\documentclass[iop]{emulateapj}
\usepackage{graphicx}
\shorttitle{M31 GC Abundances}
\shortauthors{COLUCCI, BERNSTEIN \& COHEN}

\begin{document}

\newcommand{\msol}{M_\odot}
\newcommand{\kms}{km~s$^{-1}$}
\newcommand{\rkms}{km~s$^{-1}$ \enskip}
\newcommand{\rAA}{{\AA \enskip}}
\defcitealias{m31paper}{C09}
\defcitealias{paper3}{C11}
\defcitealias{paper4}{C12}

\defcitealias{mb08}{MB08}

\title{The Detailed Chemical Properties of M31 Star Clusters I. Fe, Alpha and Light Elements\footnotemark[1]}

\footnotetext[1]{The data presented herein were obtained at the W.M. Keck Observatory, which is operated as a scientific partnership among the California Institute of Technology, the University of California and the National Aeronautics and Space Administration. The Observatory was made possible by the generous financial support of the W.M. Keck Foundation.}

\author{Janet E. Colucci\footnotemark[2]}
\affil{Observatories of the Carnegie Institution for Science, 813 Santa Barbara St., Pasadena, CA 91101}
\author{Rebecca A. Bernstein}
\affil{Observatories of the Carnegie Institution for Science, 813 Santa Barbara St., Pasadena, CA 91101} \author{\& Judith G. Cohen}
\footnotetext[2]{NSF Astronomy and Astrophysics Postdoctoral Fellow}

 \affil{Palomar Observatory, Mail Stop 105-24, California Institute of Technology, Pasadena, CA 91125 }

\email{ jcolucci@obs.carnegiescience.edu}

\begin{abstract}
 We present ages, [Fe/H] and abundances of the  alpha elements Ca I, Si I, Ti I, Ti II, and light elements Mg I, Na I, and Al I for 31 globular clusters in M31, which were obtained from high resolution, high signal-to-noise ratio (SNR$>60$) echelle spectra of their integrated light. All abundances and ages are obtained  using our original technique for high resolution integrated light abundance analysis of globular clusters.   This sample provides a never before seen  picture of the chemical history of M31.  The globular clusters are dispersed throughout the inner and outer halo, from  2.5 kpc $<$ R$_{\rm M31}$ $<$ 117 kpc.  We find a range of [Fe/H] within 20 kpc of the center of M31, and a constant [Fe/H]$\sim-1.6$ for the outer halo  clusters. We find evidence  for at least one massive globular cluster in M31 with an age between 1 and 5 Gyr. The alpha-element ratios are generally similar to Milky Way globular cluster and field star ratios. We also find chemical evidence for a late-time accretion origin for at least one  cluster, which  has a different abundance pattern than other clusters at similar metallicity.
   We find evidence for star-to-star abundance variations in Mg, Na, and Al in  the globular clusters in our sample, and find  correlations of Ca, Mg, Na, and possibly Al abundance ratios with  cluster luminosity and velocity dispersion, which can potentially be used to constrain globular cluster self-enrichment scenarios.  Data presented here were obtained with the  HIRES echelle spectrograph on the Keck I Telescope.

\end{abstract}

\keywords{galaxies: halos --- galaxies: individual (M31) --- galaxies: star clusters : general --- stars: abundances --- Local Group }

\section{Introduction}
\label{sec:intro}
\setcounter{footnote}{1}

Study of M31, the Milky Way's nearest massive neighbor, is interesting for many reasons. One of the most fundamental questions is whether the characteristics of the  M31 spiral galaxy support the assertion that the Milky Way  is a ``normal'' spiral galaxy.   This is important because we can study the properties of the Milky Way in great detail, and studies of our own galaxy by necessity are the foundation for our understanding of how galaxies in general form and evolve.   As the next closest massive galaxy, M31 is the first place to test galaxy formation theories developed from studies of the Milky Way, and in some respects is a more ideal test-case because M31 can be observed as a whole from the outside, whereas study of our own galaxy is complicated by our position within it.  

However, the distance to M31 means that we are unable to study its individual stars at the same level of detail that we can obtain in the Milky Way (MW).   For example, much of our detailed knowledge of the evolution of the MW has come from  chemical evolution studies of our Galaxy's individual stars.  Stars are ideal records of chemical evolution because their atmospheres generally retain the same chemical composition as the gas reservoir out of which they formed, and therefore  with ``fossil" chemistry of stars of all ages, one can gain unparalleled insight on the history of a galaxy.    The most precise detailed chemical abundance analyses  require high resolution, high signal-to-noise ratio (SNR) spectra, so that individual transitions of a myriad of elements can be isolated and analyzed.  Unfortunately, at a distance of 785 kpc \citep{m31distance}, the individual stars in M31 are far too faint for obtaining high resolution spectra.

With the development of our original technique for abundance analysis of high resolution {\it integrated light } (IL) spectra of globular clusters (GCs), we can now make significant advances in chemical evolution studies of distant massive galaxies.  Unresolved GCs, which are luminous, and therefore observationally accessible to large distances, can be used to learn about the chemical enrichment and formation history of other galaxies, just as they were originally used to learn about the formation of the Milky Way \citep[e.g.][]{1962ApJ...136..748E,searlezinn}.  Our technique has been developed and demonstrated on resolved GCs in the Milky Way and Large Magellanic Cloud (LMC) in a series of papers: \cite{bernstein05}, \cite{mb08} (hereafter ``MB08''), \cite{paper3} (hereafter ``C11''), and \cite{stars}.   These works demonstrate that the IL analysis provides accurate Fe abundances and [X/Fe] ratios to $\sim$0.1 dex, as well as  distinguishes ages  for GCs with a range in properties, including [Fe/H] of $-2$ to $+0$ and ages from 0.05 to 12 Gyr.  
 We also note that a detailed discussion of  potential systematic errors in high resolution IL analysis was presented for a similar technique in \cite{sakari14}. This work also  demonstrated that systematic uncertainties in GC IL analysis are small for Fe, Ca, Ti, and Ni, although
individual elements with few transitions (Ba II and Eu II) can have larger systematic uncertainties.

With this method, we have now  begun an unprecedented study of the chemical composition of the GC system of M31.   Presently, the number of confirmed, massive  GCs in M31 is $>400$;  a long history of study of  M31 GCs is embodied in the extensive  photometric and spectroscopic properties  maintained in the Revised Bologna Catalog \citep{bolognacat}.  With the  recent addition of a  large imaging survey of the outer halo of M31, the Pan-Andromeda Archaelogical Survey \citep[PAndAS,][]{2009Natur.461...66M,ibata14}, the GC system out to  projected galactocentric  radii of $\sim$150 kpc is thought to be complete to cluster magnitudes of $M_{V}=-6$ \citep{huxor14}.  
 The inner GCs of M31 have been well-studied with low resolution spectroscopy; most GCs have several metallicity estimates available from a variety of methods \citep[e.g.][]{huchra91,barmby00,perrett02,beasley05,puzia05,galleti09,caldwell11}.  As part of our ongoing project, we present the first detailed chemical abundances of GCs in M31, which now allows us to compare the detailed chemical history of old stellar populations in M31 to those in the MW for the first time.

 Detailed abundances of $\sim20$ elements were presented for a  pilot sample of 5 M31 GCs in \cite{m31paper} (hereafter ``C09'').  Here we  extend the sample of \citetalias{m31paper} and now present ages and abundances of Fe, Ca, Ti, Si, Mg, Na and Al of an additional 26 GCs in M31. In  future  papers  we will present detailed abundances of Fe-peak and r- and s-process elements in this sample of GCs. In \textsection \ref{sec:obs}, we describe the target selection, observations, data reduction and velocity measurements. In \textsection \ref{sec:analysis} we describe the equivalent width and line synthesis abundance analyses of Fe I lines, which are used to determine both [Fe/H] and age.  In \textsection \ref{sec:results} we present the results for Fe II, Ca I, Si I, Ti I, Ti II, Na I, Al I, and Mg I, and in \textsection \ref{sec:discussion} we discuss the results with respect to the star formation history of M31, formation histories of GCs in general, and previous work on the GC system of M31.

\section{Targets, Observations and Reductions}
\label{sec:obs}

\begin{figure*}
\centering
\includegraphics[scale=0.5]{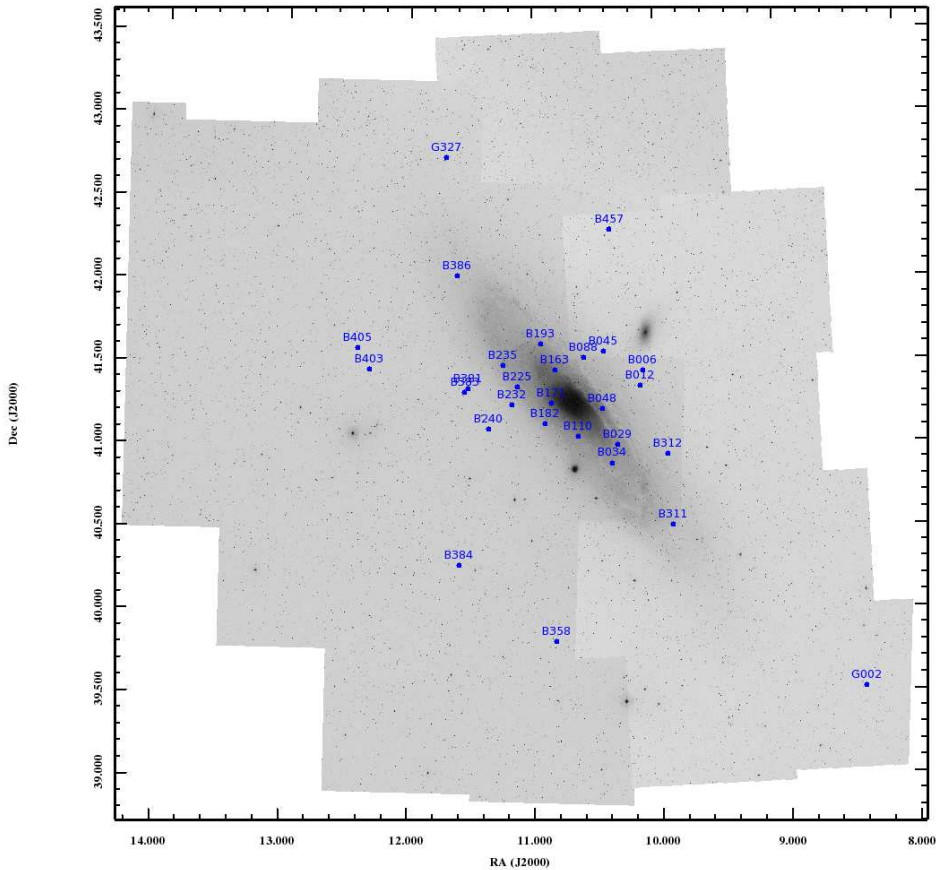}
\caption{  M31 GC  targets superimposed over a composite STScI Digitized Sky Survey image of M31 and the surrounding field.  Note that B514, MCGC5 and MGC1 are beyond the edges of the field and have galactocentric radii of  55, 79 and 117 kpc, respectively.  For comparison, the next most distant GC from the center of M31 in our sample is G002, which has a galactocentric radius of 34 kpc.}
\label{fig:field} 
\end{figure*}

\begin{deluxetable*}{lrrrrrrrr}
\scriptsize
\tablecolumns{9}
\tablewidth{0pc}
\tablecaption{Observations and Cluster Properties\label{tab:obs}}
\tablehead{
\colhead{Name}  &\colhead{RA} & \colhead{Dec}& \colhead{V} & \colhead{E(B-V)} & \colhead{${\rm R}_{{\rm M31}}$} & \colhead{Date}  &\colhead{T$_{exp}$} & \colhead{SNR (pixel$^{-1}$)} \\ \colhead{} & \colhead{(J2000)} & \colhead{(J2000)} & &\colhead{} & \colhead{(kpc)} &&\colhead{(h)} & \colhead{(6040 \AA)}  }
\startdata
 B006-G058 &    00:40:26.5 &    +41:27:26.4 & 15.46 &0.17 &   6.39 &  2008 Sep       &    3.0      &  91        \\
 B012-G064 &    00:40:32.5 &    +41:21:44.2 & 15.04 & 0.17 & 5.74 & 2008 Sep         &    2.1      &    86      \\
 B029-G090 &     00:41:17.8 &    +41:00:22.8 & 16.58 & 0.27  &6.78 & 2011 Sep        &      5.0   &     62     \\
 B034-G096 &    00:41:28.1 &    +40:53:49.6 & 15.47 & 0.16 & 6.02 &   2010 Oct       &   4.0       &   100       \\
 B048-G110 &    00:41:45.5 &    +41:13:30.7 & 16.51 &  0.36 &2.59 &   2011 Sep         & 4.0         &   76       \\
 B088-G150 &      00:42:21.1 &    +41:32:14.3 & 15.00 & 0.46$^{a}$ &  3.80 & 2009 Sep           &    1.2      &    85$^{1}$      \\
 &   &  & &&&   2011 Sep           &   2.0       &          \\

 B110-G172 &    00:42:33.1 &    +41:03:28.4 & 15.28 &  0.12& 2.93 &  2010 Oct        &    3.0     &    112      \\
B163-G217 &  00:43:17.0 &     +41:27:44.9 & 15.04  &0.21 & 3.00 &  2012 Sep  &     2.7 &   114 \\
B171-G222 &  00:43:25.0 &     +41:15:37.1 & 15.28 &0.19 & 1.77  & 2012 Sep      &   3.0 &  91   \\

B182-G233 &  00:43:36.7 &     +41:08:12.2 & 15.43  &0.33& 2.88  & 2010 Oct      &   4.0 &  105  \\

 B193-G244 &    00:43:45.5 &    +41:36:57.5 & 15.33 & 0.23 & 5.41 &   2010 Oct         &    1.1      &   56       \\
 B225-G280 &    00:44:29.8 &    +41:21:36.6 & 14.15 & 0.12&  4.68 & 2008 Sep         &   1.0       & 106        \\

 B232-G286 &    00:44:40.5 &    +41:15:01.4 & 15.65 & 0.21 & 4.96 & 2008 Sep         &   3.0       &  87        \\
 B235-G297 &     00:44:57.9 &    +41:29:23.7 & 16.27 & 0.14 & 6.46 &  2011 Sep          &    4.0      & 72         \\
 B240-G302 &    00:45:25.2 &    +41:06:23.8 & 15.18 &  0.13 &7.22 &  2008 Sep        &    3.0      &    98      \\
 B311-G033 &    00:39:33.8 &    +40:31:14.4 & 15.45 &0.36 & 13.06 & 2010 Oct &      4.0    &   84       \\
 B312-G035 &    00:39:40.1 &    +40:57:02.3 & 15.52 &  0.23 &9.02 &   2008 Sep       &     3.0     &     80     \\
 B383-G318 &     00:46:12.0 &    +41:19:43.2 & 15.30 & 0.20  &8.92 &  2009 Sep       &   2.0       &     69     \\
 B384-G319 &    00:46:21.9 &    +40:17:00.0 & 15.75 & 0.10 &16.42 &   2008 Sep       &   3.5       &     91     \\
B403-G348 &   00:49:17.0   &   +41:35:08.2 & 16.22 &0.26 &17.34 & 2012 Sep &     5.0 &   77      \\
B457-G097 &  00:41:29.0 &      +42:18:37.7  & 16.91 &0.13 &14.58  & 2012 Sep&      6.0 &   54      \\

B514-MCGC4 &    00:31:09.8 &    +37:53:59.6 & 15.76 & 0.09$^{b}$& 55.30 & 2008 Sep         &    3.5      &       72   \\
  G327-MVI &      00:46:49.6 &    +42:44:44.6 & 15.90 &0.18$^{a}$  &22.70 &    2009 Sep          &    3.0      & 68         \\
      G002 &    00:33:33.8 &    +39:31:18.5 & 15.93 &0.08$^{c}$  &33.62 &    2008 Sep      & 3.6         &   85       \\
 MCGC5-H10 &     00:35:59.8 &     +35:41:03.9 & 16.09 &0.05$^{b}$&  78.68 &  2008 Sep        &   4.3       &     77     \\
      MGC1 &    00:50:42.5 &    +32:54:58.7 & 15.50 &0.17$^{d}$& 117.05 &  2008 Sep        &  3.2        &     67     \\
\cutinhead{Clusters analyzed in C09}
 B045-G108 & 00:41:43.1 & +41:34:20.0  & 15.83  &0.18& 4.90      &  2006 Sep &4.5 &100 \\
 B358-G219 &  00:43:17.9 &   +39:49:13.2  & 15.12&0.06$^{a}$ & 19.86   &  2006 Sep& 2.9&  110 \\
 B381-G315 &  00:46:06.6 &   +41:20:58.9  & 15.76&0.24 &8.72    &   2006 Sep& 4.0  & 100 \\
 B386-G322 &  00:46:27.0 &   +42:01:52.8 & 15.64   &0.18&14.08  &   2006 Sep  &3.5&  90 \\
 B405-G351 &  00:49:39.8 &  +41:35:29.7  &   15.20 & 0.18 &18.28 &   2006 Sep & 3.0& 100\\

\enddata
\tablecomments{Cluster identifications, positions, V magnitudes, and
  projected galactocentric radii from M31 are taken from the Revised
  Bologna Catalog \citep{bolognacat}. Reddening values are taken from
  \cite{caldwell11}, with the exceptions of a.) \cite{fan08} b.)\cite{2007ApJ...655L..85M}
 c.) \cite{2007AJ....133.2764B} d.)   \cite{2010MNRAS.401..533M}
1. SNR of combined 2009 and 2011 spectrum.
}
\end{deluxetable*}

\begin{figure}
\centering
\includegraphics[trim = 19mm 0mm 10mm 0mm, clip,scale=0.55]{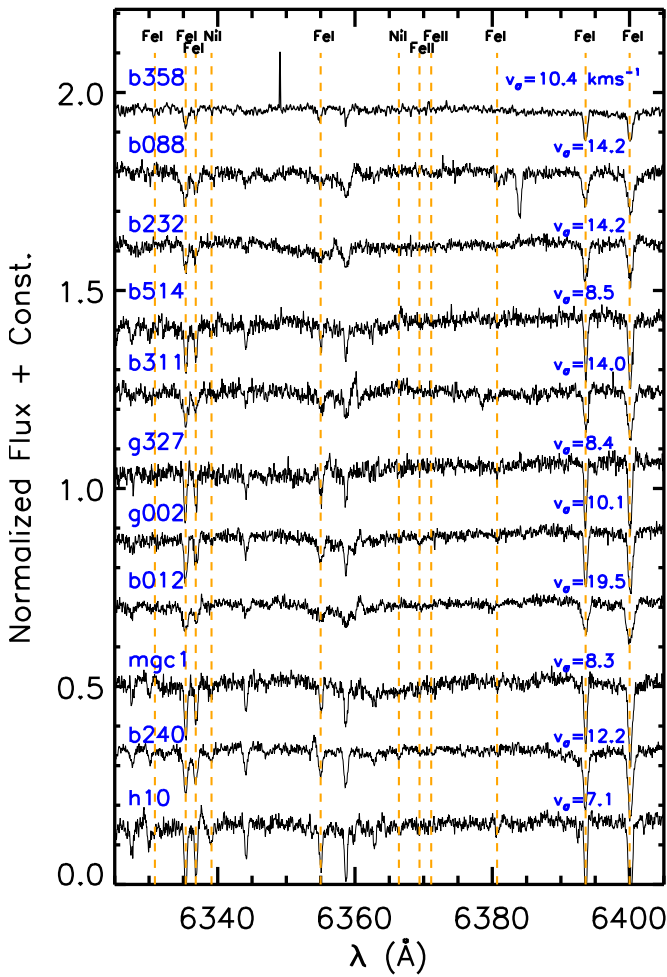}
\caption{A portion of the spectrum of each of the  11 most metal-poor GCs in our sample.  
The spectra are normalized to 1.0, with a base at 0.0,  and a constant offset has been applied to each for visualization.
  Metallicity increases from top to bottom.  The cluster names are shown, as well as the velocity dispersions we measure in \textsection \ref{sec:vdisp}.  Dashed lines correspond to Fe I, Fe II, and Ni I transitions, as noted.}
\label{fig:spec1} 
\end{figure}

\begin{figure}
\centering
\includegraphics[trim = 13mm 0mm 10mm 0mm, clip,scale=0.55]{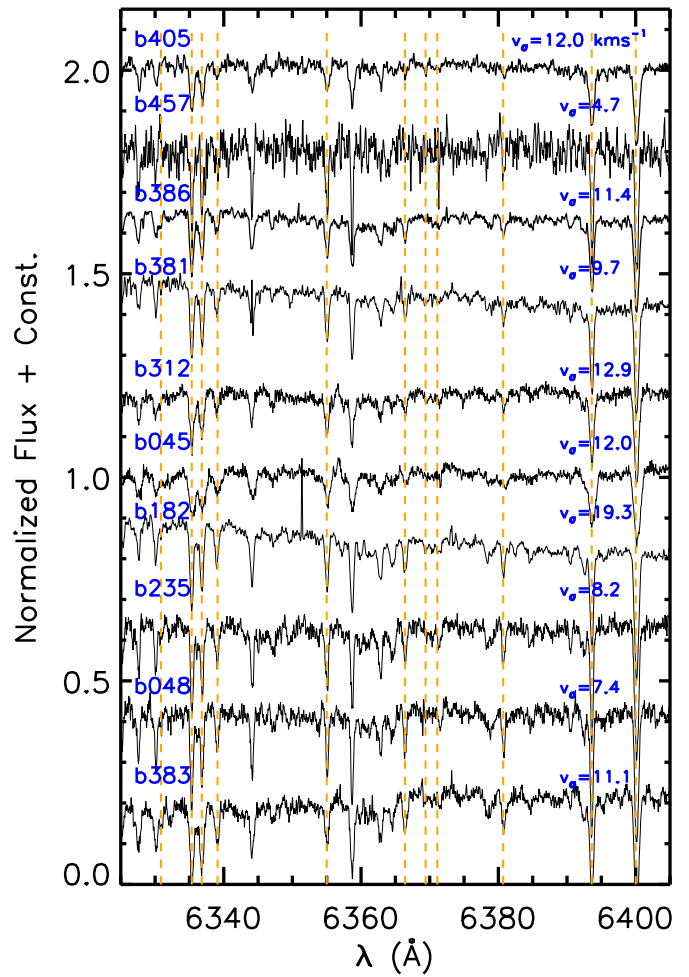}
\caption{ The same as Figure \ref{fig:spec1} for the 10 intermediate metallicity GCs in our sample.}
\label{fig:spec2} 
\end{figure}

\begin{figure}
\centering
\includegraphics[trim = 18mm 0mm 5mm 0mm, clip,scale=0.55]{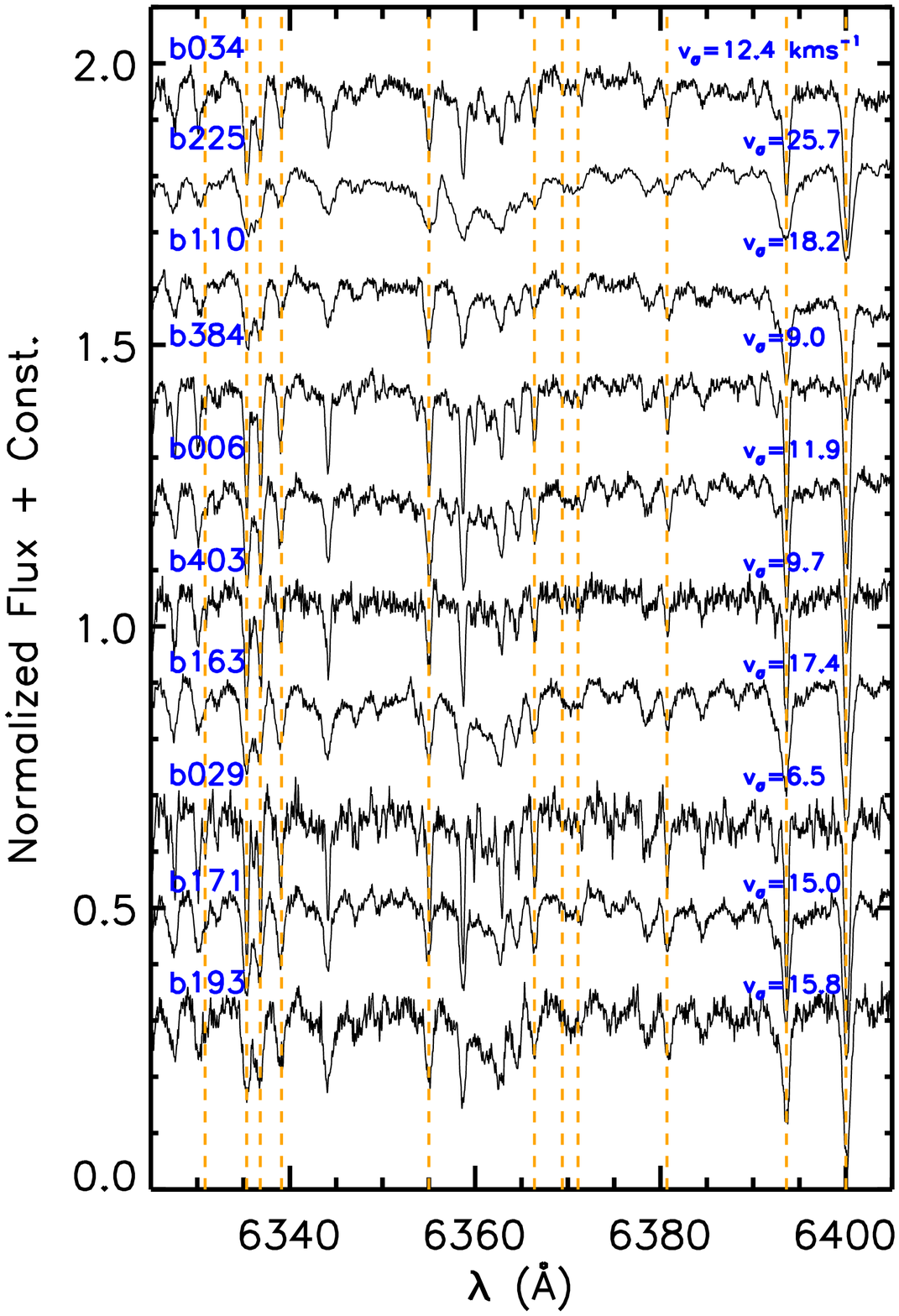}
\caption{  The same as Figure \ref{fig:spec1} for the 11 most metal-rich GCs in our sample.}
\label{fig:spec3} 
\end{figure}

Our GC targets were chosen from  the Revised Bologna Catalog \citep{bolognacat}, and have all been previously spectroscopically confirmed as members of the M31 GC system.  Our selection criteria required that the GCs be more luminous than V$\sim$17 mag, but less luminous than V$\sim$15 mag because the brightest clusters have the highest velocity dispersions (v$_{\sigma}$), leading to more line broadening and blending and are thus more difficult to analyze. We  picked GCs that are in relatively uncrowded regions and that are  not projected onto the highest surface brightness part of the M31 bulge or disk.  While our sample is obviously not complete, we have selected  GCs with a wide range in previously estimated [Fe/H], age, v$_{\sigma}$,  and   projected galactocentric distance from M31 (R$_{\rm M31}$) in order to increase our chances of surveying the range of properties present in M31 GCs.  The 
magnitudes and spatial information are listed for all of the GCs  
in Table~\ref{tab:obs}. In Figure \ref{fig:field} we show the locations of the GCs in our sample over a composite STScI Digitized Sky Survey image of M31 and the surrounding field.  Note that the three GCs with the largest projected distances from M31--- B514, MCGC5 and MGC1--- are beyond the edges of the field, which emphasizes the extensive radial coverage of our sample.

We obtained high resolution IL spectra of the M31
GCs using the  HIRES spectrograph
 \citep{1994SPIE.2198..362V} on the Keck I telescope. The data were taken over several observing runs from 2008-2012.  In all observing runs we used  identical setups that utilized  the D3 decker, which has a  slit size of $1.7"\times7.0"$ and spectral resolution of $R=24,000$, which is sufficient to resolve individual spectral lines of GCs with v$_{\sigma}\geq 7 $ \kms.  22 of the 26  GCs have previously measured half light radii \citep{2007AJ....133.2764B,peacock10,ma12,wang13} that are    between $\sim$0.6"$-$1.1" , which  means that 70-90$\%$ of the GC light was included in the $1".7\times7.0"$
slit during the observation.  We assume that the other 4 GCs have similar half light radii, with the conclusion that the GC populations are well sampled in the integrated light. 
 The wavelength coverage of the HIRES spectra is approximately
3800$-$8300 \AA.  Total exposure times were between 1$-$6 hours for each GC
and are listed in Table~\ref{tab:obs}  along with the date each GC was observed.  The total exposure for each GC was divided into 1800 or 3600 s increments to aid in cosmic ray removal.
The SNR
estimates at 6000 \rAA are also given in Table \ref{tab:obs}.  Data were reduced with standard flat  fielding, sky subtraction, and wavelength calibration routines in the
HiRes Redux pipeline.\footnote{ http://www.ucolick.org/$\sim$xavier/IDL/index.html} To remove the blaze function of HIRES we used low order polynomial fits to the spectrum of a G star taken during each run, which should approximately have the same color as the integrated light of GCs \citepalias{mb08}.  In Figures \ref{fig:spec1}-\ref{fig:spec3} we show a portion of the final spectra of the GCs in a $\sim$100 \rAA region centered at approximately 6365 \AA, which is a region that includes several spectral features used in the abundance determinations.  The GCs are shown in order of increasing metallicity from our analysis, and velocity dispersion of each GC are noted for reference.

\subsection{Velocity Dispersion Measurements}
\label{sec:vdisp}

One-dimensional velocity dispersion (v$_{\sigma}$) measurements were obtained by cross correlation with Galactic template stars, as described in \citetalias{m31paper}.  In brief, the IRAF task {\it fxcor} is used to cross correlate the GC spectra with a template star on an order by order basis. The full width at half maximum of the cross correlation peaks is then converted to a line-of-sight velocity dispersion using an empirical relation, as described in \cite{1979AJ.....84.1511T}.  The template stars used in this analysis include HD188510 (G5V), HR6757 (G8II), HR6940 (G8II-III), and HR7325 (G9III), and were observed with identical setups as the GC targets during each run. The variation in derived velocity dispersion when different  template stars were used was generally less than 1-2 \kms, which is comparable to the scatter in measurements between individual echelle orders for individual template stars. For our final measurements we average the results for all four template stars, with an uncertainty equal to the standard deviation of the mean. The results are listed in Table \ref{tab:velocities}, along with the heliocentric corrected radial velocities.  In Table \ref{tab:velocities}, we also list previously measured radial velocities and velocity dispersions for the GCs. The majority of GCs with previously measured radial velocities  agree with our results to  within 3 $\sigma$ of the quoted errors.

\begin{deluxetable*}{l|rrrr|rrlrl}
\scriptsize
\tablecolumns{10}
\tablewidth{0pc}
\tablecaption{Velocity Dispersions and Radial Velocities\label{tab:velocities}}
\tablehead{&\multicolumn{4}{c}{This Work} & \multicolumn{5}{c}{Literature}\\
\colhead{Name}  &\colhead{v$_{\sigma}$} & \colhead{Error}& \colhead{v$_{r}$} & \colhead{Error} & \colhead{v$_{\sigma,Lit}$}  &\colhead{Error} & \colhead{v$_{r,Lit}$} &\colhead{Error} &\colhead{Ref.}\\ \colhead{} & \colhead{kms$^{-1}$} &   \colhead{kms$^{-1}$} & \colhead{kms$^{-1}$}& \colhead{kms$^{-1}$} & \colhead{kms$^{-1}$} & \colhead{kms$^{-1}$}& \colhead{kms$^{-1}$} & \colhead{kms$^{-1}$}  }
\startdata
 B006-G058 &    11.93 & 0.45 & -238.1 &0.2 &11.9   & 0.7 & -236.5 &0.6    &1,3,5   \\

 B012-G064 &    19.50  &   0.76 &-359.4&0.1 &17.8 &2.2  &-360.7  &0.6 &1,5\\

 B029-G090 &  6.51 &  0.74  &-520.1&0.3&6.8  &0.6 & -505.2  &0.6&1\\
 B034-G096 &   12.38 &  0.41  & -555.5&0.4   &\nodata&\nodata   &-539   &6&2   \\
 B048-G110 &   7.42   & 0.73  &-241.7&0.1& 7.1  &0.5 &-228.1 & 0.5 &1 \\
 B088-G150 &  14.25 &   0.90 &-491.9&0.2&16.5  & 1.0  & -489.4  &0.6 &1\\
 B110-G172 &   18.20 &  0.54  &-238.4&0.6& 19.6  &1.1   & -237.0  & 0.5&1 \\
B163-G217 & 17.41 & 0.83 &-174.8&0.1& 18.8  & 1.0 & -163.5  &0.5   &1\\
B171-G222 & 15.04 &  0.81 &-287.6& 0.1& 15.6  & 0.9 & -267.5  &  0.5 &1\\
B182-G233 & 19.29 &  0.57 &-361.7&0.5 &18.4  & 1.0  & -356.6  & 0.5&1\\
 B193-G244 &  15.79 & 0.37 &-59.4& 0.5 &14.7  & 2.1   & -62.1  & 0.5 &1,5  \\

 B225-G280 &  25.73 & 1.14   &-154.2&0.2&27.2  & 1.6    &  -163.7  & 1.6&1,3,5\\

 B232-G286 &   14.24 &   0.72  &-188.3& 0.3& 13.3   &  0.8  & -191.7 & 0.6 &1\\
 B235-G297 &   8.20  &   0.59&-106.7&0.3 & 8.2  & 0.5    &-92.4   & 0.4  &1\\
 B240-G302 &   12.23&   0.48 &-53.1&0.2 & 12.4  & 0.6 & -55.8  & 2.0  &1,5\\

 B311-G033 &  14.01 &   0.50  &-514.0&0.2& \nodata &\nodata   & -469 & 10  &2\\
 B312-G035 &   12.91 &  1.24 &-172.6&0.3 &\nodata &\nodata& -174 & 11&2\\
 B383-G318 &    11.13 &  0.40 &-231.1&0.3 &\nodata& \nodata & -253 & 9 &2\\
 B384-G319 &    9.00 &   0.56 & -359.4 &0.3  &10.3 & 0.4 & -363.8 & 0.3  &2,5\\

B403-G348 &  9.70  & 0.39  &-366.0&0.3  &\nodata &\nodata& -358 & 48 &2\\
B457-G097 & 4.73 &   0.80   &-72.1& 0.3 &\nodata&\nodata &-63 & 15&2\\
B514-MCGC4 &    8.49 & 0.55 &-474.7&0.1  &\nodata&\nodata&-458  &23 &2\\
  G327-MVI &  8.43 &   0.68  &-270.3&0.2 &\nodata&\nodata&-251 & 11   &2 \\
      G002 &   10.12 & 0.52 & -349.3 &0.2&  9.7  &0.3 &  -313 & 17 &5\\
 MCGC5-H10 &    7.12& 0.80   &-354.6&0.1& 7.2  & 0.4  &-358.3 &1.9&4 \\
      MGC1 &    8.29  &  0.72 &-354.6&0.1& \nodata &\nodata&-355 & 2&2 \\

\enddata
\tablerefs{ 1. \cite{strader11}.   2.\cite{bolognacat}.  3.\cite{dubath97}    4. \cite{alvesbrito09}  5. \cite{djorgovski97}. Where two references are listed, we have calculated the average of the two results and an error equal to the standard deviation in the mean. }
\end{deluxetable*}

\begin{figure}
\centering
\includegraphics[trim = 0mm 0mm 0mm 20mm, clip,scale=0.4,angle=90]{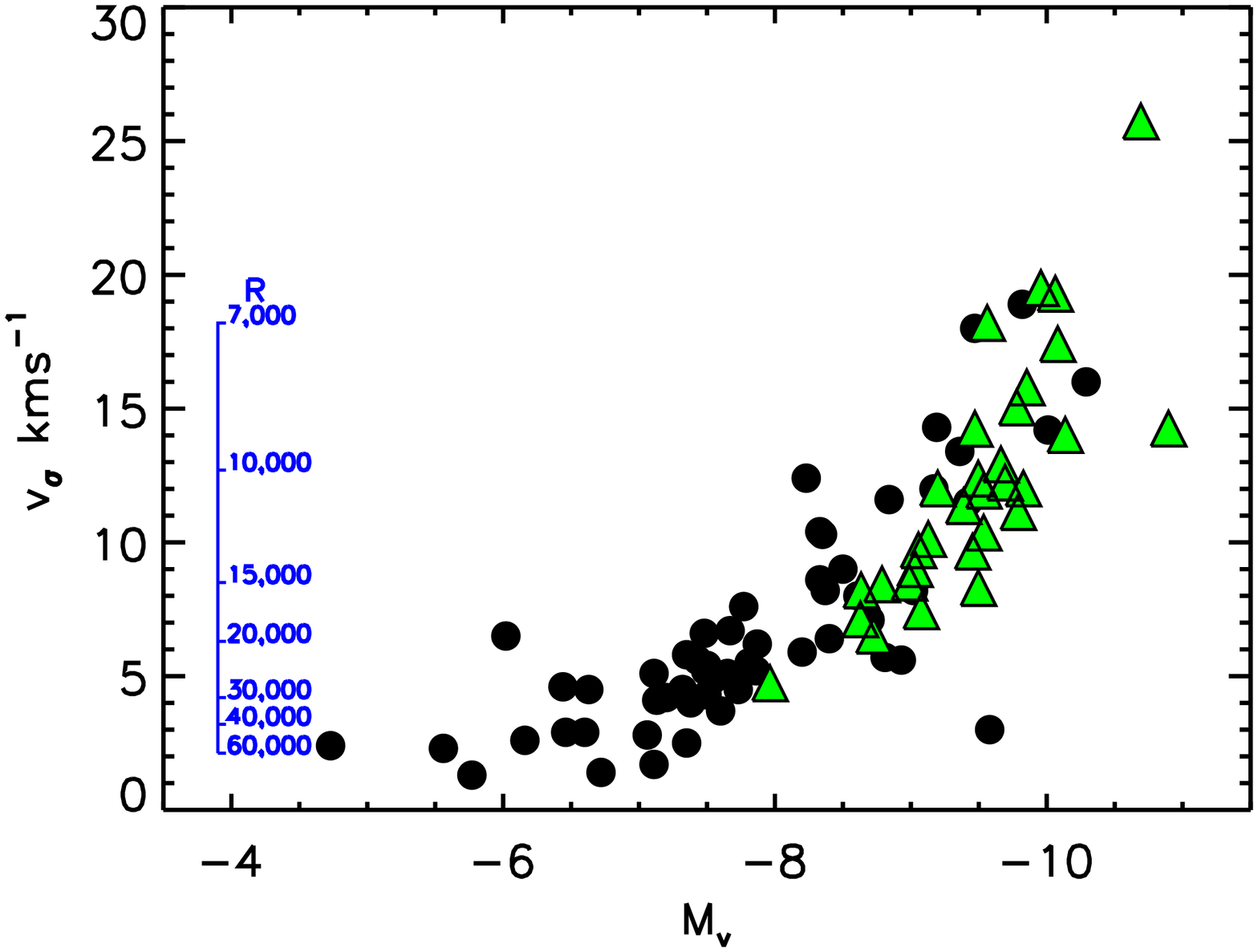}
\caption{Green triangles show our  velocity dispersion measurements for the M31 sample as a function of reddening corrected absolute V magnitude. Reddening and magnitudes are listed in Table \ref{tab:obs}, and we assume a distance modulus of 24.47 \citep{m31distance} and extinction parameter of R$_{v}$=3.1.  For comparison, a subset of the MW GC system is shown in black; data is taken from \cite{1997A&A...324..505D} and the 2010 revision of the  \cite{1996AJ....112.1487H} catalog.  The inset axis shows the effective spectral resolution corresponding to the velocity dispersion broadening. }
\label{fig:vdisp-mag} 
\end{figure}

Nearly all of the previously measured velocity dispersions agree with our results to within 1 $\sigma$, and all agree to within 2 $\sigma$.  We note that some differences between analyses are expected due to differences in apertures, but we neglect that effect here because our primary goal in measuring velocity dispersions is to use them in spectral synthesis analysis. Further discussion of the M31 GC velocity dispersions and an analysis of mass-to-light (M/L) ratios will be presented in Colucci et al. (2014, in prep).  
For 9 of the GCs, we present the first measurements of the velocity dispersion (B034, B311, B312, B383, B403, B457, B514, G327, and MGC1).
In Figure \ref{fig:vdisp-mag} we show our velocity dispersion measurements for the M31 GCs as a function of the reddening corrected absolute V-magnitudes of the GCs. 
 In this work we  use the V and E(B-V) values referenced in Table \ref{tab:obs}, a distance modulus of 24.47 \citep{m31distance} and the extinction  parameter $R_V=3.1$.
    Figure  \ref{fig:vdisp-mag} confirms that the M31 GCs generally show the same trend as Milky Way GCs  \citep[data taken from ][2010 revision]{1997A&A...324..505D, 1996AJ....112.1487H}, and that the M31 sample is mainly found in  the more luminous range because of our selection criteria.   Figure \ref{fig:vdisp-mag}  also  shows on the inset axis the corresponding effective spectral resolution of the GCs due to the velocity dispersion broadening, which can be found on  the inset axis. This shows the large range in effective resolution, and that nearly all of the GC spectra are completely resolved with an instrumental setup of $R=24,000$.

\section{Abundance and Age Analysis}
\label{sec:analysis}

Our method for obtaining detailed abundances from integrated light GC spectra was presented for a pilot sample of M31 GCs in \citetalias{m31paper}.  In that paper we performed an IL equivalent width (EW) analysis using our routine ILABUNDS \citepalias{mb08}.  In this work we initially repeat that analysis for Fe I lines, which is reviewed in \textsection \ref{sec:ews}, and then refine that analysis as needed with an additional IL spectrum synthesis analysis, as described in \textsection \ref{sec:syn}, below.

\subsection{ IL EW Analysis}
\label{sec:ews}

As described in  \citetalias{m31paper}, absorption line EWs are measured with the semi-automated program GETJOB \citep{1995AJ....109.2736M}.  The line lists used in our analysis are taken from references compiled in  \cite{1994ApJS...91..749M}, \cite{1995AJ....109.2757M},
\cite{1998AJ....115.1640M},  \citetalias{mb08} and \cite{johnson06}. 

ILABUNDS utilizes the 2010 version of MOOG  \citep{1973ApJ...184..839S}  to calculate flux-weighted IL EWs to match to the observed EWs.  In order to calculate  IL EWs, we   construct synthetic color magnitude diagrams (CMDs) from Teramo isochrones \citep{2004ApJ...612..168P,2006ApJ...642..797P,2007AJ....133..468C}.  We use  canonical isochrones  with  an extended asymptotic giant branch (AGB), $\alpha-$enhanced low$-$temperature opacities calculated according to \cite{ 2005ApJ...623..585F}, and a mass$-$loss parameter of $\eta$=0.2.  As in all of our analyses, we apply an IMF of the form in  \cite{2002Sci...295...82K}.   The synthetic CMDs are divided into $\sim$25 boxes of stars with similar
properties, with each containing $\sim$4\% of the total $V$-band
flux. The atmospheres of the average stellar types are interpolated from the 1-D, plane parallel, ODFNEW and AODFNEW model grids of Kurucz\footnote{The models are available  from R. L. Kurucz's Website at  http://kurucz.harvard.edu/grids.html}\citep[e.g.][]{2004astro.ph..5087C}. All abundances in both the EW and line synthesis analysis are calculated under the assumption of local
thermodynamic equilibrium (LTE).   All abundance ratios relative to solar are calculated with the solar abundance values of \cite{asplundreview}.

The age and [Fe/H]  solutions for  each cluster are  identified as the range in synthetic CMD ages and [Fe/H] that produce the most self-consistent results using the 10$-$80 individual Fe I lines measured in each cluster.  The best solutions have the smallest line-to-line statistical error ($\sigma_{{\rm N}}$), and minimal dependence of Fe I abundance with line excitation potential (EP), wavelength, and EW.  
The line-to-line scatter also includes systematic uncertainties between the lines themselves.
For each cluster there is a range in CMD ages that produce similarly self-consistent solutions.  For older clusters this range is typically 10$-$15 Gyr, and leads to  a systematic uncertainty in [Fe/H] of $\lesssim$ 0.05 dex, which we denote $\sigma_{\rm {Age}}$.  For younger clusters the preferred age range may be smaller, but the $\sigma_{\rm {Age}}$ can be  larger ($\sim$ 0.1 dex), due to the more rapidly changing stellar populations at younger ages.    For the total uncertainty in [Fe/H] for each cluster, we add the statistical error in the mean abundance ($\sigma_{ {\rm N}} / \sqrt{{\rm N} -1} $) (which also includes systematic errors between lines) and the systematic age uncertainty $\sigma_{\rm {Age}}$ in quadrature.  

\subsection{Refined Fe Line Synthesis Analysis}
\label{sec:syn}
 
 Since the publication of our pilot study in \citetalias{m31paper}, we have further refined our abundance  analysis techniques using the IL spectral synthesis component of ILABUNDS \citepalias{paper3}.  In \citetalias{paper3} we implemented a $\chi^2$-minimization scheme with the IL spectral synthesis in order to recover more elemental abundances from lower SNR data (SNR$\sim$40).  This type of analysis can also improve the precision of the measurements in cases where line blending is significant, because unlike EW analysis of single features, the synthesis includes the contributions from all the nearby lines in a specified region.  It also allows for more accurately establishing the ``pseudo" continuum around the lines of interest \citep[see][for a more detailed discussion of continuum in IL spectra]{sakari}.  In terms of the analysis of GC IL spectra, line blending is most significant when the overall metallicity is high, and/or when the cluster's velocity dispersion is large.  Therefore, it is  especially important to evaluate the impact of line blending on our analysis when we apply our  technique to clusters in galaxies whose GC systems are thought to reach higher overall metallicities than the MW's GC system.  In addition,  this is important for analyzing  GCs in more distant galaxies where we are observationally limited to probing the brightest, most massive, portion of the GC luminosity function.  In this case, the GCs we can observe will likely have larger velocity dispersions than the typical GCs that we can observe in the Local Group \citep[see][for massive GCs in NGC 5128]{cent}.
 
Our refined Fe line synthesis analysis first consists of an automated procedure to synthesize a region $\pm$10 \rAA around each Fe I feature in the preferred line list used for our EW analysis, which was described above.  As in \citetalias{paper3}, the wavelengths, {\it gf} values, and other atomic parameters of the  neighboring features around the preferred Fe I lines are drawn from the larger Kurucz database.\footnote{http://kurucz.harvard.edu/linelists.html} Other than the line list, to calculate the IL spectra we only need input a synthetic CMD, which has its own associated [Fe/H] that is used as a starting abundance.  The procedure then synthesizes IL spectra in each region with abundances that vary from $\pm$0.5 dex from the initial [Fe/H], in steps of 0.1 dex.   We have found that for the typical SNRs we obtain, 0.1 dex increments in [Fe/H] are sufficient for discriminating meaningful abundance differences in  Fe I lines with different atomic parameters  across the full wavelength range.    In principle, however, this choice is arbitrary and a smaller increment could be used for exceptional  quality data.   

We note that this step size is a negligible source of
	    uncertainty compared to the systematics, which are
	    demonstrated by the line-to-line scatter when multiple
	    lines are available and the uncertainty due to the unknown
	    age of the GC, which is represented in our analysis by the
	    range in CMD ages we use.   The total uncertainty for any abundance measured from
	   multiple lines is estimated from the line-to-line scatter 
 ($\sigma_{ {\rm N}}$), which will include systematic uncertainties between the
	   lines themselves.  These systematics are more difficult to quantify
	   (oscillator strengths, incomplete line lists, effect of
	   stellar population mismatch, etc.) than the statistical
	   errors associated with measuring the EW, which are much smaller.

In order to perform a meaningful $\chi^2$-minimization we must appropriately compare the observed spectra with the synthesized spectra, as described in \citetalias{paper3}. As a starting point, we automatically normalize the data to an average of the top $\sim$100 maximum values of the flux across the pixels in the 20 \rAA region of each synthesis. We bias the normalization to the highest flux values, with a reasonable allocation for the noise level, in order to account for absorption lines preferentially lowering the pseudo-continuum. The $\chi^2$-minimization can then be performed in a region that is approximately $\pm$0.25 \rAA around the Fe I line.   As pointed out in \citetalias{paper3}, in practice one must review the data-synthesis comparison around each Fe I  line by eye in order to obtain the most accurate measurements.  First, it is particularly  important to evaluate the normalization over a broad region, because subtle differences in matching the pseudo-continuum in high metallicity or highly broadened cluster spectra can greatly impact the results. Note that in our analysis we have   chosen to isolate the Fe I lines of interest, rather than performing a  $\chi^2$-minimization over the full 20 \rAA region.   We choose to do this because we are interested in specific Fe I transitions with the most accurate {\it gf} values, and we want to minimize uncertainties introduced by an (unavoidably) more uncertain extended line list in the region.  By reviewing each line we can also eliminate Fe lines that are too badly blended to provide a meaningful measurement, those that may have strong non-LTE effects,  lines that are in areas of particularly bad local noise, those that are coincident with sky absorption lines, etc.

Once the cleanest set of Fe I lines is established in this way, the rest of the [Fe/H] and age analysis proceeds automatically in the same way as in the EW analysis.  We use the mean abundance derived from the set of Fe I lines to iteratively solve for a self-consistent CMD at each age, where the input CMD abundance is equal to the final derived abundance for all lines. This results in one self consistent synthetic CMD of a given [Fe/H] at each age. Then,   because we are still isolating individual Fe lines, we use the same diagnostics to constrain the most appropriate age for each GC; namely the self-consistency of the abundance from individual Fe lines as a function of wavelength, EP, and EW.   The one difference is that in this case we must calculate a ``pseudo-EW" for each line, which is essentially the EW we would have observed in the absence of line blending.  In order to do this, we re-synthesize IL spectra with the final inferred [Fe/H] using {\it only} the single Fe I transitions in our line list, and then calculate the inferred, pseudo-EWs for each Fe line.   

This refined Fe line synthesis analysis is much more time  and user  intensive than the semi-automated EW analysis.  It is also more user intensive than other techniques that rely on automated full spectrum fitting.   However, GC IL spectra are fundamentally complex, and there are  more  measurement subtleties  than in analysis of individual stars, and these uncertainties are only compounded when full spectrum fitting over regions with poorly calibrated line lists, large non-LTE effects, or strong lines that are not on the linear region of the curve of growth.   The refined Fe line synthesis analysis outlined here makes it possible to recover precise, reliable abundance information from GC IL spectra that have SNR too low for standard EW analysis, or are significantly affected by line blending due to  high overall metallicity and/or   large velocity broadening.

Due to the intensiveness of this technique for determining  the overall [Fe/H] of a GC, we have performed tests to establish the situations where EW analysis is not sufficient for measuring accurate abundances and ages.  We highlight some examples using the M31 sample in the next section.

\subsubsection{EW vs. Synthesis Tests}
\label{sec:synthonly}

We  initially performed an EW analysis for all the GCs in our sample.   Next we performed a refined Fe line synthesis analysis for a subset of the GCs with large v$_{\sigma}$ and/or high [Fe/H].   To begin, the subset included all of the GCs with velocity dispersions greater than 15 \kms, as well as all of the GCs with [Fe/H]$> -0.5.$  Note that B225, which has the largest velocity dispersion by far, is not included in this subset because we were not able to perform an EW analysis with GETJOB.   Some of the GCs presented here satisfy both of the criteria.   In addition, we performed the synthesis for a subset of GCs that sample the rest of the range of velocity dispersion and [Fe/H] in order to establish the parameter space where EW analysis produces results as accurate as the synthesis.  The full test subset of GCs and the results from both analyses are listed in Table \ref{tab:ew_v_syn}. 

\begin{figure*}
\centering
\includegraphics[scale=0.65,angle=90]{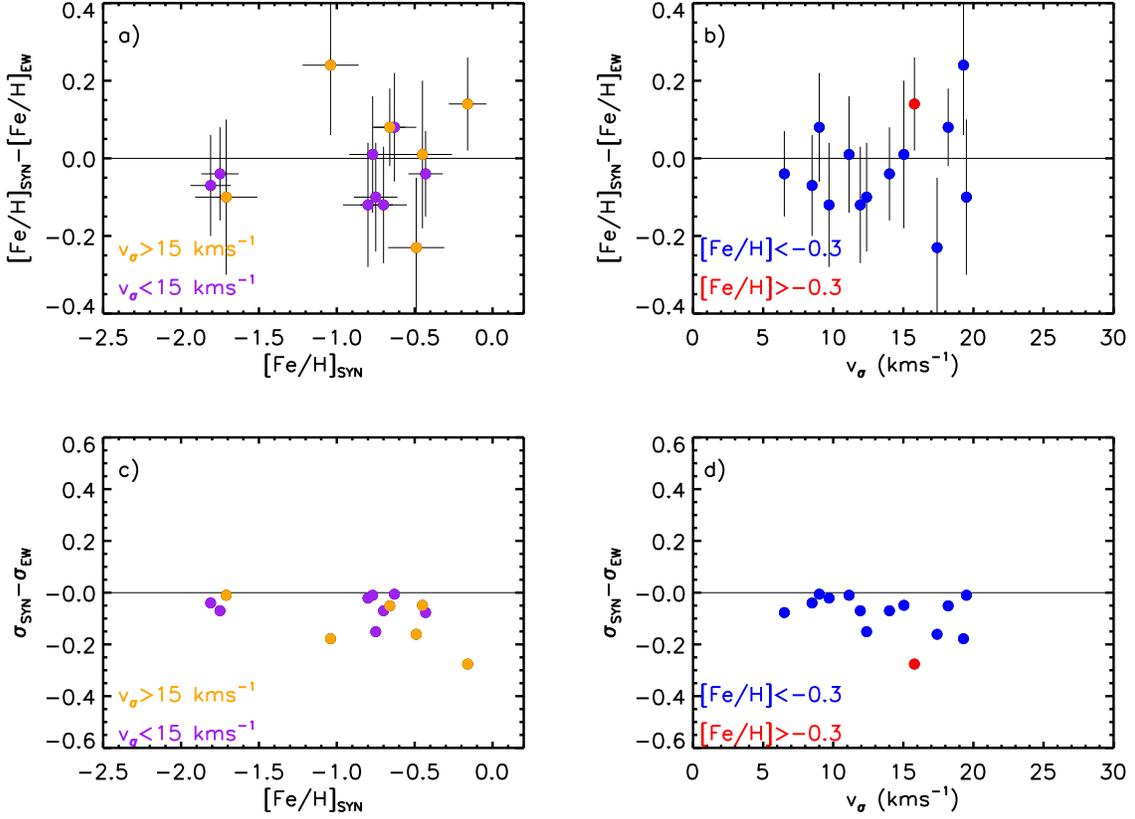}
\caption{ The results of our tests between EW and line synthesis analysis.  a) The difference in mean [Fe/H] from EWs and synthesis as a function of the [Fe/H] determined from line synthesis.  GCs with velocity dispersions $> 15$ \kms are highlighted in orange, and those with v$_{\sigma} < 15$ \kms are highlighted in purple.  b) The same as a), instead as a function of velocity dispersion.  Blue and red designate clusters with [Fe/H] below and above $-0.3$.  c). The difference in the statistical error of the mean [Fe/H] between the two analyses as a function of [Fe/H] determined from the line synthesis analysis.  d) The same as c), only as a function of velocity dispersion.    }
\label{fig:ewVSsyn} 
\end{figure*}

In Figure \ref{fig:ewVSsyn} we show the general trends in the comparison between EW and synthesis analysis.  We have quantified the differences in two ways;  first in terms of the difference in the final derived [Fe/H], and second in terms of the reduction of the statistical error ($\sigma_{{\rm N}}$), i.e. the scatter in abundance between individual Fe I lines.  We show both of these quantities as a function of [Fe/H] and v$_{\sigma}$ in Figure \ref{fig:ewVSsyn}a,c and Figure \ref{fig:ewVSsyn}b,d, respectively.  For the [Fe/H] comparison we also show the one sigma error bars ($\sigma_{{\rm N}}$) as a guide for evaluating consistent results between analyses.

   Inspection of Figure \ref{fig:ewVSsyn} seems to imply that large velocity dispersions have a bigger impact on the results than  high [Fe/H].  In general, the naive expectation is true in that line synthesis analysis  results in abundances that are slightly lower than EW analysis, however this is not necessarily the case for solutions that had large statistical errors ($\sigma_{{\rm N}}$) to begin with.    As might be expected, the most discrepant cases occur for GCs that have {\it  both} a large velocity dispersion  {\it and} a high [Fe/H].  At worst, the [Fe/H] differs by $\sim$0.25 dex, and the statistical error can be reduced by $\sim0.25$ dex, which is a change of 70 \%.   Nonetheless, in most other cases the differences are less dramatic.  Figure \ref{fig:ewVSsyn}a shows that the [Fe/H] results are consistent within 1 $\sigma$ for all of the GCs that have velocity dispersions $<$15 \kms, and for the lowest metallicity clusters even the statistical errors ($\sigma_{{\rm N}}$) are nearly identical in both cases, which means that EW abundance analysis is just as accurate as full synthesis analysis in this regime.   
    Figure \ref{fig:ewVSsyn}b more clearly shows that inconsistent [Fe/H] are not seen until GC velocity dispersions  are $>$ 15 \kms. Figure  \ref{fig:ewVSsyn}d shows that the statistical errors can be significantly improved for GCs with velocity dispersions as low as 12 \kms, even though the final [Fe/H] for these GCs are not formally inconsistent between analyses.   Our results suggest that GCs with [Fe/H] at least as high as $-0.4$ can be accurately analyzed with EWs if the velocity dispersion is not a concern.   The two GCs in our sample that have  higher  [Fe/H] also have large velocity dispersions, but the most  conservative conclusion is that the high [Fe/H] is exacerbating the problem.   Therefore, our conclusions from this analysis are that it is necessary to perform the more intensive line synthesis analysis for GCs with velocity dispersions $>$ 15 \kms, and those with [Fe/H]$> -0.3.$  Finally we note that  none of the GCs in \citetalias{m31paper} meet these criteria, and thus the Fe results and conclusions from the  EW analysis of \citetalias{m31paper} are unaffected.

 \begin{deluxetable}{lrr|rr|r}
\tablecolumns{8}
\tablewidth{0pc}
\tablecaption{EW vs. Synthesis Analysis \label{tab:ew_v_syn}}
\tablehead{
\colhead{Name} & \multicolumn{2}{c}{EW} &
\multicolumn{2}{c}{Synthesis} &\colhead{v$_{\sigma}$}\\
 & \colhead{[Fe/H]} & \colhead{$\sigma$} & \colhead{[Fe/H]}& \colhead{$\sigma$} &\colhead{kms$^{-1}$} }
\startdata

 B029-G090   &   -0.39  &    0.19    &    -0.43  &   0.11 &
    6.5 \\
 B514-MCGC4   &   -1.74  &    0.17   &     -1.81  &    0.13
  &  8.5  \\
B384-G319   &   -0.71  &    0.15   &     -0.63  &   0.14  &
9.0 \\
 B403-G348  &  	-0.68  &    0.18  &   	-0.80 &    0.16 	  & 9.7 \\

 B383-G318    &    -0.78  &    0.16  &       -0.80  &    0.14      &   11.1 \\

 B006-G058   &    -0.58  &    0.22  &   -0.70	 & 0.15		  & 	11.9 \\
 B034-G096   &   -0.65  &     0.29  &    -0.75  &   0.14  &    12.4 \\
 B311-G033  &  -1.71  &    0.19  &     -1.75  &   0.12   &  14.0 \\

 B171-G222  &  -0.46  &    0.24   &     -0.45  &   0.19   &  15.0 \\
 B193-G244   &   -0.30  &    0.40 &    -0.16  &   0.12   &   15.8 \\
 
 B163-G217    &    -0.26  &    0.34   &        -0.49   &   0.18  & 17.4 \\

 B110-G172   &   -0.74  &     0.15 &     -0.66  &   0.10   &  18.2 \\
 B182-G233   &   -1.28  &    0.36   &    -1.04  &   0.18   &  19.3 \\

   B012-G064   &     -1.61  &    0.21  &   	 -1.71 &    0.20   &  	  19.5 \\

\enddata
\end{deluxetable}

\begin{figure}
\centering
\includegraphics[angle=90,scale=0.3]{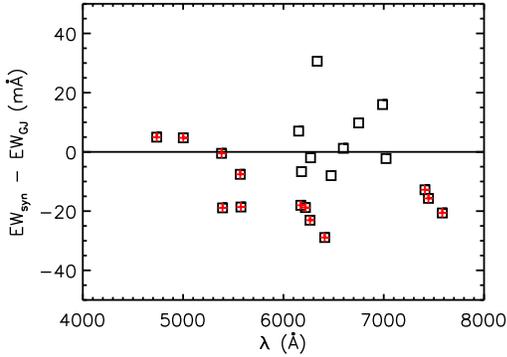}
\caption{ The difference in the EWs measured with GETJOB and the pseudo-EW measured from line synthesis for B384, as a function of wavelength.  A solid line is drawn at at 0 to guide the eye.   Fe lines with EWs $>$ 100 m\AA~ are highlighted in red.  }
\label{fig:ews} 
\end{figure}

\begin{figure}
\centering
\includegraphics[angle=90,scale=0.3]{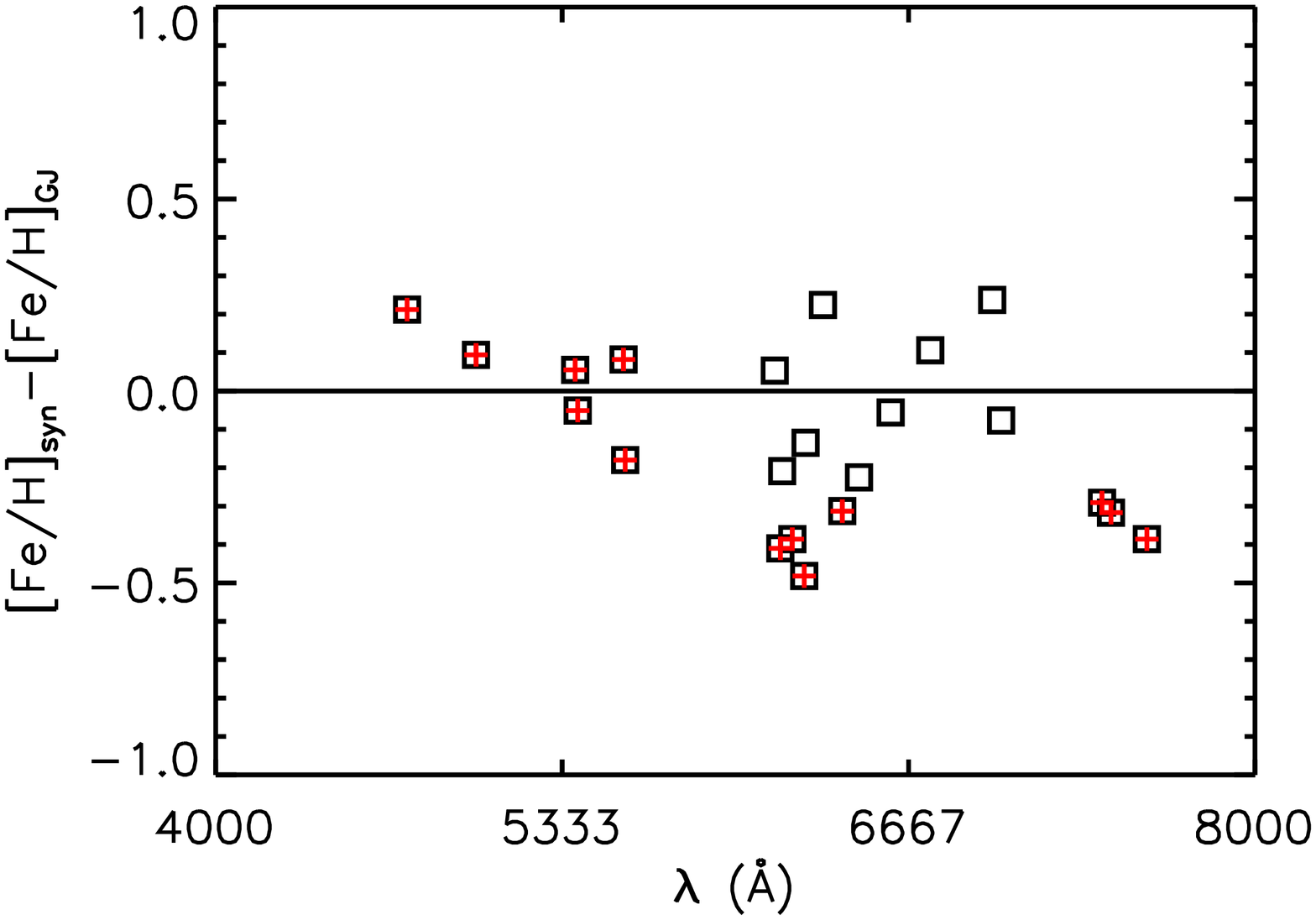}
\caption{ The difference in the [Fe/H] obtained using EWs from GETJOB and the [Fe/H] measured from line synthesis for B384, as a function of wavelength.  A solid line is drawn at at 0 to guide the eye.   Fe lines with EWs $>$ 100 m\AA~ are highlighted in red.  }
\label{fig:ewswave} 
\end{figure}

\begin{figure*}
\centering
\includegraphics[angle=90,scale=0.28]{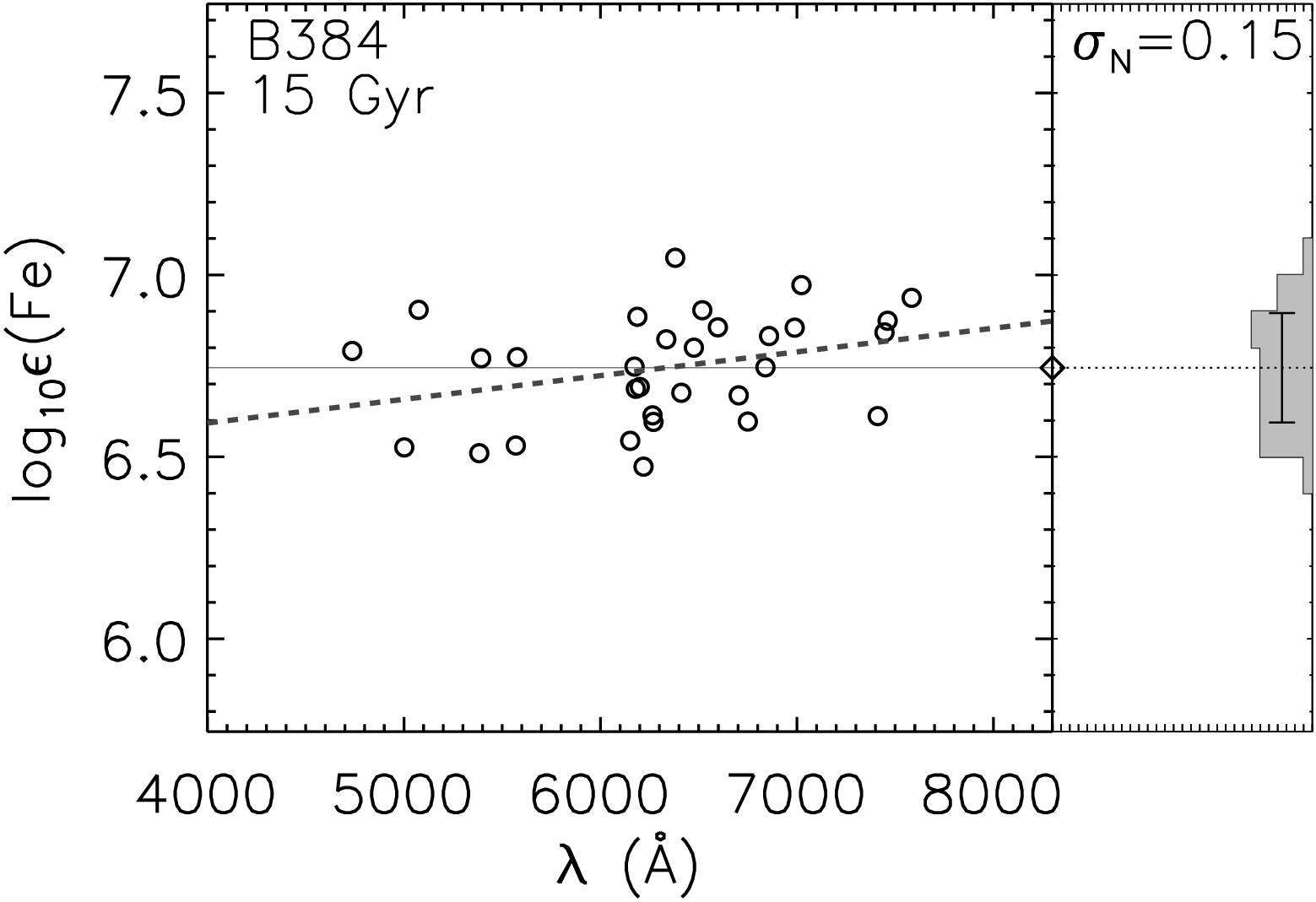}
\includegraphics[angle=90,scale=0.28]{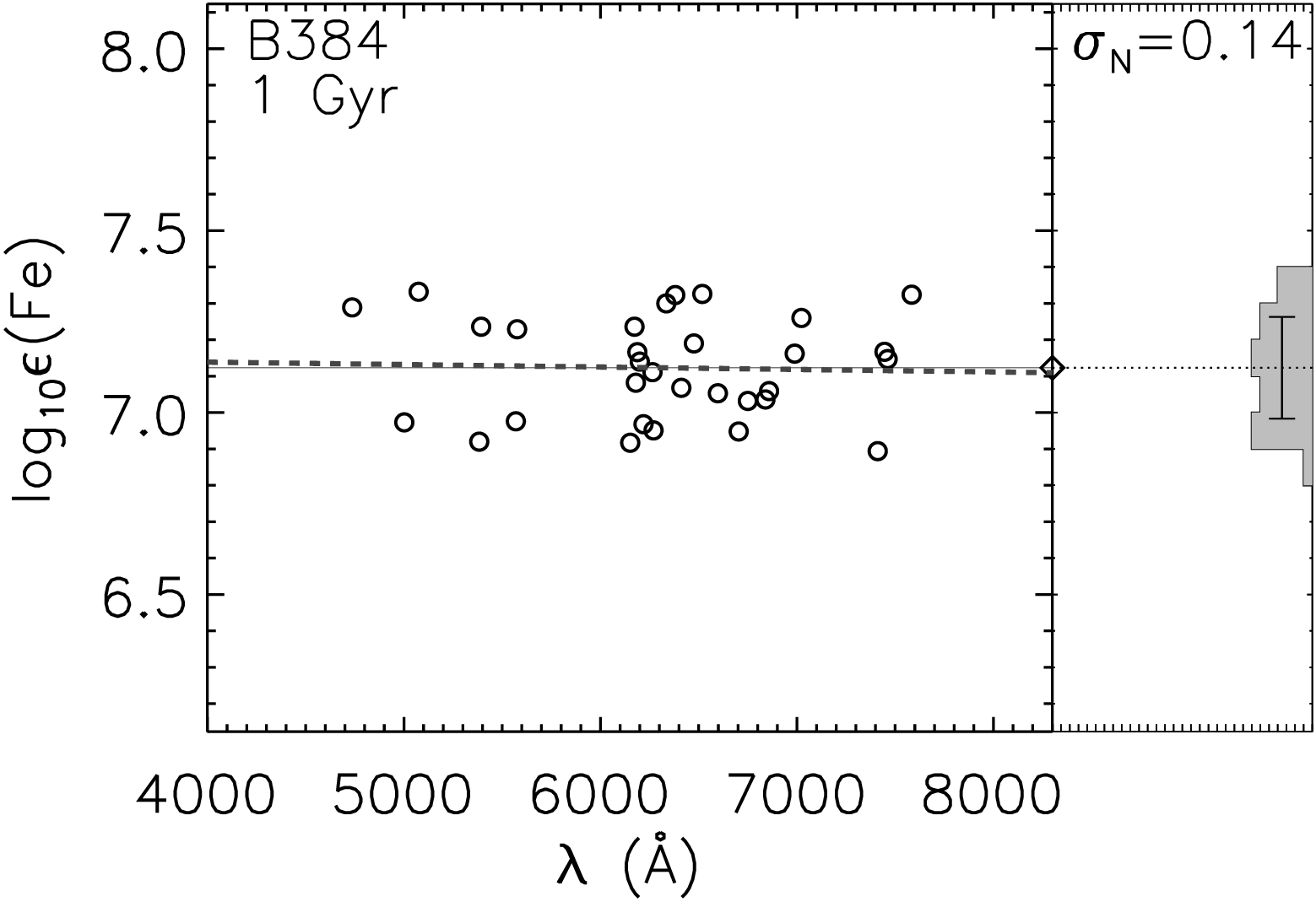}

\includegraphics[angle=90,scale=0.28]{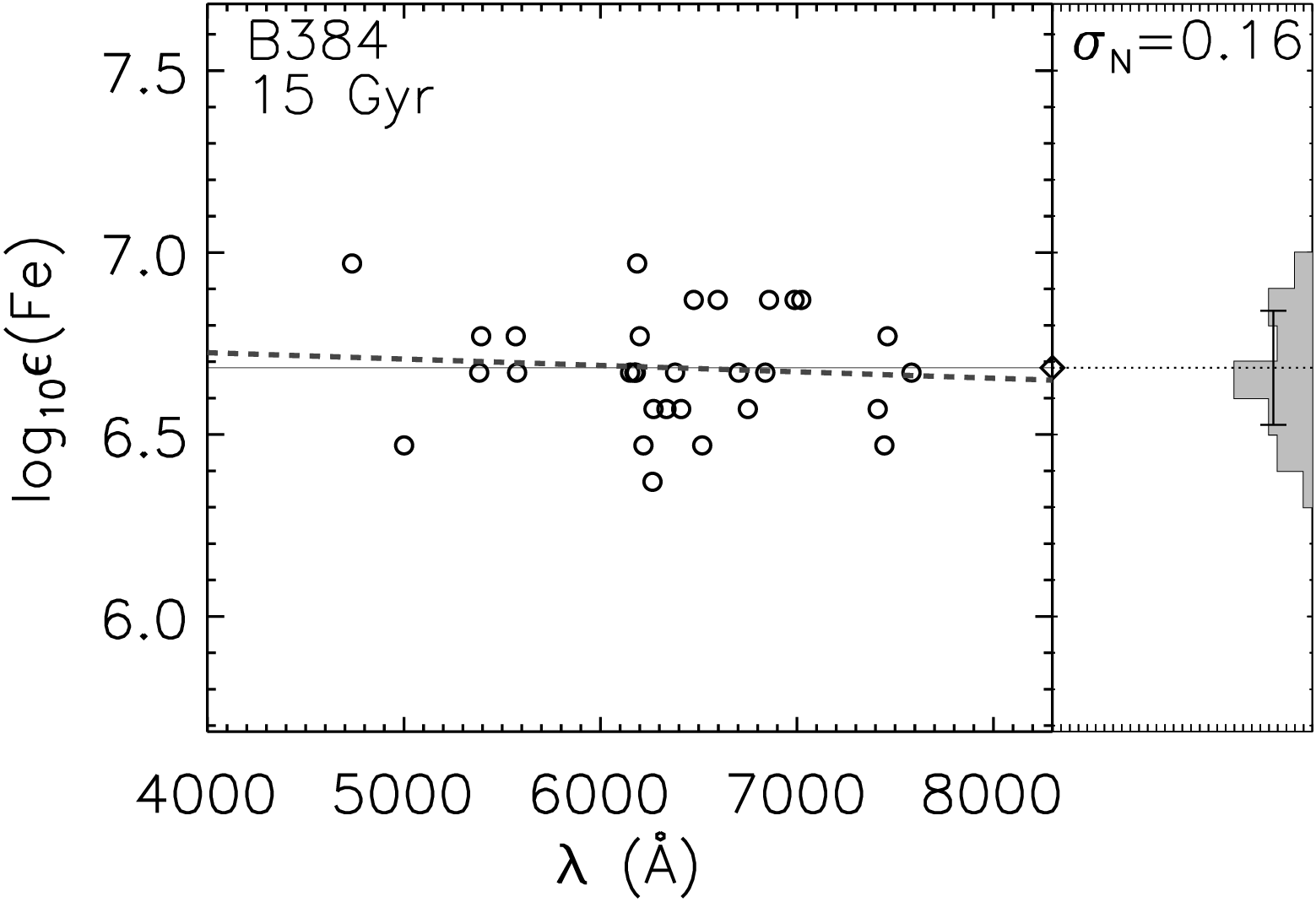}
\includegraphics[angle=90,scale=0.28]{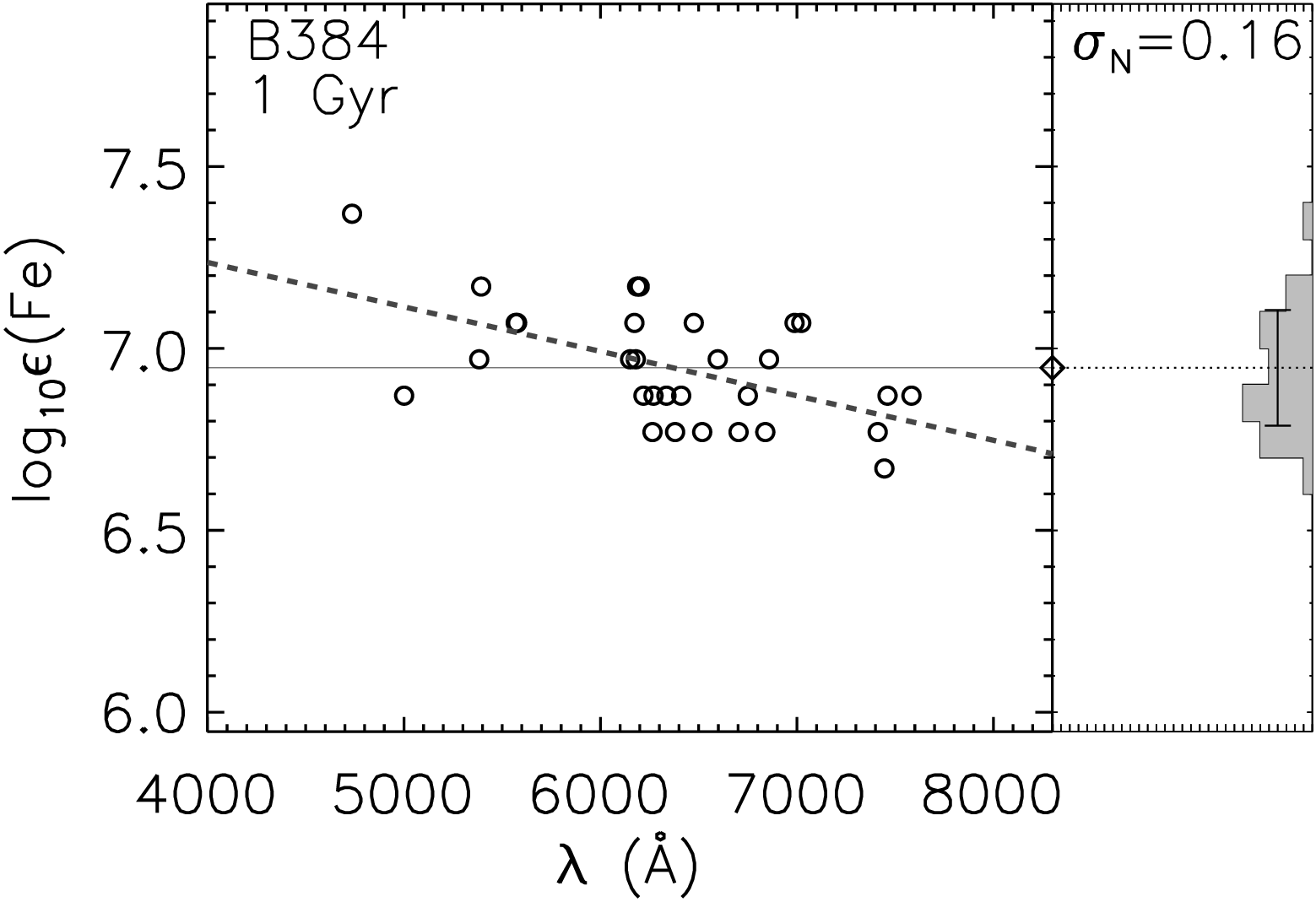}
\vskip 0.2 cm
\caption{ The individual  Fe line  abundances as a function of wavelength  for B384.  The top panels are the results using EWs measured in GETJOB, and the bottom panels show the results using line synthesis.  In both cases an old (15 Gyr) solution is shown on the left, and a young (1 Gyr) solution in shown on the right. solid lines show the mean abundance from all lines, and the dashed line shows a linear least squares fit. Using GETJOB, the young solution is preferred, while with line synthesis the old solution is preferred. More details can be found in the text.}
\label{fig:b384} 
\end{figure*}

The  v$_{\sigma}$ and [Fe/H] effects are straightforward to quantify, but we  have also found  other situations when a more intensive line synthesis is beneficial or even necessary.   The first cases are for  GCs where the data quality is low enough that the EW analysis doesn't converge to a [Fe/H] solution with a  reasonable statistical error.    
 B457, which has a relatively low velocity dispersion of 4.7 \kms, has the lowest SNR in our sample (SNR=54), and was difficult to analyze with EWs.
For that reason, we use line synthesis for the final Fe I analysis.  Note that the SNR of B193 is similar
 (SNR=56).   For that cluster, synthesis analysis was used because it satisfies our velocity dispersion criterium (v$_{\sigma}$=15.8).   

A second, perhaps more  interesting, case occurred when the EW analysis suggested a GC had a young or intermediate age.  Initially we found three GCs that appeared young when analyzed with EWs (B029, B034, B384).    These three GCs are among the more metal-rich GCs in our sample,  so we were suspicious of how higher metallicity could subtly be influencing the results.  We also note that the existence of intermediate age GCs in M31 is still heavily debated, so it is particularly important that we investigate any potential systematic effects that may cause us to measure young ages.   Therefore, we performed the more in-depth line synthesis analysis to be  as confident as possible with the results. For B029, the line synthesis analysis didn't change the age determination, which is discussed in more detail in \textsection \ref{sec:b029}.  However, for B034 and B384, we found that the subtle measurement differences in the line synthesis analysis changed the age solutions  so that they were more consistent with old ages. 
We suspect that the changes are due to the disproportionate  influence of  cool giants on  strong, red Fe lines.    
   In Figures \ref{fig:ews} and \ref{fig:ewswave} we investigate the differences between  the solutions in an attempt to determine what the cause may be. 
   Here we use B384 as an example, but the same behaviors are seen in the solution for B034.  In Figure \ref{fig:ews}, we show the change in the EW measured from GETJOB and that measured in the line synthesis analysis as a function of wavelength, and in Figure \ref{fig:ewswave} we show the change in derived abundance as a function of wavelength.  We have highlighted the strong lines, i.e. those    Fe I lines that have EWs $>$ 100 m\AA.   Unfortunately,  we  don't have any weak lines bluer than 6000 \rAA to
compare, but   above 6000 \AA, where we have both strong and weak lines, the
strong lines always have large changes in EW between analyses.  With this data, a general conclusion would be  that the large changes in EW for strong red lines, which will also result  in large changes in [Fe/H], are primarily responsible for the differences in the final solutions between analyses.  However, it is also possible that local issues (continuum, line blending) can be affecting this particular set of lines, so we will address this issue with a larger sample of metal-rich MW clusters in an upcoming paper \citep{mwpaper}.

It is helpful to  qualitatively observe how these subtle differences are manifested in the Fe abundance vs. wavelength diagnostics for different CMD ages. As an example we show the old (15 Gyr) and young (1 Gyr) solutions for B384 in Figure \ref{fig:b384}.  In the top panels the EW solutions are shown, and we see that the 1 Gyr solution is preferred because there is no dependence of the Fe abundance on wavelength, while for the 15 Gyr solution there is a trend of increasing  Fe abundance with wavelength.  In the bottom panels we show the synthesis solutions, and because in {\it both}  cases the Fe abundance of the strong, red lines decreased, the old 15 Gyr now exhibits no trend in Fe abundance with wavelength, and is therefore now the better solution;  the 1 Gyr solution now has a trend of decreasing Fe abundance with wavelength.

\begin{figure}
\centering

\includegraphics[scale=0.45]{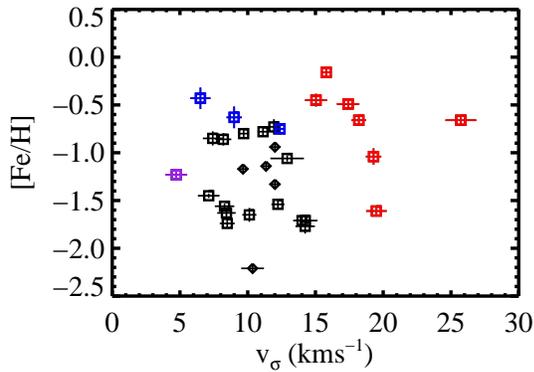}

\caption{ Final [Fe/H] plotted against velocity dispersion for the M31 sample.  Squares are new GCs in this work, diamonds show GCs from \citetalias{m31paper}. Black symbols show GCs for which the final analysis was performed with GETJOB.  Red symbols show GCs where line synthesis was used because the GCs fit the [Fe/H]$>-0.3$ or v$_{\sigma}>$ 15 \kms criteria. Purple and blue symbols show GCs where synthesis was used because of SNR or potentially young ages, respectively.}

\label{fig:fe-sigma} 
\end{figure}

 As a final summary, we show the [Fe/H] results as a function of v$_{\sigma}$ for the entire M31 sample in Figure \ref{fig:fe-sigma}.  In red we highlight GCs that meet the  criteria for requiring line synthesis analysis ([Fe/H]$> -0.3$, v$_{\sigma}>$ 15 \kms).  In purple we show the GCs where line synthesis was preferred for SNR reasons.  In blue we show the GCs where line synthesis was necessary to distinguish between young and old ages.  The 5 GCs from \citetalias{m31paper} are shown as black diamonds, demonstrating that they are outside of the regime where line synthesis is necessary.

\section{Chemical Abundance and Age Results}
\label{sec:results}

\begin{figure*}
\centering

\includegraphics[scale=0.20,angle=90]{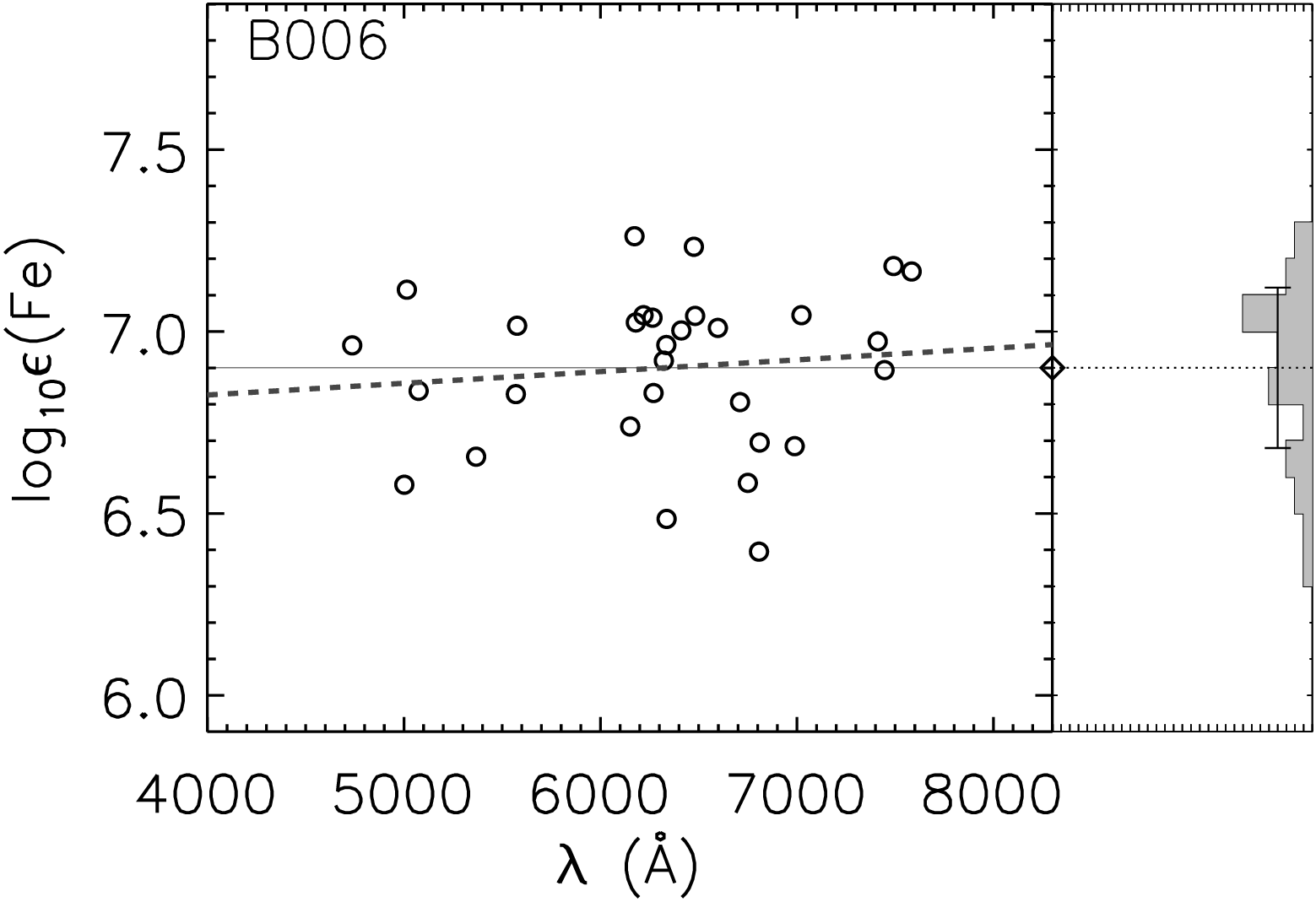}
\includegraphics[scale=0.20,angle=90]{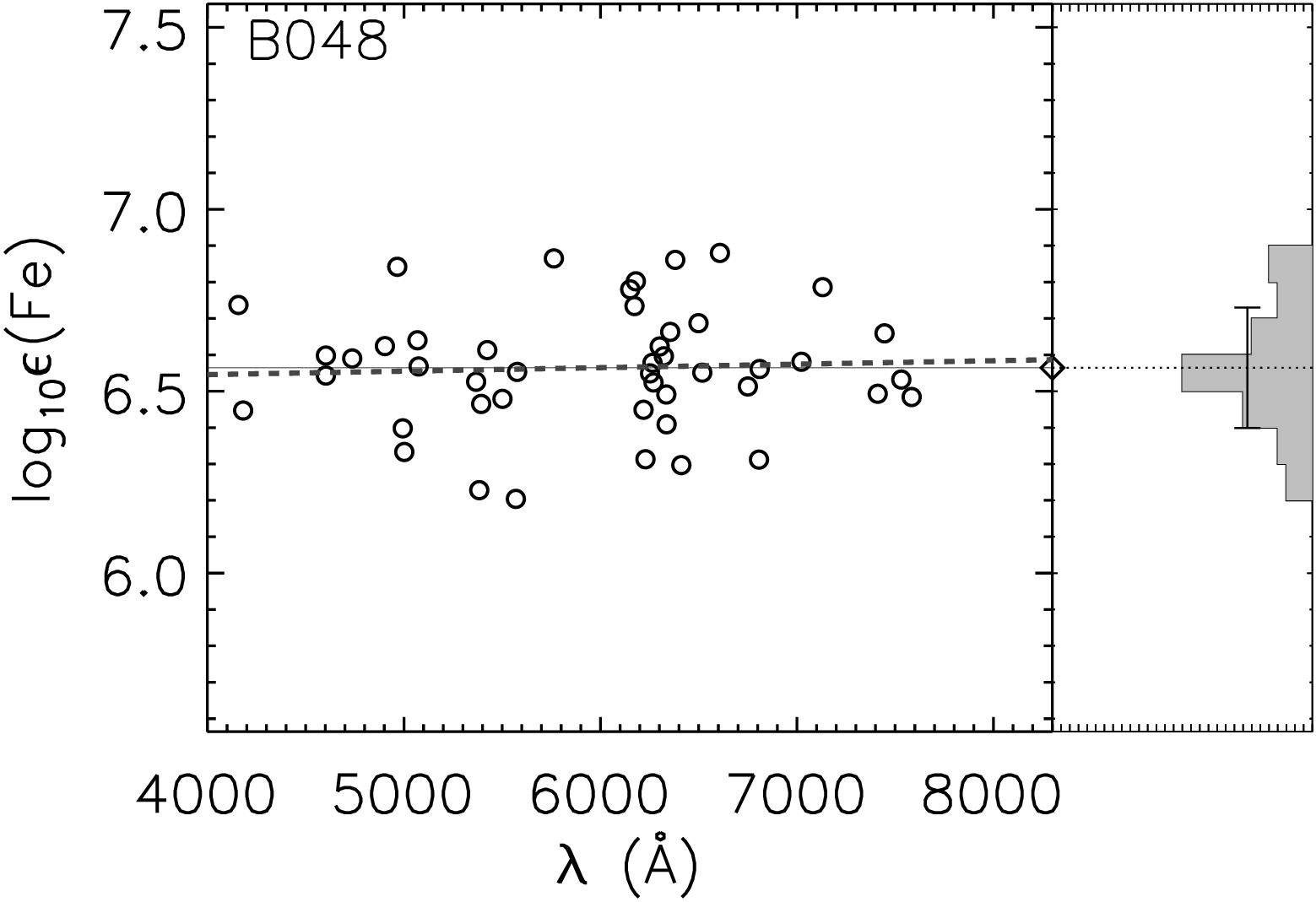}
\includegraphics[scale=0.20,angle=90]{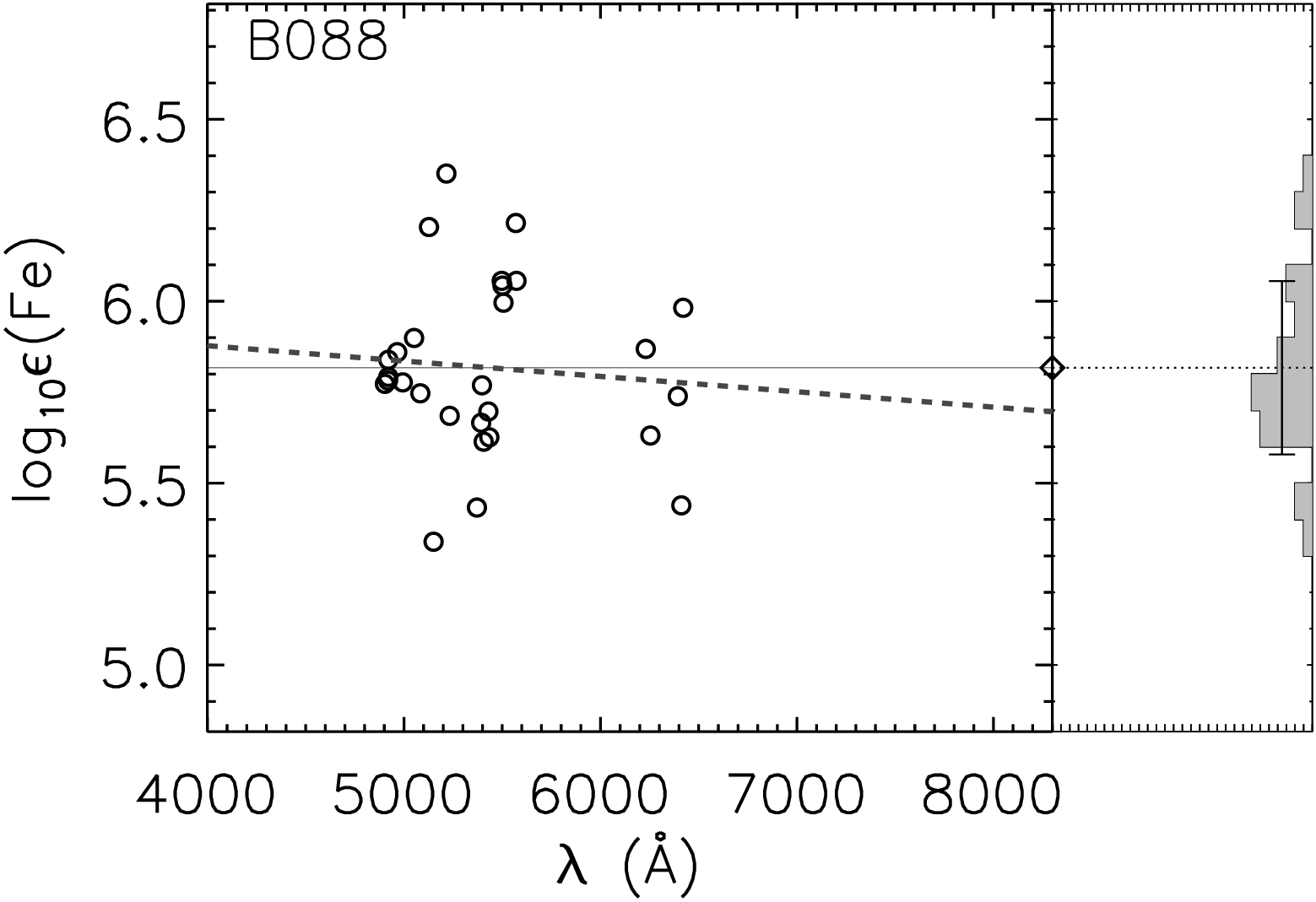}
\includegraphics[scale=0.20,angle=90]{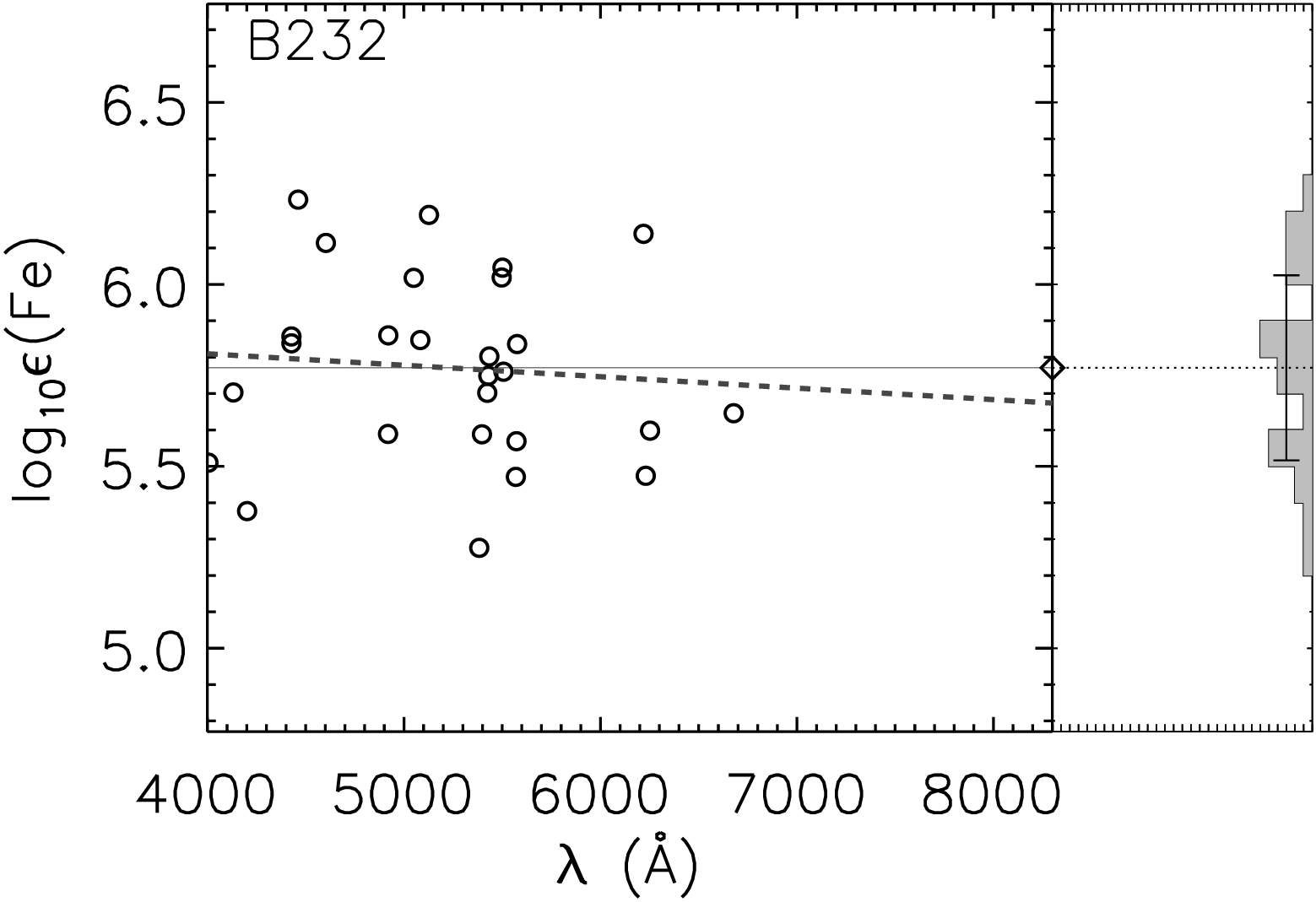}
\includegraphics[scale=0.20,angle=90]{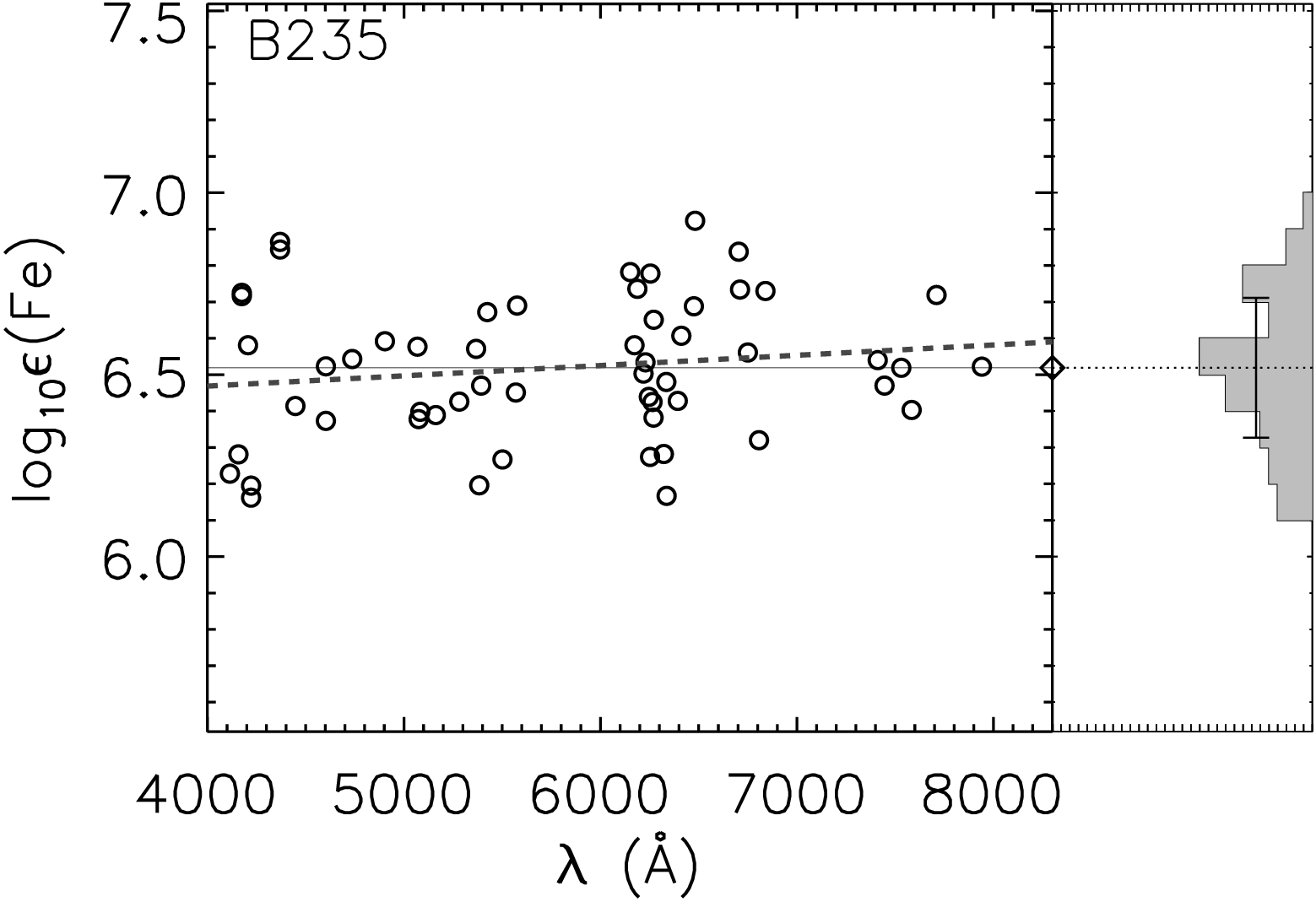}
\includegraphics[scale=0.20,angle=90]{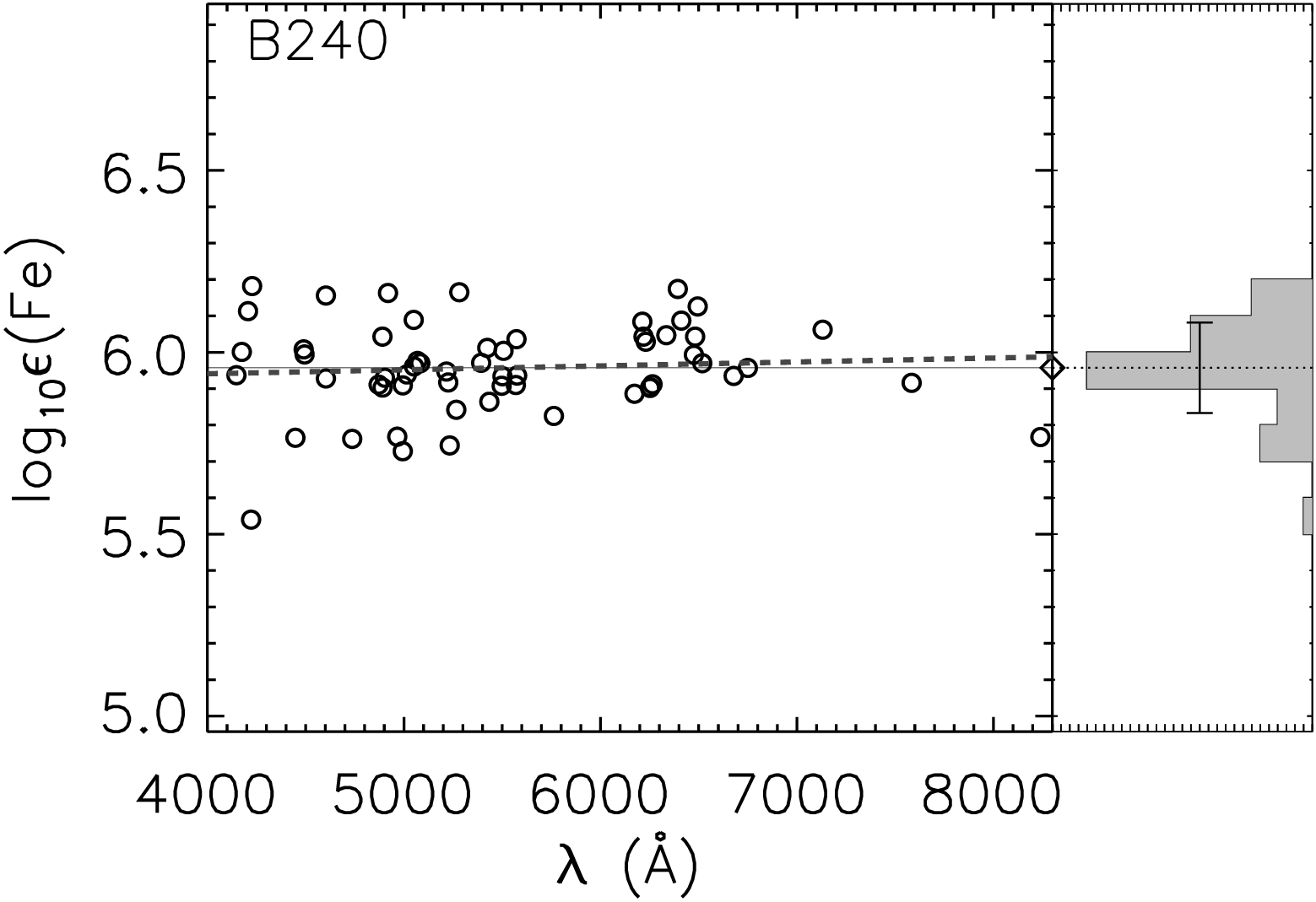}
\includegraphics[scale=0.20,angle=90]{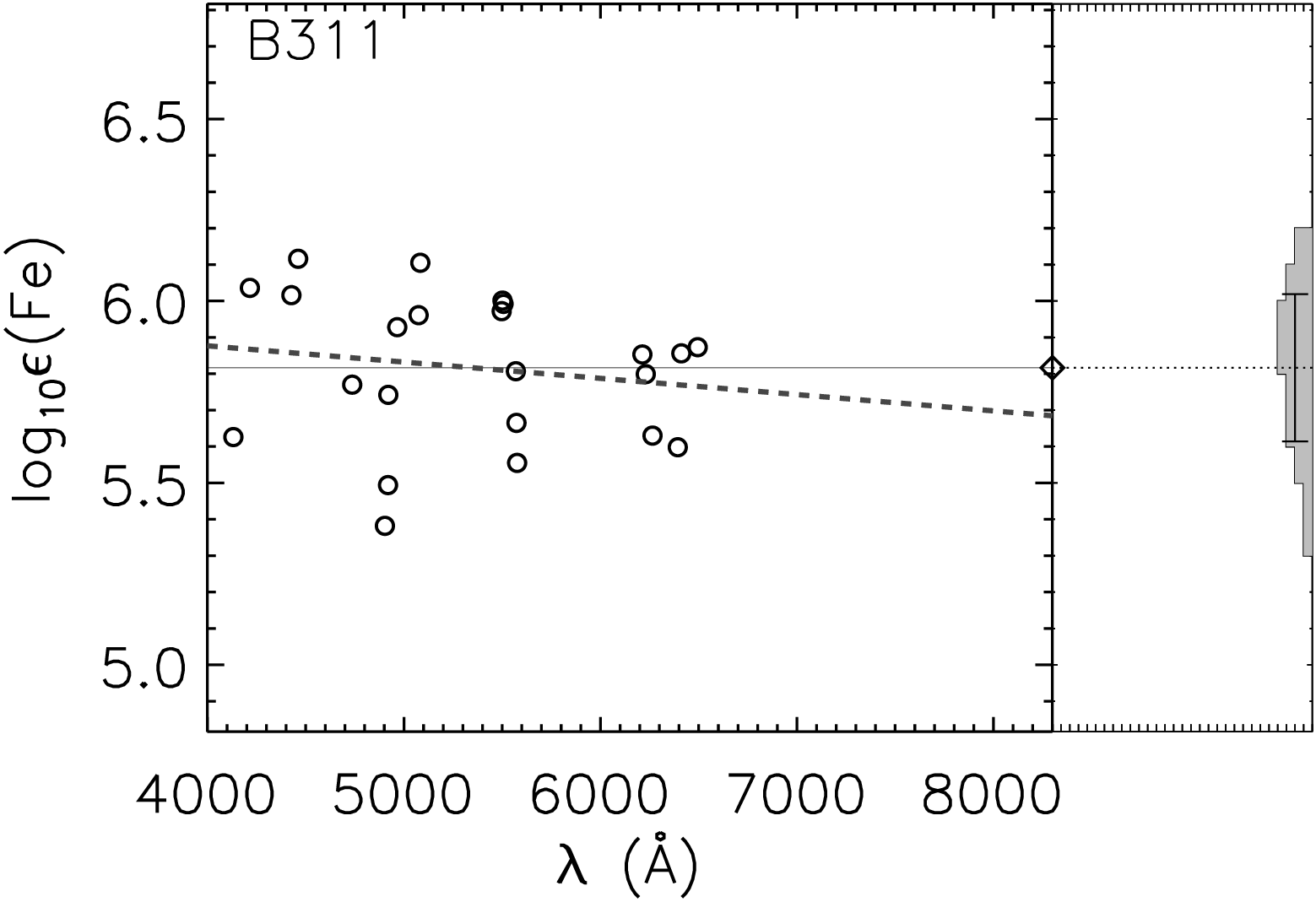}
\includegraphics[scale=0.20,angle=90]{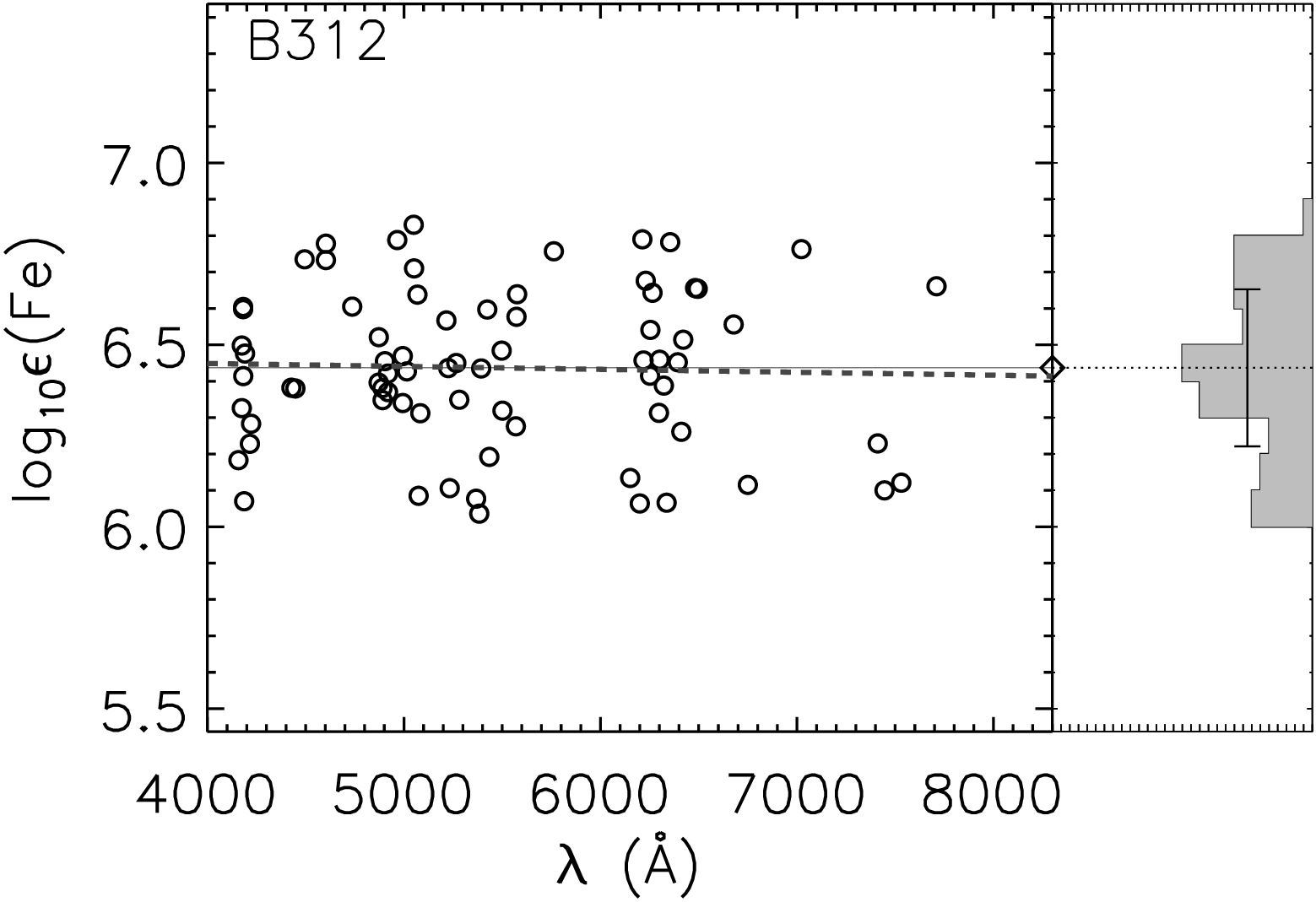}
\includegraphics[scale=0.20,angle=90]{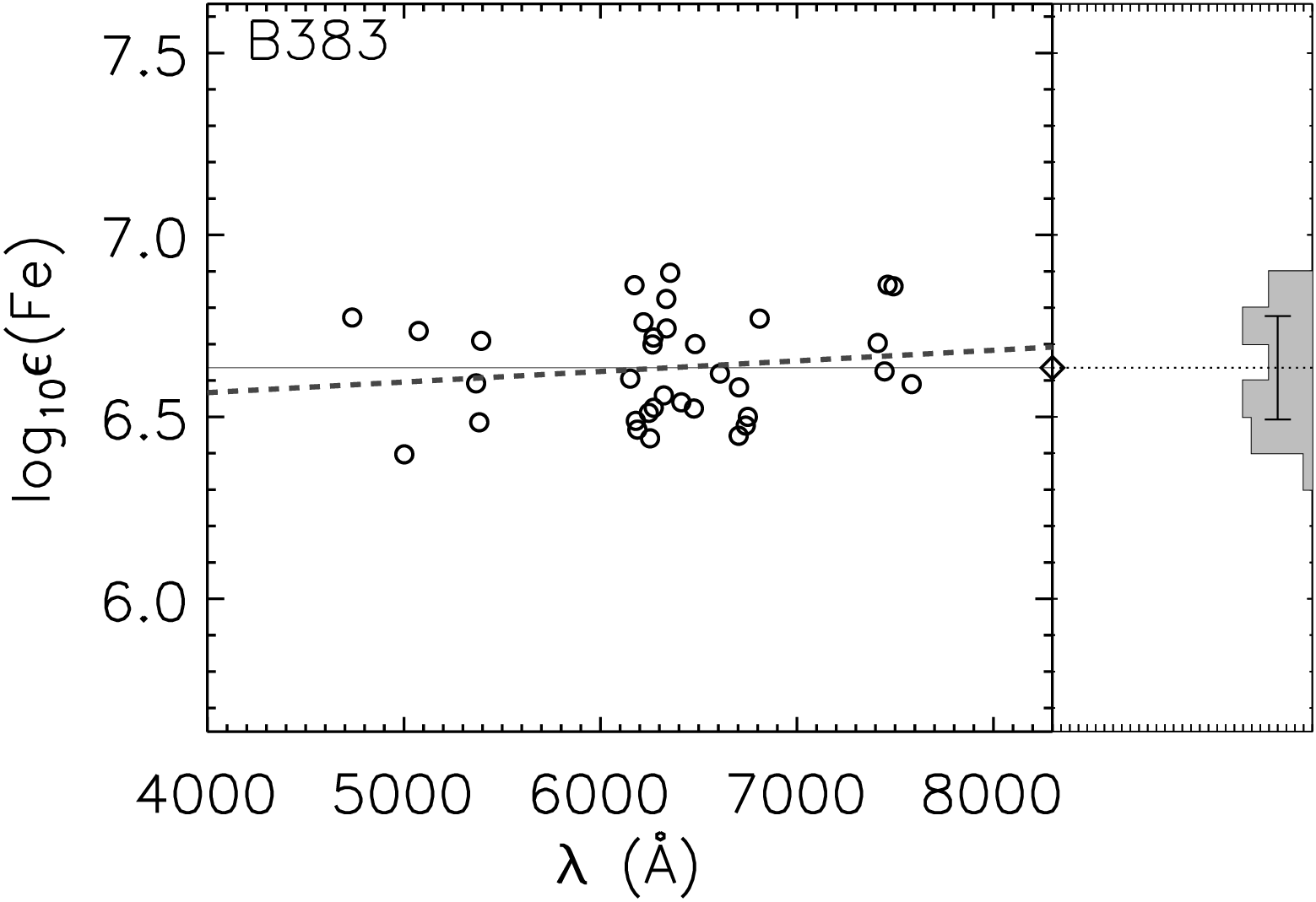}
\includegraphics[scale=0.20,angle=90]{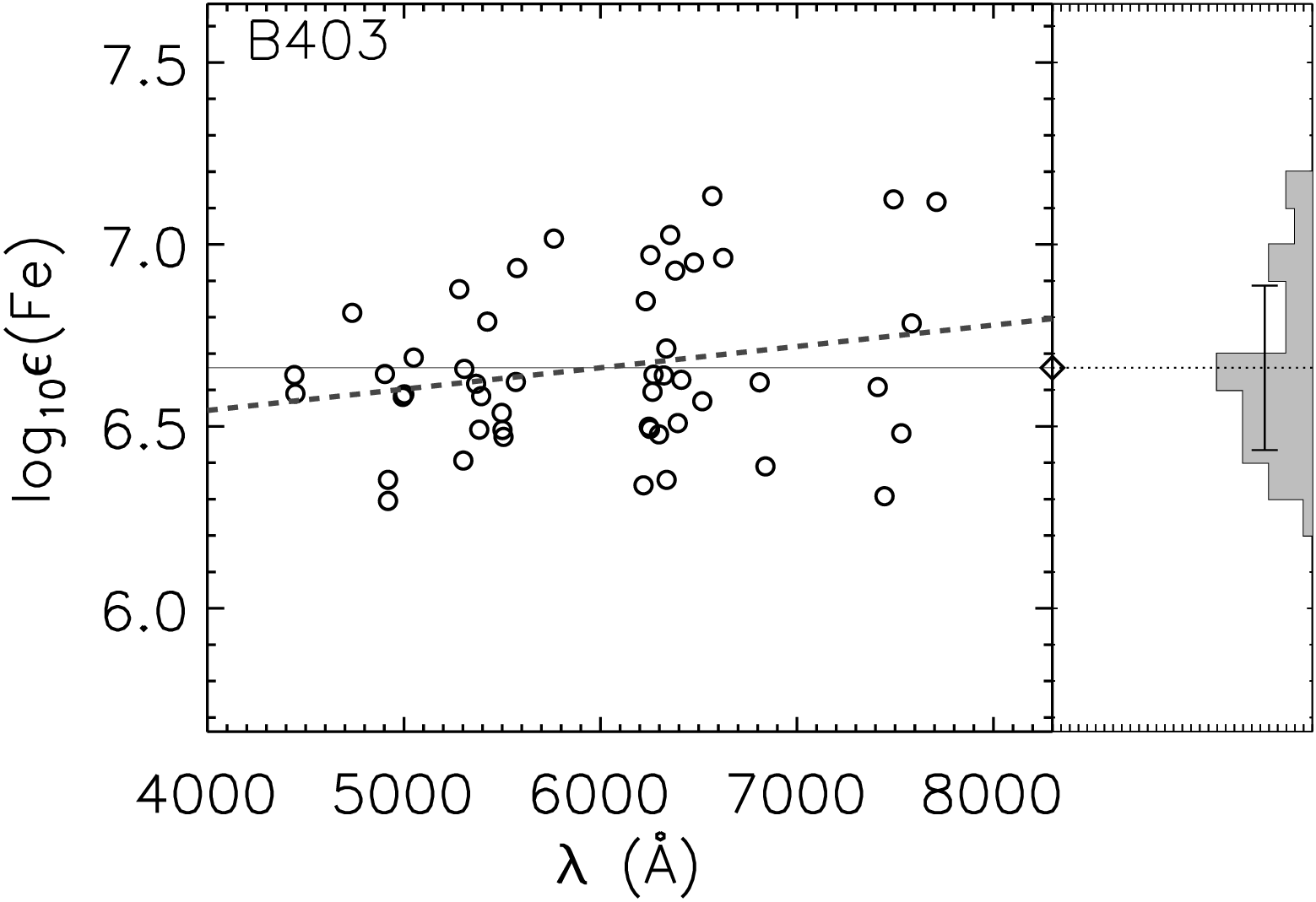}
\includegraphics[scale=0.20,angle=90]{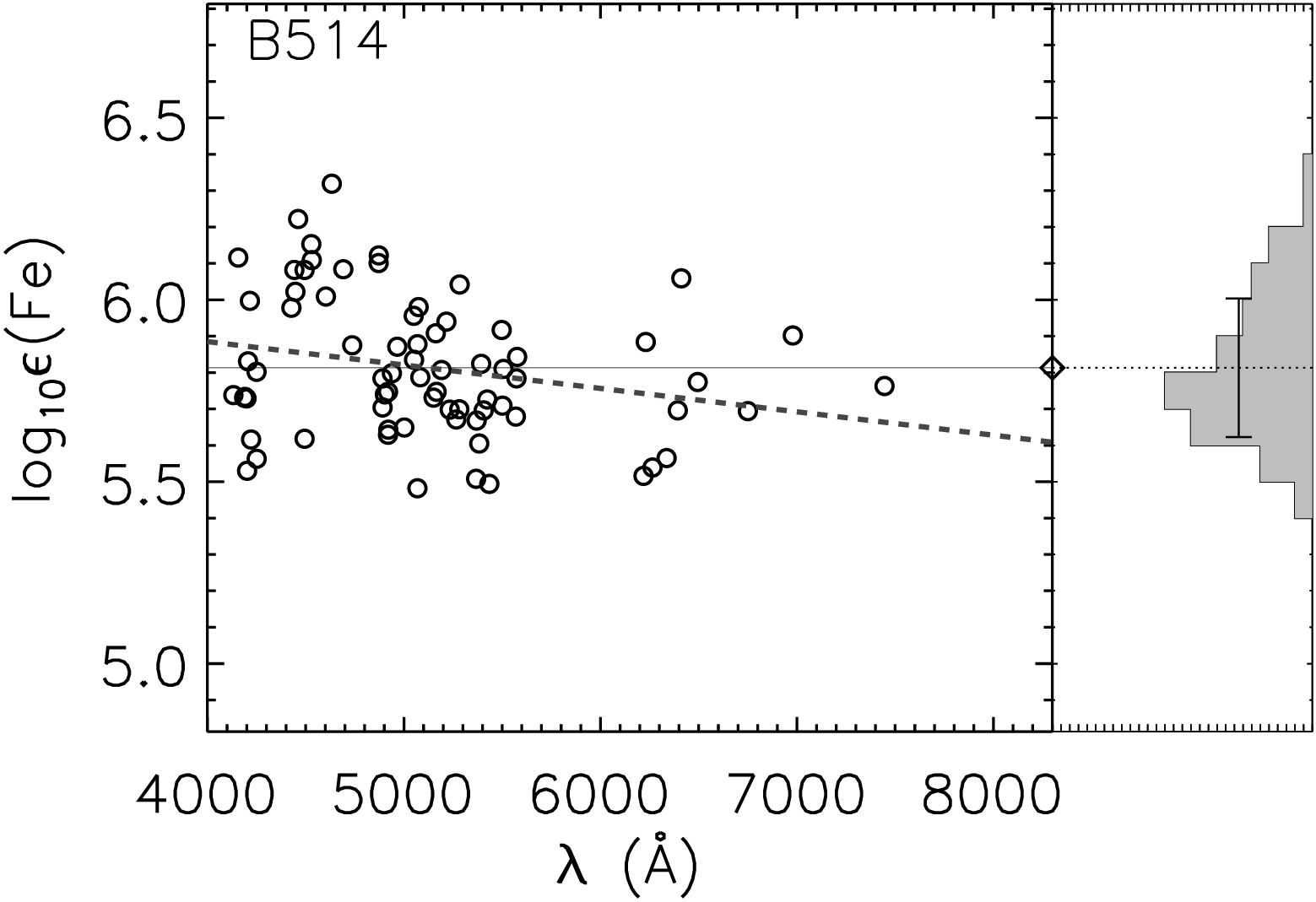}
\includegraphics[scale=0.20,angle=90]{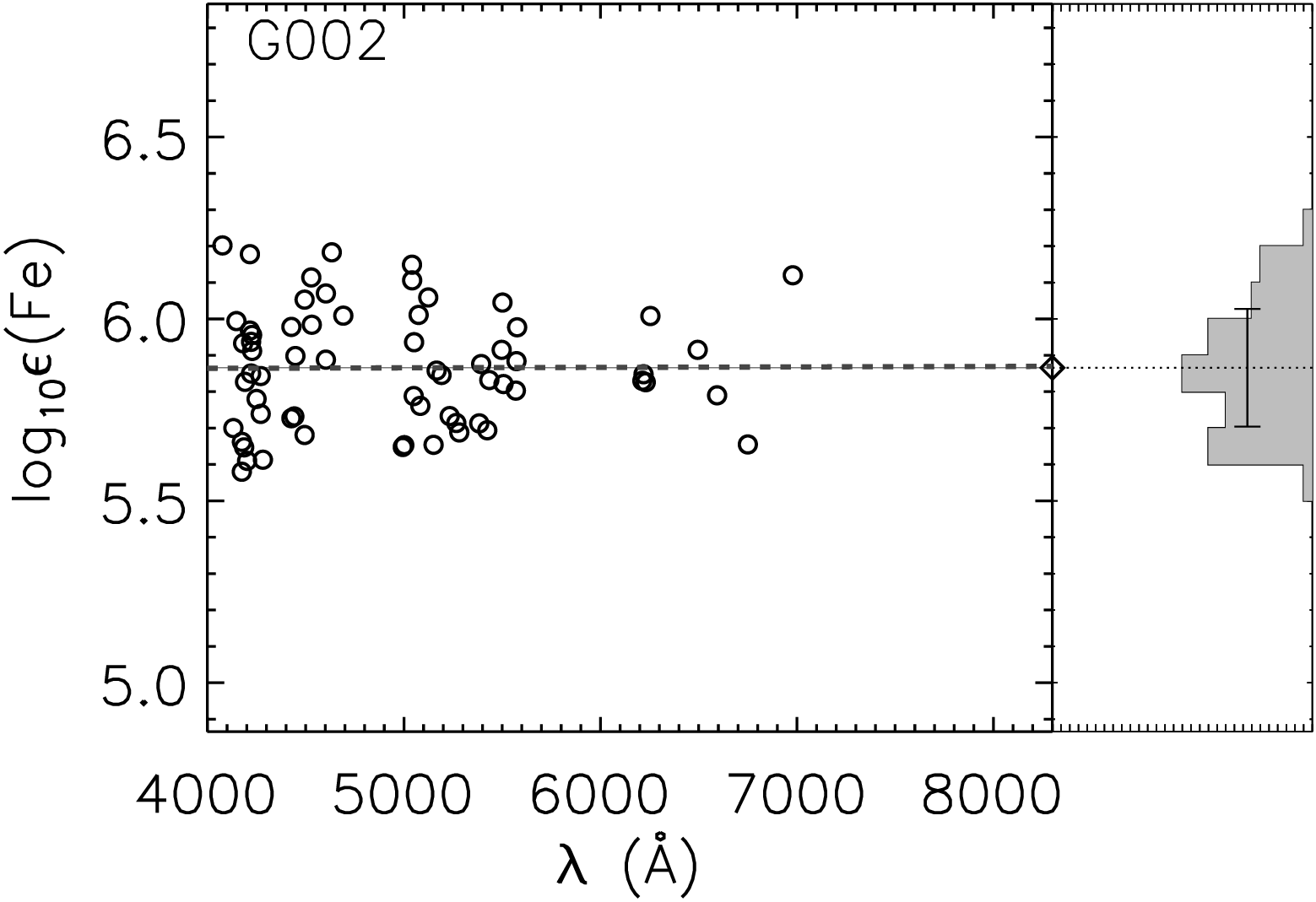}
\includegraphics[scale=0.20,angle=90]{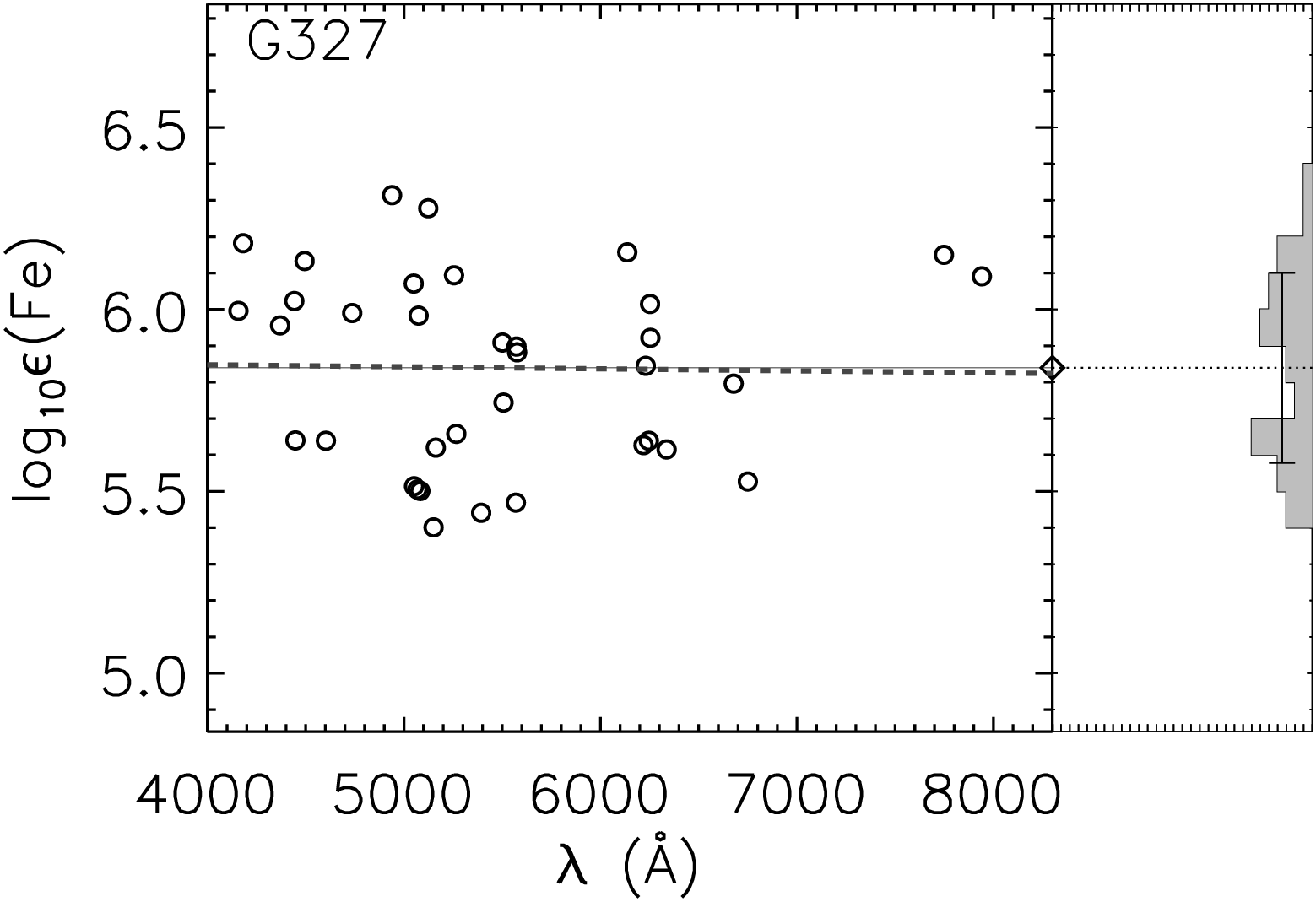}
\includegraphics[scale=0.20,angle=90]{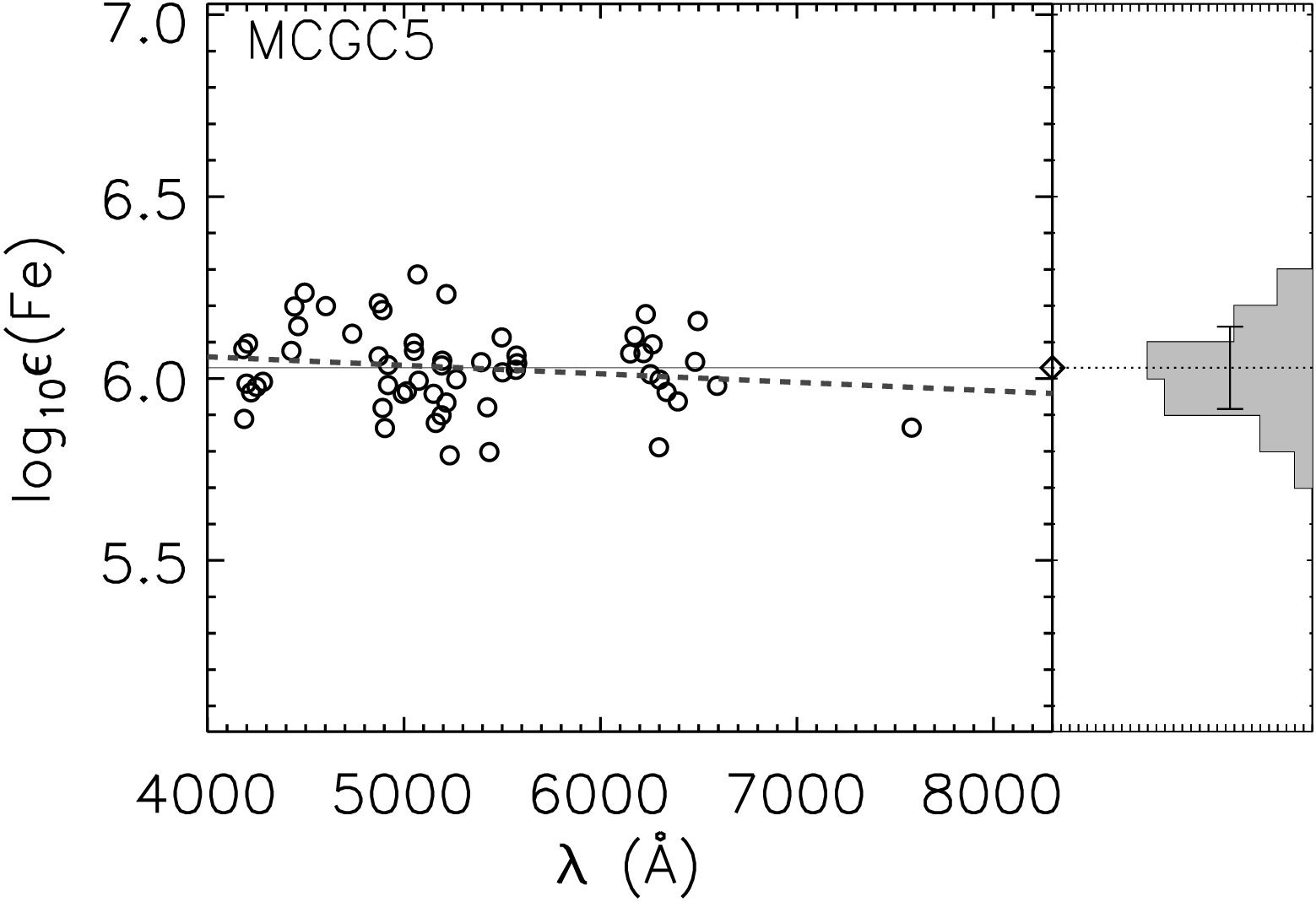}
\includegraphics[scale=0.20,angle=90]{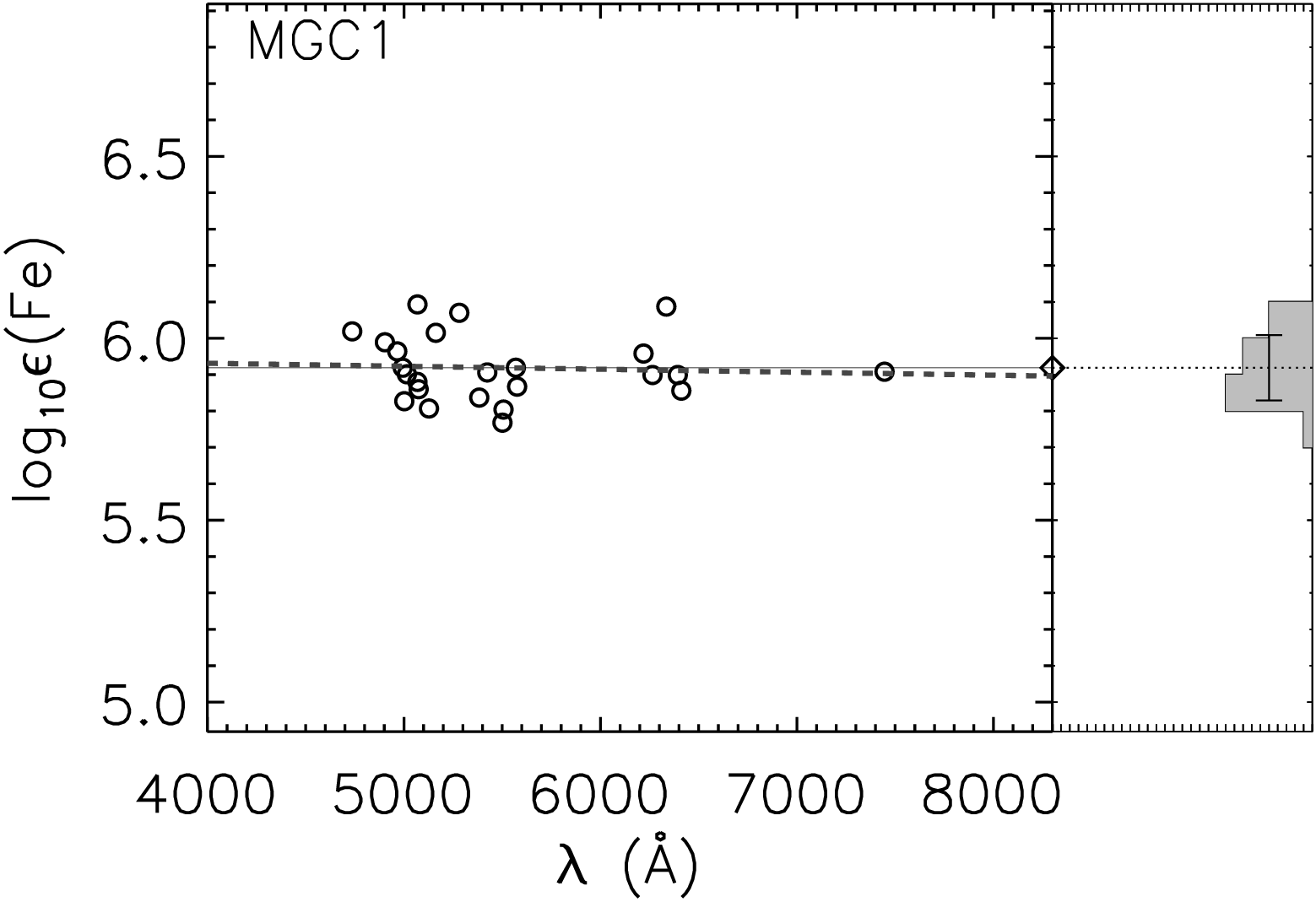}

\caption{ Fe I abundance vs. wavelength for  individual lines in GCs where final analysis was performed with  EWs measured in GETJOB. Symbols and lines are the same as in Figure \ref{fig:b384}. The total range in the y-axis is the same in all panels in order to appreciate the range in quality of the solutions.}
\label{fig:wave_ews} 
\end{figure*}

\begin{figure*}
\centering
\includegraphics[scale=0.20,angle=90]{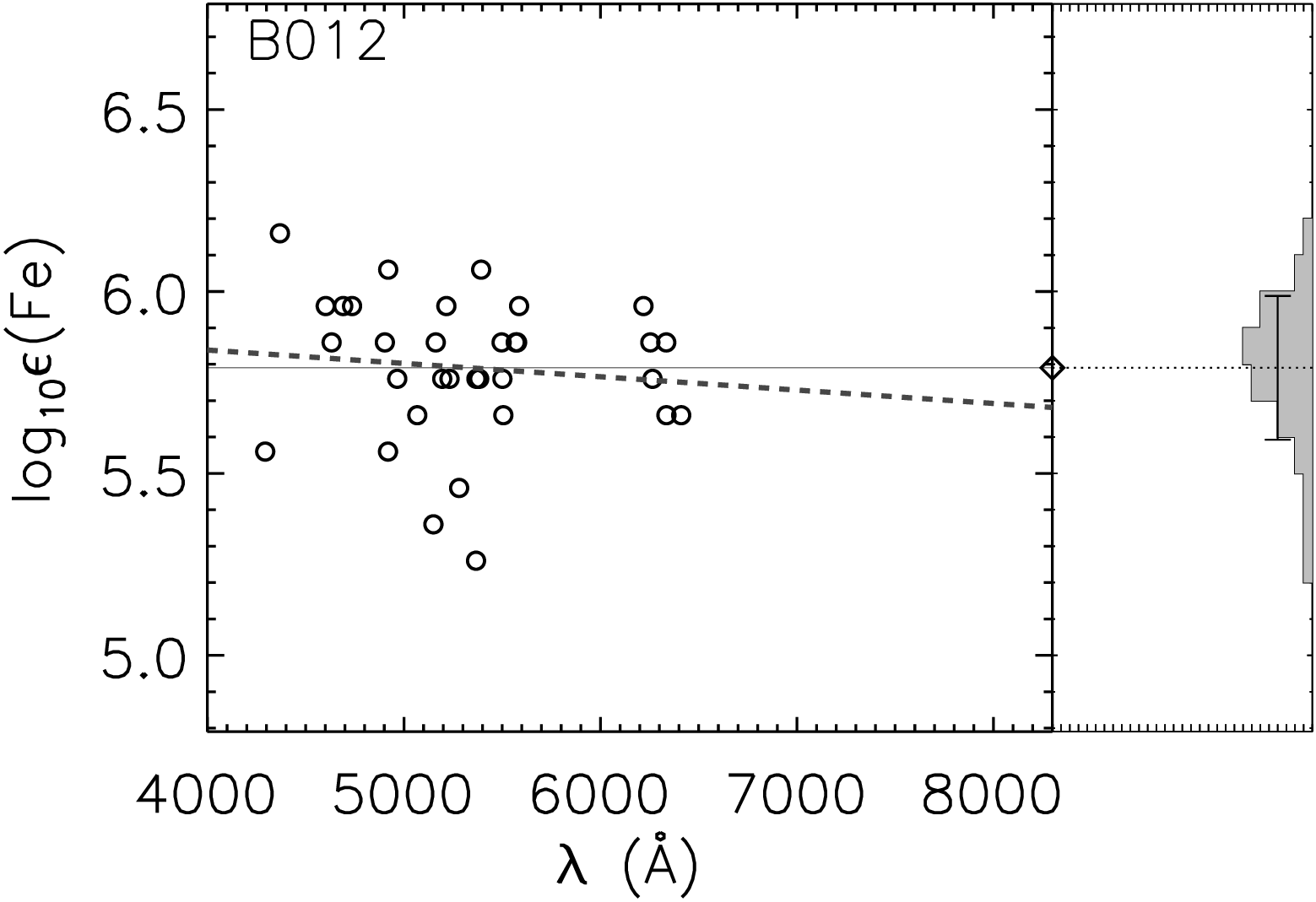}
\includegraphics[scale=0.20,angle=90]{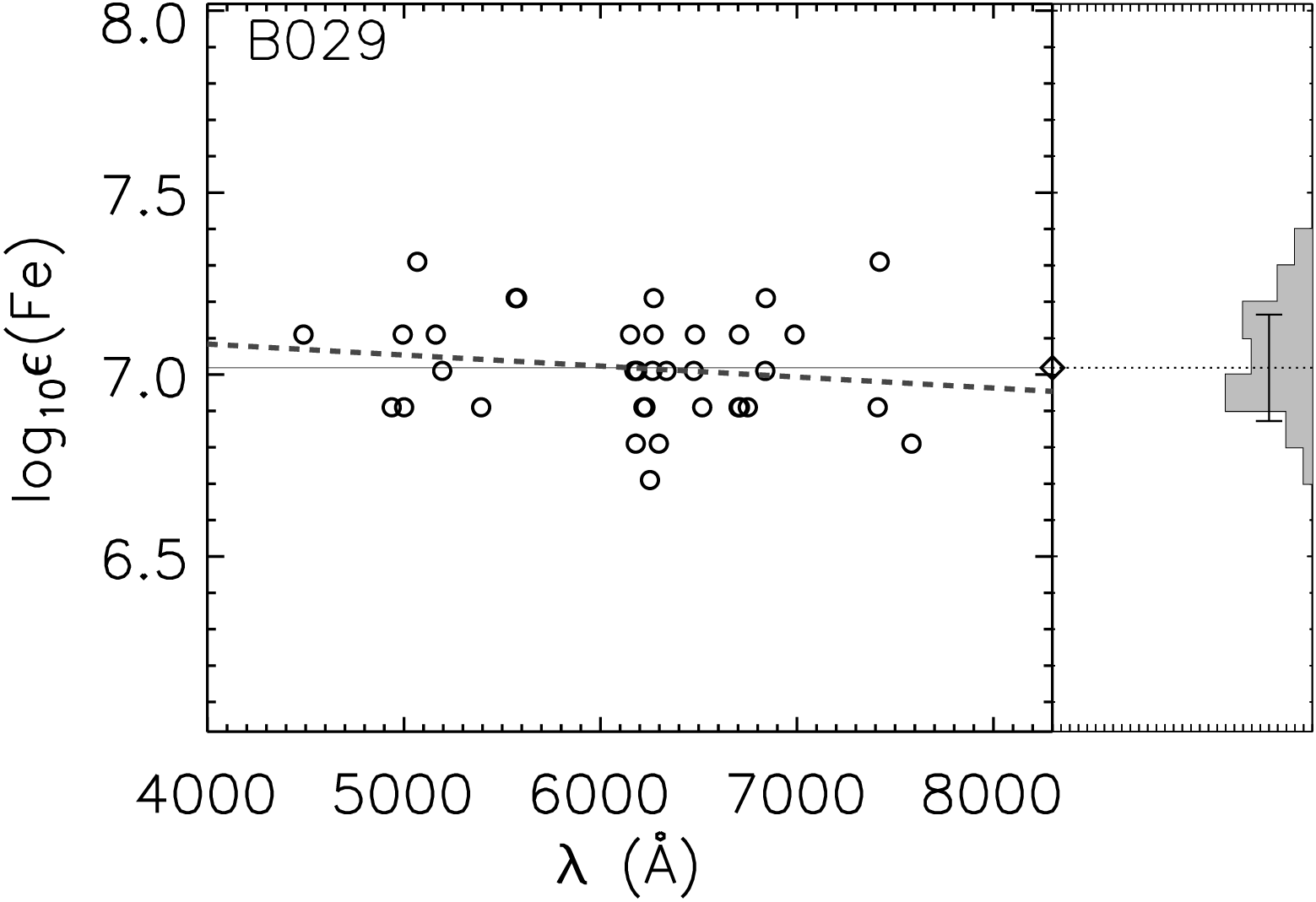}
\includegraphics[scale=0.20,angle=90]{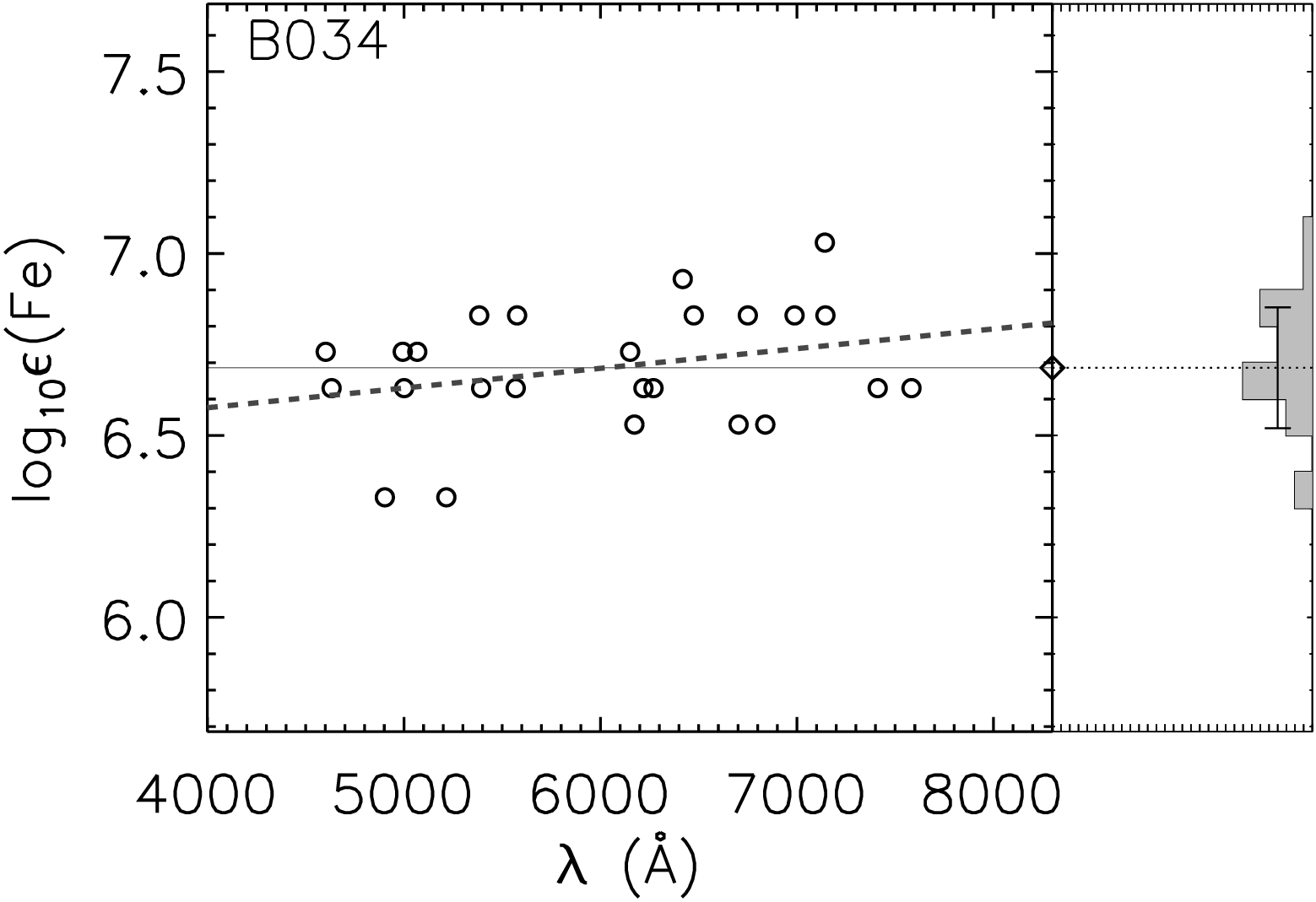}
\includegraphics[scale=0.20,angle=90]{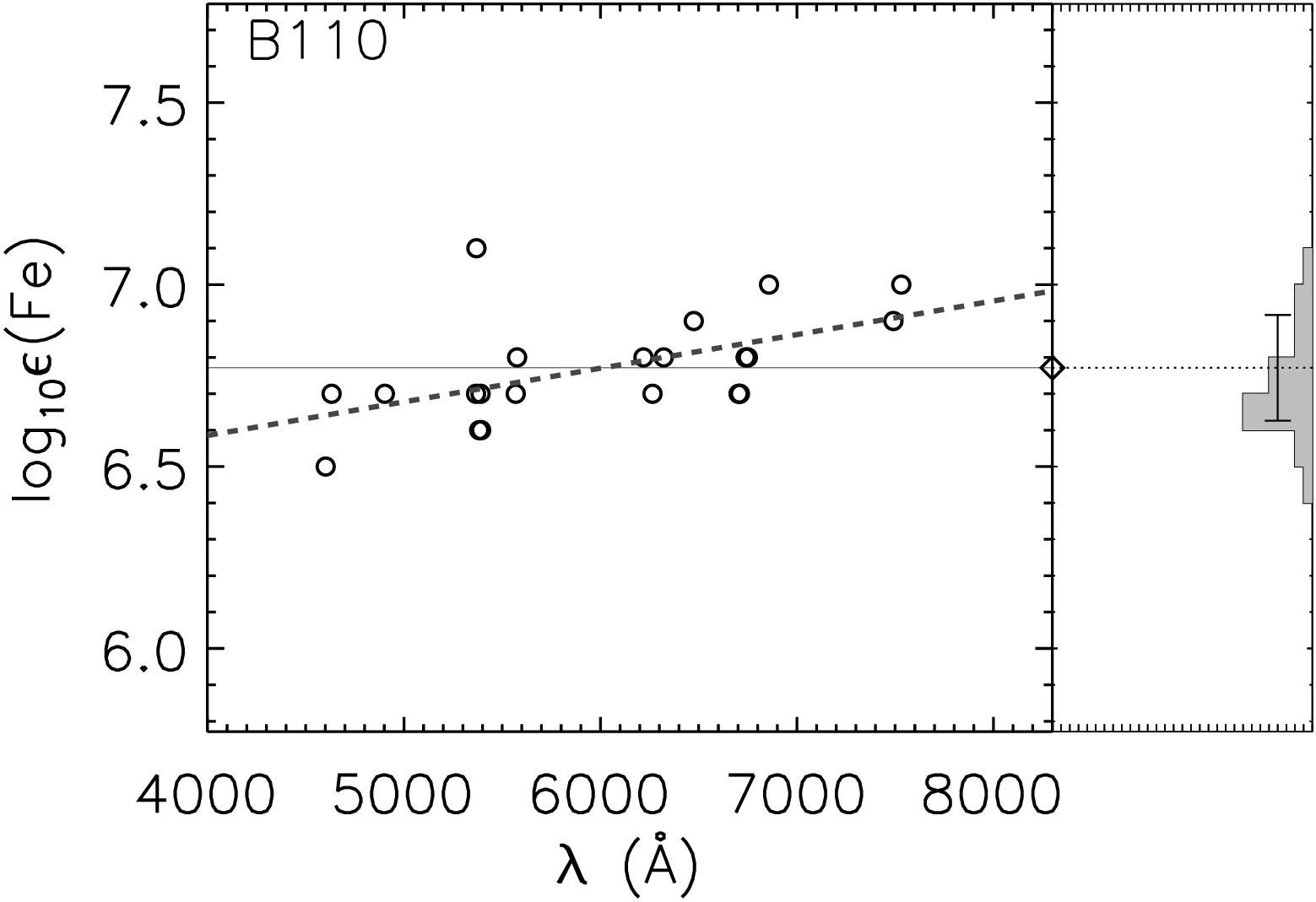}
\includegraphics[scale=0.20,angle=90]{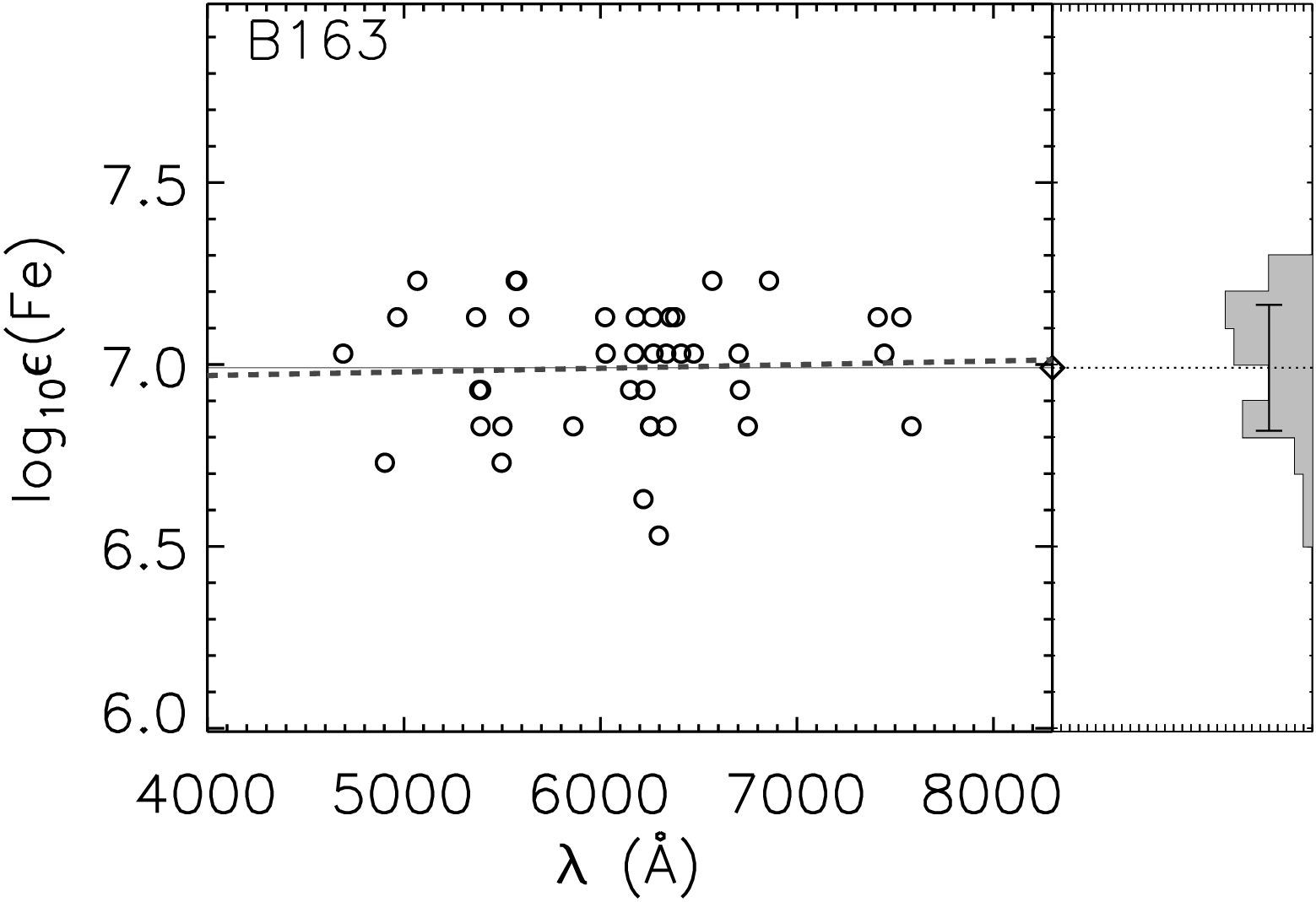}
\includegraphics[scale=0.20,angle=90]{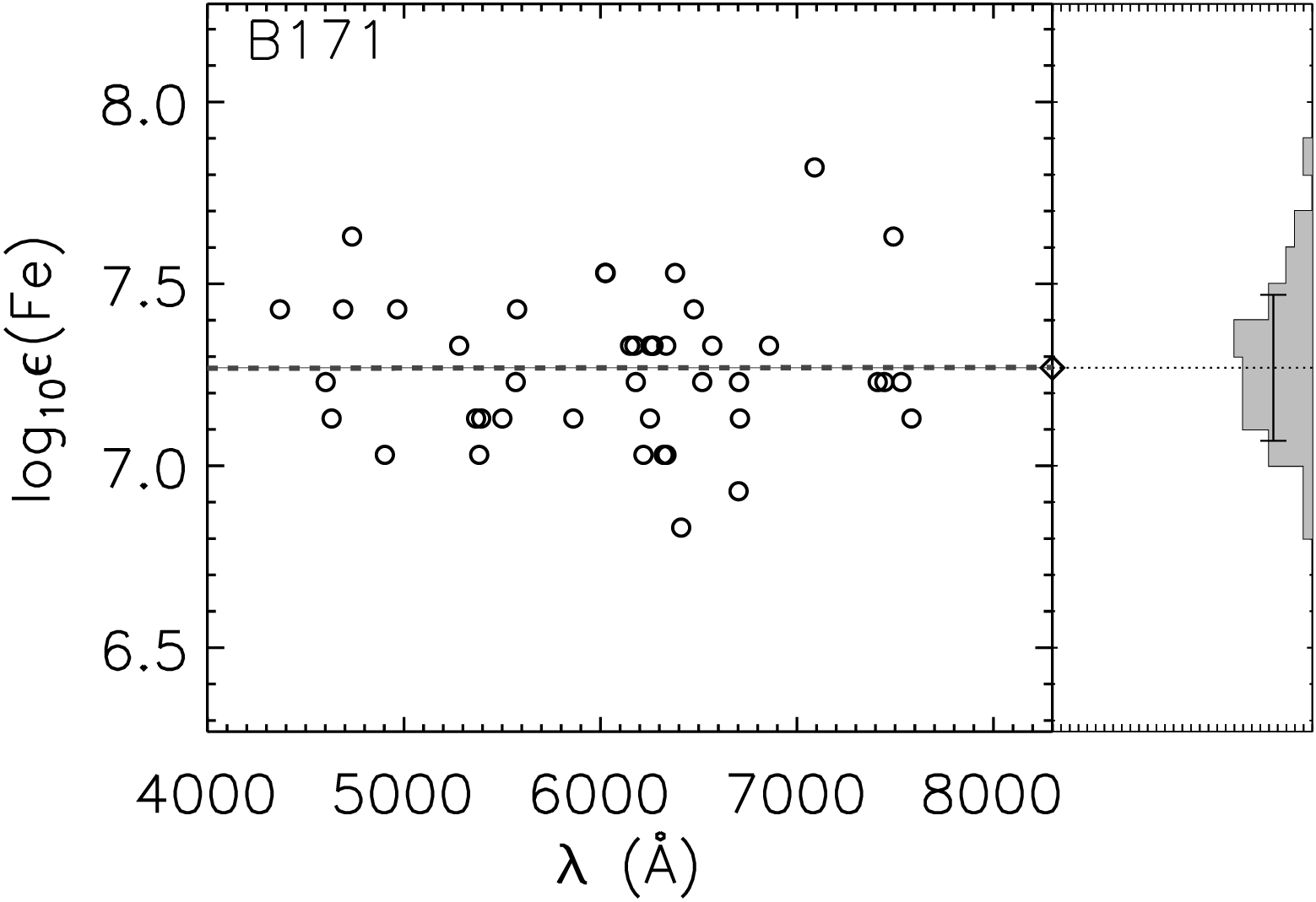}
\includegraphics[scale=0.20,angle=90]{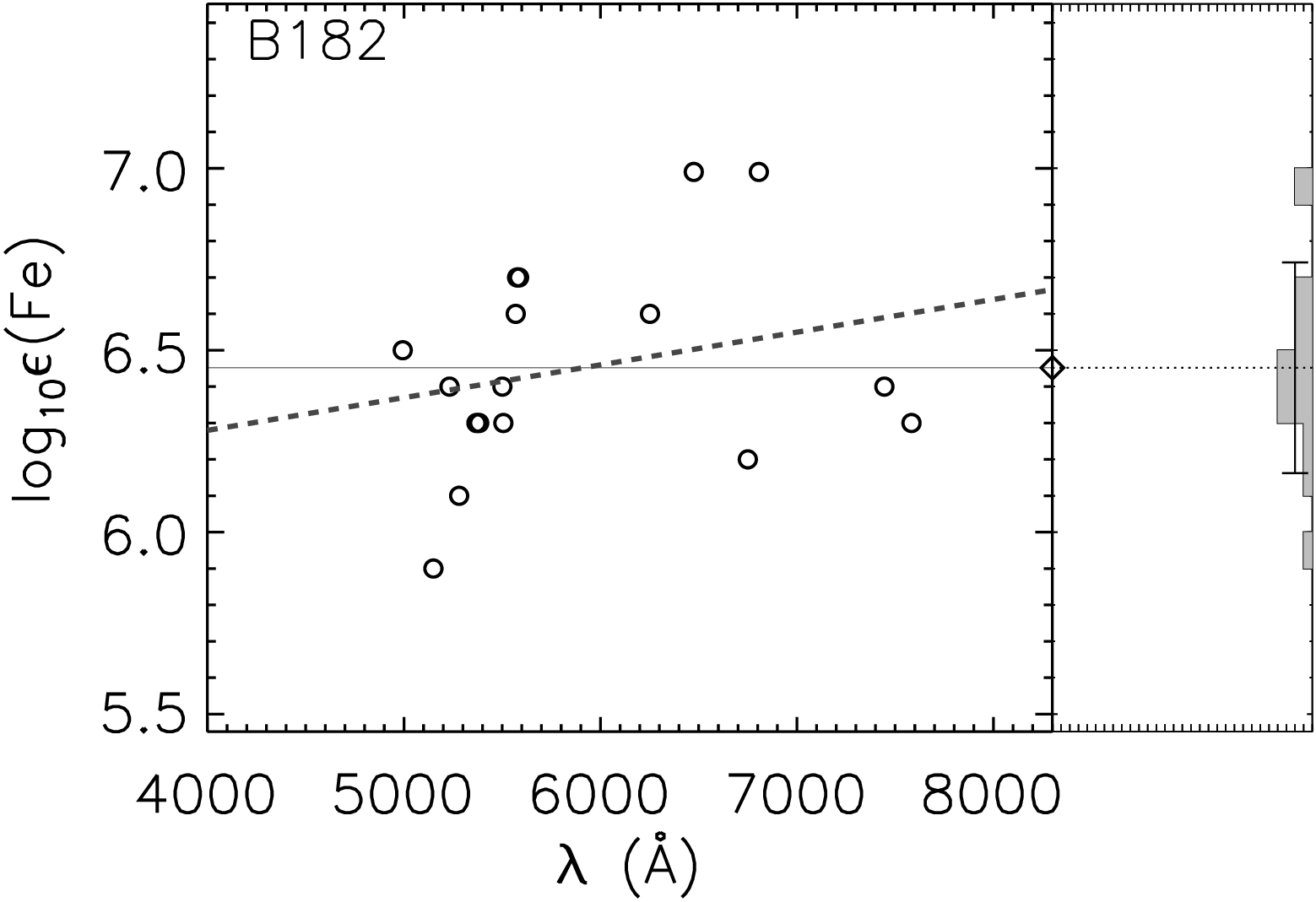}
\includegraphics[scale=0.20,angle=90]{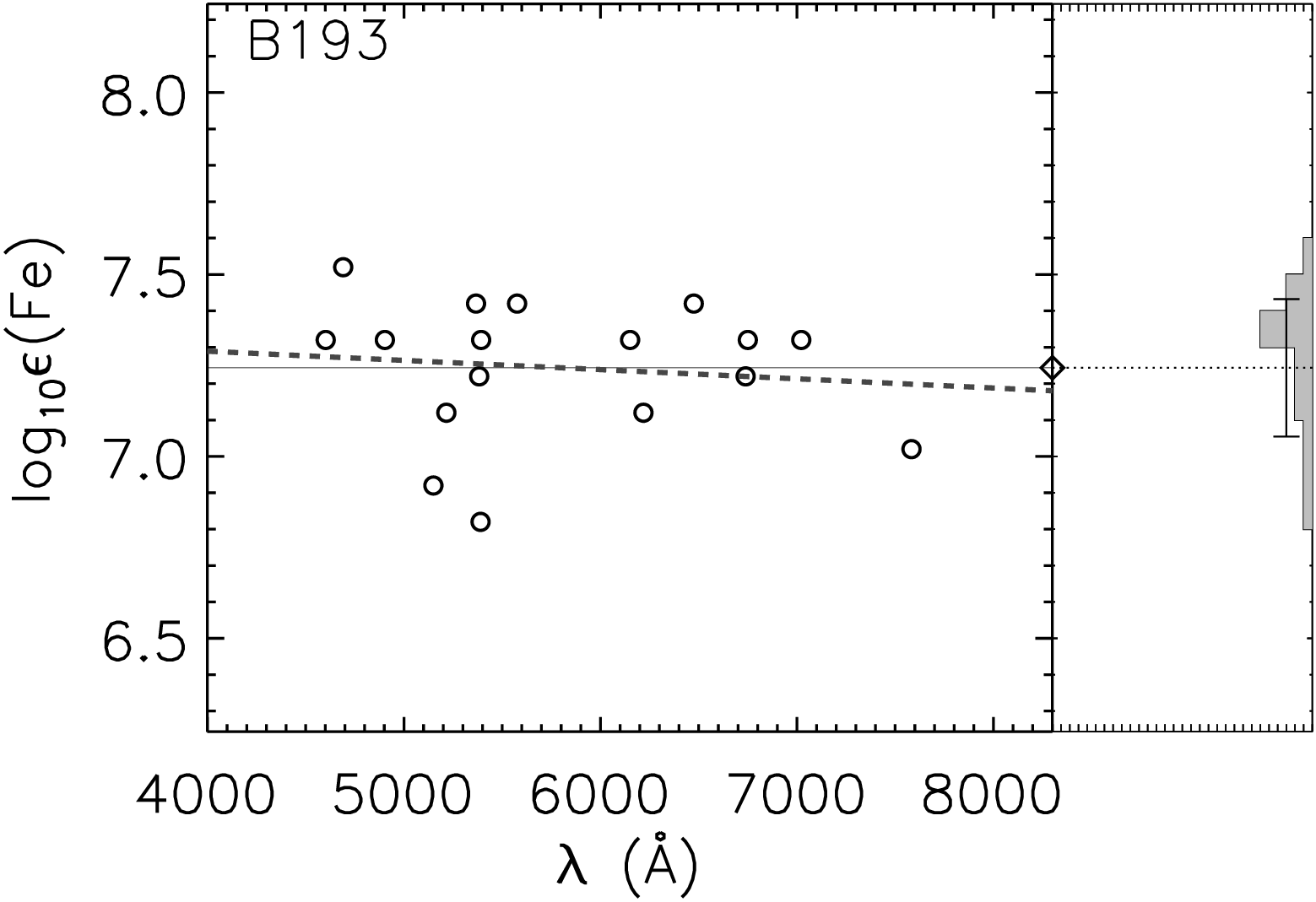}
\includegraphics[scale=0.20,angle=90]{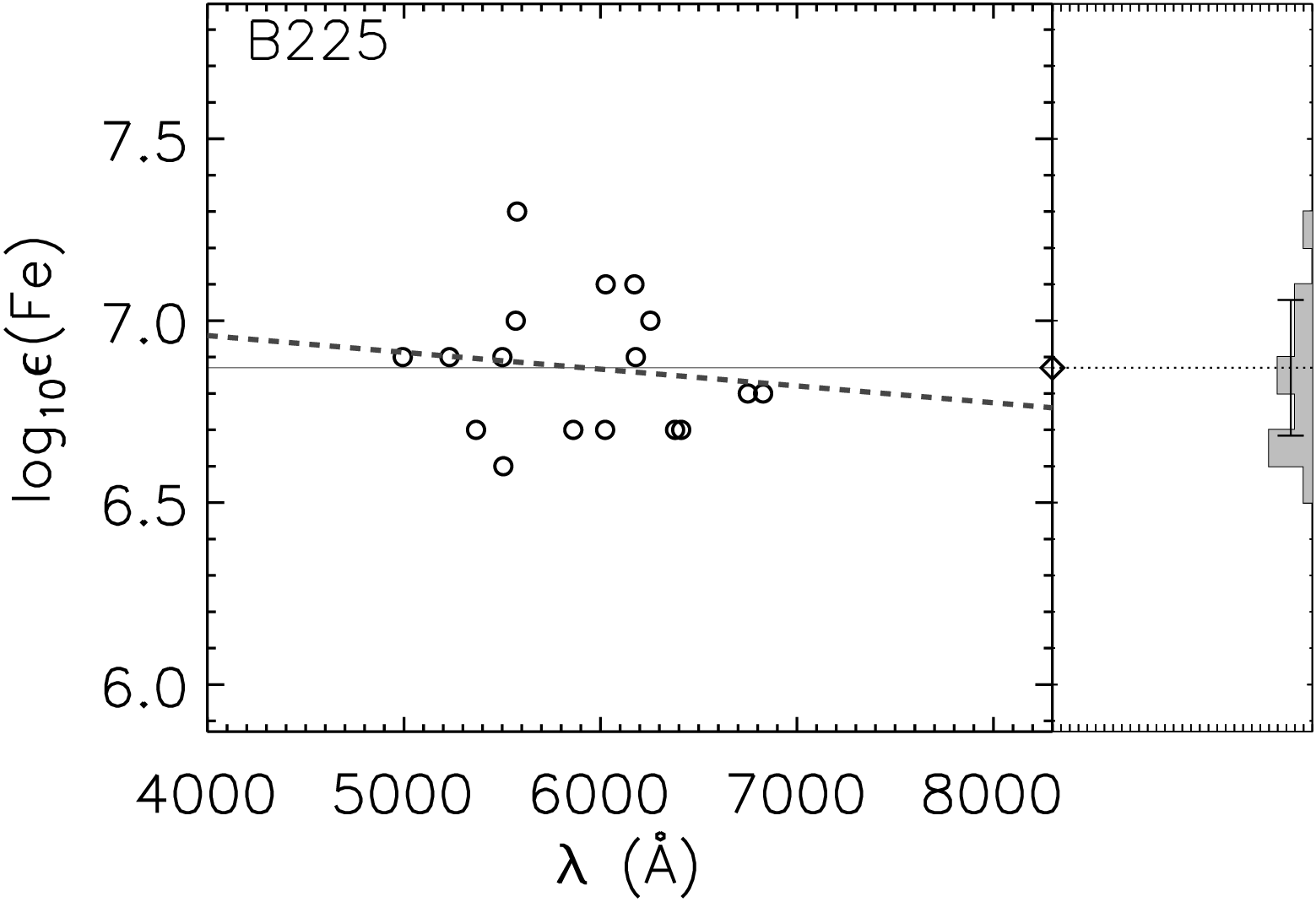}
\includegraphics[scale=0.20,angle=90]{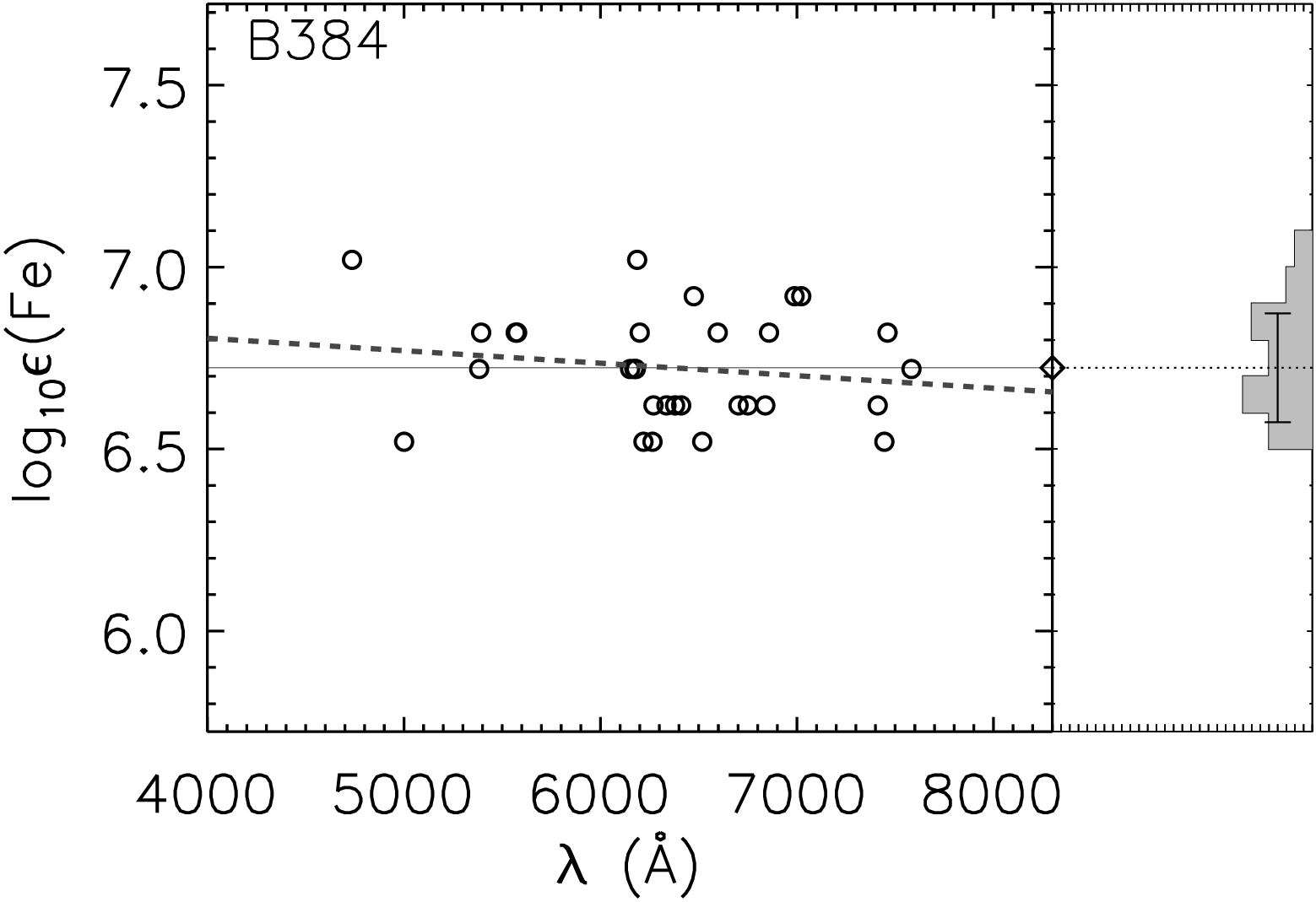}
\includegraphics[scale=0.20,angle=90]{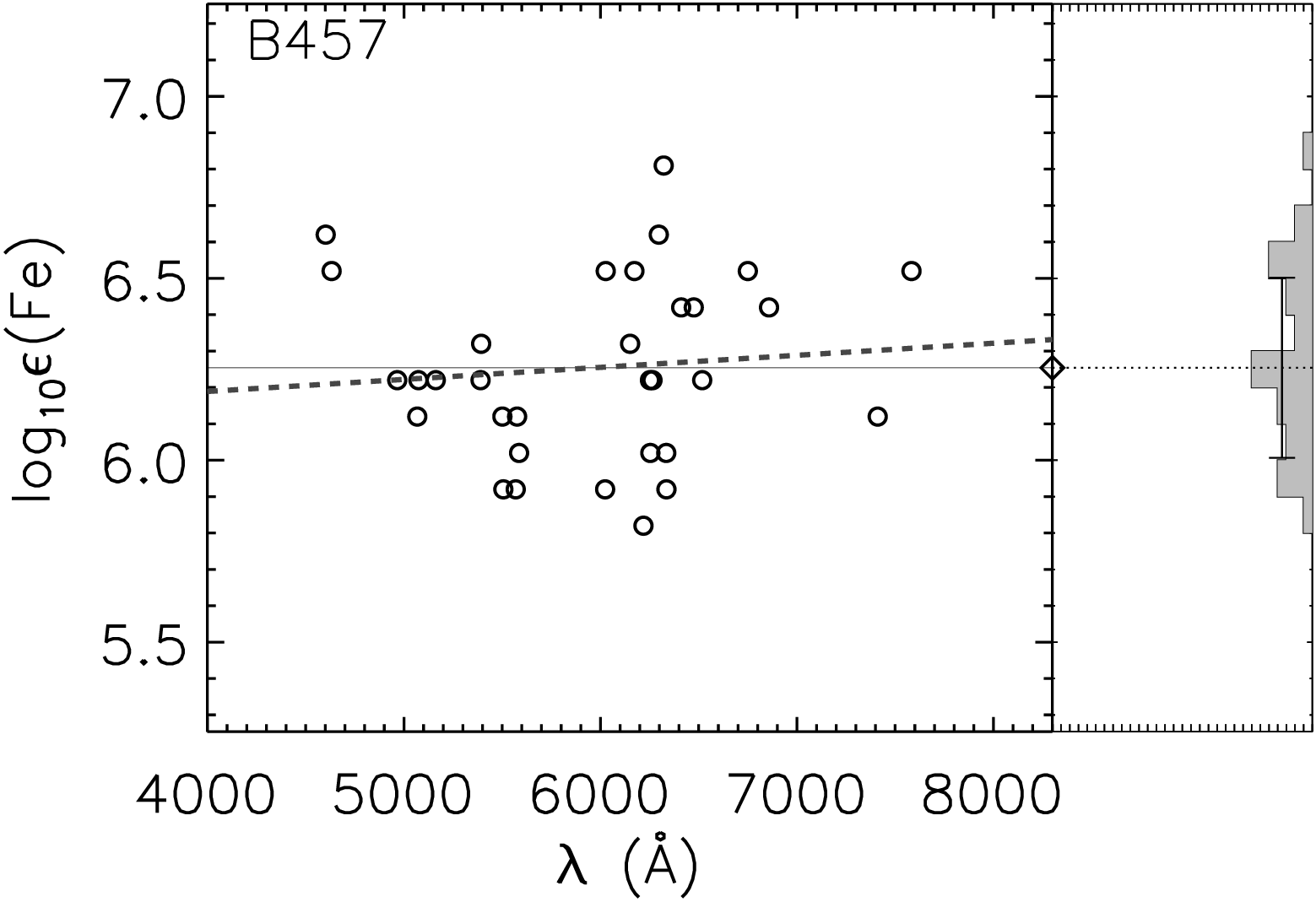}

\caption{ Fe I abundance vs. wavelength for  individual lines in GCs where final analysis was performed with line synthesis. Symbols and lines are the same as in Figure \ref{fig:b384}. The total range in the y-axis is the same in all panels in order to appreciate the range in quality of the solutions.}
\label{fig:wave_syn} 
\end{figure*}

In this work we present the Fe abundance analysis results for  26 GCs.  The EWs and line parameters of the  15 GCs for which the analysis was performed with GETJOB are listed in Table \ref{tab:fe_tab_stub}, and for the remaining 11 GCs that were analyzed with line synthesis we present the Fe abundance measured  for each line in Table \ref{tab:fe_syn_stub}.  The final [Fe/H] and age results and their accompanying uncertainties for the 31  GCs in our full sample are presented in Table \ref{tab:abund};  the 5 GCs analyzed previously  in \citetalias{m31paper} are separated at the end of the table for clarity.   Table \ref{tab:abund} also lists the number of Fe I lines used in the analysis and whether the final results were obtained with an equivalent width  analysis (EW) or synthesis analysis (SYN).  The last column  denotes whether the final abundances were calculated with solar-scaled (S) or alpha-enhanced (A) isochrones and stellar atmospheres. This decision  is based  on the alpha element abundances measured for Ca I, Si I, Ti I, and Ti II presented in \textsection \ref{sec:alpha}.  We require the averaged [X/Fe] ratio of all of the alpha elements to be $>+0.15$ for  alpha-enhanced calculations to be used instead of scaled-solar.  We also present the final Fe solutions as a function of wavelength for the 26 GCs analyzed in this work in Figure \ref{fig:wave_ews} for those analyzed with GETJOB and in Figure \ref{fig:wave_syn} for those analyzed with line synthesis.

\begin{deluxetable*}{rrr|rrrrrrrrrrrrr}
\scriptsize
\tablecolumns{16}
\tablewidth{0pc}
\tablecaption{M31 GC Fe I EWs\label{tab:fe_tab_stub}}
\tablehead{
 \colhead{$\lambda$}& \colhead{E.P.} &  \colhead{log{\it gf}} & \colhead{B006} & \colhead{B048} &  \colhead{B088}&  \colhead{B232}  & \colhead{B235} & \colhead{B240} & \colhead{B311} &  \colhead{B312}  & \colhead{B383}  & \colhead{B403}& \colhead{B514} & \colhead{G002} &  \colhead{G327} \\ \colhead{(\AA)}& \colhead{(eV)} && \colhead{(m\AA)} & \colhead{(m\AA)}& \colhead{(m\AA)}& \colhead{(m\AA)}& \colhead{(m\AA)}& \colhead{(m\AA)}& \colhead{(m\AA)}& \colhead{(m\AA)}& \colhead{(m\AA)}& \colhead{(m\AA)}& \colhead{(m\AA)}& \colhead{(m\AA)}& \colhead{(m\AA)} }
\startdata

6322.694 &  2.590 & -2.440 &  93.6 &  82.3 &  \nodata&  \nodata&  67.0 &  \nodata&  \nodata&  66.5 &  81.5 &  87.6 &  \nodata&  \nodata&  \nodata\\
6335.337 &  2.200 & -2.170 & 138.4 & 115.9 &  \nodata&  \nodata& 115.9 &  83.8 &  \nodata&  \nodata& 139.9 & 135.8 &  \nodata&  \nodata&  \nodata \\
6336.830 &  3.690 & -0.670 &  95.8 &  99.4 &  \nodata&  \nodata&  85.8 &  \nodata&  \nodata&  73.5 & 123.6 &  99.9 &  45.0 &  \nodata&  49.1 \\
6355.035 &  2.840 & -2.330 &  \nodata&  76.5 &  \nodata&  \nodata&  \nodata&  \nodata&  \nodata&  76.6 &  89.7 &  98.8 &  \nodata&  \nodata&  \nodata\\
6380.750 &  4.190 & -1.370 &  \nodata&  51.6 &  \nodata&  \nodata&  \nodata&  \nodata&  \nodata&  \nodata&  \nodata&  57.6 &  \nodata&  \nodata&  \nodata\\
6393.612 &  2.430 & -1.500 &  \nodata&  \nodata&  85.0 &  \nodata& 138.8 & 114.3 &  76.0 & 130.2 &  \nodata& 149.8 &  80.0 &  \nodata&  \nodata \\
6411.658 &  3.650 & -0.650 & 140.9 &  97.0 &  45.4 &  \nodata& 120.2 &  80.5 &  64.0 &  88.7 & 113.7 & 124.4 &  74.8 &  \nodata&  \nodata  \\
6421.360 &  2.280 & -1.980 &  \nodata&  \nodata&  83.8 &  \nodata&  \nodata&  \nodata&  \nodata& 118.1 &  \nodata&  \nodata&  \nodata&  \nodata&  \nodata\\
6475.632 &  2.560 & -2.930 &  88.5 &  \nodata&  \nodata&  \nodata&  66.9 &  32.4 &  \nodata&  \nodata&  58.5 &  82.6 &  \nodata&  \nodata&  \nodata\\
6481.878 &  2.280 & -2.980 &  94.2 &  \nodata&  \nodata&  \nodata&  94.0 &  45.3 &  \nodata&  74.8 &  83.2 &  \nodata&  \nodata&  \nodata&  \nodata \\
6494.994 &  2.400 & -1.250 &  \nodata&  \nodata&  \nodata&  \nodata&  \nodata& 132.3 & 109.6 & 179.5 &  \nodata&  \nodata& 101.8 & 112.1 &  \nodata \\
6498.945 &  0.960 & -4.670 &  \nodata&  81.7 &  \nodata&  \nodata&  \nodata&  \nodata&  \nodata&  \nodata&  \nodata&  \nodata&  \nodata&  \nodata&  46.3 \\
6518.373 &  2.830 & -2.400 &  \nodata&  68.5 &  \nodata&  \nodata&  \nodata&  38.3 &  \nodata&  \nodata&  \nodata&  72.4 &  \nodata&  \nodata&  \nodata\\
6569.224 &  4.730 & -0.380 &  \nodata&  \nodata&  \nodata&  \nodata&  \nodata&  \nodata&  \nodata&  \nodata&  \nodata&  91.6 &  \nodata&  \nodata&  \nodata  \\
6593.874 &  2.430 & -2.380 &  \nodata&  \nodata&  \nodata&  \nodata&  \nodata&  \nodata&  \nodata&  \nodata&  \nodata&  \nodata&  \nodata&  47.4 &  \nodata\\
6597.571 &  4.770 & -0.970 &  40.3 &  \nodata&  \nodata&  \nodata&  \nodata&  \nodata&  \nodata&  \nodata&  \nodata&  \nodata&  \nodata&  \nodata&  \nodata\\
6608.044 &  2.270 & -3.940 &  \nodata&  46.9 &  \nodata&  \nodata&  \nodata&  \nodata&  \nodata&  \nodata&  36.6 &  \nodata&  \nodata&  \nodata&  \nodata\\
6625.039 &  1.010 & -5.280 &  \nodata&  \nodata&  \nodata&  \nodata&  \nodata&  \nodata&  \nodata&  \nodata&  \nodata&  67.8 &  \nodata&  \nodata&  \nodata\\
6677.997 &  2.690 & -1.400 &  \nodata&  \nodata&  \nodata&  70.3 &  \nodata&  91.4 &  \nodata& 129.5 &  \nodata&  \nodata&  \nodata&  \nodata&  80.9 \\
6703.576 &  2.760 & -3.060 &  \nodata&  \nodata&  \nodata&  \nodata&  57.4 &  \nodata&  \nodata&  \nodata&  39.5 &  \nodata&  \nodata&  \nodata&  \nodata\\

\enddata
\tablecomments{This table is presented in its entirety in machine readable form in the electronic edition of the journal.  Lines listed twice were measured in two orders with overlapping wavelength coverage.}
\end{deluxetable*}

\begin{deluxetable*}{r|rrrrrrrrrrr}
\scriptsize
\tablecolumns{12}
\tablewidth{0pc}
\tablecaption{M31 GC Fe I Synthesis Abundances\label{tab:fe_syn_stub}}
\tablehead{
 \colhead{$\lambda$} & \colhead{B012} & \colhead{B029} &  \colhead{B034}&  \colhead{B110}  & \colhead{B163} & \colhead{B171} & \colhead{B182} &  \colhead{B193}  & \colhead{B225} & \colhead{B384} & \colhead{B457}  \\ \colhead{(\AA)} }
\startdata

6297.799 &  \nodata &  6.81 &  \nodata &  \nodata &  6.53 &  \nodata &  \nodata &  \nodata &  \nodata &  \nodata &  6.62 \\
6322.694 &  \nodata &  \nodata &  \nodata &  6.80 &  \nodata &  7.03 &  \nodata &  \nodata &  \nodata &  \nodata &  6.81 \\
6335.337 &  5.86 &  \nodata &  \nodata &  \nodata &  7.03 &  7.33 &  \nodata &  \nodata &  \nodata &  \nodata &  6.02 \\
6336.830 &  5.66 &  7.01 &  \nodata &  \nodata &  6.83 &  7.03 &  \nodata &  \nodata &  \nodata &  6.62 &  5.92 \\
6355.035 &  \nodata &  \nodata &  \nodata &  \nodata &  7.13 &  \nodata &  \nodata &  \nodata &  \nodata &  \nodata &  \nodata \\
6380.750 &  \nodata &  \nodata &  \nodata &  \nodata &  7.13 &  7.53 &  \nodata &  \nodata &  6.70 &  6.62 &  \nodata \\
6411.658 &  5.66 &  \nodata &  \nodata &  \nodata &  7.03 &  6.83 &  \nodata &  \nodata &  6.70 &  6.62 &  6.42 \\
6419.956 &  \nodata &  \nodata &  6.93 &  \nodata &  \nodata &  \nodata &  \nodata &  \nodata &  \nodata &  \nodata &  \nodata \\
6475.632 &  \nodata &  7.01 &  6.83 &  6.90 &  7.03 &  7.43 &  6.99 &  7.42 &  \nodata &  6.92 &  6.42 \\
6481.878 &  \nodata &  7.11 &  \nodata &  \nodata &  \nodata &  \nodata &  \nodata &  \nodata &  \nodata &  \nodata &  \nodata \\
6518.373 &  \nodata &  6.91 &  \nodata &  \nodata &  \nodata &  7.23 &  \nodata &  \nodata &  \nodata &  6.52 &  6.22 \\
6569.224 &  \nodata &  \nodata &  \nodata &  \nodata &  7.23 &  7.33 &  \nodata &  \nodata &  \nodata &  \nodata &  \nodata \\
6597.571 &  \nodata &  \nodata &  \nodata &  \nodata &  \nodata &  \nodata &  \nodata &  \nodata &  \nodata &  6.82 &  \nodata \\
6703.576 &  \nodata &  6.91 &  6.53 &  6.70 &  7.03 &  6.93 &  \nodata &  \nodata &  \nodata &  6.62 &  \nodata \\

\enddata
\tablecomments{This table is presented in its entirety in machine
  readable form in the
  electronic edition of the journal. Abundances for individual lines
  are measured in steps of 0.1 dex from the mean abundance using all
  lines.  The typical statistical measurement uncertainty of
  individual lines is 0.05 dex. Note that these statistical uncertainties do not dominate
	the total uncertainty.  See text for further discussion.  Abundances for each GC are calculated using its CMD solution
  that has  the
  oldest age.}
\end{deluxetable*}

\begin{deluxetable*}{lrrrrrrrlr}

\tablecolumns{10}
\tablewidth{0pc}
\tablecaption{M31 GC Ages and Fe I Abundances \label{tab:abund}}
\tablehead{
\colhead{Name}   & \colhead{Age} & \colhead{$\Delta$Age}&\colhead{[Fe I/H]} & \colhead{{\rm N}$^{1}$} & \colhead{$\sigma_{ {\rm N}}$$^{2}$}& \colhead{$\sigma_{ {\rm Age}}$$^{3}$} & \colhead{$\sigma_{ {\rm T}}$$^{4}$}&\colhead{EW/SYN$^{5}$} & \colhead{$\alpha^{6}$} \\  & \colhead{(Gyrs)} &  \colhead{(Gyrs)}  }
\startdata

 B006-G058 & 12.5 &  2.5 &  -0.73 &  30 &   0.22 &  0.07 &  0.08 & EW & A \\
 B012-G064 & 11.5 &  1.5 &  -1.61 &  30 &   0.21 &  0.01 &  0.04 & SYN & A \\
 B029-G090 &  2.1 &  0.9 &  -0.43 &  33 &   0.11 &  0.11 &  0.11 &  SYN &A \\
 B034-G096 & 12.5 &  2.5 &  -0.75 &  24 &   0.14 &  0.05 &  0.06 & SYN & A \\
 B048-G110 & 12.5 &  2.5 &  -0.85 &  46 &   0.17 &  0.07 &  0.07 & EW & A \\
 B088-G150 & 14.0 &  1.0 &  -1.71 &  28 &   0.22 &  0.02 &  0.05 & EW & A \\
 B110-G172 &  6.5 &  3.5 &  -0.66 &  18 &   0.10 &  0.04 &  0.05 &  SYN &A \\
 B171-G222 &  12.5 &  2.5 &  -0.45 &  43 &   0.19 &  0.06 &  0.07 & SYN & A \\
 B163-G217 & 11.5 &  1.5 &  -0.49 &  41 &   0.18 &  0.03 &  0.04 & SYN & A \\ 
 B182-G233 & 12.5 &  2.5 &  -1.04 &  14 &   0.18 &  0.07 &  0.09 & SYN & A \\
B193-G244 &  8.0 &  5.0 &  -0.16 &  14 &   0.12 &  0.04 &  0.05 &  SYN &A \\
 B225-G280 & 10.0 &  3.0 &  -0.66 &  16 &   0.13 &  0.03 &  0.05 &  SYN& A \\

B232-G286 & 14.0 &  1.0 &  -1.77 &  27 &   0.23 &  0.06 &  0.08 & EW & A \\
 B235-G297 & 12.5 &  2.5 &  -0.86 &  55 &   0.19 &  0.06 &  0.07 & EW & A \\
 B240-G302 & 12.5 &  2.5 &  -1.54 &  58 &   0.13 &  0.04 &  0.04 & EW & A \\
 B311-G033 & 14.0 &  1.0 &  -1.71 &  23 &   0.19 &  0.04 &  0.06 & EW & A \\
 B312-G035 & 12.5 &  2.5 &  -1.06 &  74 &   0.22 &  0.02 &  0.03 & EW & A \\
 B383-G318 & 12.5 &  2.5 &  -0.78 &  34 &   0.16 &  0.04 &  0.05 &  EW &A \\
 B384-G319 &  6.5 &  3.5 &  -0.63 &  29 &   0.14 &  0.11 &  0.11 &  SYN &A \\
 B403-G348 & 11.0 &  1.0 &  -0.80 &  49 &   0.16 &  0.03 &  0.04 & EW & A \\
 B457-G097 & 11.0 &  4.0 &  -1.23 &  32 &   0.25 &  0.04 &  0.06 & SYN & A \\
B514-MCGC4 & 14.0 &  1.0 &  -1.74 &  72 &   0.17 &  0.04 &  0.04 &  EW& A \\
  G327-MVI & 11.5 &  1.5 &  -1.65 &  36 &   0.25 &  0.06 &  0.07 &  EW &A \\
      G002 & 11.5 &  1.5 &  -1.63 &  64 &   0.16 &  0.04 &  0.04 &  EW &S \\
 MCGC5-H10 & 12.5 &  2.5 &  -1.45 &  57 &   0.10 &  0.02 &  0.02 & EW&   A \\
      MGC1 & 11.5 &  1.5 &  -1.56 &  24 &   0.09 &  0.02 &  0.03 &  EW &A
      \\
\hline
\cutinhead{GCs Analyzed in \citetalias{m31paper}}
\\
 B045-G108 & 12.5 &  2.5 &  -0.94 &  49 &   0.22 &  0.03 &  0.04 & EW&   A \\

 B358-G219 & 12.5 &  2.5 &  -2.21 &  47 &   0.21 &  0.03 &  0.04 &EW &   A \\
 B381-G315 & 12.5 &  2.5 &  -1.17 &  61 &   0.17 &  0.03 &  0.04 &EW &   A \\
 B386-G322 & 11.0 &  4.0 &  -1.14 &  35 &   0.16 &  0.03 &  0.04 &EW &
 A \\
 B405-G351 & 12.5 &  2.5 &  -1.33 &  42 &   0.26 &  0.03 &  0.05 & EW&   A \\

\enddata

\tablecomments{Results from \citetalias{m31paper} are reproduced here
  for completeness. 1. Number of Fe I lines measured. 2. Statistical
  error of the mean Fe I abundance. 3. Dependence of the final
  abundance on the age of the CMD. 4. Final total uncertainty in
  abundance, $\sigma_{ {\rm T}}= ((\sigma_{ {\rm N}} / \sqrt{{\rm N} -1}
  )^{2} + \sigma_{\rm {Age}}^{2})^(1/2)$. 5. Denotes whether the final
analysis was performed with GETJOB (EW) or line synthesis
(SYN). 6.  Denotes whether solar-scaled (S) or alpha-enhanced (A)
isochrones and stellar atmospheres were used in the final analysis.}

\end{deluxetable*}

\begin{figure}
\centering

\includegraphics[scale=0.5]{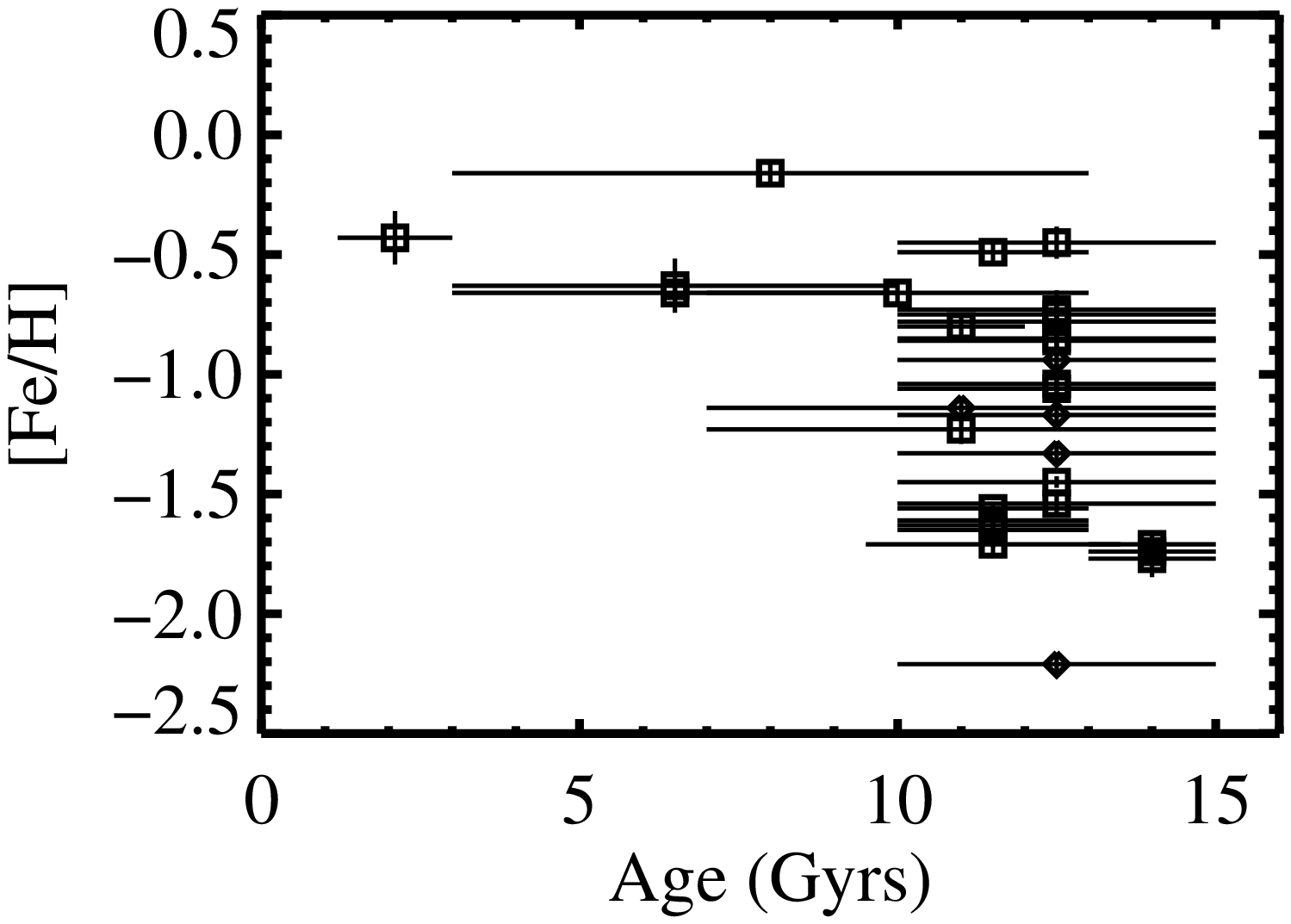}

\caption{ The age-metallicity relationship for the M31 GC sample.  GCs from \citetalias{m31paper} are shown as diamonds. All GCs, with the exception of one, have solutions consistent with ages of $\geq$ 10 Gyr. }
\label{fig:fe-age} 
\end{figure}

\subsection{Age-Metallicity Relationship}
\label{sec:age-met}

 In Figure \ref{fig:fe-age} we show the age-metallicity relationship for our current sample of M31 GCs.  With the exception of B029, all of the GCs are consistent with having ages at least as old as 10 Gyr.  A handful of the higher metallicity GCs (B110, B193, B384) are  consistent with either having ages between 5 and 10 Gyr   or having a strong contribution to the IL from hot blue HB stars.  However,  the colors for these GCs  suggest that it is more likely that these GCs are old with blue HBs \citep{kang12}.    Finally, we find an intermediate age for one GC (B029) as mentioned in \textsection \ref{sec:synthonly}, which we discuss in more detail in the following section.

\subsubsection{B029: Evidence for an Intermediate Age }
\label{sec:b029}
B029 is the only GC in our current sample for which we find an intermediate age; in this work we define intermediate age as having formed between 1 and 5 Gyrs ago. As discussed in \textsection \ref{sec:synthonly}, to check this result, we have performed the Fe I analysis with both EWs and synthesis and we find consistency between the two methods.  In addition, we have performed experiments with adding  very hot, completely blue horizontal branches to the synthetic CMDs.  In performing these tests, we replaced all of the red horizontal branch (HB) stars with  blue HB stars, while conserving the total V flux on the HB.  We can find no ``extreme" HB cases that give a more self consistent Fe I solution across all of the diagnostics than the one obtained with an intermediate age. The diagnostic patterns we find for B029 are similar to what we found for  the $\sim$2 Gyr LMC clusters we analyzed in  \citetalias{paper3}, where we demonstrated that the Fe line diagnostics are successful in identifying clusters that have  ages $<$5 Gyr.

\begin{figure*}
\centering

\includegraphics[scale=0.20,angle=90]{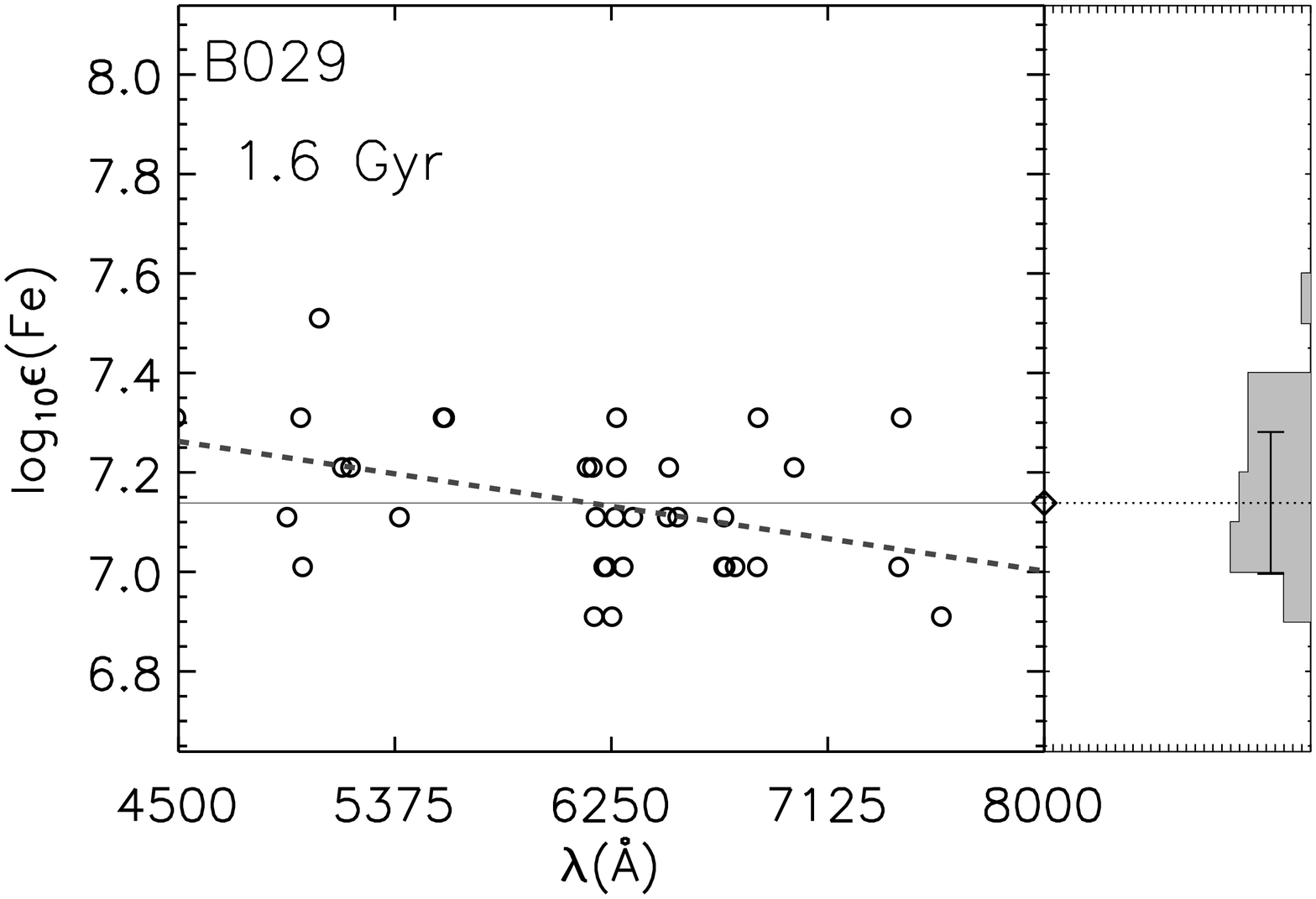}
\includegraphics[scale=0.20,angle=90]{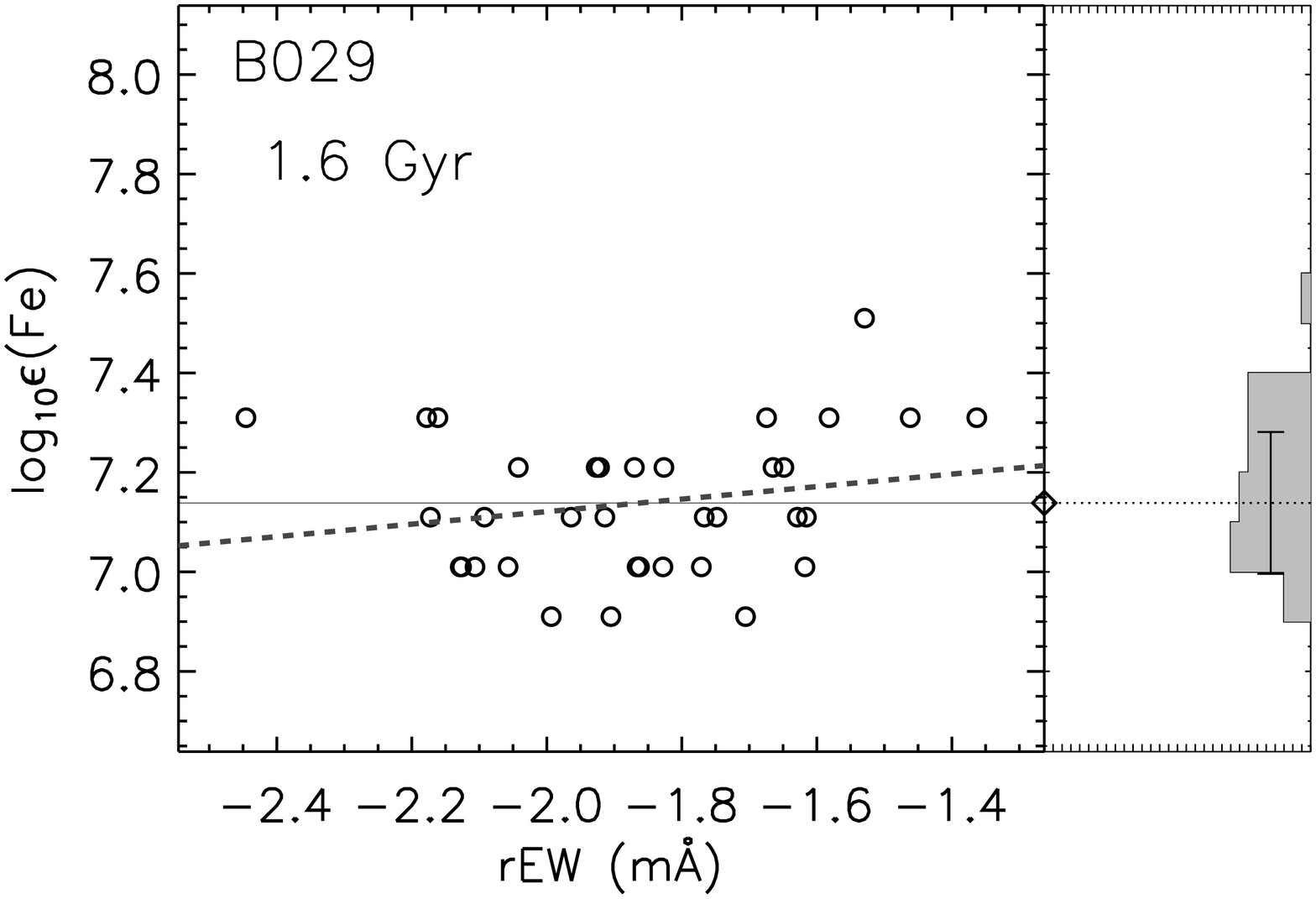}
\includegraphics[scale=0.20,angle=90]{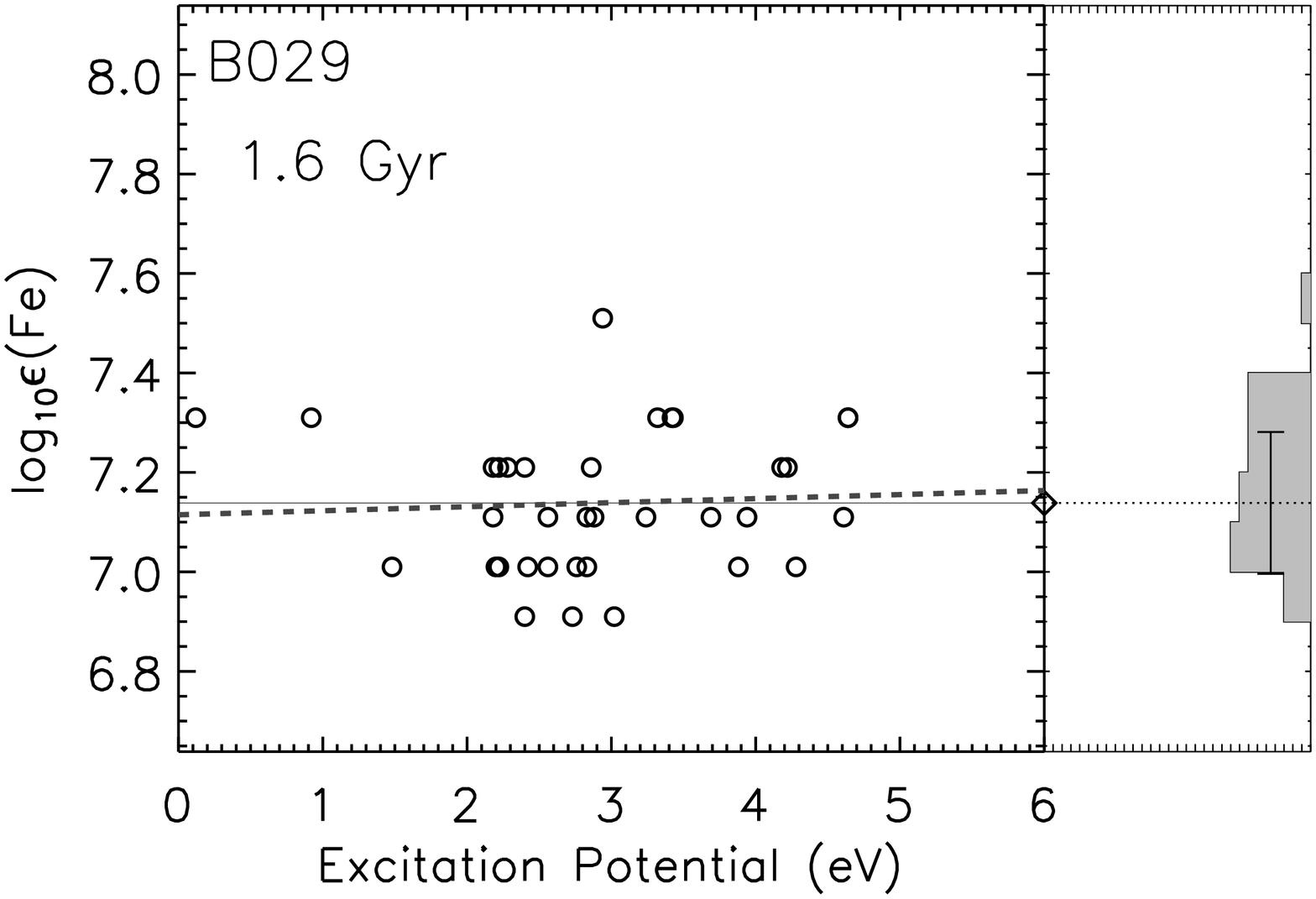}

\includegraphics[scale=0.20,angle=90]{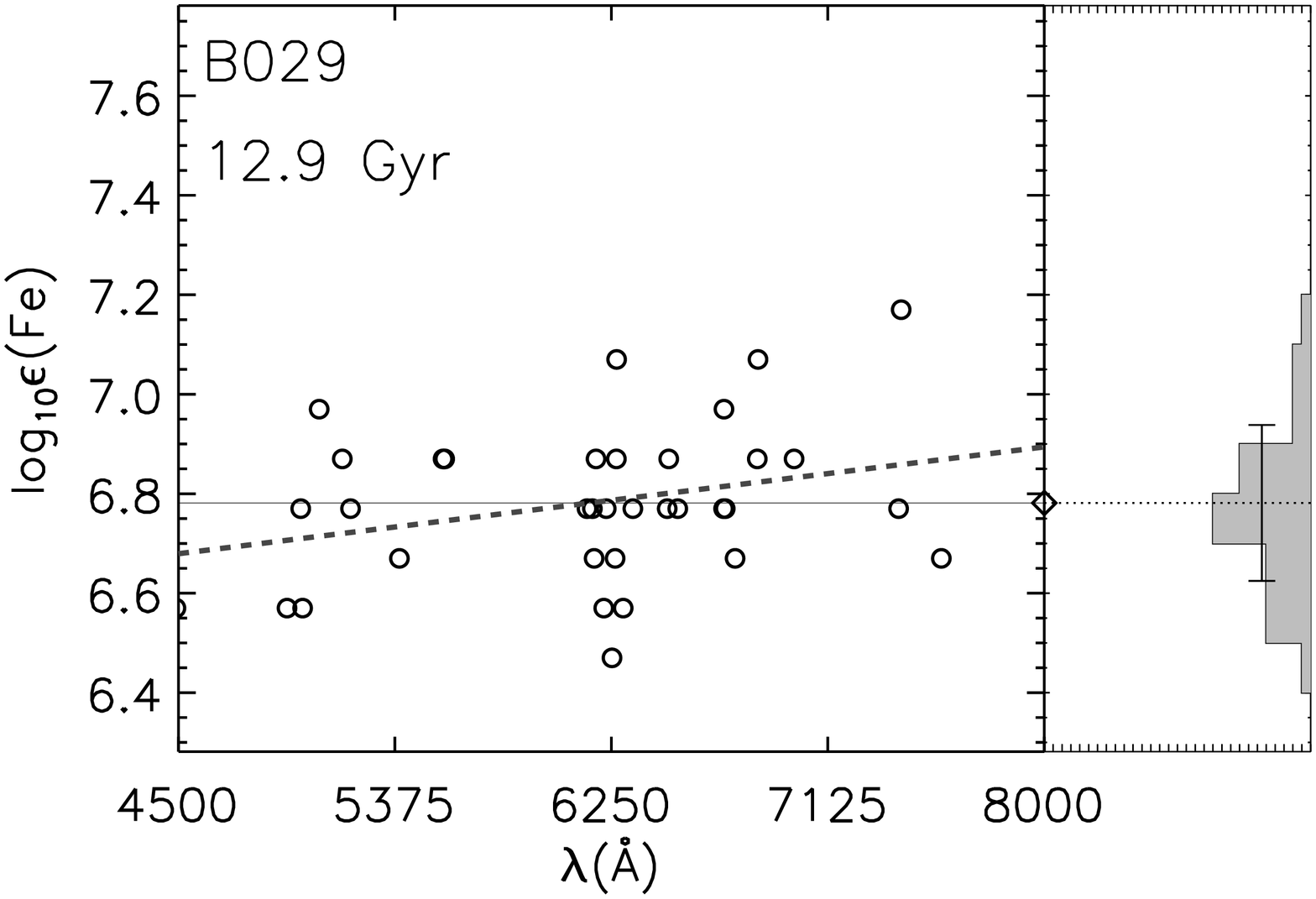}
\includegraphics[scale=0.20,angle=90]{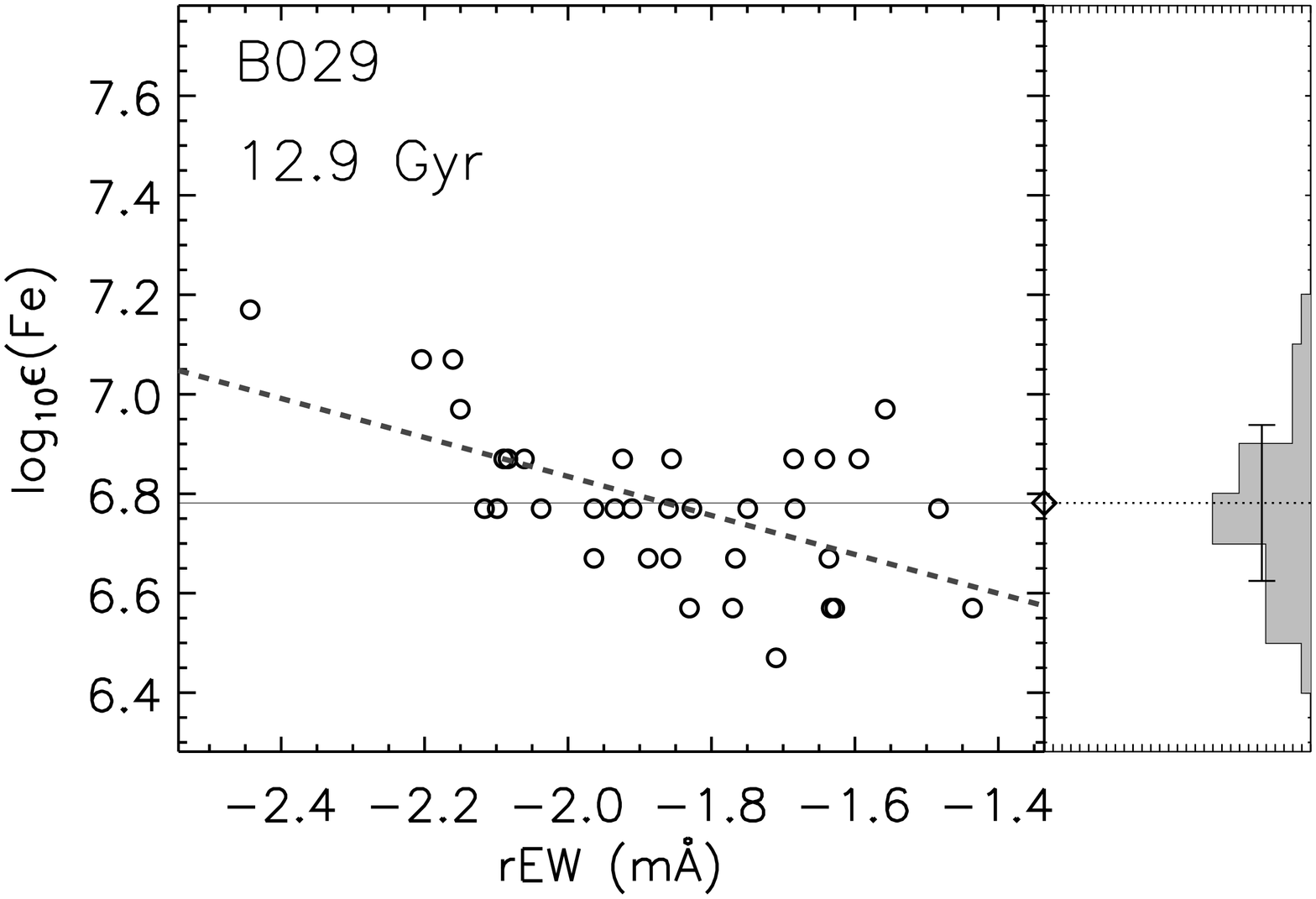}
\includegraphics[scale=0.20,angle=90]{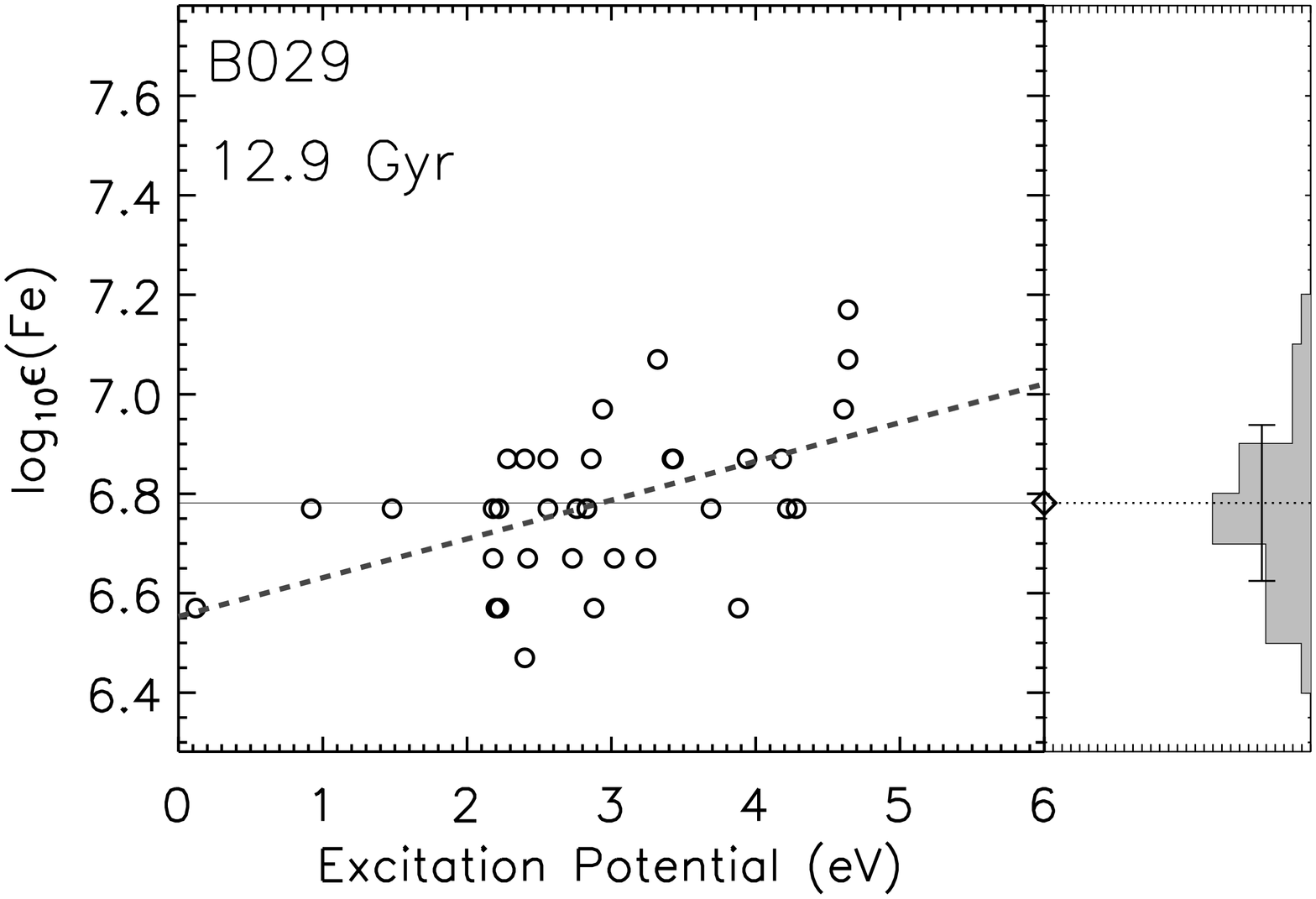}

\includegraphics[scale=0.20,angle=90]{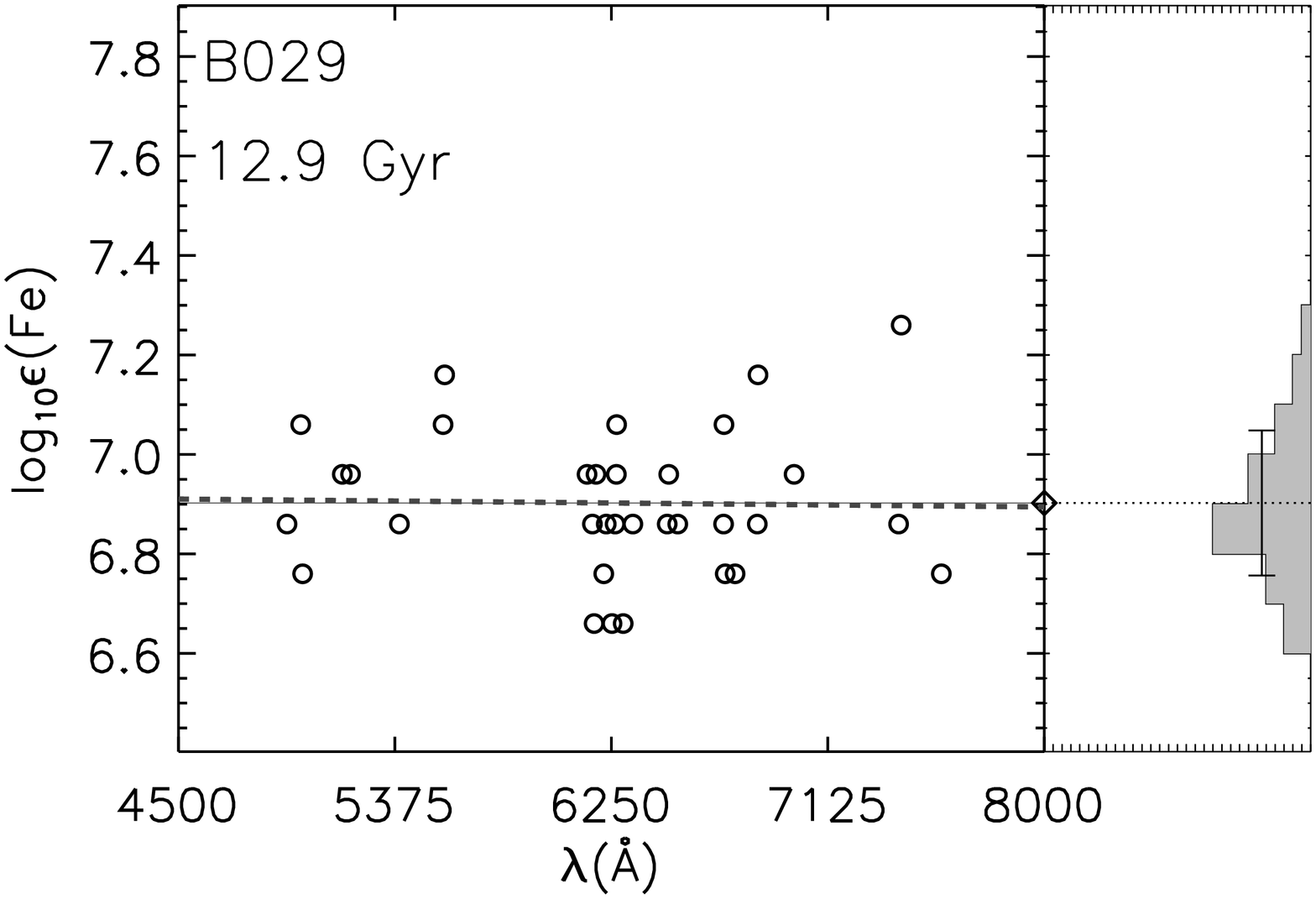}
\includegraphics[scale=0.20,angle=90]{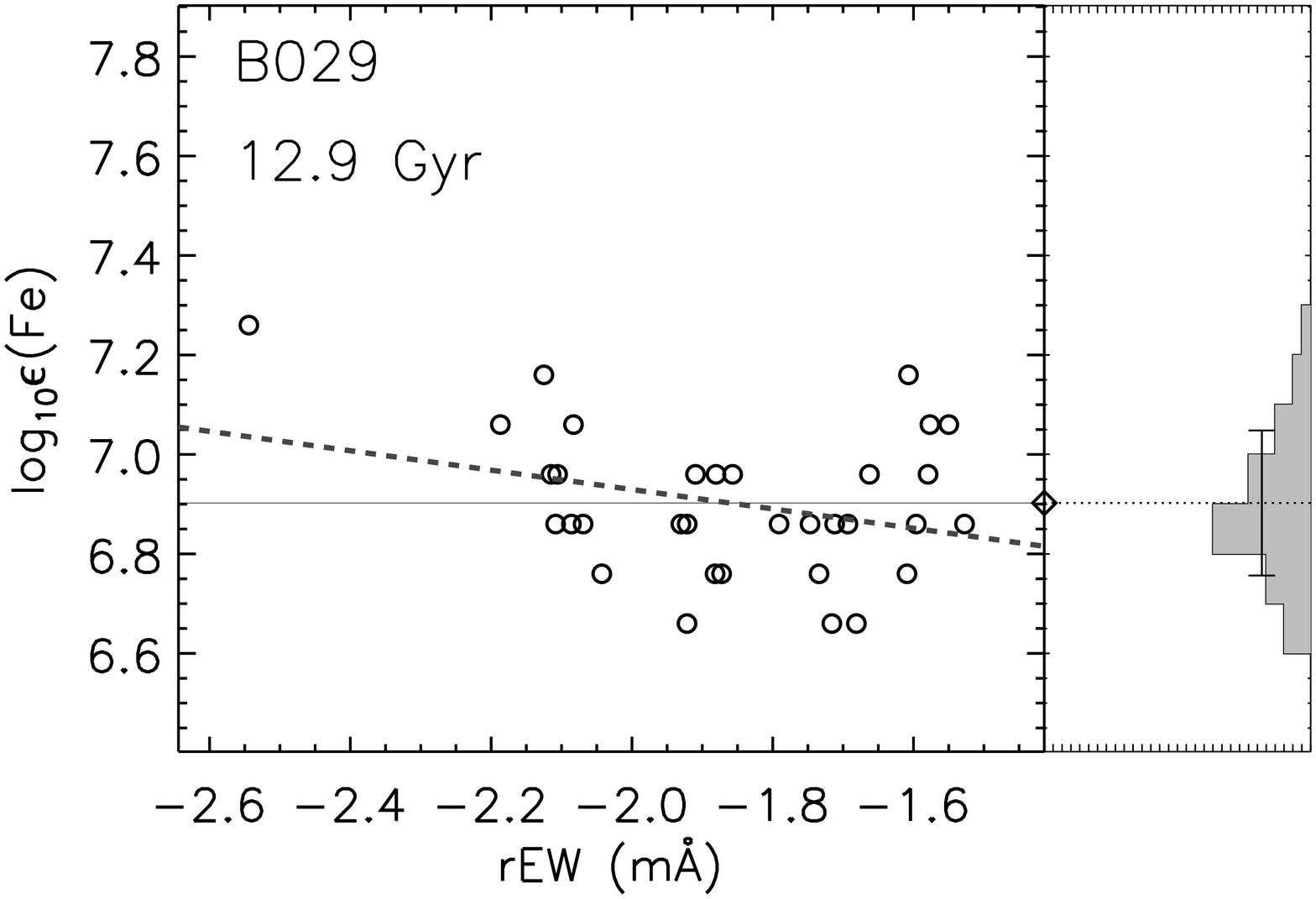}
\includegraphics[scale=0.20,angle=90]{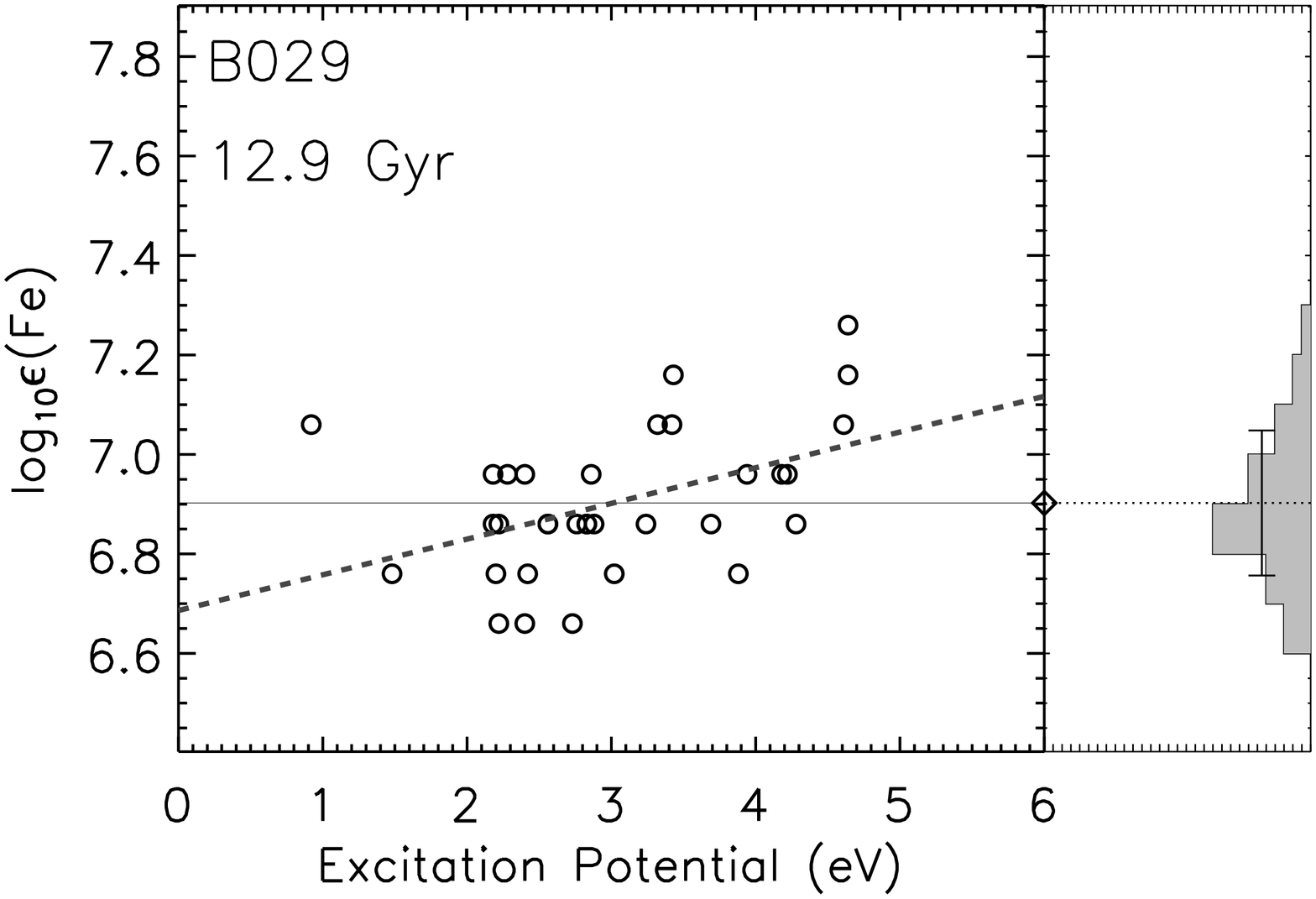}

\caption{ Fe line diagnostics (behavior with wavelength, reduced EW (rEW$\equiv$log$_{10}$(EW/$\lambda$)), and EP from left to right) for B029. Symbols and lines are the same as in Figure \ref{fig:b384}.  
The top panels show the line synthesis solutions for an age of  1.6 Gyr. The middle panels show the  solutions for an age of  12.9  Gyr. The bottom panels show the  solutions for an age of 12.9 Gyr and the ad hoc addition of an extreme blue HB. The best overall solutions are the top panels, which have an age of 1.6 Gyr.
}

\label{fig:b029} 
\end{figure*}

To illustrate the difference evident in our analysis, in Figure \ref{fig:b029} we show the analysis diagnostics for an age of 1.6 Gyr, an age of $\sim$13 Gyr, and an age of $\sim$13 Gyr with an extreme, hot HB.  There are several points to take away from Figure \ref{fig:b029}. First, the Fe solutions in the top panels, for an age of 1.6 Gyr, clearly have a smaller statistical scatter and smaller trends with wavelength, EW, and EP than the solutions in the middle panels for an age of 13 Gyrs.    We note that it is possible to eliminate the trend in [Fe/H] with wavelength by using a 13 Gyr CMD with a completely blue HB, as shown in the bottom panels.   However,  the statistical scatter and behavior of [Fe/H] with EW and EP in the extreme blue HB case are still not as good as in the 1.6 Gyr case.  Prior experience with our training set  GCs suggest low statistical scatter is a critical characteristic of a good solution \citep{mb08,scottphd,paper3,mwpaper}. Therefore, we conclude that overall the best solutions for B029 are for ages of $\sim$2 Gyr, and a relatively high overall metallicity of [Fe/H]=$-0.43$.

\begin{figure}
\centering
\includegraphics[scale=0.5]{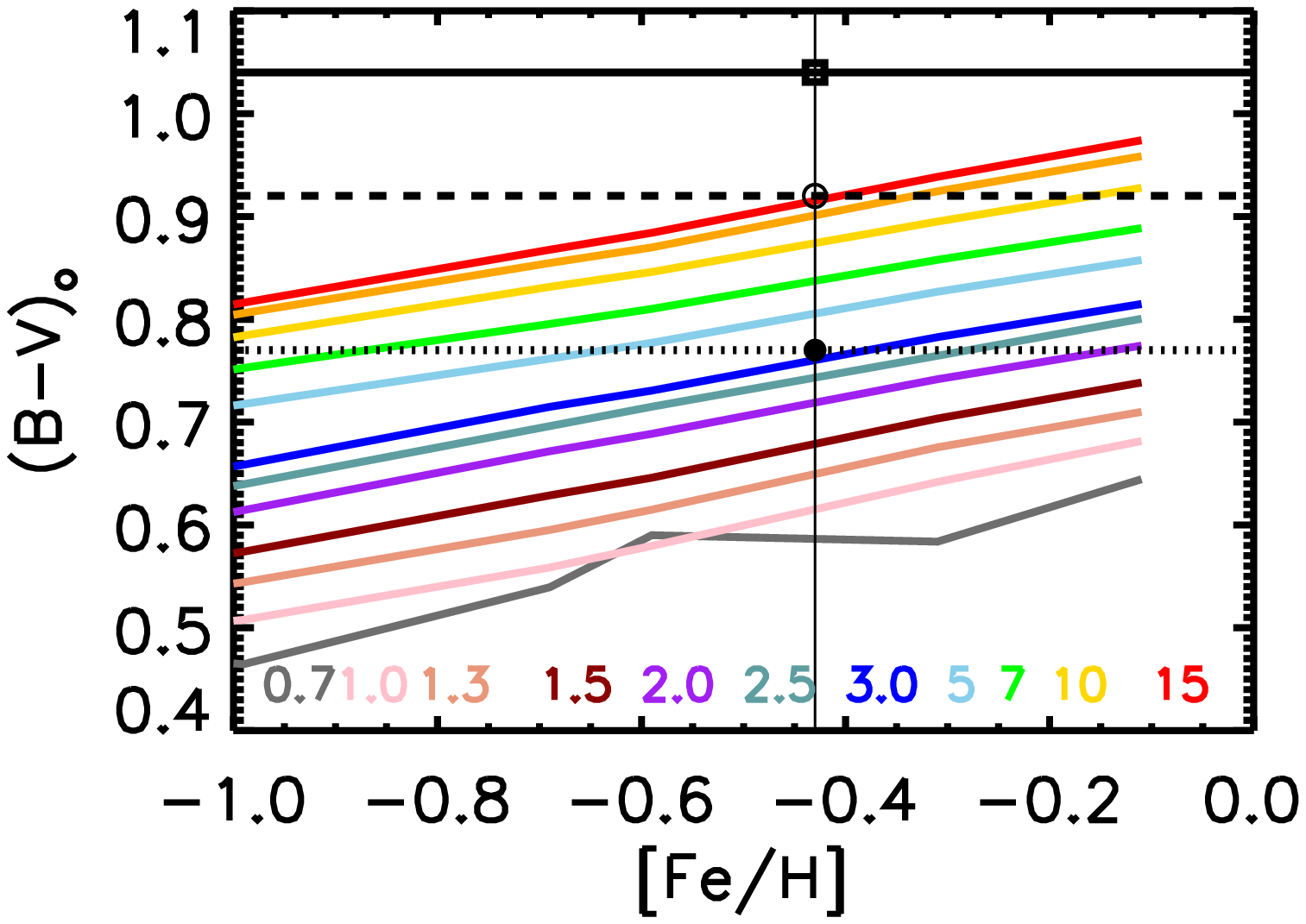}
\caption{ Predicted integrated color as a function of [Fe/H] and age, as calculated from the Teramo isochrones we use in our analysis. To guide the eye, a vertical solid line marks the [Fe/H] solution we find for B029. The observed B-V color of B029 from the Revised Bologna Catalog \citep{bolognacat} is shown as a horizontal solid line.  The dashed line shows the reddening corrected color of B029, using the E(B-V)=0.12 from \cite{fan08}.  The dotted line shows the corrected color using the E(B-V)=0.27 from \cite{caldwell11}.  The latter is consistent with the age range we derive.
 }
\label{fig:b029color} 
\end{figure}

The only definitive way to confirm an intermediate age for B029 is to obtain a deep HST CMD that reaches the turnoff, which unfortunately is not available, and would require several days of integration time \citep[e.g.][]{brown2004}. There are other indirect, non-spectroscopic strategies for arguing for or against intermediate ages for  extragalactic GCs when CMDs that reach the turnoff are unavailable.  
We consider three common consistency checks in this section, but note that age dating unresolved GCs has a long, extensive  and sometimes conflicting history.

First, the simplest consistency check is to consider integrated optical colors.    Unfortunately,  B029 is significantly reddened, likely because it is  projected onto the outer disk of M31.   The  E(B-V) in the literature vary quite a bit;  \cite{fan08} find E(B-V)=0.12, while \cite{caldwell11} find E(B-V)=0.27.  The differences in these E(B-V) values alone make conclusions drawn from colors ambiguous.  As an example, in Figure \ref{fig:b029color}, we show the predicted (B-V)$_{{\rm o}}$ from the Teramo isochrones used in our analysis compared to the observed B-V for B029.   Figure \ref{fig:b029color} shows that the lower E(B-V) value of \cite{fan08} is consistent with an old age for B029, but that the higher  of \cite{caldwell11} is consistent with the younger age we determine (however we note that \cite{caldwell11} find an old age for B029 from low resolution spectroscopy).  We conclude that  an intermediate age for B029 cannot be ruled out by the  observed optical colors.

  Second, mass-to-light (M/L) ratios are often  investigated, under the assumption that intermediate age GCs will have larger mass-to-light ratios than old GCs, because the stellar populations are less evolved.    \cite{strader11} calculated M/L ratios for a large sample of M31 GCs, and for B029 found M/L=1.5 in V and M/L =0.5 in K, which is comparable to the rest of their old,  high [Fe/H]  GCs in M31.  However,   \cite{strader11} also find that the behavior of the M/L ratios in GCs does not follow that expected from stellar population models.  Similarly, \cite{zaritsky1} and \cite{zaritsky2} also discuss how M/L ratios of well-studied, resolved GCs  do not consistently match  current theoretical expectations.  Therefore, it's hard  to draw definitive conclusions about GC properties using M/L ratios until their behavior  is better understood theoretically.
  
  Third, near-ultraviolet (NUV) and  far-ultraviolet (FUV) colors are   used in order to distinguish between  GCs with young ages and GCs that have very hot HB stars.  An old GC with significant numbers of hot (T$_{eff}>$10,000 K) HB stars will be detectable in the FUV, with progressively older ages producing stronger FUV flux \citep[eg.][]{lee02}.  Very young GCs ($\leq$ 0.5 Gyr) will also have significant NUV and FUV flux due to hot main sequence stars, but these GCs can be identified by  significantly bluer optical colors than what is observed for B029 \citep{reygalex}.  True intermediate age GCs are faint or undetectable in the FUV  \citep{leeworthey05}.  \cite{reygalex}, with an update by \cite{kang12}, present the GALEX FUV and NUV catalog of M31 GCs.  B029 is not detected in either the NUV or FUV, which is consistent with having an intermediate age.   However, the  high reddening remains a possible explanation for the lack of blue flux, so even this evidence is not  particularly definitive.

  In summary, we do not find compelling independent evidence to rule out an intermediate age for B029.  
  Because Balmer lines are temperature sensitive, they have historically been the primary spectroscopic age indicator.
However,  ages derived from Balmer line strengths are notoriously problematic because the same line depths can be achieved for true  intermediate age clusters and old clusters that have significant numbers of blue HB stars \citep[e.g.][]{1994ApJS...95..107W, 2004ApJ...608L..33S, 2011MNRAS.412.2445P}.  A strength of our analysis compared to other spectroscopic techniques is that we  use only unsaturated  Fe lines to constrain the ages of the clusters so that we can avoid the degeneracies between hot main sequence and hot HB stars.  Instead,  unsaturated Fe lines are  principally sensitive to the temperature of the red giant branch (RGB), AGB,  and turnoff stars \citepalias[see discussions in][]{mb08,m31paper,paper3}. Therefore, this is the first evidence for the existence of an intermediate age  cluster in M31 that does not solely rely on integrated colors or Balmer line absorption from line indexes \citep{barmby00,2003AJ....125..727J,beasley04,puzia05,beasley05}. Additional high resolution abundance analyses of GCs that are not detected in GALEX would be interesting to further investigate the intermediate age population of GCs in M31.

\subsection{Abundances from Fe II Lines}
\label{sec:feii}
\begin{figure}
\centering
\includegraphics[scale=0.5]{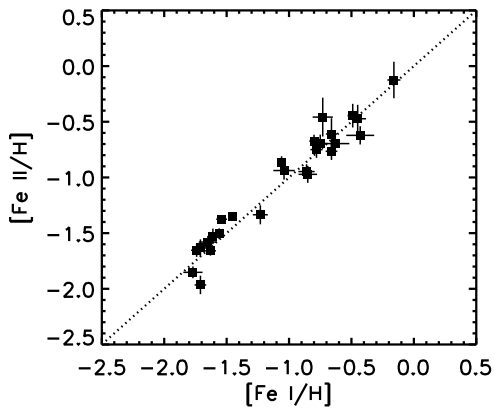}
\caption{ The comparison of the mean abundances measured from Fe II lines and the mean abundances measured from Fe I lines. The dotted line shows 1:1 correspondence. }
\label{fig:feii} 
\end{figure}

In addition to abundance measurements from Fe I lines, which we use to constrain the age and metallicity of the GCs, we are also able to measure abundances from Fe II lines.  We have measured all of the Fe II abundances using spectral synthesis and  $\chi^{2}$-minimization, since there are fewer trusted transitions, and they are often significantly more blended than the Fe I lines.  
For consistency, we have  re-measured the Fe II lines for GCs from \citetalias{m31paper} using spectral synthesis.  The final abundance of each Fe II line, as well as the adopted line parameters are given in Table \ref{tab:stub_ele_linetable}.  The mean [Fe II/H] abundances are listed in Table \ref{tab:nfabund}, along with the statistical error and the number of Fe II lines used in the analysis.  With line synthesis, on average we are able to measure $\sim$10 Fe II lines for each GC, which is an improvement over what we were able to measure in \citetalias{m31paper}, where we averaged 3-4 Fe II lines per GC.   We find that the abundances measured from Fe II lines are  consistent with the abundances measured from Fe I lines, as shown in Figure \ref{fig:feii}.  The Fe II and Fe I abundances all agree to within 2 $\sigma$, with most agreeing within 1 $\sigma$.  The excellent agreement of the Fe II abundances is  independent confirmation of the accuracy of our measurements.

\subsection{Alpha Elements}
\label{sec:alpha}

One goal of this work is to study the alpha element abundance pattern of M31 GCs, as sampled by the elements Ca, Si, and Ti.  These alpha elements are produced and ejected into the interstellar medium (ISM) in large quantities by type II supernovae (SN II), whose progenitors are massive stars that live and die on short timescales on the order of millions of years \citep[e.g.][]{woosley95}.  They build up quickly in active star formation epochs.  The particular utility of alpha elements for learning about star formation histories comes from the comparison of their abundances to elements formed on  different timescales.  The standard comparison is to Fe or Fe-peak elements, which are contributed to the ISM in large quantities on comparatively longer timescales in supernovae type Ia (SN Ia), and in the simplest sense the behavior of the [$\alpha$/Fe] in a galaxy lets us infer the rates and timescales of star formation throughout its history.  Here we aim to compare the [$\alpha$/Fe] abundance pattern in M31 GCs to GCs and stars in the MW in order to constrain the relative strengths and durations of  star formation in M31.

\begin{figure}
\centering
\includegraphics[trim = 10mm 0mm 0mm 20mm, clip,angle=90,scale=0.46]{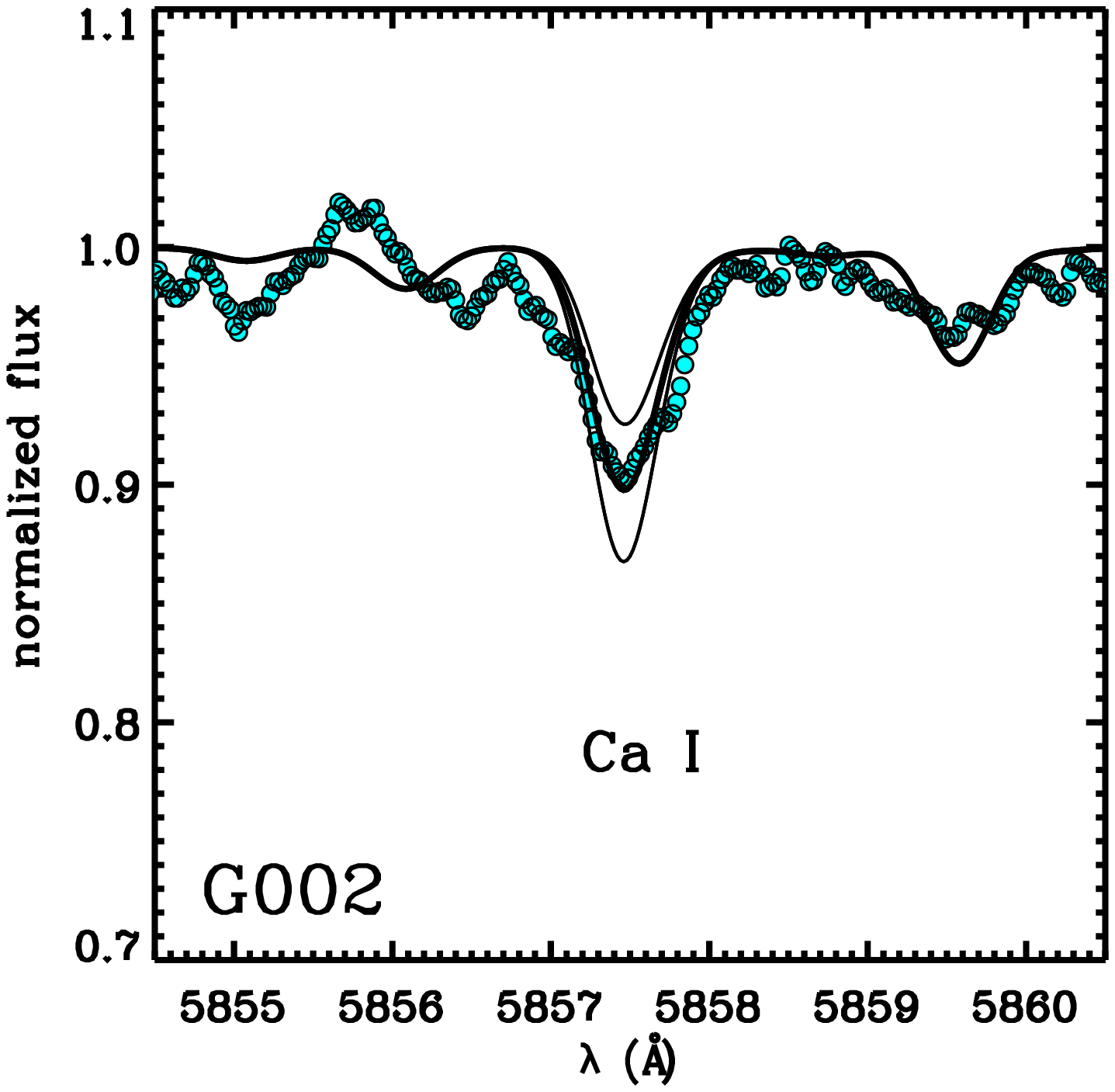}
\includegraphics[trim = 0mm 10mm 0mm 20mm, clip,angle=90,scale=0.46]{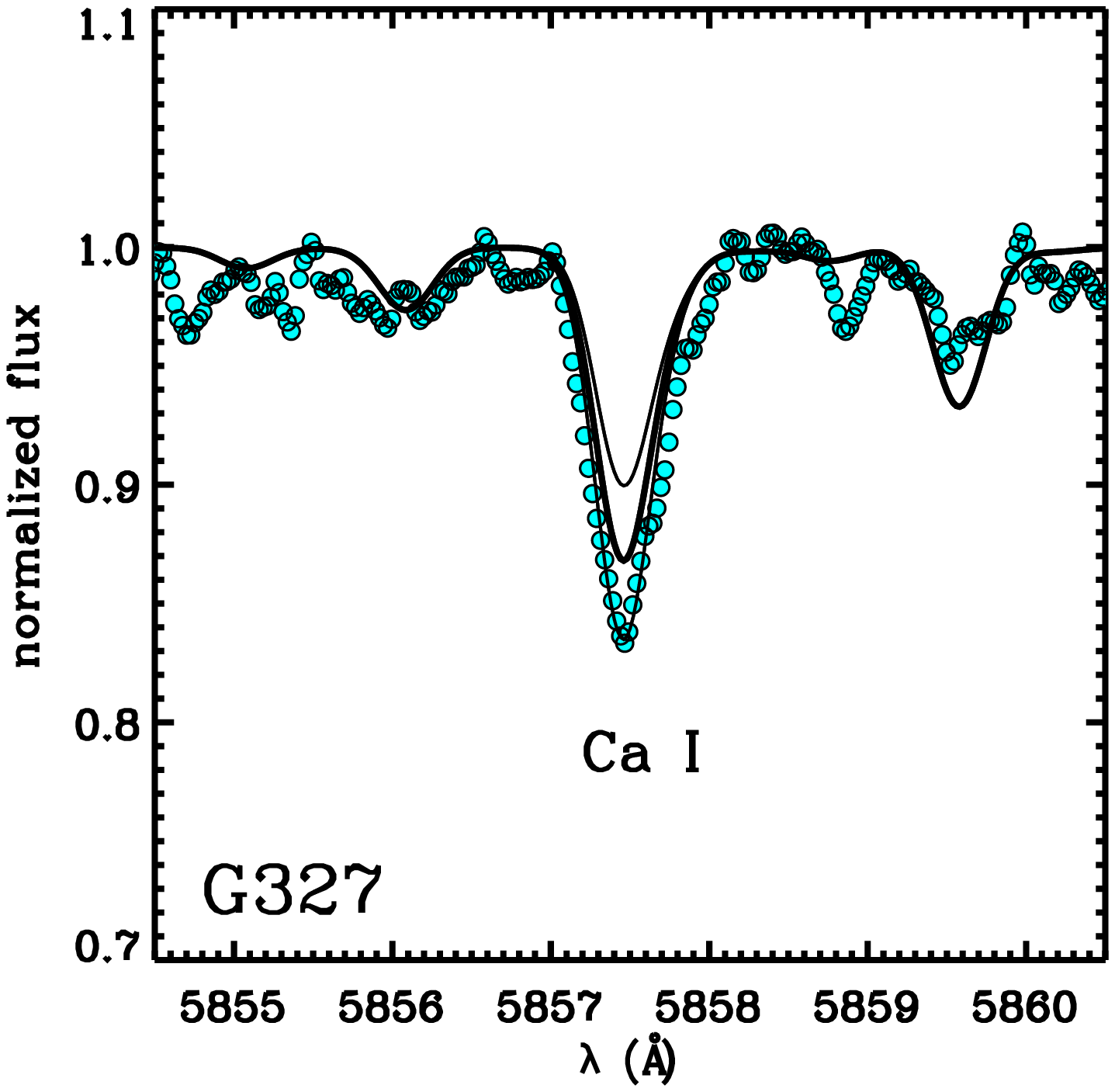}

\caption{ Example Ca I synthesis fits for G002 (top) and G327 (bottom), which have similar metallicities.  Points show the data, which has been smoothed by 5 pixels for presentation. Solid lines correspond to ratios of [Ca/Fe]=$-0.3, +0.0, +0.30$.  The best fitting abundance for G002 is [Ca/Fe]=$+0.0$, while for G327 is is [Ca/Fe]$=+0.3$. }
\label{fig:ca} 
\end{figure}

\begin{figure}
\centering
\includegraphics[trim = 10mm 0mm 0mm 20mm, clip,angle=90,scale=0.46]{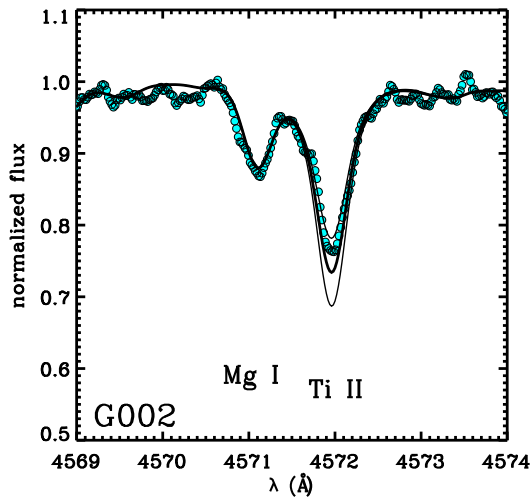}
\includegraphics[trim = 0mm 10mm 0mm 20mm, clip,angle=90,scale=0.46]{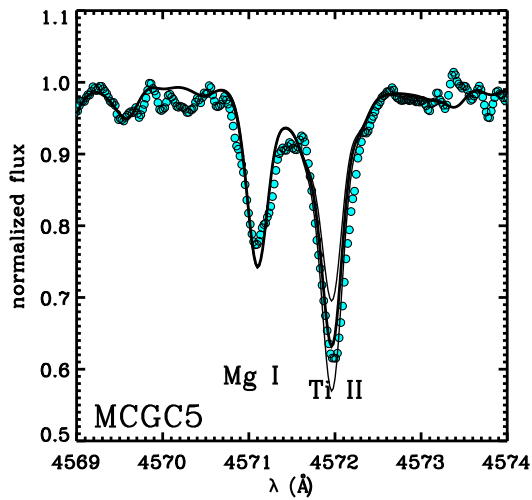}

\caption{ Example Ti II synthesis fits for G002 (top) and MCGC5 (bottom). Points show the data, which has been smoothed by 5 pixels for presentation. Solid lines correspond to ratios of [Ti/Fe]=$-0.5, +0.0, +0.50$, and the mean [Mg/Fe] for each GC ([Mg/Fe]=$-0.5$ for G002 and [Mg/Fe]=$-0.1$ for MCGC5) has been adopted to account for blending with the Mg I feature.   G002 has a lower [Ti/Fe] than MCGC5. }
\label{fig:ti2} 
\end{figure}

We measure abundances for the alpha-elements Ca I, Si I, Ti I, and Ti II.  Note that we also measure Mg I, however, we do not include it here because it is not necessarily mono-metallic in GCs and so is instead  included in  \textsection \ref{sec:light}. 
  All of the alpha element measurements were made using IL spectral synthesis and  $\chi^{2}$-minimization, which enables us to recover more individual lines for each element and with higher precision than the EW analysis we employed in \citetalias{m31paper}.     The  abundances measured for each line  are presented in Table \ref{tab:stub_ele_linetable}.  The final  averaged measurements for the alpha elements for the full M31 sample are listed in Table \ref{tab:tnfabund2}.   In Figure \ref{fig:ca} and Figure \ref{fig:ti2} we show examples of the quality of the synthesis fits to Ca I and Ti II for three GCs.

The uncertainties in the abundances of the non-Fe elements are calculated in the same way as for Fe I when multiple lines are available, which gives an indication of the systematic errors between lines, as well as the uncertainty due to the age of the CMD. A special case occurs when we have only a single measurement of a given species, and therefore no indication of the line-to-line scatter.  As an example, we have only one measurement of Si I for B012, and one measurement of Ti I for B088.  In these cases we have adopted a typical line-to-line scatter of 0.1 dex as an estimate of the systematic error in the abundance, which is then added in quadrature with the systematic age uncertainty  to obtain a total uncertainty.  
 
 The lines we measure in the GC IL spectra are all lines that are commonly measured in individual stars.  Here we include some additional notes on their behavior in the IL spectra.  Our Ca I line list includes  17 lines;  the majority of which are strong and isolated enough for clean measurement in most GCs.    The 6169 \rAA line is a blend of two Ca I features and is only included when the abundance can be cleanly measured.  The 6455 \AA, 6464 \AA, and 6508 \rAA lines  are weak and usually only measured in the more metal-rich GCs that have low velocity dispersions.  
 
 The Si I line list includes 11 lines, which are all weaker features than most of the Ca I lines.    The 5793 \AA, 5948 \AA, 6721 \AA, and 7932 \rAA are the weakest and are only measured in a handful of GCs that are metal rich and have low velocity dispersions.  The 7415 \rAA and 7423 \rAA are blended with Ni I lines, and were only measured when the velocity dispersion was low enough that the line profiles were sufficiently separated.   
 
 Our Ti I line list includes 22 lines,  most of which are blended with other features to some extent.  We include measurements of the blended lines when the line profiles are sufficiently separated that we are confident with the abundance measurement. Figure \ref{fig:ti2} shows a typical case where the line is blended with a nearby feature but the abundance is well determined.  The Ti I lines that are most isolated are 4991 \AA, 5866 \AA, and 6743 \AA.  The redder Ti I lines ($\lambda > 5200$ \AA) are generally weaker, and so are not usually cleanly measured in more metal-poor GCs.  The 5173 \rAA and 5192 \rAA are in the Mg b region, and are usually measured in more metal-poor GCs where the continuum is well determined. Likewise, the 6554 \rAA and 6556 \rAA Ti I lines are near H alpha and are only included when the wings of H alpha are not significant, so that the continuum is well determined. 
 
 17 lines are included in the Ti II line list;  most are somewhat blended, similar to Ti I.  The most isolated Ti II lines are 4501 \AA, 4589 \AA, and 5381 \AA.  The 5185 \rAA line is in the Mg b region, and was only included when the continuum was well determined.  The 4865 \rAA line is near H beta and was only included when the wings of H beta did not affect the continuum near the line.

\begin{figure}
\centering
\includegraphics[scale=0.50]{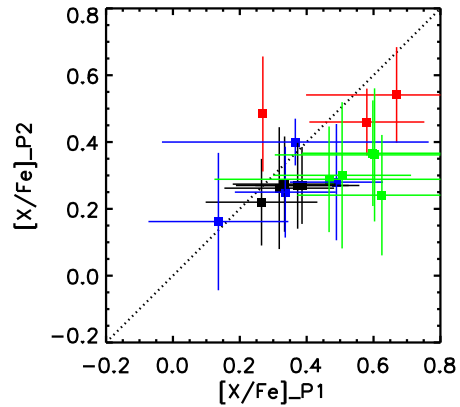}
\caption{A comparison of alpha element ratios measured using EWs in \citetalias{m31paper} ([X/Fe]\_P1) to those measured using spectral synthesis in this work ([X/Fe]\_P2).  Black, red, blue and green show Ca I, Si I, Ti I and Ti II ratios, respectively. The dotted line shows 1:1 correspondence.}
\label{fig:comparealphas} 
\end{figure}

 For consistency, we have re-measured the abundances of the alpha elements with IL synthesis for the GCs  presented in \citetalias{m31paper}.  In Figure \ref{fig:comparealphas}, we compare the results for the EW analysis in \citetalias{m31paper} to the results measured using line synthesis in this work. In general the results are in good agreement,  and we find that when using synthesis the number of lines recoverable is generally larger, and the statistical error is reduced.  Only Ti II abundances appear to be significantly different, which may be because most of the Ti II lines suffer from line blending.   This highlights the better precision obtained using line synthesis, which is especially important for elements that have only a handful of analyzable spectral features, such as Si I.  We also note that we were able to add  measurements for Si I for B405 and  Ti I for B358, which we were not able to measure with EWs in \citetalias{m31paper}.

The results in Table \ref{tab:nfabund} and Table \ref{tab:tnfabund2} show that the total uncertainty in the abundance ratios, which is determined by adding the  $\sigma_{{\rm X}}$ and $\sigma_{{\rm A,X}}$ in quadrature, is usually $<$0.1 dex, and the age uncertainty, $\sigma_{{\rm A,X}}$, is usually less than or comparable to the statistical uncertainty, $\sigma_{{\rm X}}$.   However, in some cases, the $\sigma_{{\rm A,X}}$ dominates the abundance ratio  uncertainty.   It is expected that this would be the case for GCs with larger age uncertainties, like B193 and B457, which have uncertainties of 5 and 4 Gyr, respectively. However, these GCs also have the lowest SNR data of the current sample.  If the age uncertainty was solely due to a stellar population mismatch in the true and theoretical CMDs, this uncertainty would remain even if better quality, higher SNR data were obtained.   However, since we determine the age using the behavior of the abundances measured from different Fe I lines, it is possible that reducing the line to line scatter with better quality data could also have an impact on the age determination.

\begin{figure}
\centering
\includegraphics[scale=0.50]{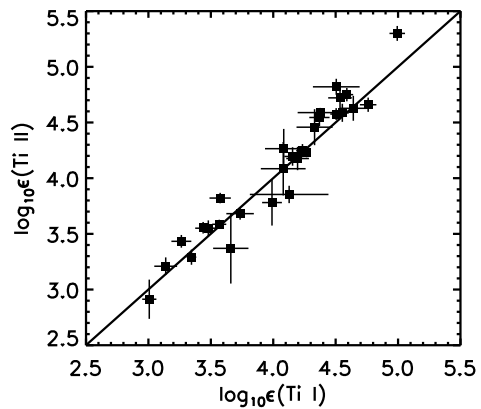}
\caption{  The same as Figure \ref{fig:feii} for Ti I and Ti II abundances. }
\label{fig:ti_comparison} 
\end{figure}

Because we have a large sample of both Ti I and Ti II abundances, it is interesting to compare the results for neutral and ionized species, as we did for Fe I and Fe II in \textsection \ref{sec:feii}.  This comparison is  shown for the absolute Ti I and Ti II abundances  in Figure \ref{fig:ti_comparison}.  We find that the abundances we obtain for  Ti I are consistent with those of  Ti II, just as the  Fe I abundances were consistent with the Fe II abundances.

\begin{figure*}
\centering
\includegraphics[scale=0.45]{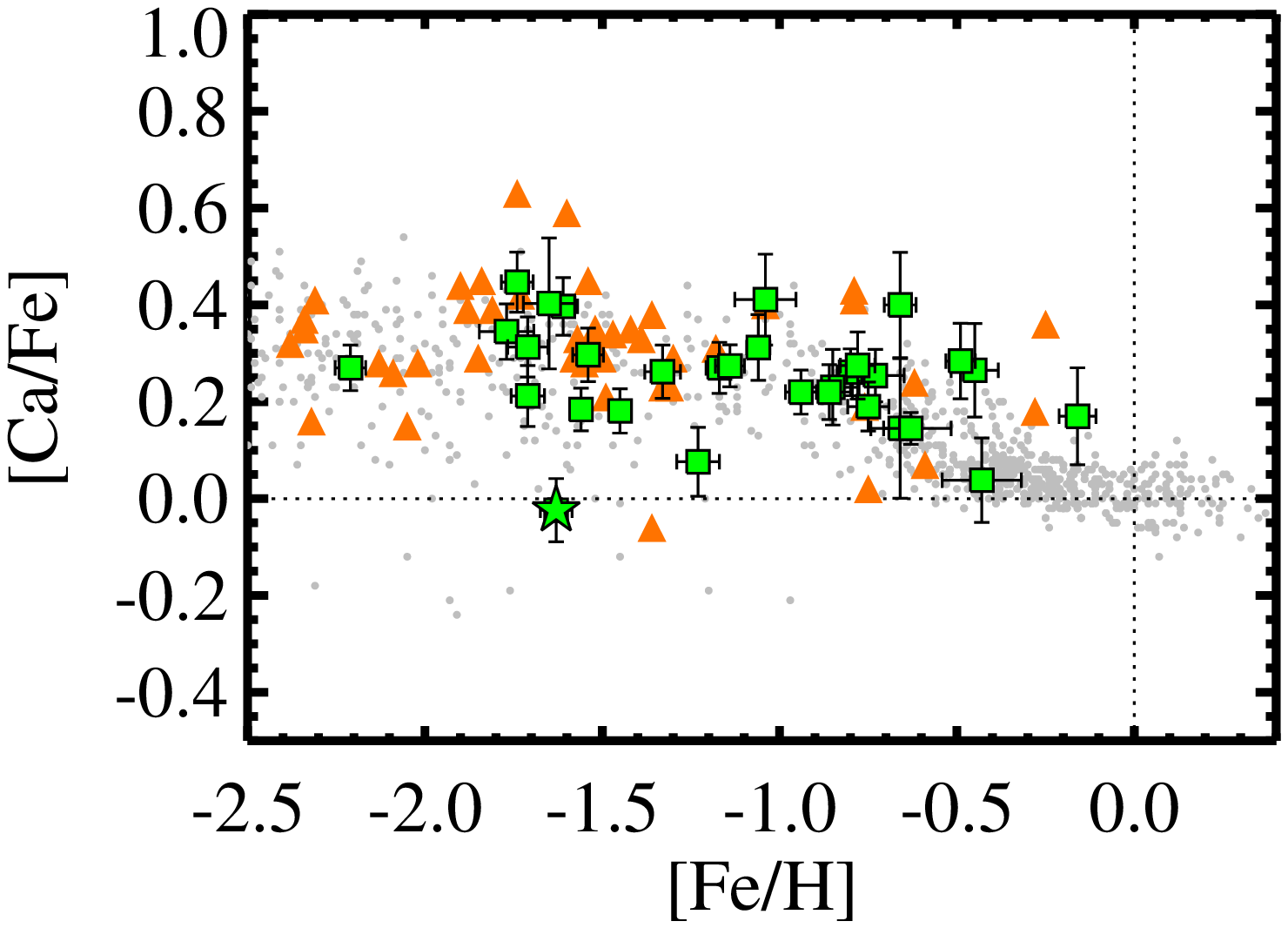}
\includegraphics[scale=0.45]{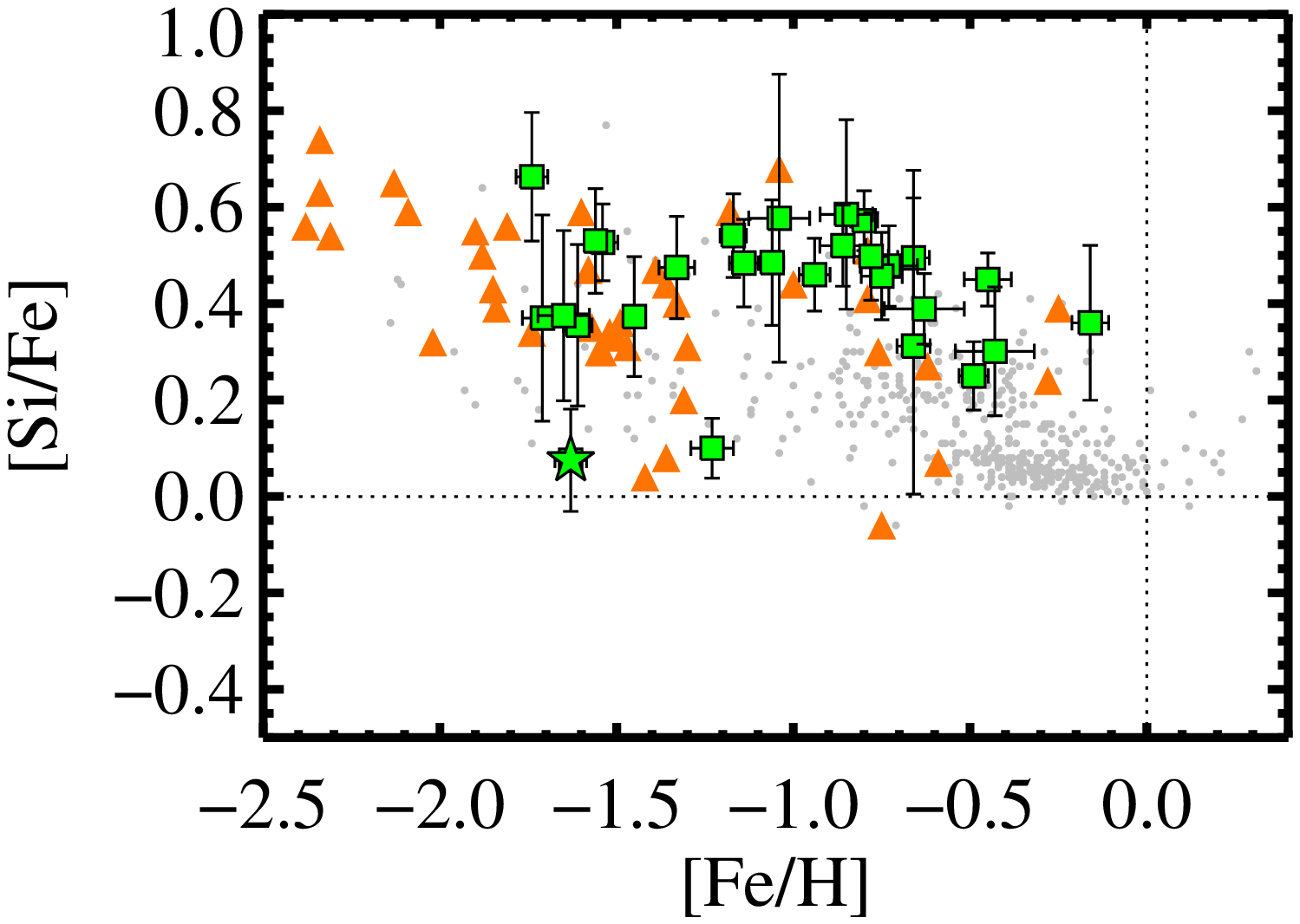}
\includegraphics[scale=0.45]{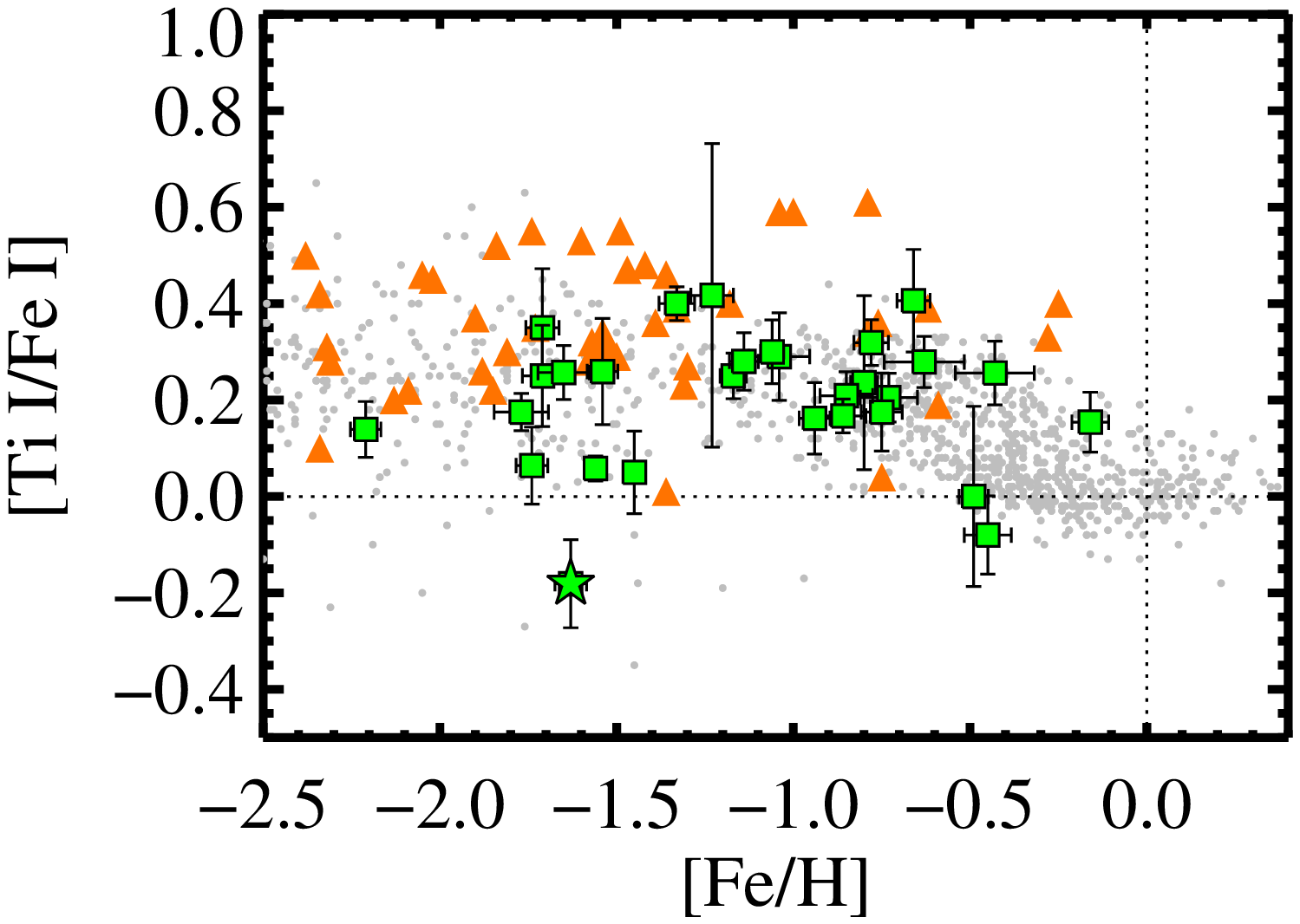}
\includegraphics[scale=0.45]{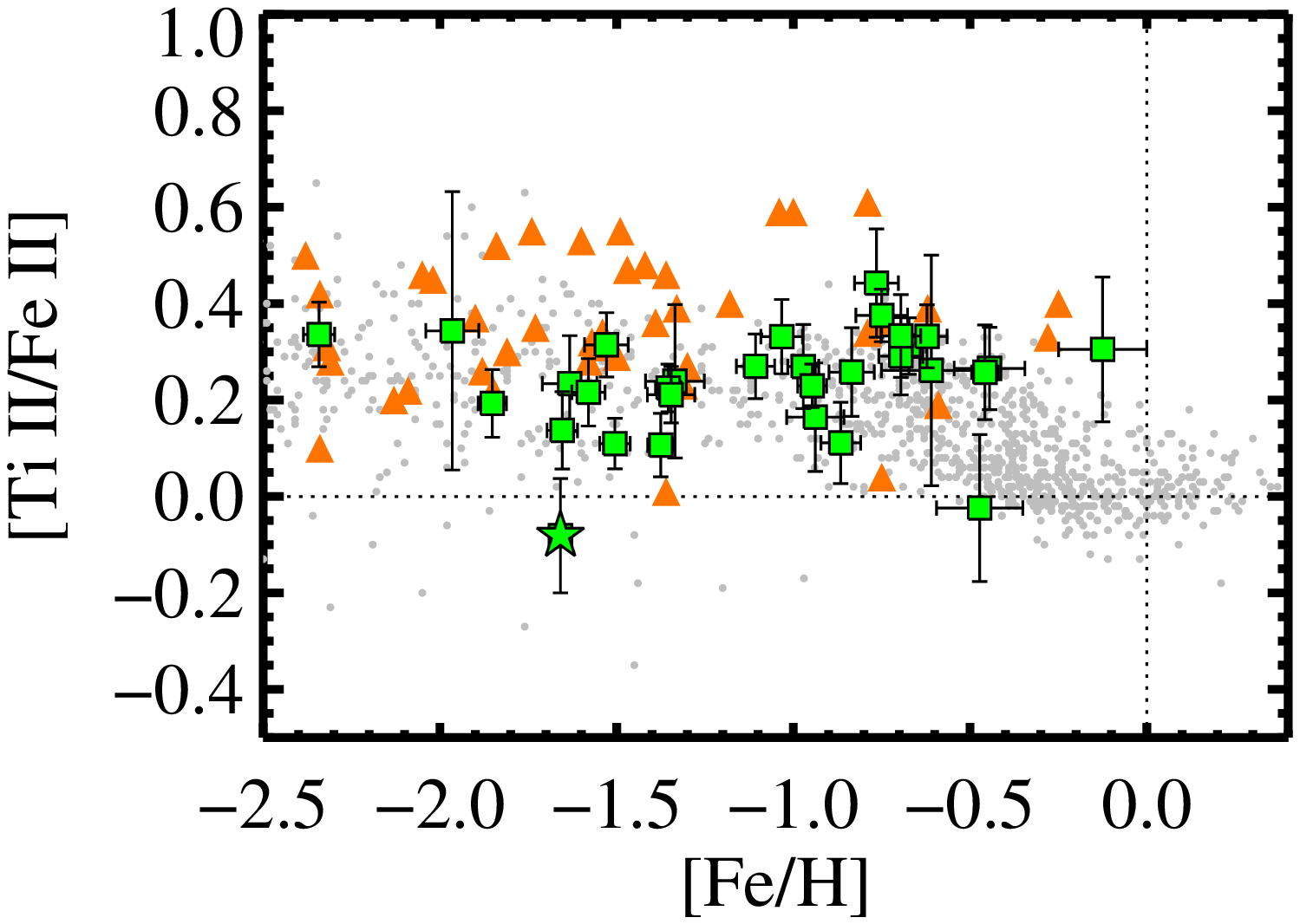}

\caption{ Abundance ratios for the alpha elements Ca I, Si I, Ti I,  and Ti II. Ratios of neutral species are taken with respect to Fe I and ionized species with respect to Fe II.   Green symbols show the M31 GC IL abundances. Green stars show G002; the GC where solar-scaled atmospheres and isochrones were applied.  Stellar MW GC abundances from the compilation of \cite{pritzl05} are shown as orange triangles, and MW field star abundances from \cite{reddy06}, \cite{fulbright07}, \cite{barklem05}, and the compilation of \cite{venn04} are shown as gray circles.  
}
\label{fig:alpha} 
\end{figure*}

\begin{figure*}
\centering
\includegraphics[scale=0.70]{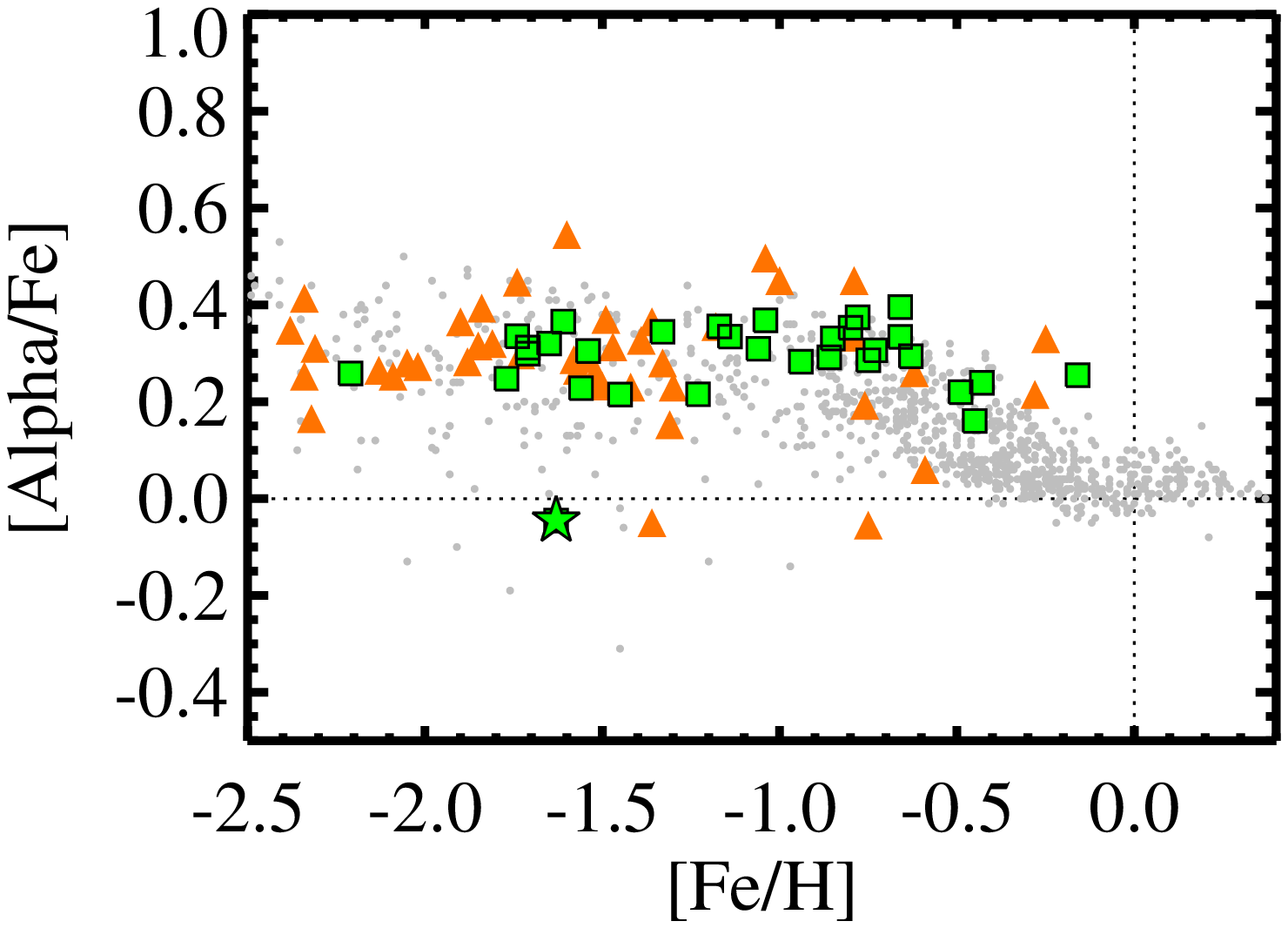}
\caption{ Mean alpha abundances  from  Ca I, Si I, Ti I , and Ti II ratios.  Symbols and references are the same as in Figure \ref{fig:alpha}.}
\label{fig:mean} 
\end{figure*}

In Figure \ref{fig:alpha},  we compare the alpha element ratios as a function of [Fe/H] for  M31 and the MW. 
For the MW comparison we use the compilation of GC mean abundances in \cite{pritzl05},  which were obtained from homogenizing the abundances of individual stars by different authors. 
In general, the [Ca/Fe], [Si/Fe], [Ti I/Fe], and [Ti II/Fe] abundance patterns of the M31 GCs are similar to the MW abundance patterns. 
It is interesting to compare the [$\alpha$/Fe] plateau values, because the plateau is in principle sensitive to the relative numbers of high mass stars, and contains information about the early halo stellar population history.
We  calculate the mean plateau values using all GCs with [Fe/H]$<-0.7$; for easy comparison, the mean values for the  M31 sample and the stellar GC data from \cite{pritzl05} are listed in Table \ref{tab:alphas}.   We find that the mean plateau values for the four individual elements, as well as a mean of the sample as a whole, agree very well. 
In addition, we show the  mean obtained from the four elements  for each individual GC in both samples  in Figure \ref{fig:mean}, where the agreement for the M31 GCs and MW GCs in the plateau region is clear. 
Interestingly, we also note that the [Ca/Fe] abundances in Figure \ref{fig:alpha} also follow the MW field star abundances very closely in the ``knee" region from [Fe/H]$\sim -1$ to [Fe/H]$\sim 0$, perhaps more closely than the MW's own GCs.  The decrease in [Ca/Fe] in the knee region may be interpreted as  an increasingly dominant contribution of SN Ia enrichment over SN II enrichment in the ISM, since SN Ia produce Fe in larger amounts than SN II.

\begin{deluxetable*}{rrrrr|rrrrrrrrr}
\scriptsize
\tablecolumns{14}
\tablewidth{0pc}
\tablecaption{M31 GC Line Abundances\label{tab:stub_ele_linetable}}
\tablehead{
 \colhead{Species} && \colhead{$\lambda$}& \colhead{E.P.} &  \colhead{log{\it gf}} &\multicolumn{9}{c}{12+log(X/H)}\\  && \colhead{(\AA)}& \colhead{(eV)} & & \colhead{B006} & \colhead{B012}& \colhead{B029}& \colhead{B034}& \colhead{B045}& \colhead{B048} &  \colhead{B088}& \colhead{B110} & \colhead{B163}  }
\startdata
  Na   &   I   &5682.650 &  2.100 & -0.700  &      6.22  &      4.90  &      6.24  &      5.89  &      5.43  &   \nodata  &   \nodata  &   \nodata  &      6.37  \\
  Na   &   I   &5688.220 &  2.100 & -0.460  &      6.32  &      5.20  &      6.24  &      6.09  &      5.53  &   \nodata  &   \nodata  &   \nodata  &      6.57  \\
  Na   &   I   &6154.230 &  2.100 & -1.570  &      6.12  &      5.39  &      6.34  &      5.89  &      5.53  &   \nodata  &   \nodata  &   \nodata  &   \nodata  \\
  Na   &   I   &6160.753 &  2.100 & -1.270  &   \nodata  &      5.39  &   \nodata  &   \nodata  &   \nodata  &   \nodata  &   \nodata  &   \nodata  &   \nodata  \\
\\  
Mg   &   I   &4167.277 &  4.346 & -0.995  &   \nodata  &   \nodata  &   \nodata  &      7.10  &   \nodata  &      6.84  &   \nodata  &   \nodata  &   \nodata  \\
  Mg   &   I   &4351.921 &  4.346 & -0.520  &   \nodata  &      6.55  &   \nodata  &   \nodata  &   \nodata  &   \nodata  &   \nodata  &   \nodata  &   \nodata  \\
  Mg   &   I   &4571.102 &  0.000 & -5.569  &      7.38  &   \nodata  &   \nodata  &      7.00  &      6.89  &      6.54  &      5.36  &   \nodata  &   \nodata  \\
  Mg   &   I   &4703.003 &  4.346 & -0.377  &      7.38  &   \nodata  &   \nodata  &   \nodata  &      6.59  &      6.44  &      5.46  &   \nodata  &   \nodata  \\

\enddata
\tablecomments{This table is presented in its entirety in machine readable format in the electronic edition of the journal.}

\end{deluxetable*}

\begin{deluxetable*}{l|rrrr|rrrr|rrrr|rrrr}
\addtolength{\tabcolsep}{-25pt}
\scriptsize
\tablecolumns{17}
\tablewidth{0pc}
\tablecaption{M31 GC  Mean  Abundances \label{tab:nfabund}}
\tablehead{ 
\colhead{\scriptsize Name}  & \colhead{\scriptsize[FeII/H]}& \colhead{\scriptsize$\sigma_{ {\rm FeII}}$} &  \colhead{\scriptsize{\rm N$_{{\rm FeII}}$}} & \colhead{\scriptsize$\sigma_{ {\rm A,FeII}}$} &\colhead{\scriptsize[NaI/Fe]} & \colhead{\scriptsize$\sigma_{ {\rm NaI}}$} &  \colhead{\scriptsize{\rm N$_{{\rm NaI}}$}}  & \colhead{\scriptsize$\sigma_{ {\rm A,Na}}$} & \colhead{\scriptsize[MgI/Fe]} & \colhead{\scriptsize$\sigma_{ {\rm MgI}}$} &  \colhead{\scriptsize{\rm N$_{{\rm MgI}}$}}  & \colhead{\scriptsize$\sigma_{ {\rm A,Mg}}$} & \colhead{\scriptsize[AlI/Fe]} & \colhead{\scriptsize$\sigma_{ {\rm AlI}}$} &  \colhead{\scriptsize{\rm N$_{{\rm AlI}}$}}  & \colhead{\scriptsize$\sigma_{ {\rm A,Al}}$}\\ \colhead{(1)} & \colhead{(2)} &\colhead{(3)} &\colhead{(4)} &\colhead{(5)} &\colhead{(6)} &\colhead{(7)} &\colhead{(8)} &\colhead{(9)} &\colhead{(10)} &\colhead{(11)} &\colhead{(12)} &\colhead{(13)} &\colhead{(14)} &\colhead{(15)} &\colhead{(16)} &\colhead{(17)}
 }
\startdata
 B006-G058 &  -0.46 &  0.04 &  12 &  0.17 &   0.52 &  0.04 &   3 &  0.02 &   0.32 &  0.03 &   5 &  0.03 &   0.62 &  0.13 &   2 &  0.10 \\

B012-G064 &  -1.53 &  0.06 &   9 &  0.03 &   0.62 &  0.13 &   4 &  0.00 &   -0.08 &  0.07 &   2 &  0.04 &   \nodata &   \nodata &   \nodata &   \nodata \\

 B029-G090 &  -0.62 &  0.06 &  11 &  0.06 &   0.48 &  0.08 &   3 &  0.07 &   0.17 &  0.08 &   3 &  0.05 &   0.38 &  0.14 &   2 &  0.11 \\

 B034-G096 &  -0.70 &  0.06 &  11 &  0.05 &   0.47 &  0.08 &   3 &  0.00 &   0.21 &  0.08 &   5 &  0.04 &   0.50 &  0.10 &   1 &  0.00 \\
 
B045-G108 &  -0.84 &  0.06 &  11 &  0.07 &   0.12 &  0.04 &   3 &  0.07 &   0.04 &  0.15 &   4 &  0.05 &   0.10 &  0.07 &   2 &  0.07 \\
 B048-G110 &  -0.97 &  0.04 &  16 &  0.06 &   \nodata &   \nodata &   \nodata &   \nodata &   0.10 &  0.08 &   5 &  0.11 &   0.40 &  0.07 &   2 &  0.07 \\
 B088-G150 &  -1.97 &  0.07 &   6 &  0.04 &   \nodata &   \nodata &   \nodata &   \nodata &  -0.48 &  0.00 &   3 &  0.02 &   \nodata &   \nodata &   \nodata &   \nodata \\
 B110-G172 &  -0.77 &  0.06 &   8 &  0.05 &   \nodata &   \nodata &   \nodata &   \nodata &   0.40 &  0.10 &   1 &  0.00 &   \nodata &   \nodata &   \nodata &   \nodata \\
 B163-G217 &  -0.45 &  0.10 &   4 &  0.04 &   0.72 &  0.07 &   2 &  0.04 &   0.25 &  0.10 &   1 &  0.07 &   0.50 &  0.00 &   2 &  0.00 \\
 B171-G222 &  -0.47 &  0.12 &   6 &  0.03 &   0.73 &  0.08 &   3 &  0.00 &   0.18 &  0.17 &   4 &  0.02 &   0.38 &  0.21 &   2 &  0.04 \\
 B182-G233 &  -0.94 &  0.08 &   7 &  0.01 &   \nodata &   \nodata &   \nodata &   \nodata &   0.35 &  0.10 &   1 &  0.07 &   \nodata &   \nodata &   \nodata &   \nodata \\
 B193-G244 &  -0.12 &  0.12 &   5 &  0.11 &   \nodata &   \nodata &   \nodata &   \nodata &   0.40 &  0.10 &   1 &  0.00 &   \nodata &   \nodata &   \nodata &   \nodata \\
 B225-G280 &  -0.61 &  0.14 &   2 &  0.01 &   \nodata &   \nodata &   \nodata &   \nodata &   0.45 &  0.10 &   1 &  0.07 &   \nodata &   \nodata &   \nodata &   \nodata \\
 B232-G286 &  -1.85 &  0.04 &   9 &  0.03 &   0.35 &  0.07 &   2 &  0.00 &  -0.10 &  0.04 &   2 &  0.00 &   \nodata &   \nodata &   \nodata &   \nodata \\
 B235-G297 &  -0.95 &  0.04 &  15 &  0.05 &   0.43 &  0.04 &   3 &  0.05 &  -0.01 &  0.09 &   4 &  0.02 &   0.35 &  0.10 &   1 &  0.07 \\
 B240-G302 &  -1.38 &  0.04 &  14 &  0.01 &   0.23 &  0.07 &   2 &  0.04 &   -0.03 &  0.05 &   3 &  0.08 &   \nodata &   \nodata &   \nodata &   \nodata \\
 B311-G033 &  -1.63 &  0.08 &   6 &  0.00 &   \nodata &   \nodata &   \nodata &   \nodata &  -0.07 &  0.12 &   3 &  0.05 &   \nodata &   \nodata &   \nodata &   \nodata \\
 B312-G035 &  -0.87 &  0.06 &  10 &  0.01 &   0.38 &  0.00 &   2 &  0.04 &   0.00 &  0.14 &   2 &  0.00 &   \nodata &   \nodata &   \nodata &   \nodata \\

 B358-G219 &  -2.35 &  0.04 &   8 &  0.04 &   \nodata &   \nodata &   \nodata &   \nodata &  -0.03 &  0.21 &   2 &  0.00 &   0.69 &  0.05 &   2 &  0.05 \\

 B381-G315 &  -1.03 &  0.06 &  11 &  0.00 &   0.23 &  0.11 &   3 &  0.00 &   0.15 &  0.07 &   5 &  0.01 &   \nodata &   \nodata &   \nodata &   \nodata \\

 B383-G318 &  -0.75 &  0.07 &  13 &  0.01 &   0.55 &  0.00 &   3 &  0.07 &   0.24 &  0.07 &   4 &  0.01 &   0.35 &  0.14 &   2 &  0.07 \\
 B384-G319 &  -0.70 &  0.06 &  14 &  0.08 &   0.33 &  0.04 &   3 &  0.05 &   0.15 &  0.06 &   4 &  0.00 &   0.30 &  0.10 &   1 &  0.00 \\

 B386-G322 &  -1.11 &  0.05 &  11 &  0.03 &   0.38 &  0.07 &   2 &  0.04 &   0.13 &  0.05 &   5 &  0.01 &   \nodata &   \nodata &   \nodata &   \nodata \\
 B403-G348 &  -0.67 &  0.05 &  15 &  0.03 &   0.38 &  0.07 &   3 &  0.02 &   0.18 &  0.05 &   6 &  0.00 &   0.20 &  0.00 &   2 &  0.00 \\

 B405-G351 &  -1.35 &  0.07 &   9 &  0.04 &   0.25 &  0.07 &   2 &  0.00 &  -0.03 &  0.07 &   5 &  0.01 &   \nodata &   \nodata &   \nodata &   \nodata \\
 B457-G097 &  -1.33 &  0.08 &  10 &  0.02 &   0.07 &  0.14 &   4 &  0.04 &  -0.42 &  0.15 &   5 &  0.11 &   \nodata &   \nodata &   \nodata &   \nodata \\
\scriptsize
B514-MCGC4 &  -1.65 &  0.04 &  12 &  0.04 &   0.45 &  0.10 &   1 &  0.07 &  -0.16 &  0.07 &   4 &  0.12 &   \nodata &   \nodata &   \nodata &   \nodata \\
      G002 &  -1.66 &  0.03 &  12 &  0.03 &   \nodata &   \nodata &   \nodata &   \nodata &  -0.55 &  0.07 &   3 &  0.07 &   \nodata &   \nodata &   \nodata &   \nodata \\
  G327-MVI &  -1.58 &  0.05 &  10 &  0.03 &   \nodata &   \nodata &   \nodata &   \nodata &   0.02 &  0.10 &   4 &  0.02 &   \nodata &   \nodata &   \nodata &   \nodata \\
\scriptsize
 MCGC5-H10 &  -1.35 &  0.04 &  14 &  0.00 &  -0.03 &  0.07 &   2 &  0.11 &  -0.09 &  0.06 &   5 &  0.02 &   \nodata &   \nodata &   \nodata &   \nodata \\
      MGC1 &  -1.50 &  0.04 &  13 &  0.02 &   0.10 &  0.00 &   2 &  0.00 &   0.09 &  0.15 &   3 &  0.08 &   \nodata &   \nodata &   \nodata &   \nodata \\

\enddata

\tablecomments{All abundances are obtained from line synthesis. Column 2 is the mean [Fe II/H] abundance. Columns  6, 10, and 14 indicate the mean abundance ratios ([X/Fe]) with respect to Fe I. Columns 3, 7, 11, and 15 are the statistical error in the mean for each species ($\sigma_{ {\rm X}} = \sigma / \sqrt{{\rm N_{X}} -1})$. For species with only one line measurement   $\sigma_{ {\rm X}}$  is set to a value of  0.1 dex, which is  a typical line-to-line uncertainty in our analysis. Columns 4,  8, 12, and 16 are the number of lines of each species used in the analysis (N$_{{\rm X}}$). Columns 5, 9, 13, and 17 are the uncertainty in the abundance due to the age of the CMD ($\sigma_{ {\rm A,X}}$).  
}

\end{deluxetable*}

\begin{deluxetable*}{l|rrrr|rrrr|rrrr|rrrr}
\scriptsize
\tablecolumns{17}
\addtolength{\tabcolsep}{-35pt}
\tablewidth{0pc}
\tablecaption{M31 GC  Abundances Continued \label{tab:tnfabund2}}
\tablehead{
\colhead{\scriptsize
Name}  & \colhead{\scriptsize
[CaI/Fe]}& \colhead{\scriptsize
$\sigma_{ {\rm CaI}}$} &  \colhead{\scriptsize
{\rm N$_{{\rm CaI}}$}} & \colhead{\scriptsize
$\sigma_{ {\rm A,CaI}}$}  & \colhead{\scriptsize
[SiI/Fe]} & \colhead{\scriptsize
$\sigma_{ {\rm SiI}}$} &  \colhead{\scriptsize
{\rm N$_{{\rm SiI}}$}}  & \colhead{\scriptsize
$\sigma_{ {\rm A,SiI}}$} & \colhead{\scriptsize
[TiI/FeI]} & \colhead{\scriptsize
$\sigma_{ {\rm TiI}}$} &  \colhead{\scriptsize
{\rm N$_{{\rm TiI}}$}}  & \colhead{\scriptsize
$\sigma_{ {\rm A,TiI}}$} & \colhead{\scriptsize
[TiII/FeII]} & \colhead{\scriptsize
$\sigma_{ {\rm TiII}}$} &  \colhead{\scriptsize
{\rm N$_{{\rm TiII}}$}}  & \colhead{\scriptsize
$\sigma_{ {\rm A,TiII}}$}\\ \colhead{(1)} & \colhead{(2)} &\colhead{(3)} &\colhead{(4)} &\colhead{(5)} &\colhead{(6)} &\colhead{(7)} &\colhead{(8)} &\colhead{(9)} &\colhead{(10)} &\colhead{(11)} &\colhead{(12)} &\colhead{(13)} &\colhead{(14)} &\colhead{(15)} &\colhead{(16)} &\colhead{(17)} }
\startdata
\scriptsize 
B006-G058 &   0.25 &  0.05 &   6 &  0.02 &   0.48 &  0.05 &   6 &  0.06 &   0.20 &  0.05 &  14 &  0.02 &   0.29 &  0.06 &  10 &  0.08 \\
\scriptsize
 B012-G064 &   0.40 &  0.06 &   9 &  0.01 &   0.35 &  0.10 &   1 &  0.13 &   \nodata &   \nodata &   \nodata &   \nodata &   0.34 &  0.07 &   8 &  0.01 \\
 \scriptsize
B029-G090 &   0.04 &  0.05 &  11 &  0.07 &   0.30 &  0.05 &   7 &  0.12 &   0.26 &  0.05 &  16 &  0.05 &   0.36 &  0.06 &  11 &  0.01 \\
\scriptsize
 B034-G096 &   0.19 &  0.04 &  10 &  0.03 &   0.46 &  0.04 &   8 &  0.08 &   0.17 &  0.06 &  12 &  0.06 &   0.32 &  0.08 &  10 &  0.01 \\
 \scriptsize
B045-G108 &   0.22 &  0.04 &  11 &  0.02 &   0.46 &  0.05 &   5 &  0.06 &   0.16 &  0.06 &  11 &  0.04 &   0.33 &  0.05 &  11 &  0.08 \\
\scriptsize 
B048-G110 &   0.23 &  0.03 &  10 &  0.07 &   0.58 &  0.05 &   6 &  0.19 &   0.21 &  0.04 &  15 &  0.03 &   0.30 &  0.06 &  13 &  0.06 \\
\scriptsize
 B088-G150 &   0.21 &  0.05 &   6 &  0.04 &   \nodata &   \nodata &   \nodata &   \nodata &   0.35 &  0.10 &   1 &  0.07 &   0.37 &  0.10 &  10 &  0.27 \\
\scriptsize
 B110-G172 &   0.14 &  0.03 &   4 &  0.14 &   0.31 &  0.07 &   3 &  0.30 &   0.41 &  0.04 &   8 &  0.10 &   0.47 &  0.11 &   7 &  0.01 \\

\scriptsize
 B163-G217 &   0.28 &  0.06 &  11 &  0.04 &   0.25 &  0.07 &   2 &  0.00 &   0.00 &  0.19 &   3 &  0.00 &   0.30 &  0.07 &   6 &  0.05 \\
\scriptsize
 B171-G222 &   0.27 &  0.09 &  10 &  0.04 &   0.45 &  0.06 &   6 &  0.00 &  -0.08 &  0.08 &   5 &  0.03 &   0.01 &  0.15 &   7 &  0.02 \\
\scriptsize 
B182-G233 &   0.41 &  0.09 &   8 &  0.01 &   0.58 &  0.28 &   2 &  0.10 &   0.29 &  0.06 &   5 &  0.07 &   0.19 &  0.10 &   7 &  0.04 \\
\scriptsize
 B193-G244 &   0.17 &  0.07 &   2 &  0.07 &   0.36 &  0.08 &   5 &  0.14 &   0.15 &  0.06 &   6 &  0.00 &   0.33 &  0.14 &   4 &  0.06 \\
\scriptsize 
B225-G280 &   0.40 &  0.04 &   5 &  0.10 &   0.49 &  0.12 &   3 &  0.13 &   \nodata &   \nodata &   \nodata &   \nodata &   0.29 &  0.10 &   4 &  0.22 \\
\scriptsize 
B232-G286 &   0.34 &  0.04 &   8 &  0.04 &   \nodata &   \nodata &   \nodata &   \nodata &   0.17 &  0.04 &   6 &  0.01 &   0.22 &  0.07 &  12 &  0.02 \\
\scriptsize 
B235-G297 &   0.22 &  0.04 &   8 &  0.04 &   0.52 &  0.06 &   4 &  0.06 &   0.17 &  0.04 &  12 &  0.00 &   0.26 &  0.04 &  12 &  0.00 \\
\scriptsize
B240-G302 &   0.30 &  0.04 &  13 &  0.04 &   0.53 &  0.03 &   4 &  0.07 &   0.26 &  0.09 &  10 &  0.07 &   0.14 &  0.05 &  15 &  0.04 \\
\scriptsize
 B311-G033 &   0.31 &  0.06 &   7 &  0.00 &   0.37 &  0.21 &   2 &  0.03 &   0.25 &  0.10 &   4 &  0.04 &   0.26 &  0.07 &   7 &  0.07 \\
\scriptsize
B312-G035 &   0.31 &  0.03 &  12 &  0.06 &   0.48 &  0.05 &   5 &  0.12 &   0.30 &  0.07 &   9 &  0.00 &   0.14 &  0.08 &  10 &  0.01 \\
\scriptsize
 B358-G219 &   0.27 &  0.05 &   7 &  0.00 &   \nodata &   \nodata &   \nodata &   \nodata &   0.14 &  0.04 &   6 &  0.04 &   0.34 &  0.05 &  11 &  0.04 \\
\scriptsize
 B381-G315 &   0.27 &  0.04 &  12 &  0.04 &   0.54 &  0.06 &   6 &  0.06 &   0.25 &  0.04 &  11 &  0.02 &   0.37 &  0.06 &  11 &  0.04 \\
\scriptsize 
B383-G318 &   0.28 &  0.04 &  12 &  0.14 &   0.45 &  0.05 &   6 &  0.10 &   0.27 &  0.06 &  10 &  0.17 &   0.30 &  0.07 &  12 &  0.23 \\
\scriptsize
 B386-G322 &   0.27 &  0.04 &  11 &  0.01 &   0.48 &  0.08 &   7 &  0.05 &   0.28 &  0.06 &  13 &  0.01 &   0.41 &  0.07 &  11 &  0.01 \\
\scriptsize 
B384-G319 &   0.14 &  0.03 &  10 &  0.02 &   0.39 &  0.05 &   8 &  0.05 &   0.28 &  0.05 &  14 &  0.01 &   0.36 &  0.08 &  12 &  0.03 \\
\scriptsize 
B403-G348 &   0.26 &  0.04 &  11 &  0.03 &   0.57 &  0.06 &   8 &  0.00 &   0.24 &  0.07 &  12 &  0.17 &   0.34 &  0.04 &   9 &  0.04 \\
\scriptsize
 B405-G351 &   0.26 &  0.06 &  12 &  0.00 &   0.48 &  0.10 &   2 &  0.04 &   0.40 &  0.04 &   5 &  0.00 &   0.23 &  0.06 &  11 &  0.01\\
\scriptsize
 B457-G097 &   0.08 &  0.04 &   8 &  0.06 &   0.10 &  0.04 &   3 &  0.05 &   0.42 &  0.07 &  10 &  0.31 &   0.27 &  0.08 &  12 &  0.14 \\
\scriptsize

B514-MCGC4 &   0.45 &  0.04 &   9 &  0.05 &   0.66 &  0.10 &   1 &  0.09 &   0.06 &  0.04 &   7 &  0.07 &   0.17 &  0.06 &  15 &  0.06 \\
\scriptsize 
     G002 &  -0.02 &  0.02 &   8 &  0.06 &   0.07 &  0.10 &   2 &  0.04 &  -0.18 &  0.07 &   8 &  0.06 &  -0.05 &  0.08 &  12 &  0.09 \\
\scriptsize 
 G327-MVI &   0.40 &  0.07 &   6 &  0.12 &   0.38 &  0.14 &   2 &  0.11 &   0.26 &  0.04 &   7 &  0.04 &   0.25 &  0.04 &  11 &  0.06 \\
\scriptsize

 MCGC5-H10 &   0.18 &  0.04 &   9 &  0.02 &   0.37 &  0.12 &   4 &  0.03 &   0.05 &  0.09 &   7 &  0.01 &   0.25 &  0.05 &  14 &  0.01 \\
\scriptsize    
  MGC1 &   0.18 &  0.04 &   7 &  0.00 &   0.53 &  0.10 &   2 &  0.04 &   0.06 &  0.02 &   6 &  0.01 &   0.14 &  0.05 &  14 &  0.01 \\

\enddata

\tablecomments{All abundances are obtained from line synthesis. Columns  2, 6, 10, and 14 indicate the mean abundance ratios ([X/Fe]) with respect to Fe I. Columns 3, 7, 11, and 15 are the statistical error in the mean for each species ($\sigma_{ {\rm X}} = \sigma / \sqrt{{\rm N_{X}} -1})$. For species with only one line measurement   $\sigma_{ {\rm X}}$  is set to a value of  0.1 dex, which is  a typical line-to-line uncertainty in our analysis.  Columns 4,  8, 12, and 16 are the number of lines of each species used in the analysis (N$_{{\rm X}}$). Columns 5, 9, 13, and 17 are the uncertainty in the abundance due to the age of the CMD ($\sigma_{ {\rm A,X}}$).}

\end{deluxetable*}

One GC, G002, has  alpha-element abundance ratios that are lower than those of the other GCs at similar metallicity ([Fe/H]$\sim-1.6$).
 We have highlighted this  GC with a different symbol in Figure \ref{fig:alpha} for clarity.   The difference in Ca I line strength between G002 and another GC with  similar metallicity and velocity dispersion, G327,  can be seen in Figure \ref{fig:ca}, which also visually demonstrates the accuracy of the synthesis fits.  
We also find other GCs, such as B457 and B171, that have  two species with  [$\alpha$/Fe] that are lower than the average at their respective metallicities. The  two GCs also have slightly different alpha element behaviors from each other. B457 only has lower than average [Ca/Fe] and  [Si/Fe], while B171 has lower than average [Ti I/Fe] and [Ti II/Fe].   In general,  lower alpha element abundance ratios can potentially  indicate  late-time accretion from dwarf galaxies.  For example,   similar deficits in [$\alpha$/Fe] are found in the MW GC Ruprecht 106 \citep{2013ApJ...778..186V}, which is thought to have formed outside the MW.
Lower [$\alpha$/Fe] abundances are also seen in   Sagittarius GCs  that are recently or currently being accreted into the MW GC system \citep[e.g.][]{2010MNRAS.404.1203F,2005A&A...437..905S,cohenpal12}, although these GCs have higher [Fe/H] than G002.   We discuss insights into the accretion history of M31 in more detail in \textsection \ref{sec:capture}.

\begin{deluxetable}{l|r|r}
\tablecolumns{3}
\tablewidth{0pc}
\tablecaption{Mean Alpha Element Abundance Comparison \label{tab:alphas}}
\tablehead{
 & \colhead{M31 ILS}  &\colhead{MW, GC stars$^{1}$} \\ & \colhead{[Fe/H]$<-0.7$} & \colhead{[Fe/H]$<-0.7$}   }
\startdata

${\rm [Ca/Fe]}$  & $+0.26 \pm 0.10$      & $+0.27 \pm 0.13 $ \\  
${\rm [Si/Fe]}$  & $+0.45  \pm 0.14$       & $+0.38
\pm 0.18$ \\ 
${\rm [Ti~ I/Fe]}$  & $+0.21 \pm 0.13$     & $+0.23
\pm 0.16$ \\ 
${\rm [Ti~II/Fe]} $ & $+0.26 \pm 0.10$    &  $+0.27
\pm 0.18$\\  
\\
\hline
\\
${\rm [Alpha/Fe]} $ & $+0.28 \pm 0.16$     & $+0.29
\pm 0.17$\\

\enddata

\tablecomments{1. Using mean GC abundances from \cite{pritzl05}.}
\end{deluxetable}

Figure \ref{fig:alpha} also shows that, as in the MW, in M31 the different alpha elements show subtle differences in their abundance patterns. For example, [Si/Fe] ratios are typically higher than [Ca/Fe] or [Ti/Fe] ratios, which is a reflection of their different formation channels.  This is important to keep in mind, because in extragalactic studies a ``mean'' [$\alpha$/Fe] ratio is often the only measurement that can be obtained, but as Figure \ref{fig:alpha} demonstrates this can potentially wash out the interesting and constraining behavior of these elements.  In Figure \ref{fig:mean}, we average the abundances we obtain for [Ca/Fe], [Si/Fe], [Ti I/Fe], and [Ti II/Fe] to show how some information is lost when examining a mean [$\alpha$/Fe] alone, although we note that one would come to the same the general conclusion that the abundances in M31 are similar to the MW.  In terms of details that are lost, Figure \ref{fig:mean} shows that  the mean [$\alpha$/Fe] for B457 and B171 now look more similar to the bulk of the GCs, even though half of their [$\alpha$/Fe] ratios are low.  In the mean, G002 is the only GC that clearly has a different [$\alpha$/Fe] abundance pattern.   We also find that  the highest metallicity cluster, B193, is at higher mean  [$\alpha$/Fe] than the MW disk stars, but if we separate the individual elements we find that B193 only appears to be particularly inconsistent in [Si/Fe] and [Ti II/Fe].   Together, this shows that the mean [$\alpha$/Fe] is certainly a useful tool for learning about the overall chemical evolution in other galaxies, but the extra information from having precise measurements of several alpha elements shows promise for new insights in  the details of galaxy evolution.

In summary, we measure   Ca I, Si I, Ti I, and Ti II abundances  in the majority of the  GCs in our current M31 sample, providing a novel look at the evolution of [$\alpha$/Fe] over a wide range of metallicity in this galaxy.  We find that the M31 GCs generally have properties similar to MW GCs and disk stars, and discuss further implications on the star formation history of M31 in \textsection \ref{sec:discussion}.

\begin{figure}
\centering
\includegraphics[scale=0.46]{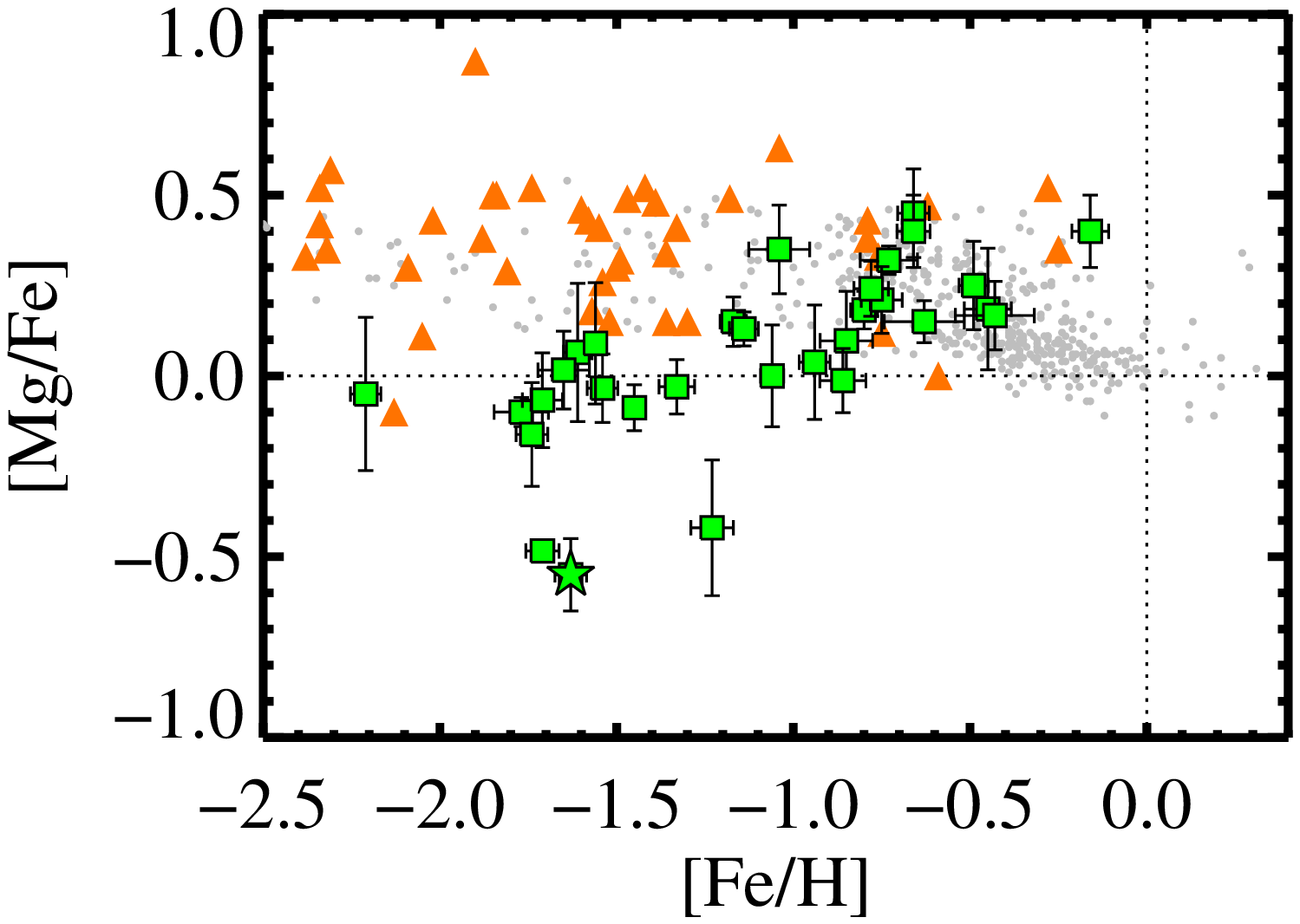}
\includegraphics[scale=0.46]{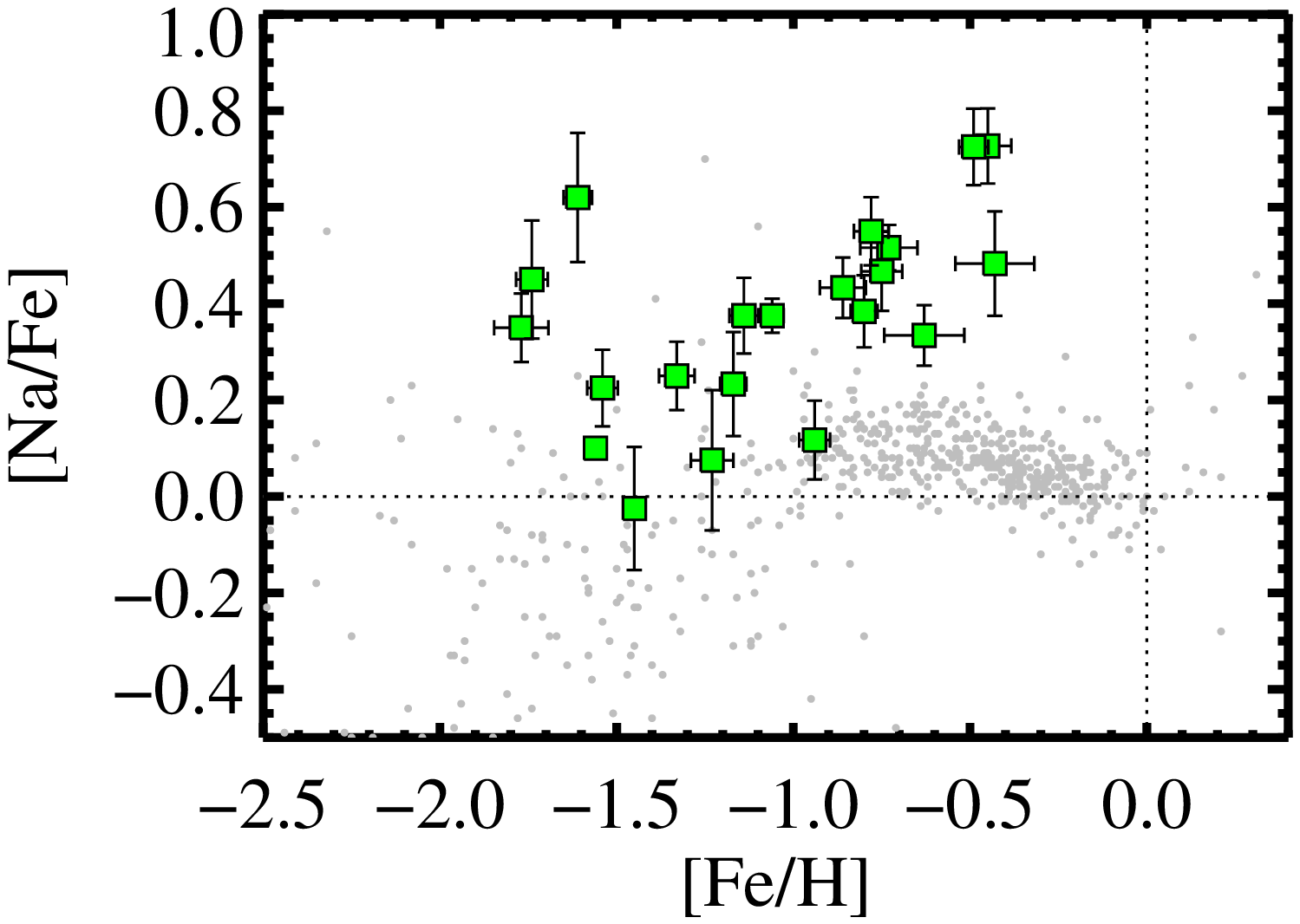}
\includegraphics[scale=0.46]{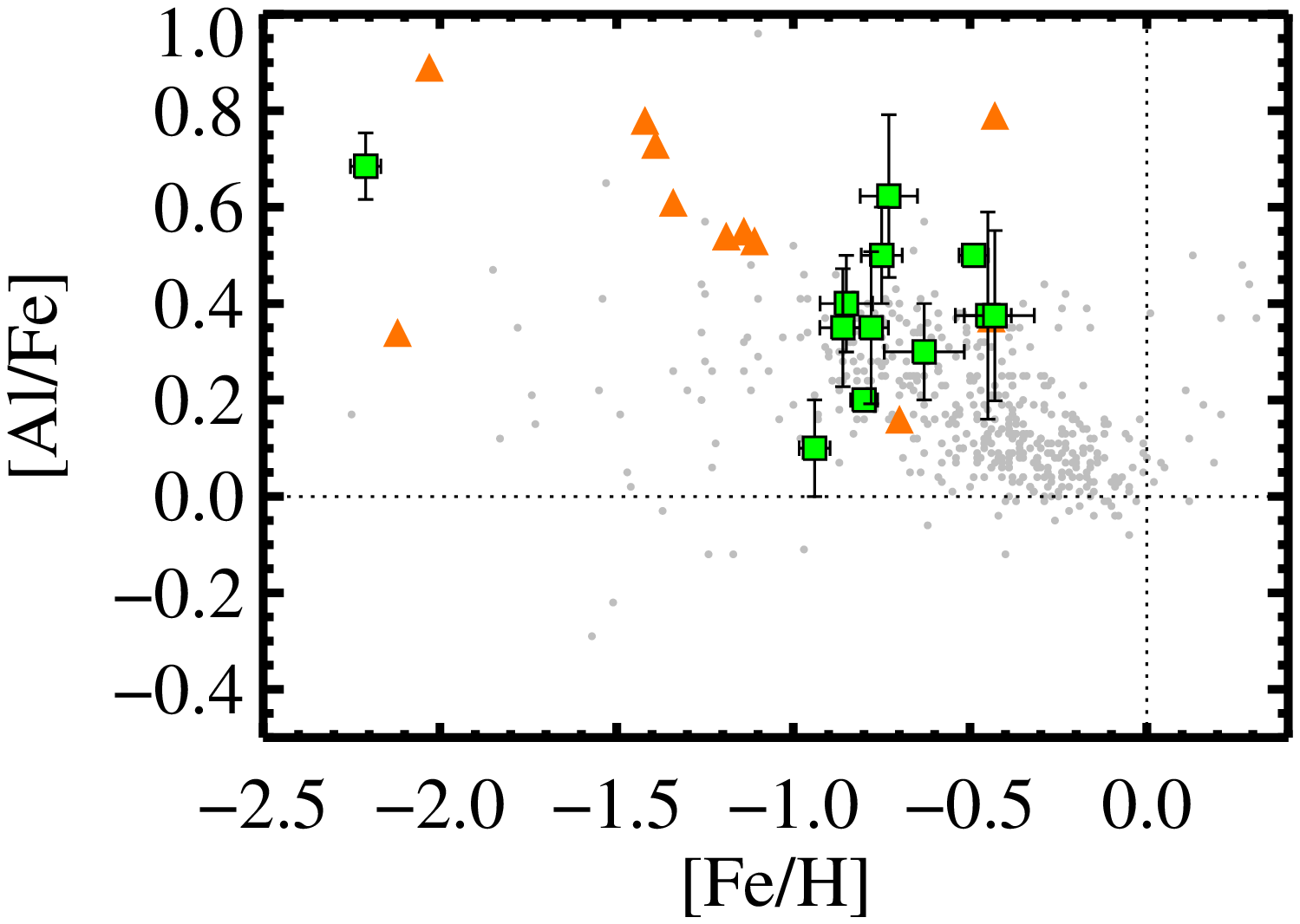}
\caption{  Light element abundance ratios  for  Mg I, Na I, and Al I.  Symbols and references are the same as in Figure \ref{fig:alpha}, with the addition of Na and Al stellar abundances from \cite{2000AJ....120.1841F}, \cite{2004A&A...425..671B}, and \cite{2006AJ....131.1766C}.}
\label{fig:light} 
\end{figure}

\subsection{Light Elements}
\label{sec:light}

We measure light element abundances for Mg I, Na I, and Al I, which are presented  in Table \ref{tab:nfabund}. Like the alpha elements, these abundances were measured with IL spectral synthesis.  The abundance ratios for these elements as a function of [Fe/H] are shown in Figure \ref{fig:light}.  These light elements are interesting because they are not necessarily monometallic in GCs, unlike [Fe/H] and the alpha elements discussed in \textsection \ref{sec:alpha}.  Instead, they have been observed to vary star-to-star in patterns that reflect high temperature proton capture nucleosynthesis \citep[see review by ][and references therein]{grattonrev}.  When these patterns are present [Na/Fe] and [Al/Fe] are enhanced above the normal values of MW field stars, and [O/Fe] and [Mg/Fe] are depleted with respect to normal values, which has led these patterns to be dubbed the  `Na-O' and `Mg-Al anticorrelations.'   The range of the star-to-star abundance variations is different between MW GCs and is thought to be related to the mechanism that causes this behavior in GCs, although a consensus on a complete theoretical understanding has yet to be found \citep[e.g.][]{2006AJ....131.1766C,2009A&A...505..727D,carretta10,2012MNRAS.423.1521D,2012ApJ...758...21C,2013MNRAS.436.2398B,2014MNRAS.439.2043M}.  The full extent of the Na-O anti correlation is observed to depend on present day GC mass, and there observationally appears to be  a minimum GC mass for correlations to be present \citep[e.g.][]{carretta10}.   Variations in the light elements have been directly observed in GC stellar samples in both the MW and external galaxies such as the LMC \citep[e.g.][]{mucc09}.  They have also been indirectly observed in the high resolution IL of GCs in M31 in \citetalias{m31paper} and \cite{italyproc}, the LMC \citepalias{paper4},  Fornax \citep{larsen12} and in MW GCs in \cite{sakari}.  The effect of the variations on the abundances obtained from IL spectra is thought to be a bias in the IL abundance toward higher ratios in [Na/Fe] and [Al/Fe], and toward lower values in [Mg/Fe].   The abundance variations are also believed to bias measurements of [Mg/Fe] in low resolution GC IL spectra, as shown in \cite{schiavon13} for a large sample of M31 GCs.

\begin{figure}
\centering
\includegraphics[trim = 10mm 0mm 0mm 20mm, clip,angle=90,scale=0.46]{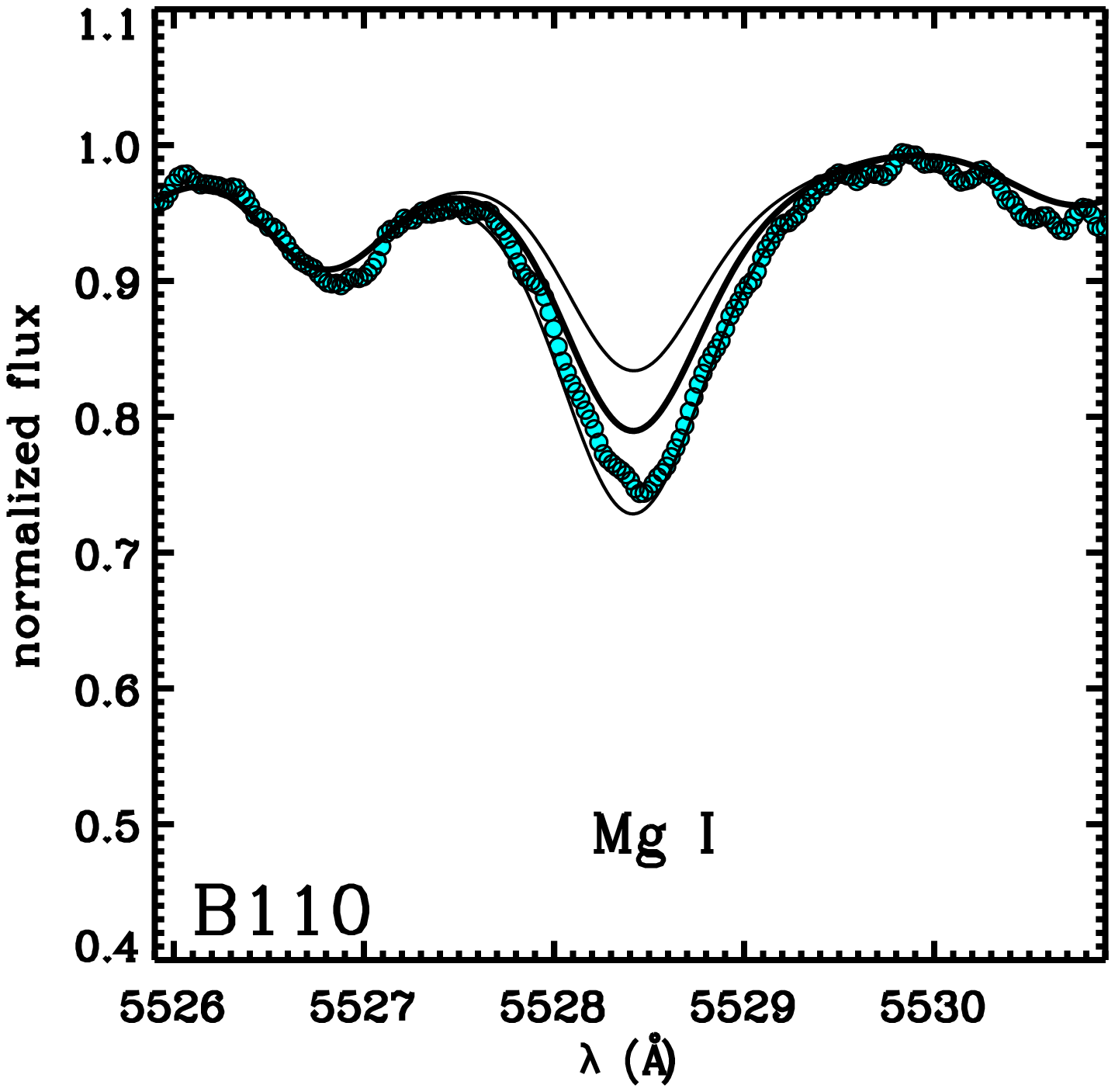}
\includegraphics[trim = 0mm 10mm 0mm 20mm, clip,angle=90,scale=0.46]{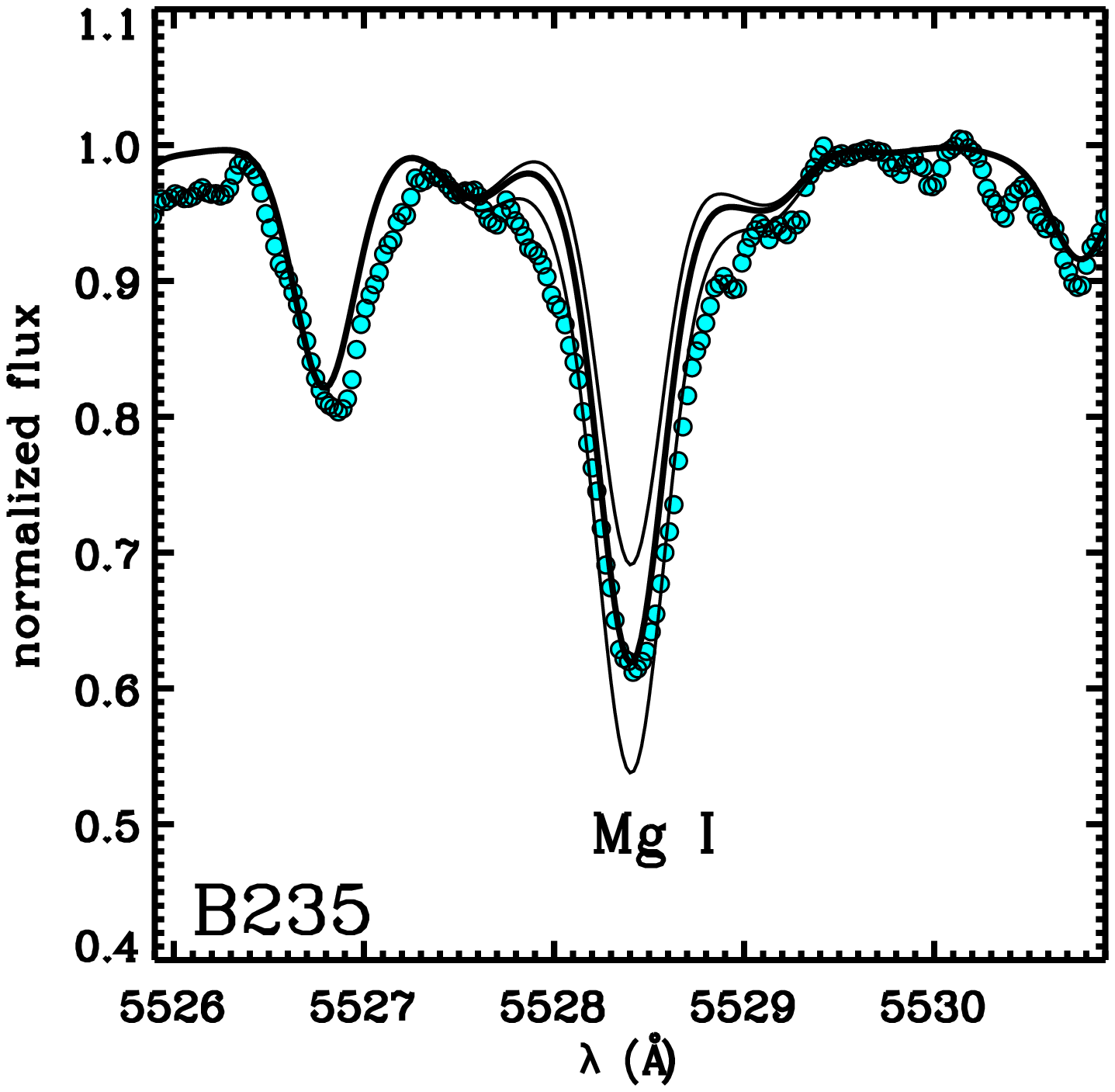}

\caption{ Example Mg I synthesis fits for B110 (top) and B235 (bottom). Points show the data, which has been smoothed by 5 pixels for presentation. Solid lines correspond to ratios of [Mg/Fe]=$-0.5, +0.0, +0.5$.  The two GCs have similar metallicities of [Fe/H]$\sim-0.7$ but different velocity dispersions; B235 has a lower [Mg/Fe] than B110. }
\label{fig:mg} 
\end{figure}

\begin{figure}
\centering
\includegraphics[trim = 10mm 0mm 0mm 20mm, clip,angle=90,scale=0.46]{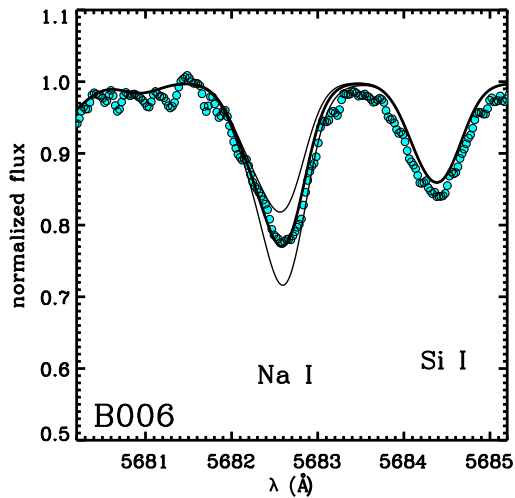}

\caption{ Example Na I synthesis fits for B006. Points show the data, which has been smoothed by 5 pixels for presentation. Solid lines correspond to ratios of [Na/Fe]=$+0.0, +0.5, +1.0$.  The mean [Si/Fe]=$+0.48$ has been adapted to account for blends.   }
\label{fig:na} 
\end{figure}

\begin{figure}
\centering
\includegraphics[trim = 0mm 10mm 0mm 20mm, clip,angle=90,scale=0.46]{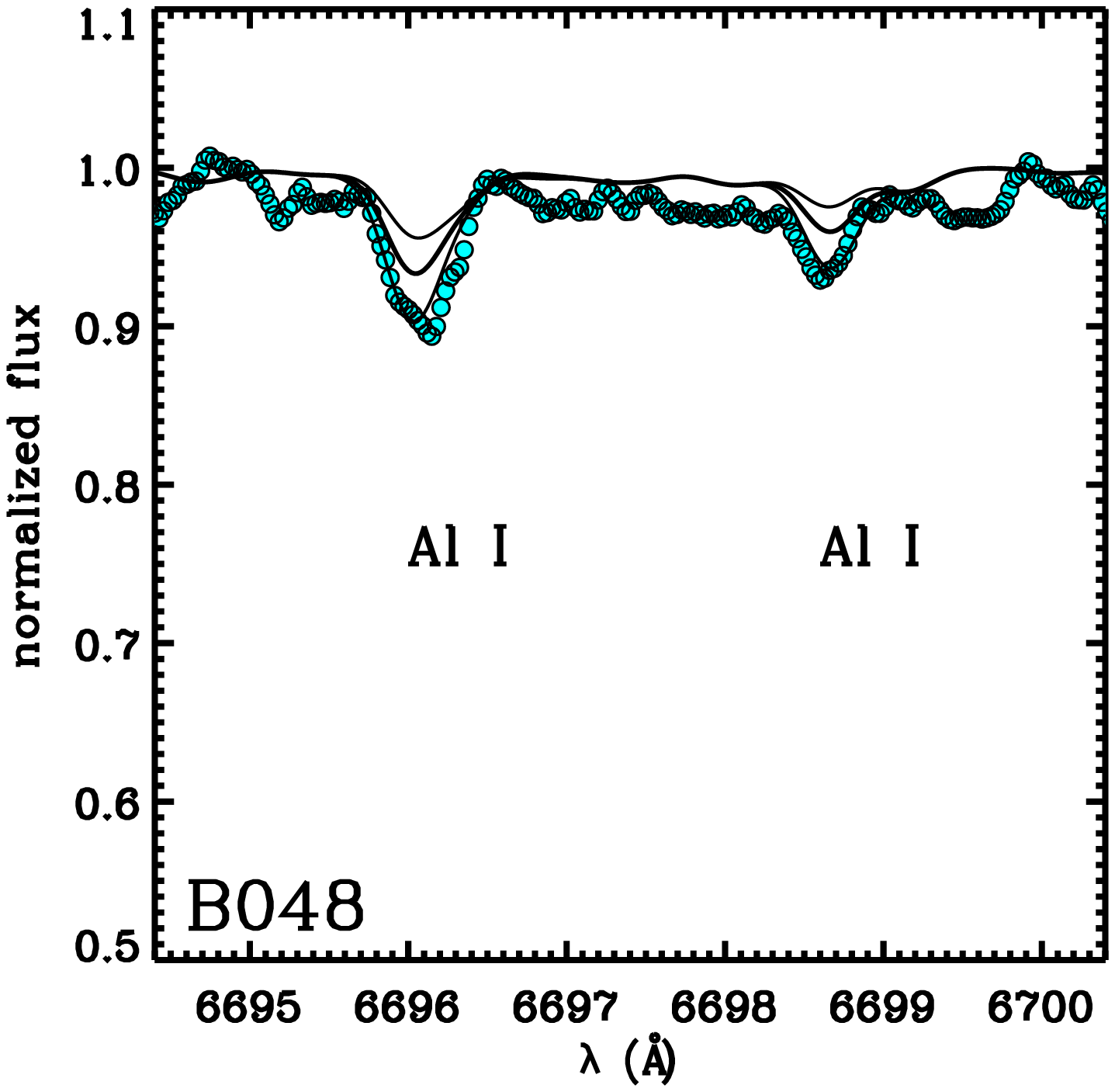}

\caption{ Example Al I synthesis fits for B048. Points show the data, which has been smoothed by 5 pixels for presentation. Solid lines correspond to ratios of [Al/Fe]=$-0.3, +0.0, +0.3$.    }
\label{fig:al} 
\end{figure}

 Figure \ref{fig:light} shows further indirect evidence for star-to-star abundance variations in M31 GCs. First, we find [Mg/Fe] abundances that are lower on average than the field star plateau values, particularly at [Fe/H]$<-0.7$.  Our Mg I measurements are obtained from a handful of weak lines across the full wavelength coverage of HIRES (4167 \AA, 4571 \AA, 4703 \AA,  5528 \AA, 5711 \AA, 7387 \AA) in order to explicitly avoid the strongest Mg I lines in the Mgb region so as to limit non-LTE or saturation problems.    The fact that the more metal-poor GCs are particularly biased to low [Mg/Fe] abundances is very interesting, although not entirely understood.   We note that \cite{larsen12} also find particularly low values of [Mg/Fe] in very metal-poor Fornax GCs, which is consistent with the picture we find here in M31. The lowest [Mg/Fe] we find is for G002, of [Mg/Fe]=$-0.5$, and an example of a clean measurement with that abundance for the 4571 \rAA line is shown in Figure \ref{fig:ti2}.   We show two examples of the Mg I 5528 \rAA line in Figure \ref{fig:mg} for GCs with higher metallicities of  [Fe/H]$\sim-0.7$.  B110 has a [Mg/Fe]$\sim+0.4$, which is similar to the other alpha element plateau values, but B235 clearly has a lower ratio of [Mg/Fe]$\sim0.0$.   We note that [Mg/Fe] abundances as low as these have also been  measured in a subset of stars in the MW outer halo GC NGC 2419 by \cite{cohen2419}.

\begin{figure}
\centering
\includegraphics[scale=0.50]{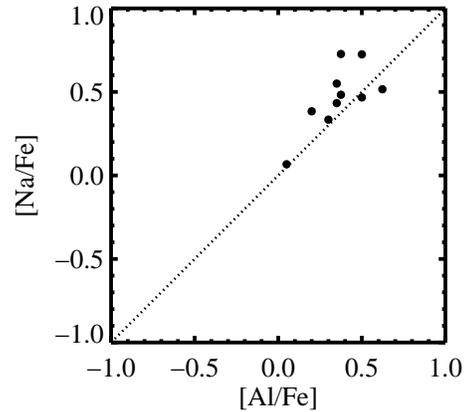}
\includegraphics[scale=0.50]{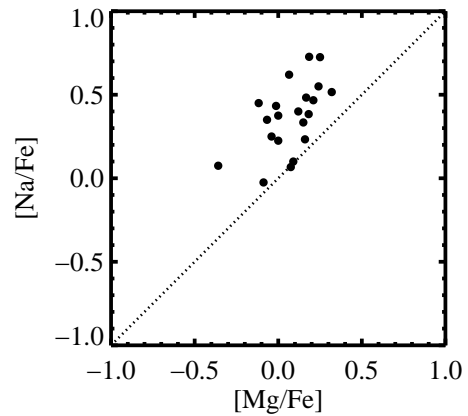}
\includegraphics[scale=0.50]{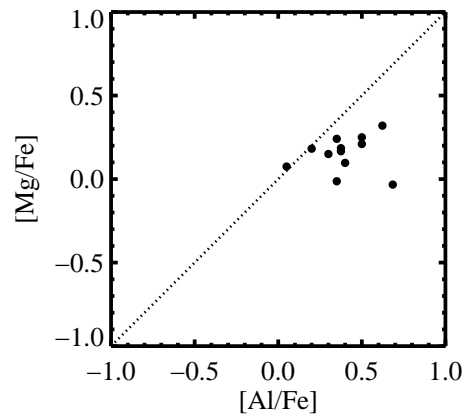}
\caption{ Possible correlations in   the light element abundance ratios in M31 GCs. The dotted line shows a 1:1 correspondence. We find initial evidence that [Na/Fe] may be loosely correlated with [Al/Fe] and [Mg/Fe], but [Al/Fe] does not appear correlated with [Mg/Fe]. }
\label{fig:light2} 
\end{figure}

We also find elevated [Na/Fe], which can be seen at all [Fe/H] in Figure \ref{fig:light}.  The Na I abundances are measured from the unsaturated 5582 \rAA and 6160 \rAA doublets, which minimizes non-LTE effects.  Although we only have 4 transitions to work with, the Na I lines are moderately strong and the 5582 \rAA doublet in particular is only partially blended, so we are able to make Na I measurements for most of the range in [Fe/H] of the sample. An example of an enhanced [Na/Fe] measurement is shown in Figure \ref{fig:na} for B006, which has a mean of [Na/Fe]=$+0.52$.  

 The [Al/Fe] abundances that we measure are also significantly super-solar, however they are primarily measured at higher metallicities and have more  overlap with abundances in normal MW field stars.  With the exception of the metal-poor GC B358 in \citetalias{m31paper}, the Al I abundances are entirely measured from the 6696 \rAA doublet, which should not require substantial non-LTE corrections.  The Al I measurements are from some of the weakest lines that we are able to cleanly measure.  An example is shown in Figure \ref{fig:al} for B048, where we also show  the subtle differences between syntheses with [Al/Fe]=$-0.3$, $+0.0$, and $+0.3$.  B048 has an [Fe/H]=$-0.89$, which is approximately the lowest metallicity for which we are able to cleanly measure an Al I abundance from these weak features, which are also not usually possible to measure for GCs with large velocity dispersions.

In Figure \ref{fig:light2} we compare the behaviors of the light element abundance ratios with each other in order to look for possible correlations in the measurements.  It is important to remember, however, that the comparison of IL flux weighted average values of the light element ratios are fundamentally different comparisons from the  standard Na-O, Mg-Al anti-correlations of {\it individual stars} within a GC.  We find evidence that [Na/Fe] is loosely correlated with [Al/Fe] and [Mg/Fe], but [Mg/Fe] and [Al/Fe] do not appear correlated.  Note that not all GCs have measurements of all three elements, and Al I is only measured in the highest metallicity GCs, which may contribute to the lack of a correlation with [Mg/Fe].
 Naively we may conclude that the possibility of  correlations between some  of  these elements simply shows that there is some consistency to   whatever mechanism is driving the star-to-star abundance variations, although  it is unclear why [Mg/Fe] and [Al/Fe] would not be correlated if their relative amounts are only determined by high temperature {\it p} capture reaction chains.  Indeed,   \cite{2006AJ....131.1766C} also find that stars within MW GCs  do not always show a clearly defined Mg-Al relationship, and that large changes in Al are only accompanied by small changes in Mg, which is consistent with what we find here in the IL abundances.

In summary, all three of the light elements we measure, Mg I, Na I, and Al I,  appear to confirm the widespread existence of star-to-star abundance variations  in M31.

\section{Discussion}
\label{sec:discussion}
In the following sections we first examine possible relationships of the M31 GC abundances with other GC properties, such as luminosity, mass and velocity dispersion.  We then discuss the abundances with respect to properties of the host galaxy, M31, such as radius. Next,  we discuss implications that our measurements have for the star formation and accretion history of M31. Finally, we present comparisons of the  high resolution measurements with previous studies using other techniques.

\subsection{Behavior With GC Properties}
\label{sec:gcbehavior}

While our sample is smaller than those in photometric or low resolution spectra studies of M31 GCs, and obviously not complete, it is interesting to look for relationships of the high precision abundances with other GC properties.   Relationships of GC luminosity with [Fe/H], or mass-metallicity relationships,  have been extensively searched for in GC systems in other galaxies \citep[e.g.][]{harris06,2008AJ....136.1828S,mieske10}.  While correlations with mass and metallicity have not been found in the MW (\cite{2008AJ....136.1828S} with data from \cite{1996AJ....112.1487H}) or M31 GC systems previously \citep[e.g.][]{2000AJ....119..727B,caldwell11}, they are commonly found in the metal-poor component of GC systems in early-type galaxies \citep{2009ApJ...699..254H}. The low-metallicity GC mass-metallicity relationship, or ``blue tilt,'' is thought to be created by GC ``self-enrichment'' \citep[e.g.][]{bailin09,mieske10,goudfrooij14}, where massive GCs are able to retain a significant fraction of their early SN II ejecta in order to incorporate it into their late forming stars. It is not yet clear how the mass-metallicity relationship of GCs seen in early type galaxies may be related to the multiple generations of stars seen in MW and LMC GCs
 \citep[e.g.][]{2009IAUS..258..233P,2009A&A...497..755M}, or the possible mechanism for star-to-star light element variations.

\begin{figure}
\centering
\includegraphics[scale=0.5]{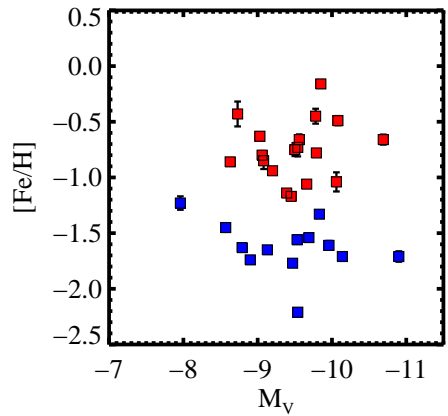}
\caption{ Luminosity-metallicity relationship for the M31 GC sample.  We divide the populations at [Fe/H]$=-1.2$.  
Magnitudes are listed in Table \ref{tab:obs}, and are taken from \cite{bolognacat},   E(B-V) and associated references are found in Table \ref{tab:obs}, and we assume a distance modulus of 24.47 \citep{m31distance} and extinction parameter $R_V=3.1$.  
}
\label{fig:fe-mag} 
\end{figure}

In Figure \ref{fig:fe-mag} we show our  [Fe/H] measurements as a function of reddening-corrected absolute magnitude in V, and we have visually separated  ``metal-poor"  and ``metal-rich" subpopulations with blue and red symbols, respectively.  We use a value of [Fe/H]$=-1.2$ to divide the sample, which  is approximately the location of the minimum between peaks  in the MW's [Fe/H]  distribution \citep[][2010 revision]{1996AJ....112.1487H}. Although our GC sample is small compared to the total GC population in M31, it is nevertheless interesting to separate behaviors of the  metal-poor and metal-rich subpopulations because of the long history of study of GC subpopulations in general \citep[e.g.][]{brodierev06}.   Approximately half of our sample is genuinely more metal poor than [Fe/H]$=-1$. We do not find statistically significant evidence for a mass-metallicity relationship in either subpopulation, or in the sample as a whole, although this would be much better addressed with a  larger sample. Even so,  this is consistent with the results  from larger M31 GC samples  of \cite{2000AJ....119..727B} and \cite{caldwell11}.

\begin{figure*}
\centering
\includegraphics[scale=0.45]{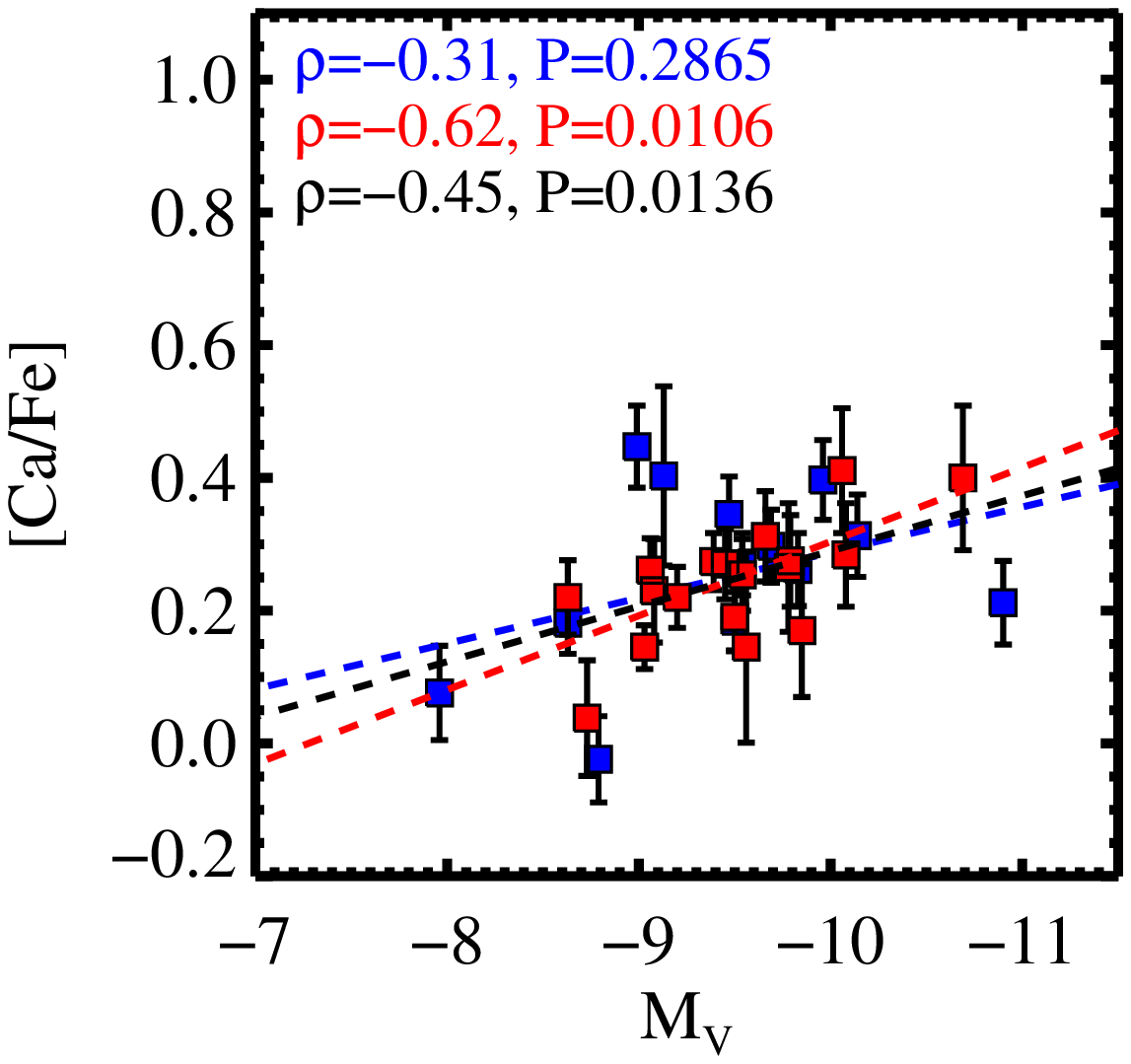}
\includegraphics[scale=0.45]{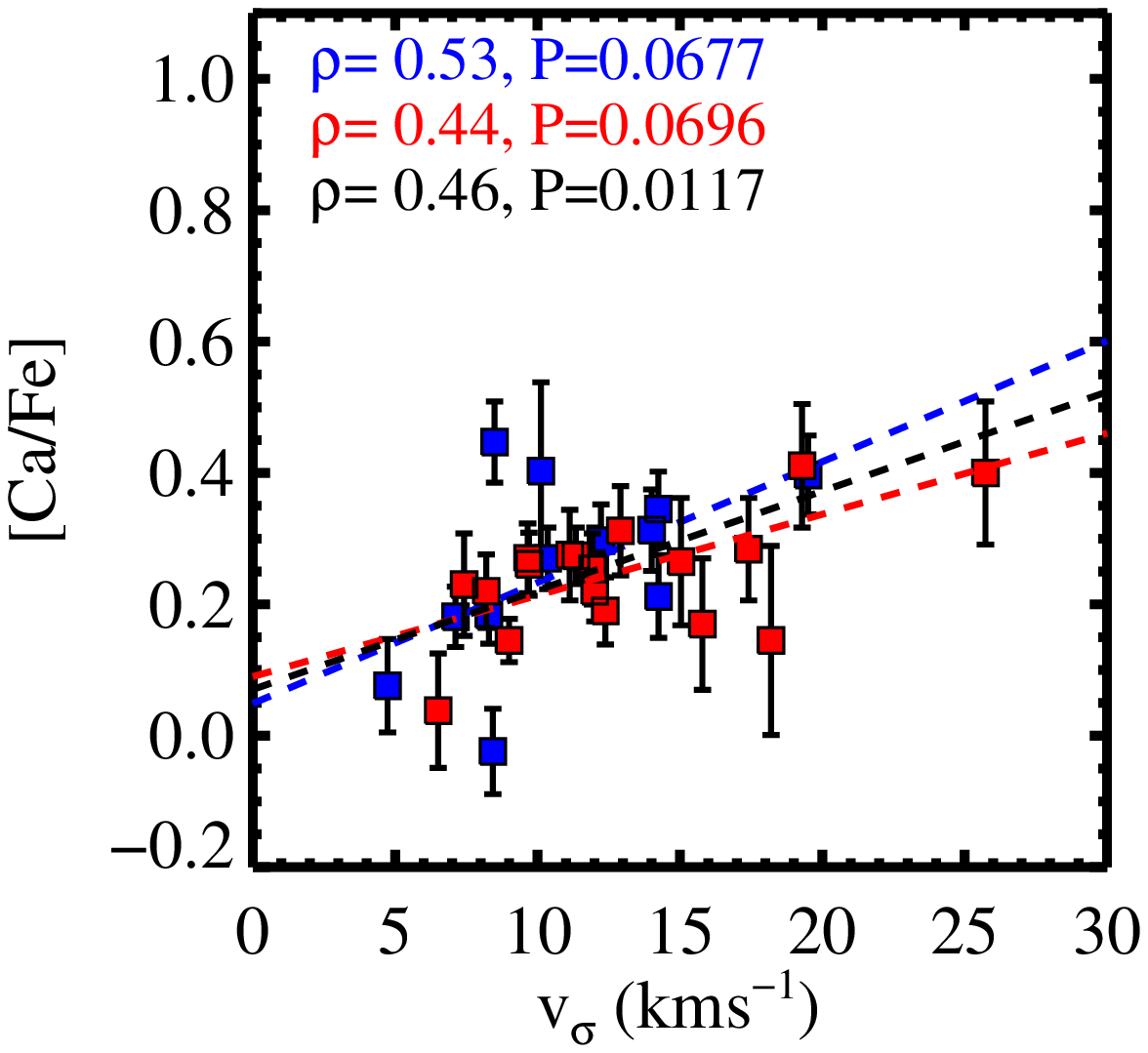}
\includegraphics[scale=0.45]{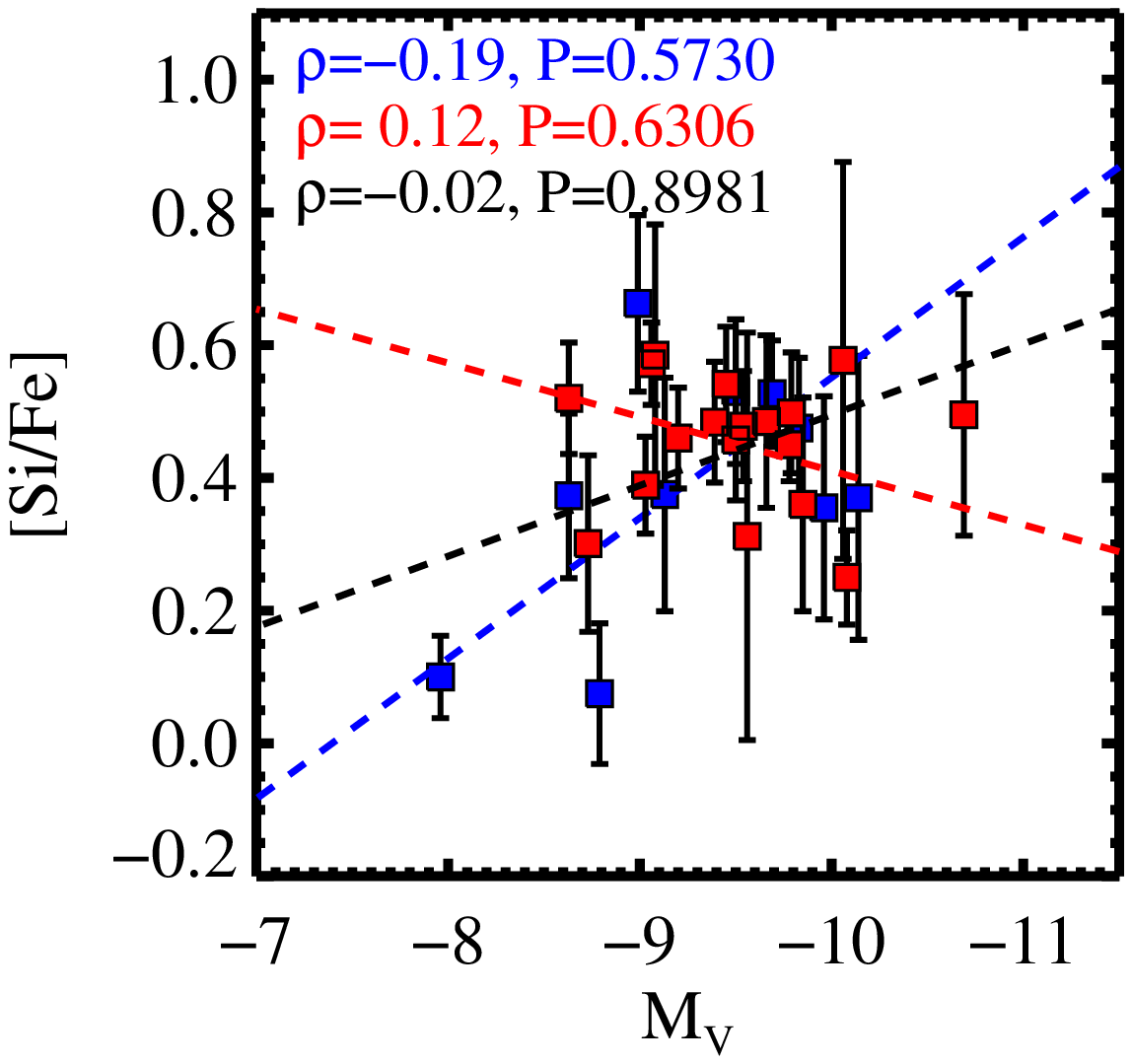}
\includegraphics[scale=0.45]{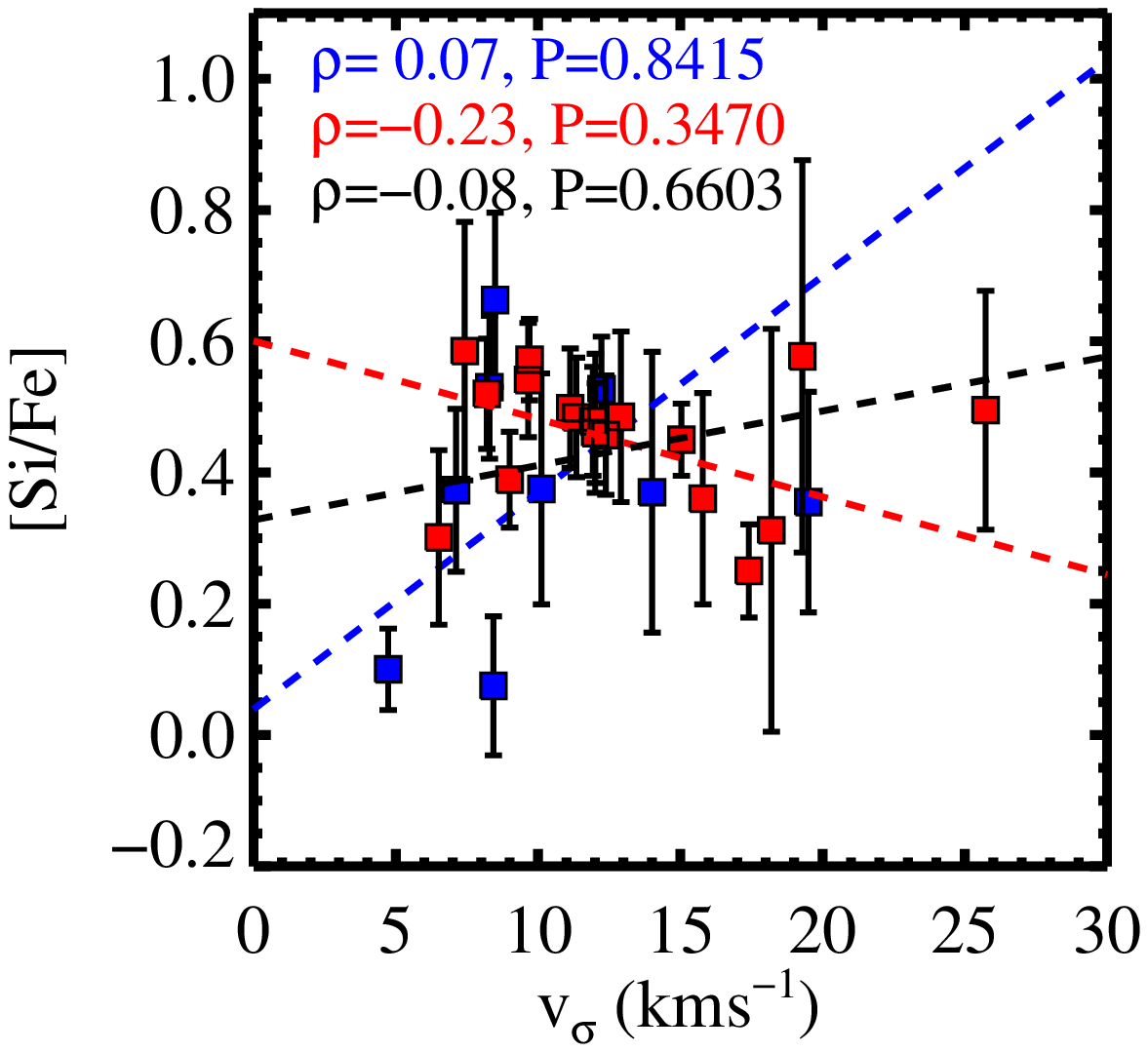}

\caption{ Behavior of [Ca/Fe] and [Si/Fe] with v$_{\sigma}$ and M$_{V}$.  GCs with [Fe/H]$<-1.2$ are shown in blue, and GCs with [Fe/H]$>-1.2$ are shown in red. Dashed lines show linear least squares fits to the metal-poor, metal-rich, and whole sample in blue, red, and black, respectively.  Corresponding color-coded  numbers in the upper left corner of each panel show the Spearman rank correlation coefficient ($\rho$) and the probability associated with the null hypothesis (P).}
\label{fig:alpha-trends} 
\end{figure*}

\begin{figure*}
\centering
\includegraphics[scale=0.45]{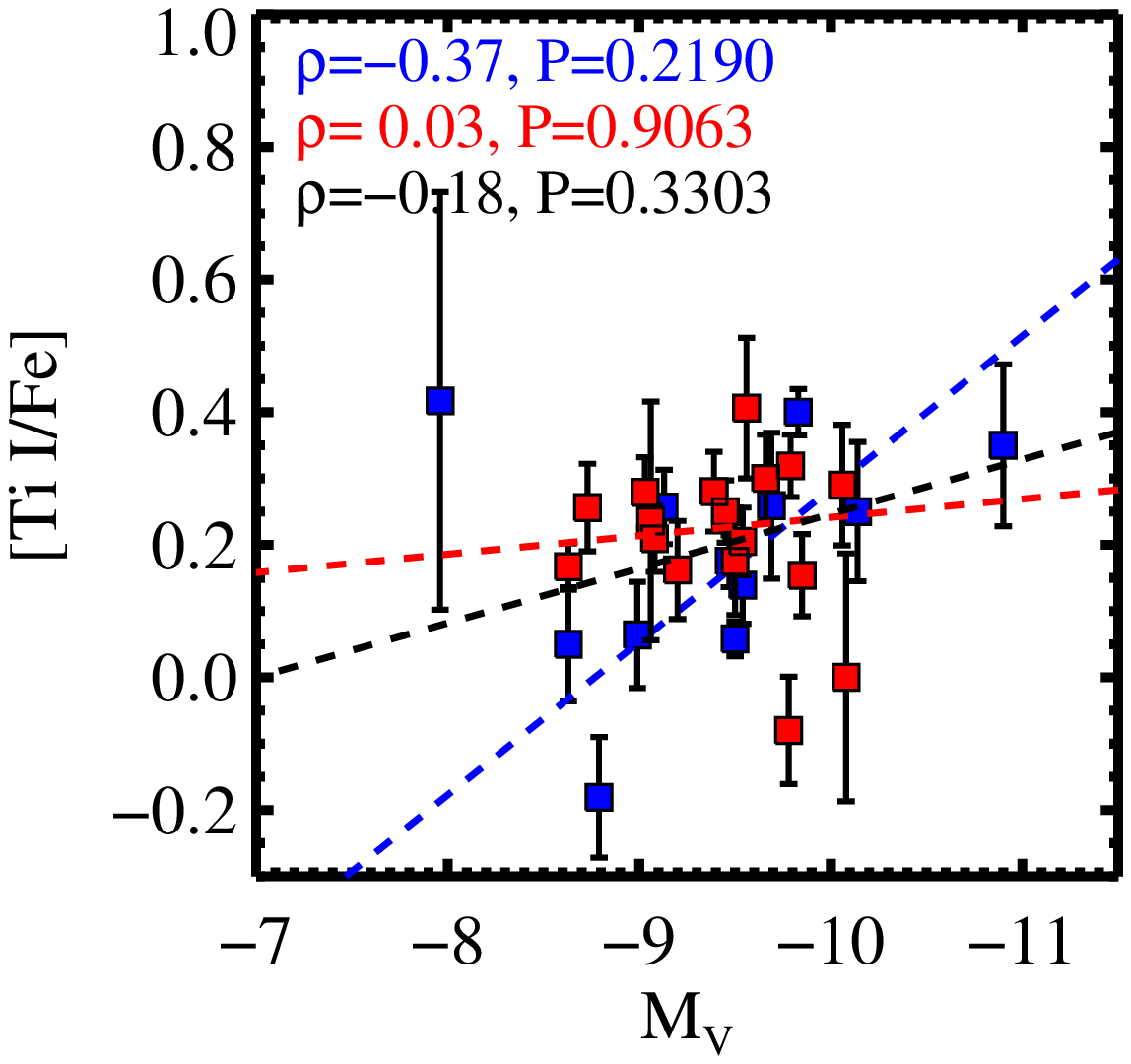}
\includegraphics[scale=0.45]{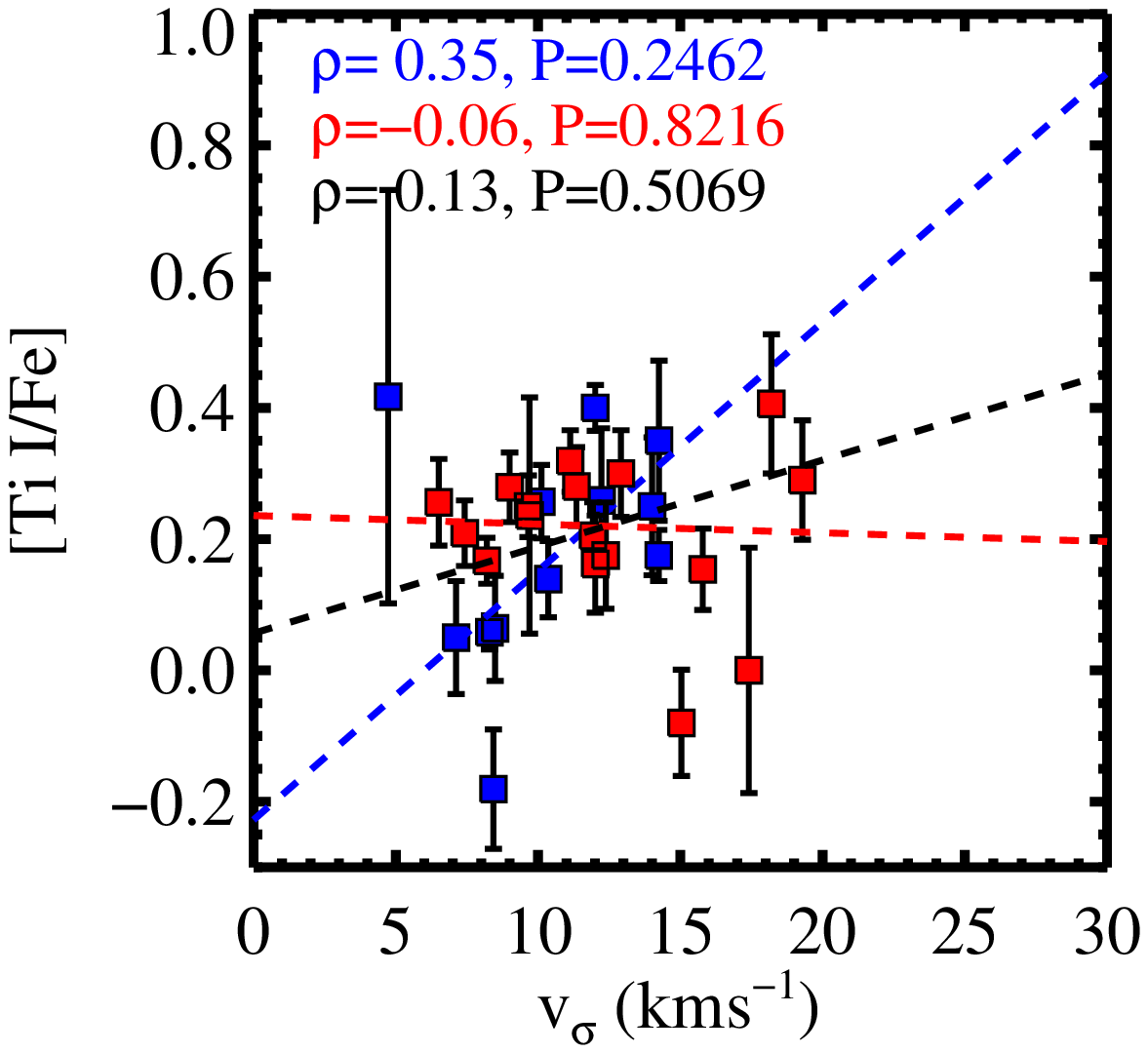}
\includegraphics[scale=0.45]{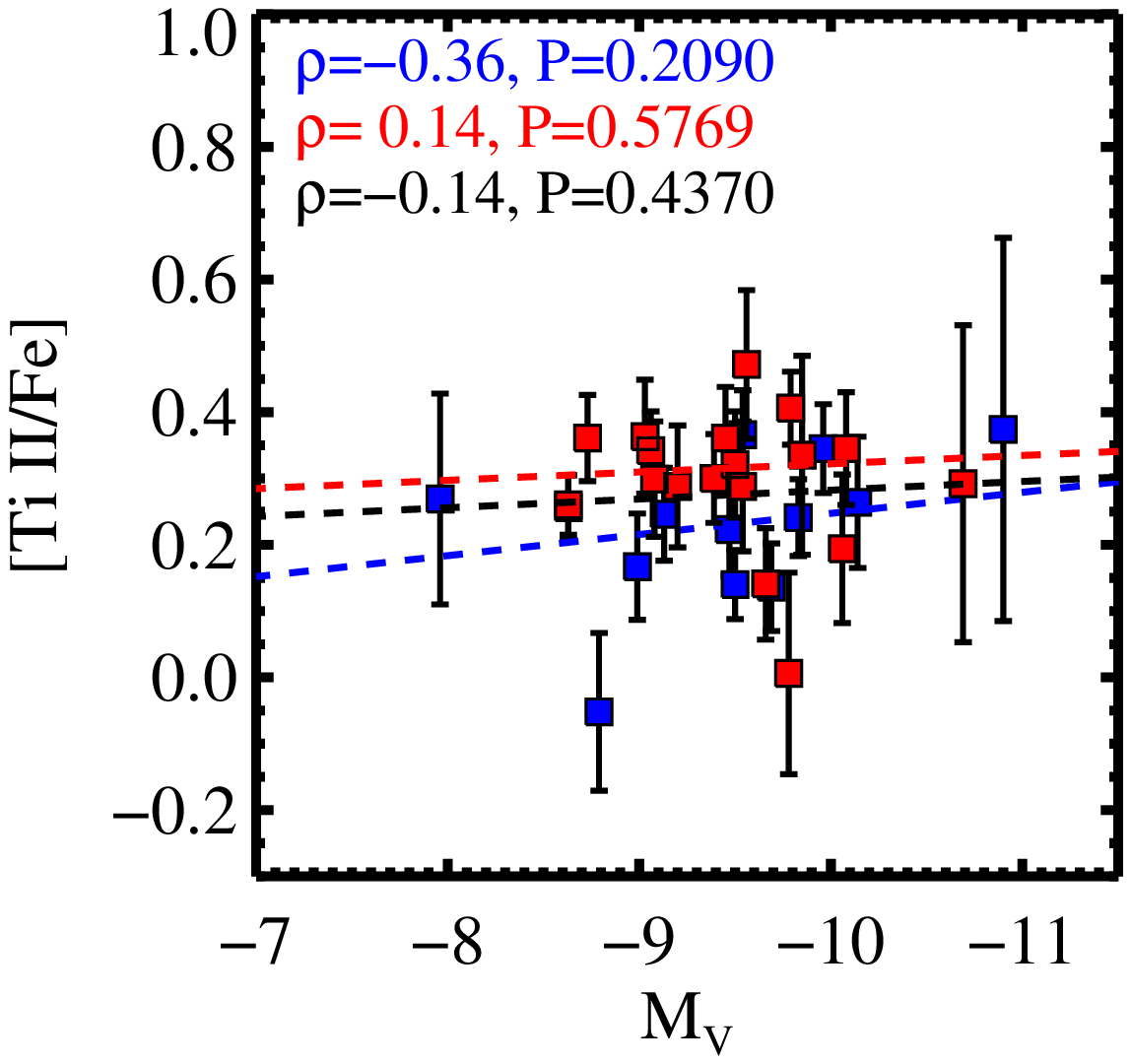}
\includegraphics[scale=0.45]{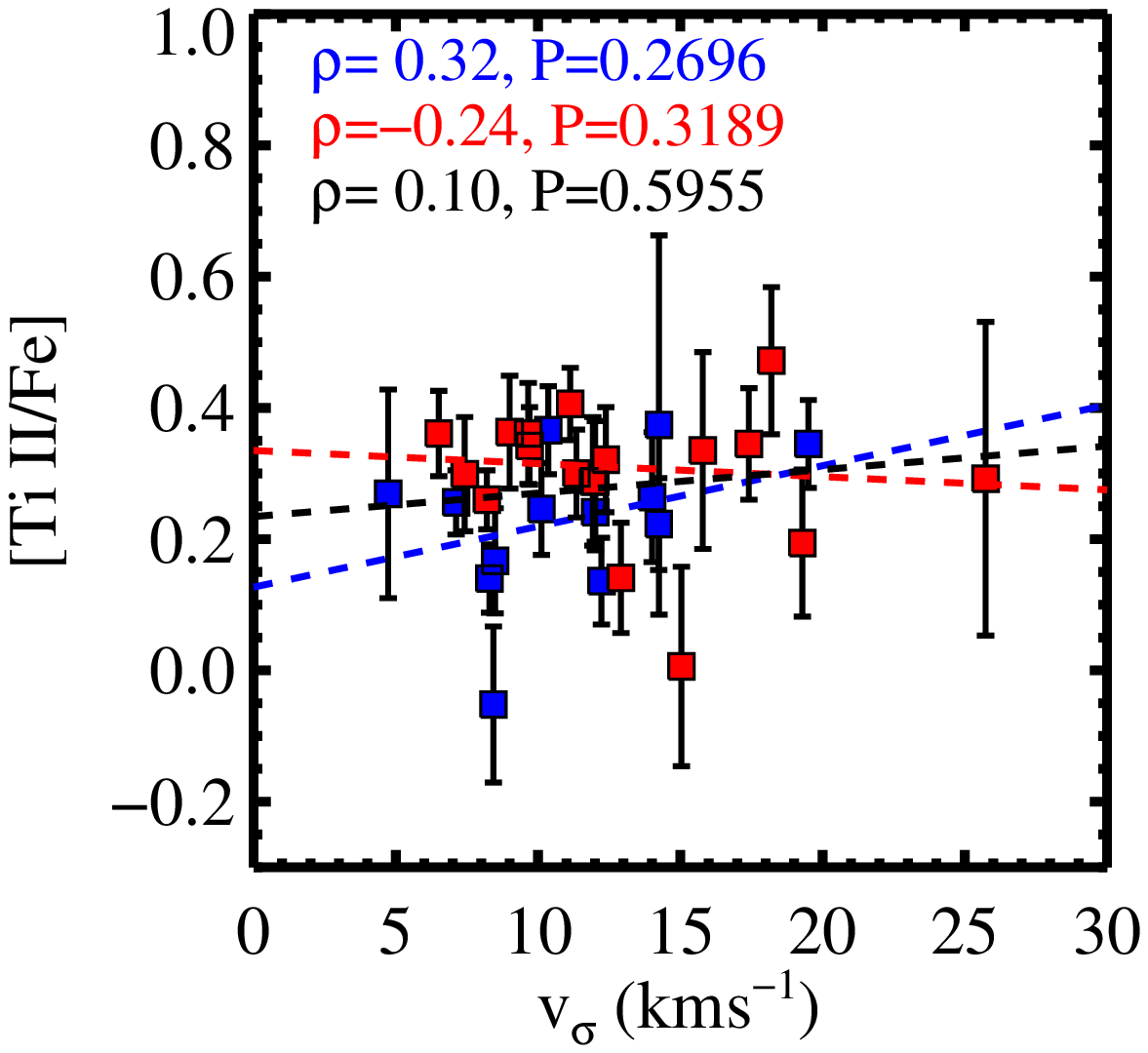}

\caption{ The same as Figure \ref{fig:alpha-trends} for [Ti I/Fe] and [Ti II/Fe]. }
\label{fig:alpha2-trends} 
\end{figure*}

\begin{figure*}
\centering
\includegraphics[scale=0.450]{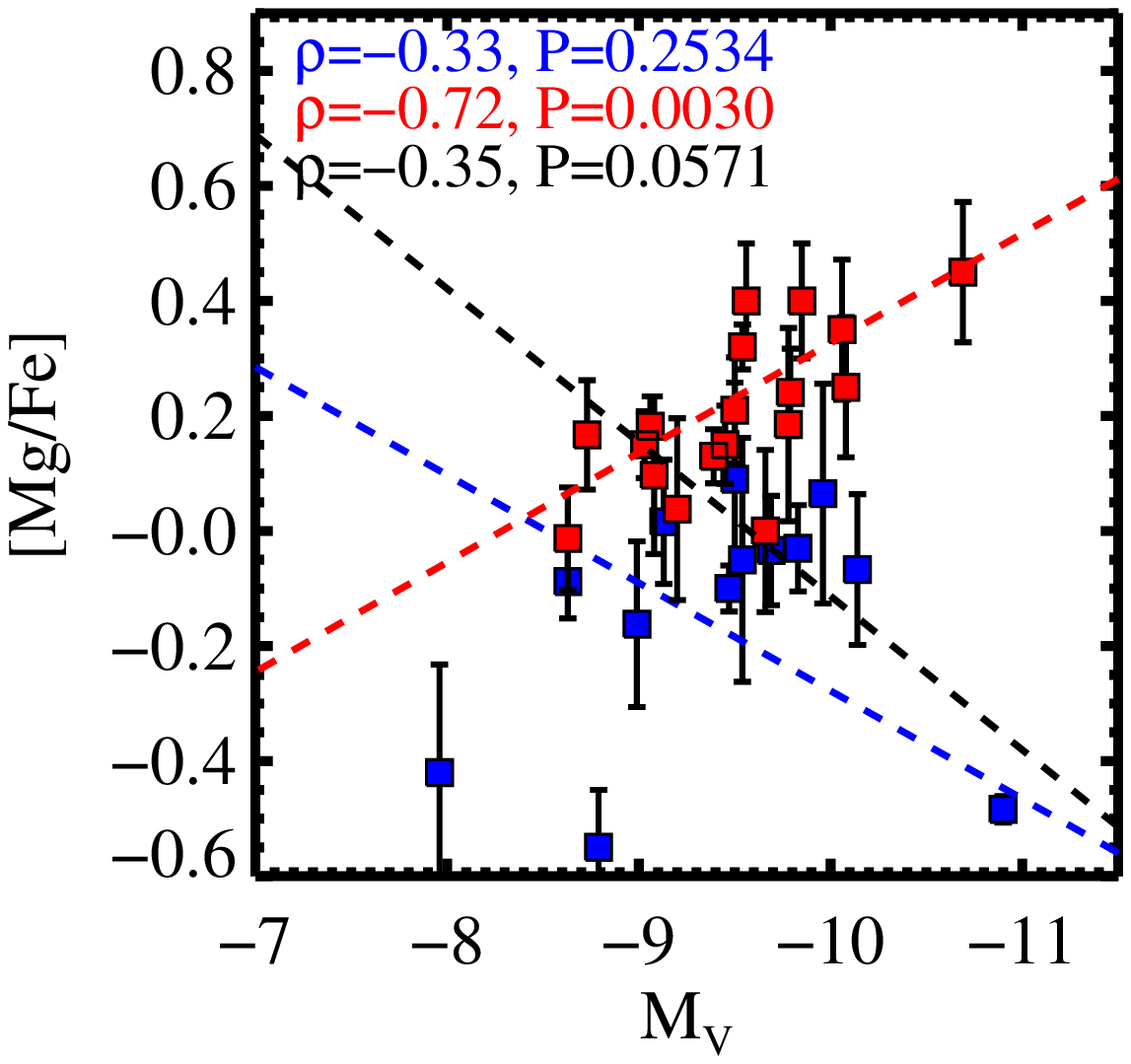}
\includegraphics[scale=0.450]{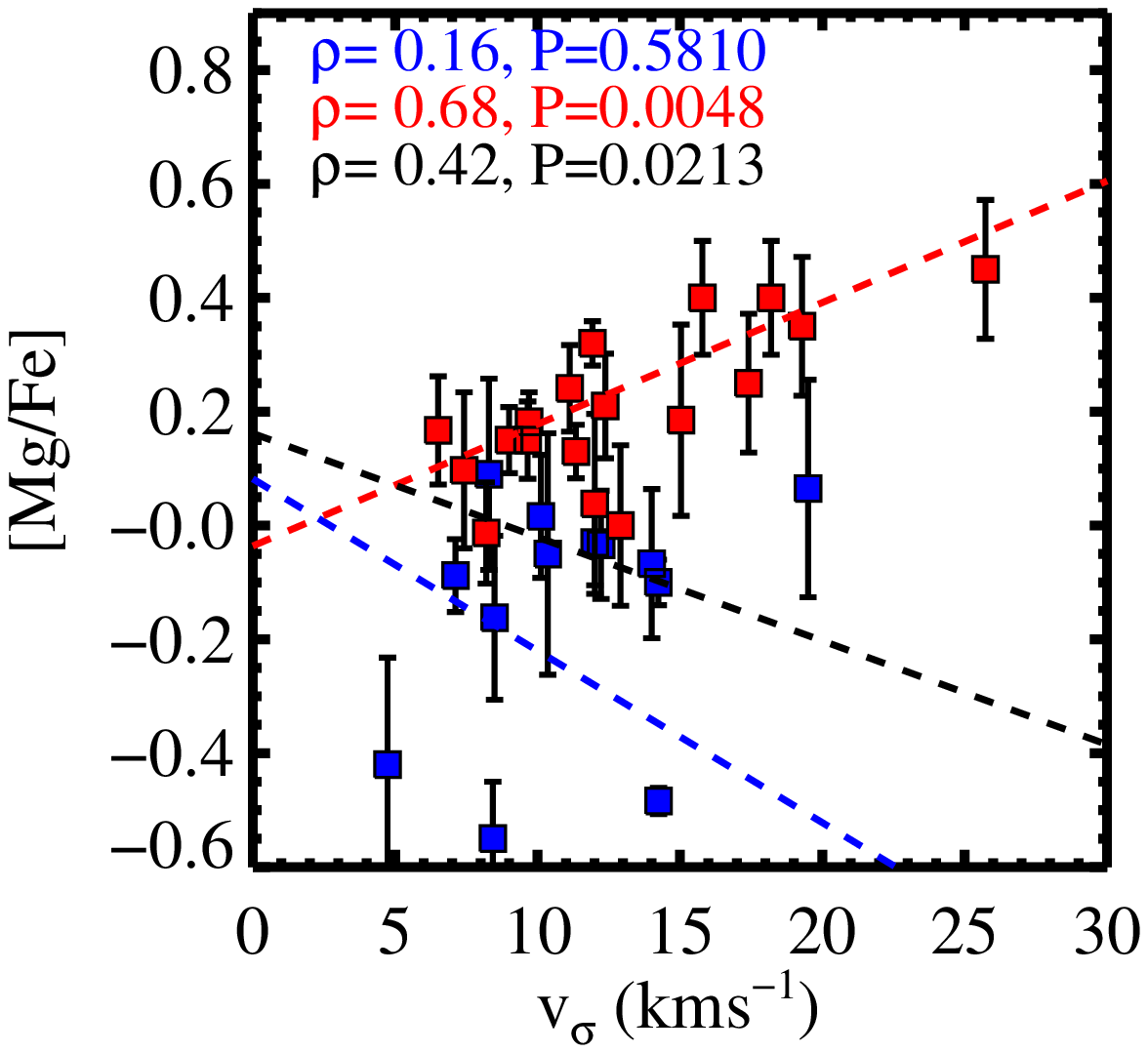}
\includegraphics[scale=0.450]{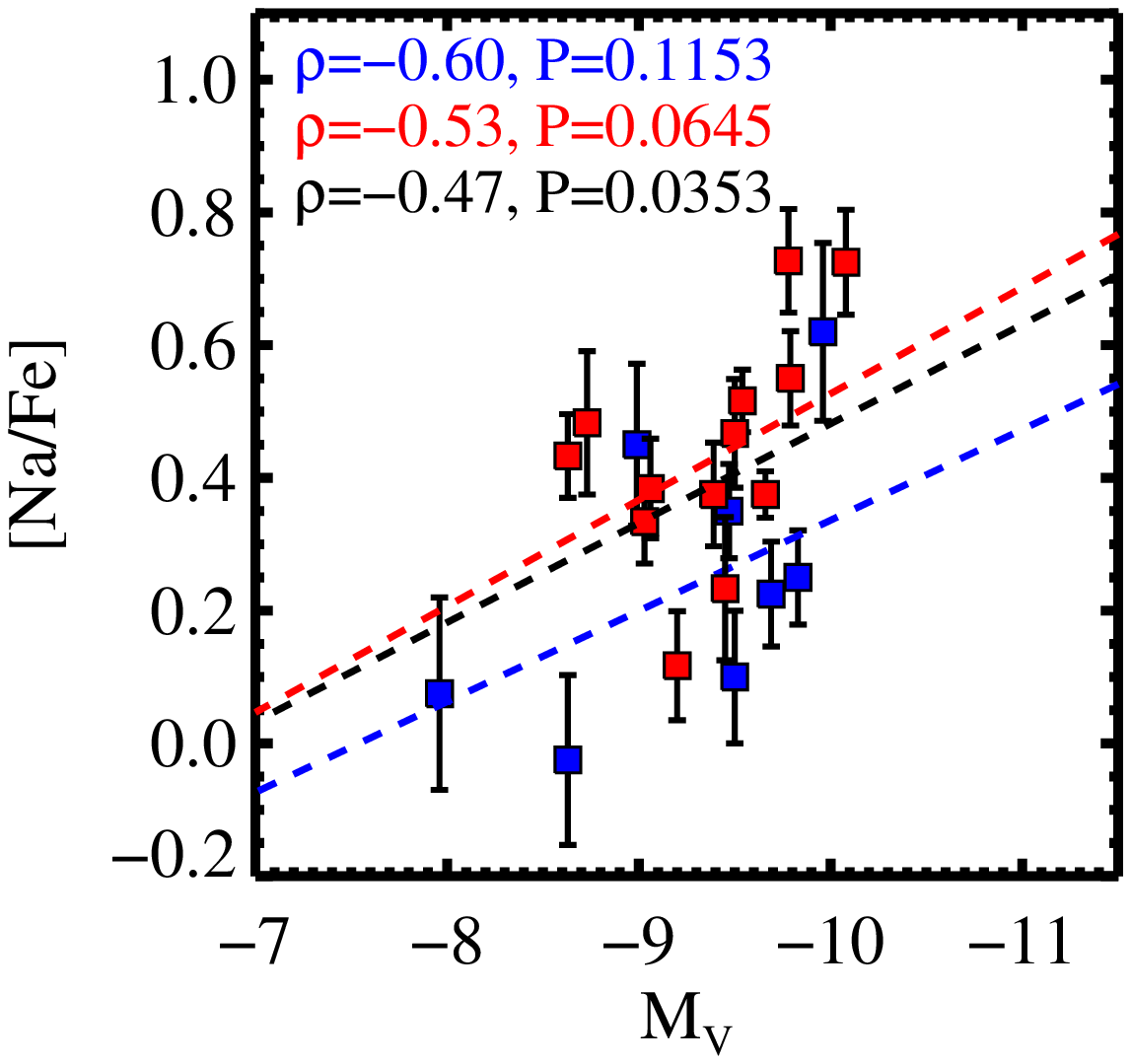}
\includegraphics[scale=0.450]{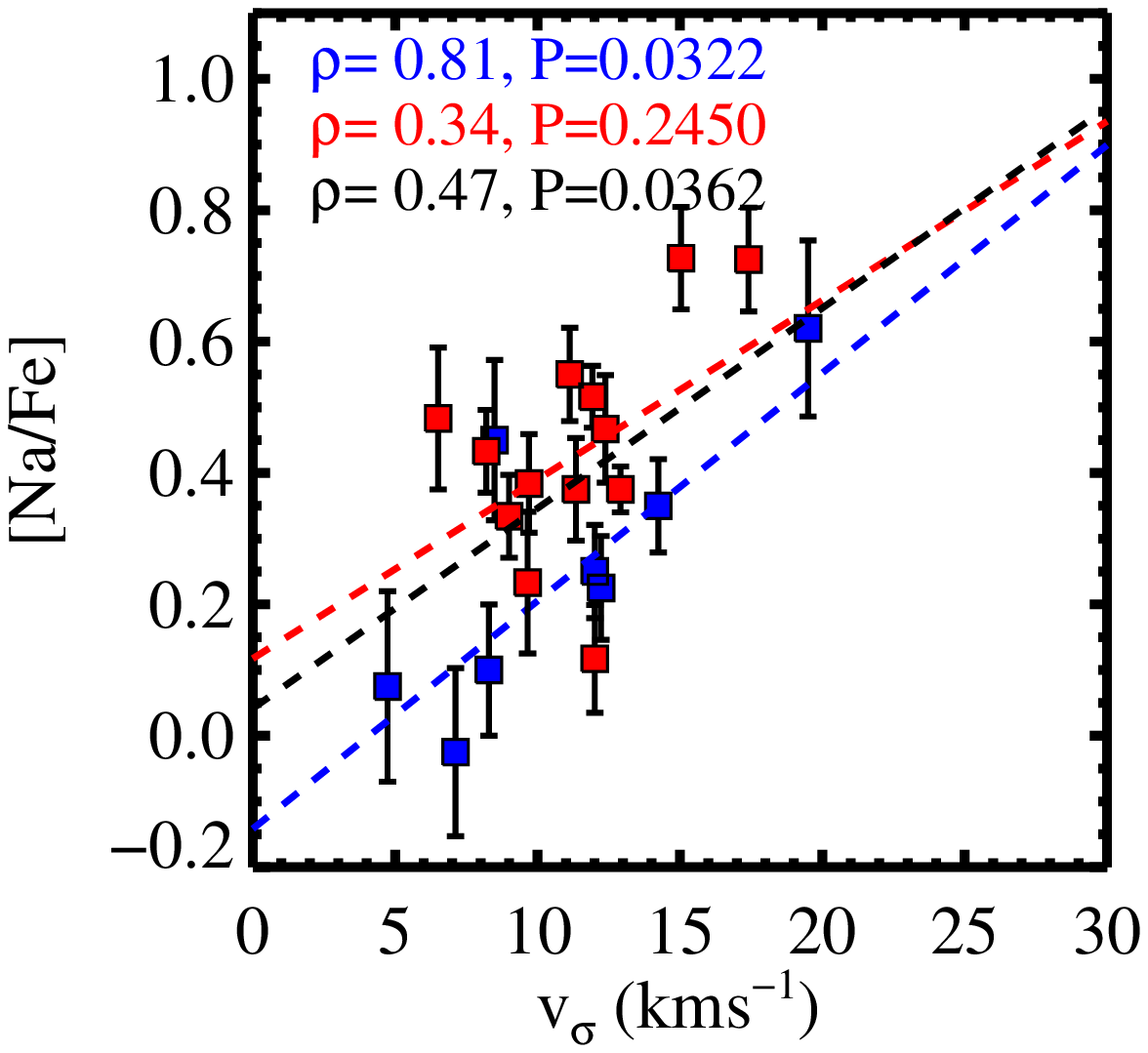}
\includegraphics[scale=0.450]{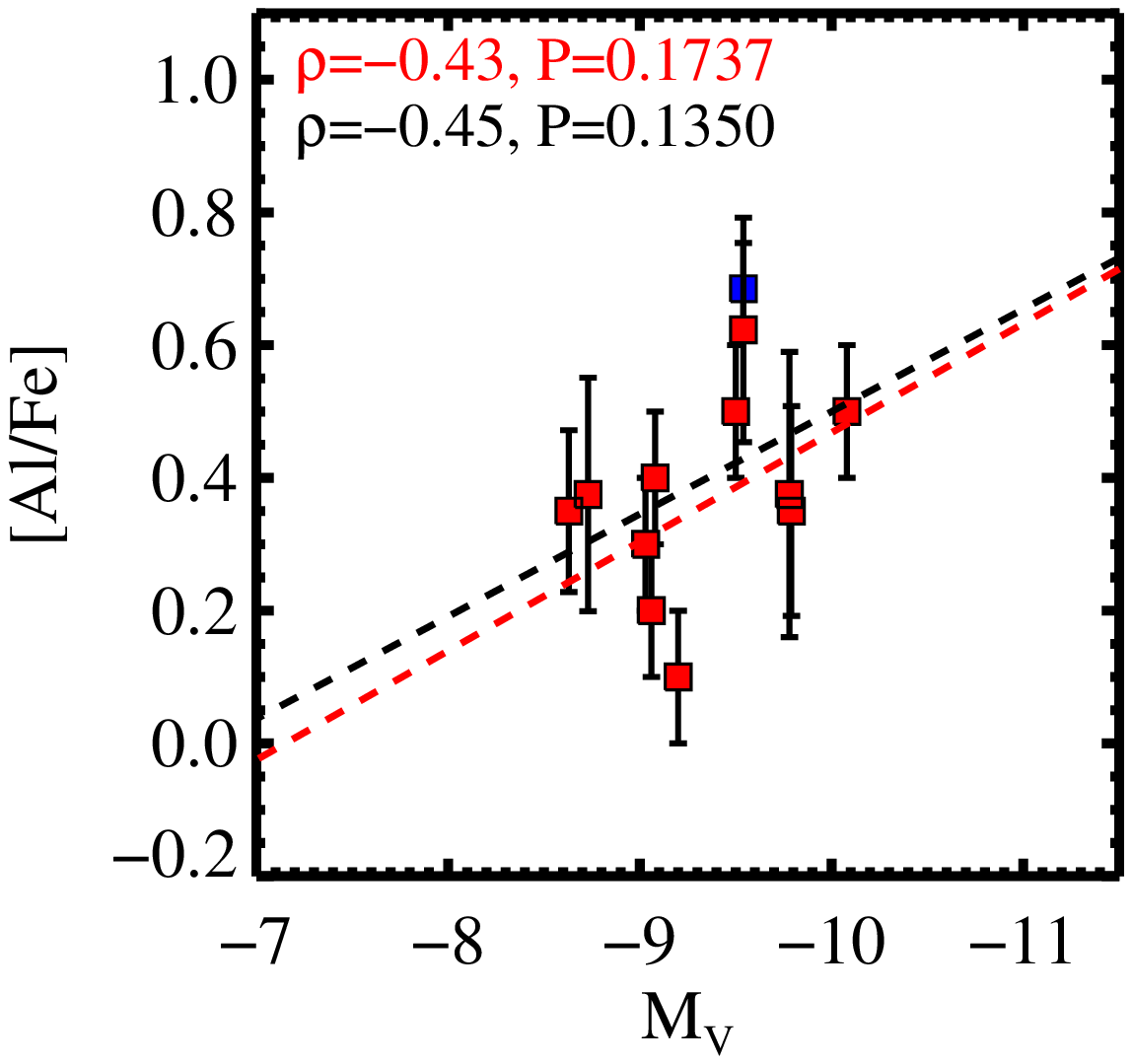}
\includegraphics[scale=0.450]{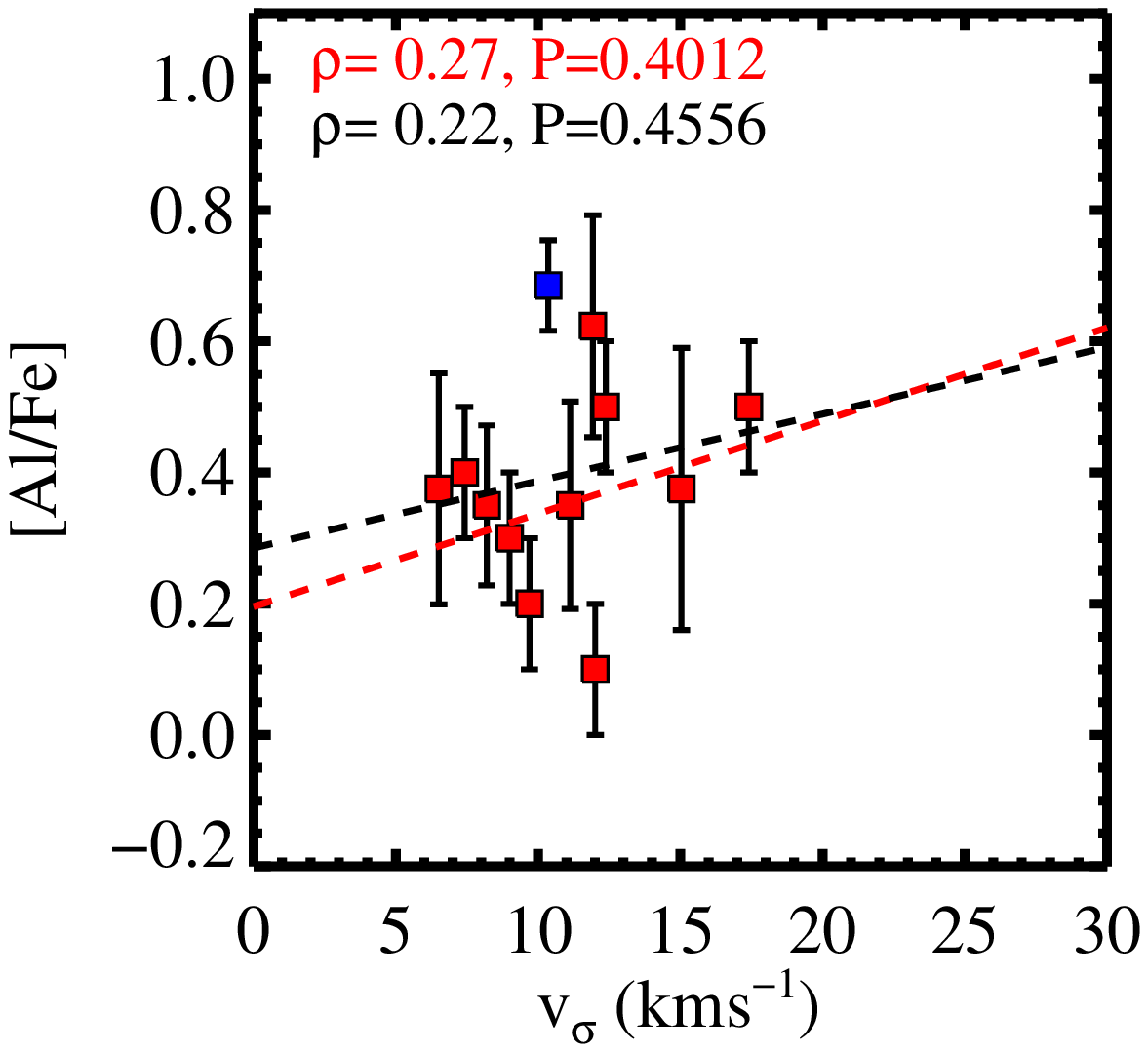}

\caption{ The same as Figure \ref{fig:alpha-trends} for [Mg/Fe], [Na/Fe], and [Al/Fe].}
\label{fig:light-trends} 
\end{figure*}

Observations of MW and M31 GCs show that the majority of old GCs have a roughly
constant M/L ratio \citep[e.g.][]{2005ApJS..161..304M,2007AJ....133.2764B,strader11}. 
 Under the assumption
of constant M/L, relationships with integrated luminosity should
reflect trends with GC mass.  Velocity dispersion should also scale
with GC mass according to the virial theorem. 
In Figure \ref{fig:alpha-trends}  we examine the behavior of the alpha element ratios as a function of absolute V magnitude (M$_{V}$) and velocity dispersion (v$_{\sigma}$), in order to investigate relationships of the abundance ratios with proxies for GC mass.  It is interesting to look for trends with mass in order to
 gain insight on GC formation, particularly the role that gas
 retention may play in multiple populations or star-to-star abundance variations.
As in Figure \ref{fig:fe-mag}, in Figure \ref{fig:alpha-trends} we show metal-poor and metal-rich GCs with blue and red symbols. We also perform  linear least squares fits to the metal-poor, metal-rich, and whole sample, which we show with blue, red and black dotted lines, respectively.    Interestingly, we find positive correlations with luminosity and velocity dispersion for the [Ca/Fe] and [Si/Fe] in the metal-poor populations. The metal-rich subpopulation also shows correlations for [Ca/Fe], but not for [Si/Fe].   For each relationship we also calculate the Spearman's rank correlation coefficient ($\rho$) and the probability (P) associated with the null hypothesis, which are noted in the upper left corner of each panel.  The Spearman's rank calculation is most useful for the [Ca/Fe] abundances, which have small, comparable errors for all of the GCs.  The Spearman's rank calculation confirms a moderate correlation with luminosity and velocity dispersion of the entire sample with [Ca/Fe] to high probability.

 We show the same analysis for the [Ti I/Fe] and [Ti II/Fe] abundances in Figure \ref{fig:alpha2-trends}.
 Both ratios have a larger scatter with M$_{V}$ and v$_{\sigma}$, and while the trend in the metal-poor population is still present, it is much less pronounced. The Spearman's rank correlation calculation does not support a strong relationship of either Ti I or Ti II with these quantities.

The same analysis for the light element ratios is shown in Figure \ref{fig:light-trends}.  We find more dramatic correlations, particularly for the metal-rich  subpopulation for   [Mg/Fe], [Na/Fe], and possibly [Al/Fe].  Moreover, we find that the trends with luminosity and velocity dispersion for the metal-rich subpopulation are offset to higher abundance ratios than the metal-poor subpopulation in [Mg/Fe], although we also note that the behavior of the metal-poor subpopulation is skewed by the three GCs with the  lowest [Mg/Fe], which span the entire range in GC mass.  Accordingly, the Spearman's rank correlation coefficient is largest, with the smallest probability for the null hypothesis, for the metal-rich GCs and [Mg/Fe].  The [Na/Fe] abundances show more moderate correlations  with luminosity and velocity dispersion.  The [Al/Fe] abundances also appear to correlate with luminosity and velocity dispersion, although we have fewer measurements for Al I, so the trends are less robust.

We also found trends of [Na/Fe] and [Al/Fe] with cluster mass  for the LMC in \citetalias{paper4}, however in that work we could not determine if the correlations were driven primarily by the age or the mass of the clusters.  In this case, our M31 GC sample is almost entirely old, so we  do find that for old GCs in  M31 the [Na/Fe] and [Mg/Fe]  abundances correlate with proxies for cluster mass.

Our results and other recent GC IL work have interesting, although perhaps puzzling, implications for models of GC formation and GC self enrichment.    A correlation of integrated light GC nitrogen abundances with mass  was recently reported by  \cite{schiavon13}, for a larger sample of metal-rich ([Fe/H]$>-1$) M31 GCs using low resolution high SNR spectra and line indexes.  However, \cite{schiavon13} did not find correlations of GC mass with Mg and Ca, contrary to what we find  here, but this could be because our high resolution measurements are more sensitive to small changes in the Mg and Ca features than the broader line indexes.

One question is whether abundance correlations with GC mass are related to multiple generations of star formation in GCs, and a second is  whether the correlations give insight into the polluters that caused the present day star-to-star abundance variations in GCs.  
There are empirically motivated theoretical predictions for light element correlations with GC mass \citep{carretta10,2012ApJ...758...21C}, that tie these two questions together.  In particular, the models of  \cite{carretta10} and \cite{2012ApJ...758...21C} favor early AGB stars as the polluters that cause the Na-O and Mg-Al anticorrelations. \cite{schiavon13} propose that   the nitrogen-mass correlation and lack of Mg- and Ca-mass correlations in their observations support AGB stars as the polluters \citep[e.g.][]{renzini08,2012MNRAS.423.1521D}. 
The Na-  and Al-mass correlations that we find here may also be consistent with  an increased fraction of ejecta from AGB stars retained in more massive GCs.
   On the other hand, it is not clear that this would explain the Mg- and Ca-mass correlations that we find, since Mg is depleted in AGB stars and Ca should not be affected.     However,  we find the most extreme and lowest [Mg/Fe] abundances  in the metal-poor subpopulation only, which do {\it not} seem to correlate with the proxies for GC mass. Therefore,  the metal-poor subpopulation may still be consistent with self enrichment by AGB ejecta. 
   
The Mg-mass correlation in the metal-rich subpopulation  may be indicative of a different form of self-enrichment, which may be more related to the  Ca-mass correlation. The metal-rich subpopulation abundance correlations may   instead be related to the self-enrichment scenario that forms the  blue tilt, or mass-metallicity relationship, seen in the most massive GCs in early type galaxies.  Further modeling is needed to determine how the abundance ratio trends we observe may be related to different GC self-enrichment scenarios.

\subsection{Behavior with Galactocentric Radius}
\label{sec:radius}

\begin{figure}
\centering
\includegraphics[angle=90,scale=0.35]{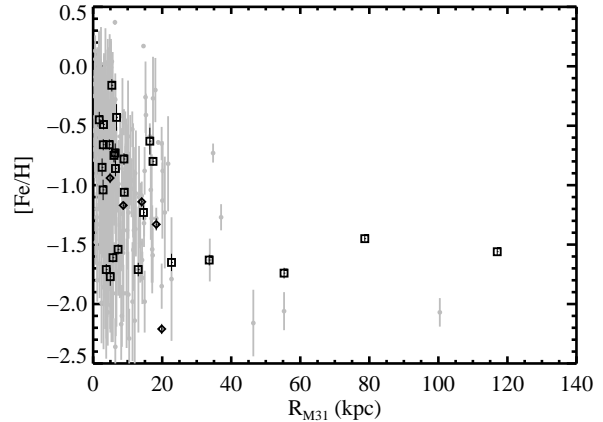}
\caption{ Behavior of [Fe/H] with galactocentric radius from M31 (R$_{\rm M31}$).  Squares show the new M31 GCs from this work and diamonds show the GCs from \citetalias{m31paper}.  Grey points show the low resolution spectra [Fe/H] measurements of the larger sample of \cite{caldwell11}, as well as the large radii GCs of \cite{alvesbrito09}. }
\label{fig:fe-r} 
\end{figure}

We show the [Fe/H] behavior with galactocentric radius from M31 in Figure \ref{fig:fe-r}, using radii tabulated in the Bologna Catalog \citep{bolognacat}. For context, we also show the larger sample of low resolution [Fe/H] measurements of \cite{caldwell11}, as well as the low resolution measurements of outer halo GCs from \cite{alvesbrito09}.  We find a spread in [Fe/H] for GCs within R$_{\rm M31}$$\sim$20 kpc, and a fairly constant [Fe/H] for GCs at R$_{\rm M31}$$>$20 kpc.  The mean value for the 5 GCs in our sample that lie outside 20 kpc from M31 is [Fe/H]$=-1.63 \pm 0.10$, which is consistent with the metallicity of outer MW halo GCs of [Fe/H]$=-1.7$ \citep[][ 2010 revision]{1996AJ....112.1487H}.

A nearly constant [Fe/H] for M31 outer halo GCs has also been found in previous works \citep{perrett02,2007ApJ...655L..85M,alvesbrito09,huxor11}.  Like our analysis, \cite{alvesbrito09} found  a flat metallicity distribution for GCs outside 30 kpc, with a mean metallicity  of  [Fe/H]=$-1.6$, which was estimated   from line indexes. Using CMD derived metallicities,  \citep{huxor11} found an outer halo metallicity of [Fe/H]$\sim-1.9$, which is lower than what we find, but the offset can probably be ascribed to a systematic offset between spectroscopic and photometric measurements. 

Interestingly, \cite{mackey13} studied two newly discovered GCs from the PAndAS survey \citep{huxor14}, which have projected galactocentric radii of $\sim$85 kpc from M31. From photometric CMD-based metallicities, \cite{mackey13} estimate that these GCs are both more metal-rich than other halo GCs, with [Fe/H]$\sim-1.35$, and possibly a few Gyr younger in age.  However these GCs are projected onto major stellar substructure in the M31 halo, and have radial velocities offset from the systemic velocity of M31, which makes it likely that these GCs were recently accreted into the M31 halo from a satellite galaxy.   \cite{mackey13} suggest that these M31 GCs are analogs to the ``young" MW halo GCs of \cite{zinn93}, which later studies  proposed were accreted at late times \citep[e.g.][]{1995AJ....109.2533D}.

Thus, evidence has been mounting that,  like the MW, M31 has both a ``young" and ``old" halo, where the young GCs are likely GCs that were accreted  late times and which may be younger and show chemical similarities to the dwarf galaxies in which they originated.   The young and old halo GCs are likely related to the stream-like and smooth halo stellar populations, which are also associated with  metal-rich ([Fe/H]$\sim-0.7$) and metal-poor ([Fe/H]$\sim-1.5$) abundances, respectively \citep{ibata14}.    The old halo GCs appear to be uniformly old and metal-poor,  consistent with the `proto-galactic' fragment formation scenario of old MW GCs first popularized by \cite{searlezinn}.  Our current sample of M31 GCs with detailed chemical abundances appears to be drawn from the old halo population, in that all of our GCs at R$_{\rm M31}$$>$20 kpc are old and metal-poor. 
A more complete high resolution census of the outer halo GCs will be interesting to study the younger  metal-rich GCs at large projected radius to determine their detailed abundance patterns.

\begin{figure*}
\centering
\includegraphics[angle=90,scale=0.3]{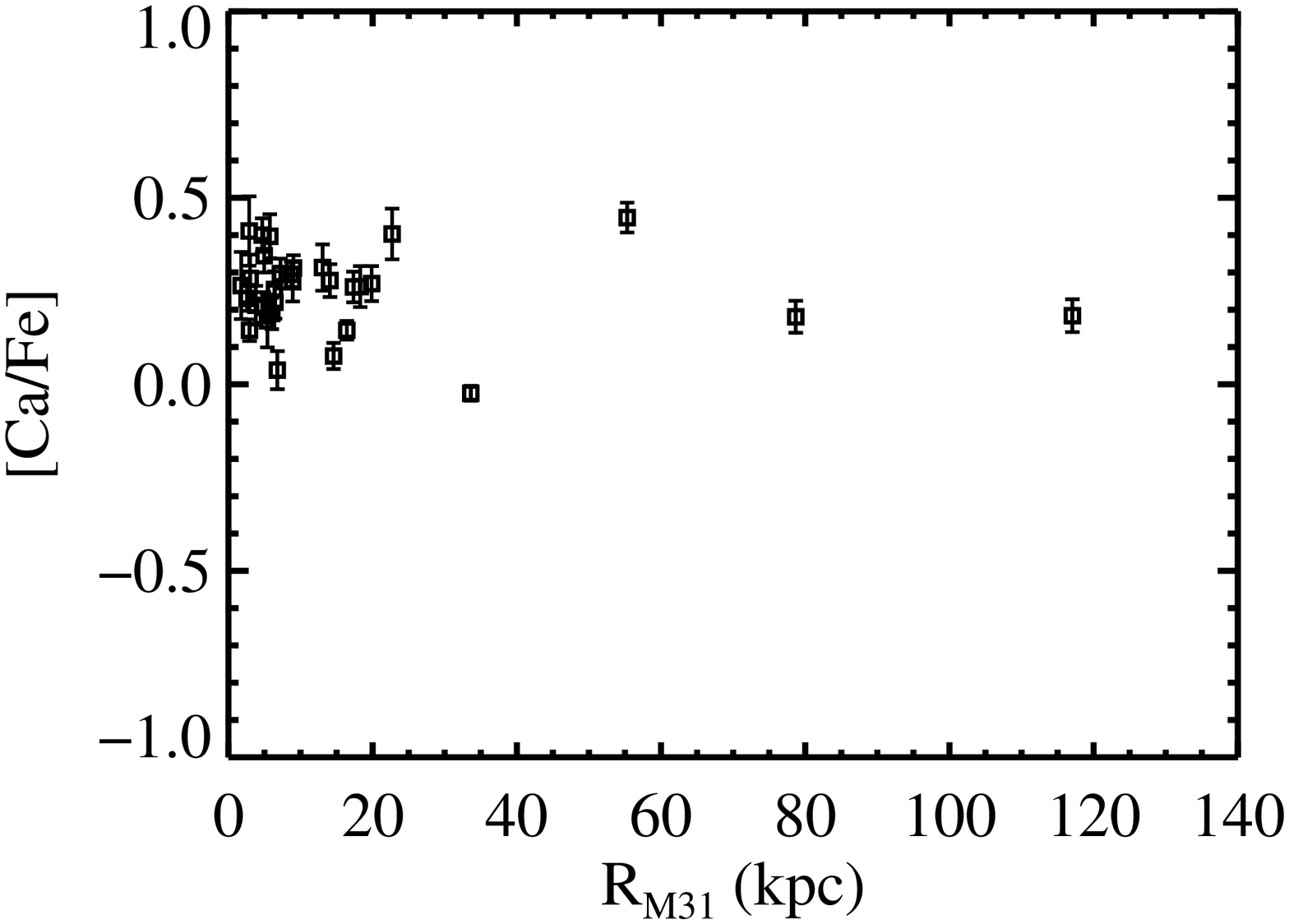}
\includegraphics[angle=90,scale=0.3]{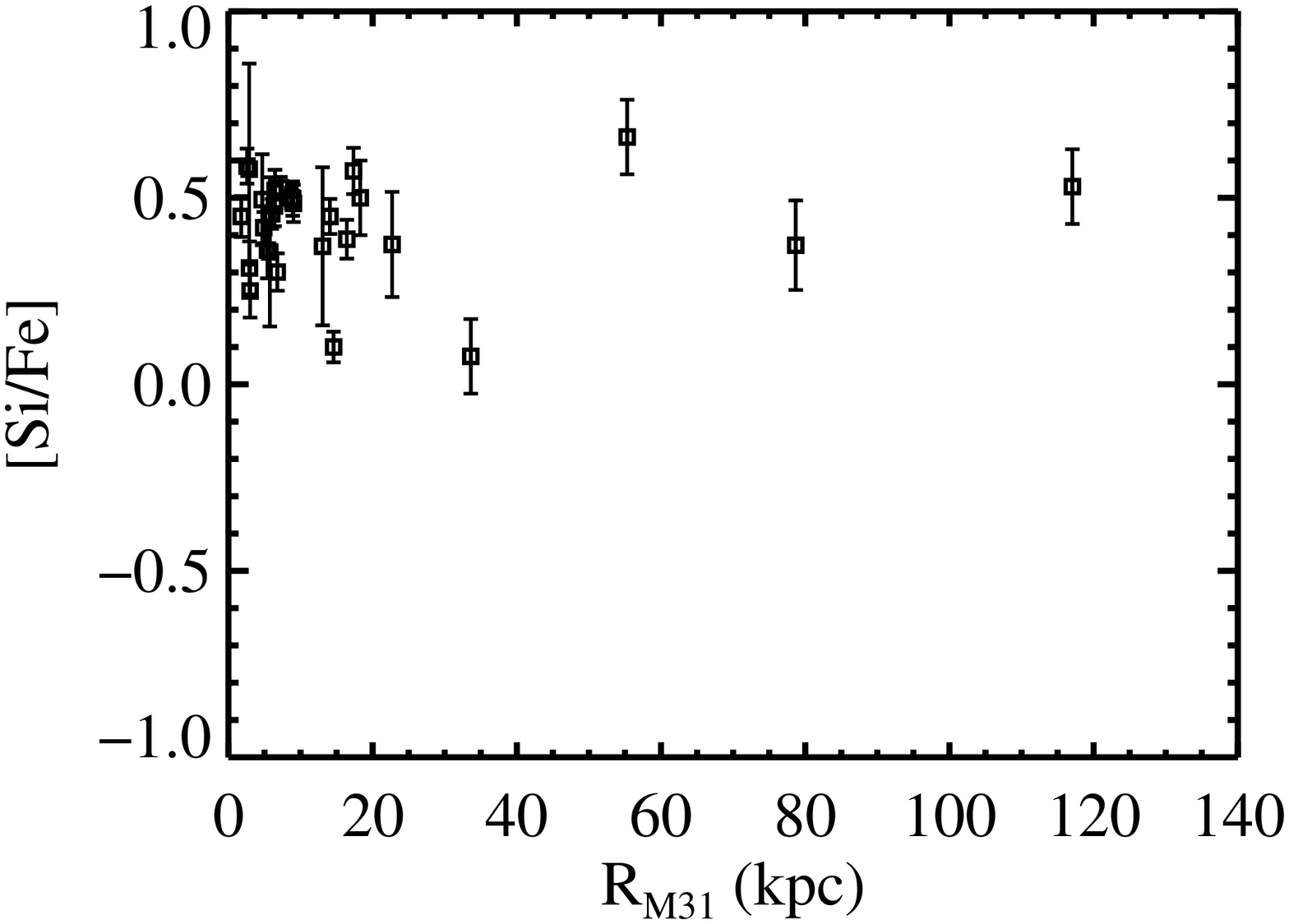}
\includegraphics[angle=90,scale=0.3]{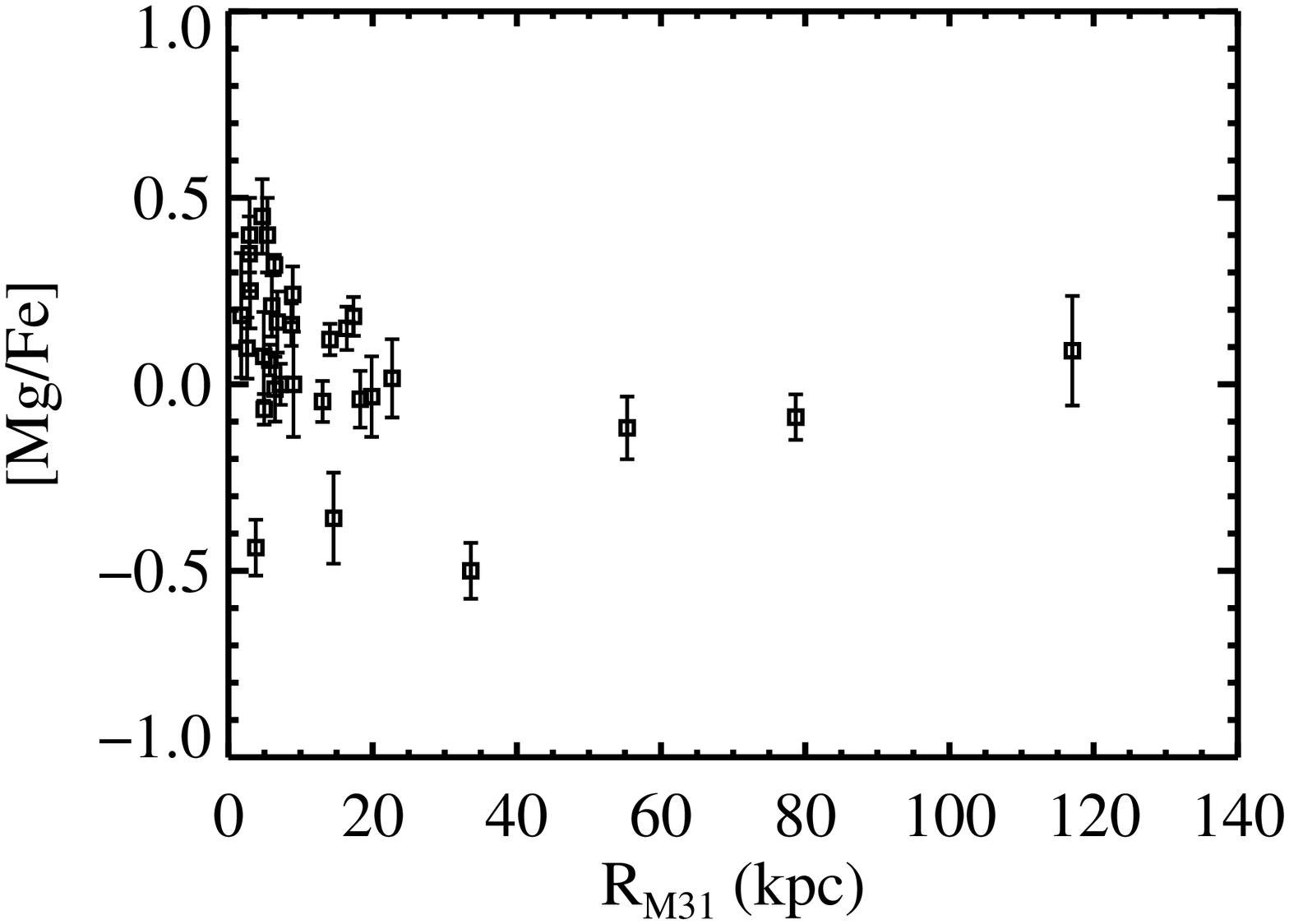}
\includegraphics[angle=90,scale=0.3]{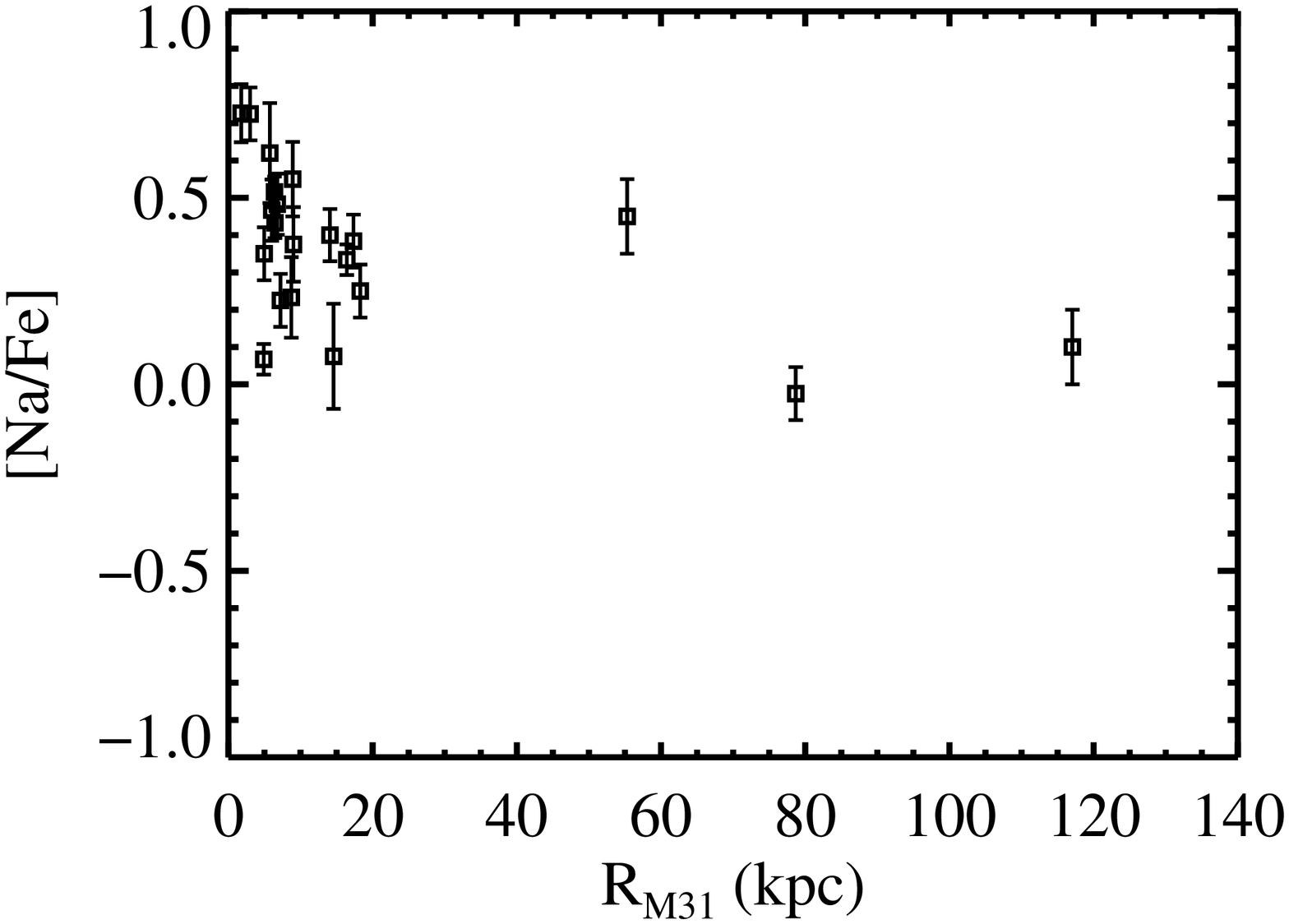}
\caption{ Behavior of abundance ratios [Ca/Fe], [Si/Fe], [Mg/Fe] and [Na/Fe]  with galactocentric radius from M31 (R$_{\rm M31}$).}
\label{fig:elements-r} 
\end{figure*}

In addition to the behavior of [Fe/H] with galactocentric radius, we are able to examine the behavior of the abundance ratios with radius, as shown for [Ca/Fe], [Si/Fe], [Mg/Fe] and [Na/Fe] in Figure \ref{fig:elements-r}. 
In the current sample, there are no compelling trends of the abundance ratios with radius, although it will be interesting to increase the sample size of GCs at large radii in the future in order to further investigate the chemical composition of the  outer halo of M31.

\subsection{Star Formation and Accretion History of M31}
\label{sec:capture}

As presented in \textsection \ref{sec:alpha}, we find that the alpha element abundances of this sample of  M31 GCs strongly resemble the average properties of  MW GCs and field stars. We find a plateau of [$\alpha$/Fe]$\sim+0.3$ that extends from [Fe/H]$=-2$ to at least [Fe/H]$=-0.7$, which indicates that at early times the chemical enrichment of M31 was dominated by the ejecta of SN II.  There is some indication from the [Ca/Fe], [Si/Fe] and [Ti/Fe] abundances that the highest metallicity GCs in our sample begin to decline at high [Fe/H], which may indicate the increasing contribution of SN Ia enrichment when these GCs formed. Continued study of the highest metallicity GCs in M31 would be interesting to confirm this behavior.  

A handful of the GCs in our sample have a scatter in the [$\alpha$/Fe] ratios obtained from different elements. This may indicate subtle differences in the ISM mixing  when the  M31 GCs were formed.At least one GC in our sample, G002,  has a noticeably different [$\alpha$/Fe] enrichment pattern than the other GCs at similar [Fe/H].
 In particular, the alpha element abundances of this
cluster are lower than for other low-metallicity GCs  ([Fe/H]$\sim-1.6$),
which is similar to the situation of the MW GC Rup 106 (\citep{1997AJ....114..180B,2013ApJ...778..186V}, [Fe/H]$\sim-1.5$ and [$\alpha$/Fe]$\sim0.0$. Because of the different alpha element abundance
pattern, it is speculated that Rup 106 formed outside the MW and was
later accreted. We compare the abundance ratios for Rup106 from the recent
analysis by Villanova et al. (2013) to both G002 and a more ?halo-
like? GC, B240, in Figure 34 for the elements in common (Fe, Ca, Si,
Ti I, and Mg).  With the exception of [Mg/Fe], it is clear that the
abundance pattern of G002 closely resembles the abundance pattern of
Rup 106, and neither Rup 106 or G002 are similar to B240.  In general, lower [$\alpha$/Fe] abundances are found in
dwarf galaxies than in the MW , although dwarf galaxies can also show
a wide range in [$\alpha$/Fe] within themselves. For example, LMC GCs
have mean [$\alpha$/Fe] ranging from $+0.0$ to $+0.4$ at [Fe/H]$<-1.5$
\citep{johnson06,mucc09, mucc10old,paper4};  Fornax GCs have [$\alpha$/Fe]$\sim+0.15$ \citep{letarte}, and the possible Sagittarius member clusters Terzan 8, Arp 2, and M54 have [$\alpha$/Fe] ranging from $+0.2$ to $+0.4$ \citep{mottini08,carrettaM54,carretta14}.  We therefore suggest that G002 was
likely accreted into the M31 GC system at late times from a massive
dwarf galaxy.
 We will  further investigate similarities of the abundances of  heavy elements in G002 with GC like Ruprecht 106 and those in dwarf galaxies in a future paper.

\begin{figure}
\centering
\includegraphics[scale=0.5]{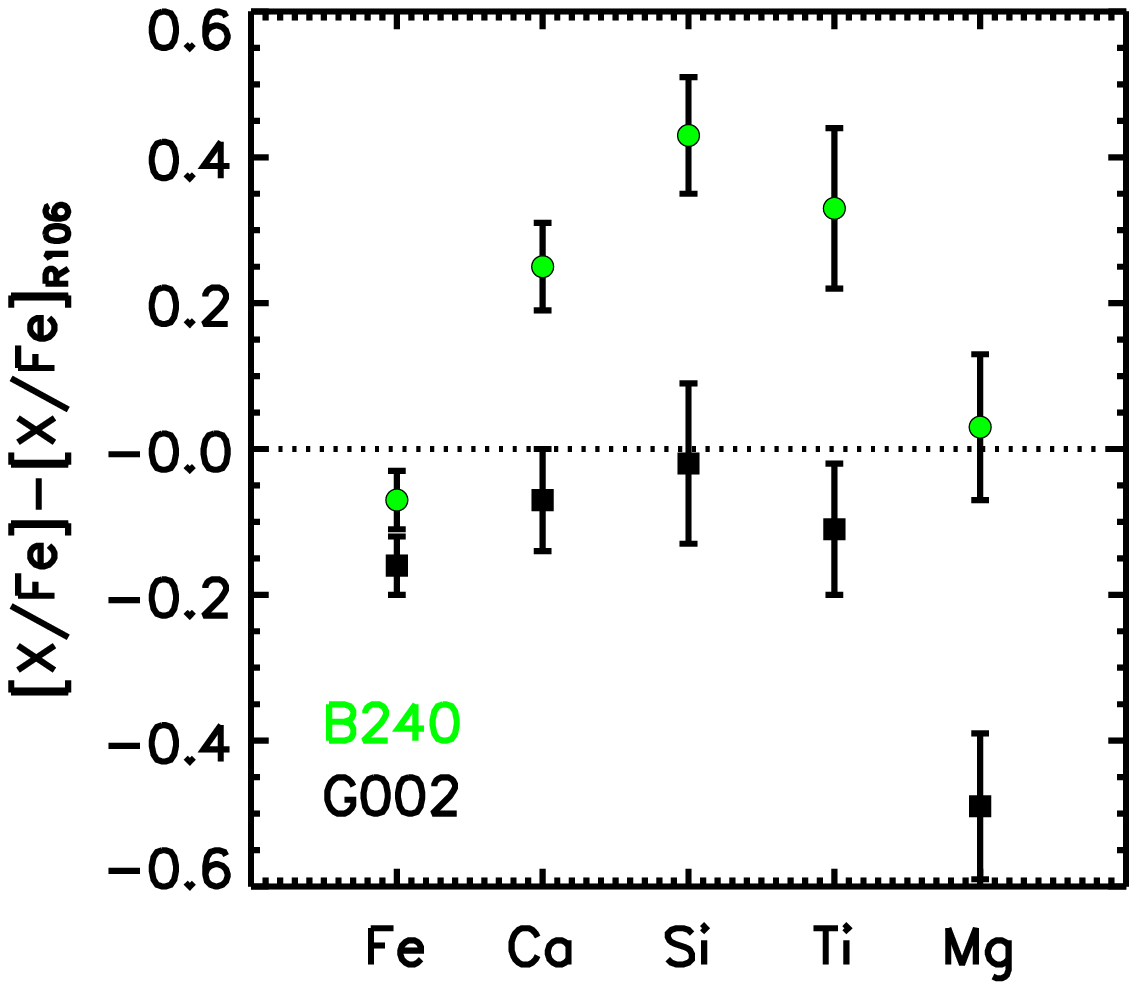}

\caption{ A comparison of the abundance pattern of G002 (black squares) and B240 (green circles) to that of Ruprecht 106 from \cite{2013ApJ...778..186V}.  The abundance ratios ([X/Fe]) are compared for the elements (X) shown, with the exception of Fe, for which we show [Fe/H].   We have adjusted the abundances of  \cite{2013ApJ...778..186V} so that all abundance ratios use the same solar abundance pattern.  For Ti, we compare our Ti I abundances, since the [Ti/Fe] in \cite{2013ApJ...778..186V} is measured solely from Ti I lines. }
\label{fig:rup106} 
\end{figure}

\begin{figure*}
\centering
\includegraphics[scale=0.450]{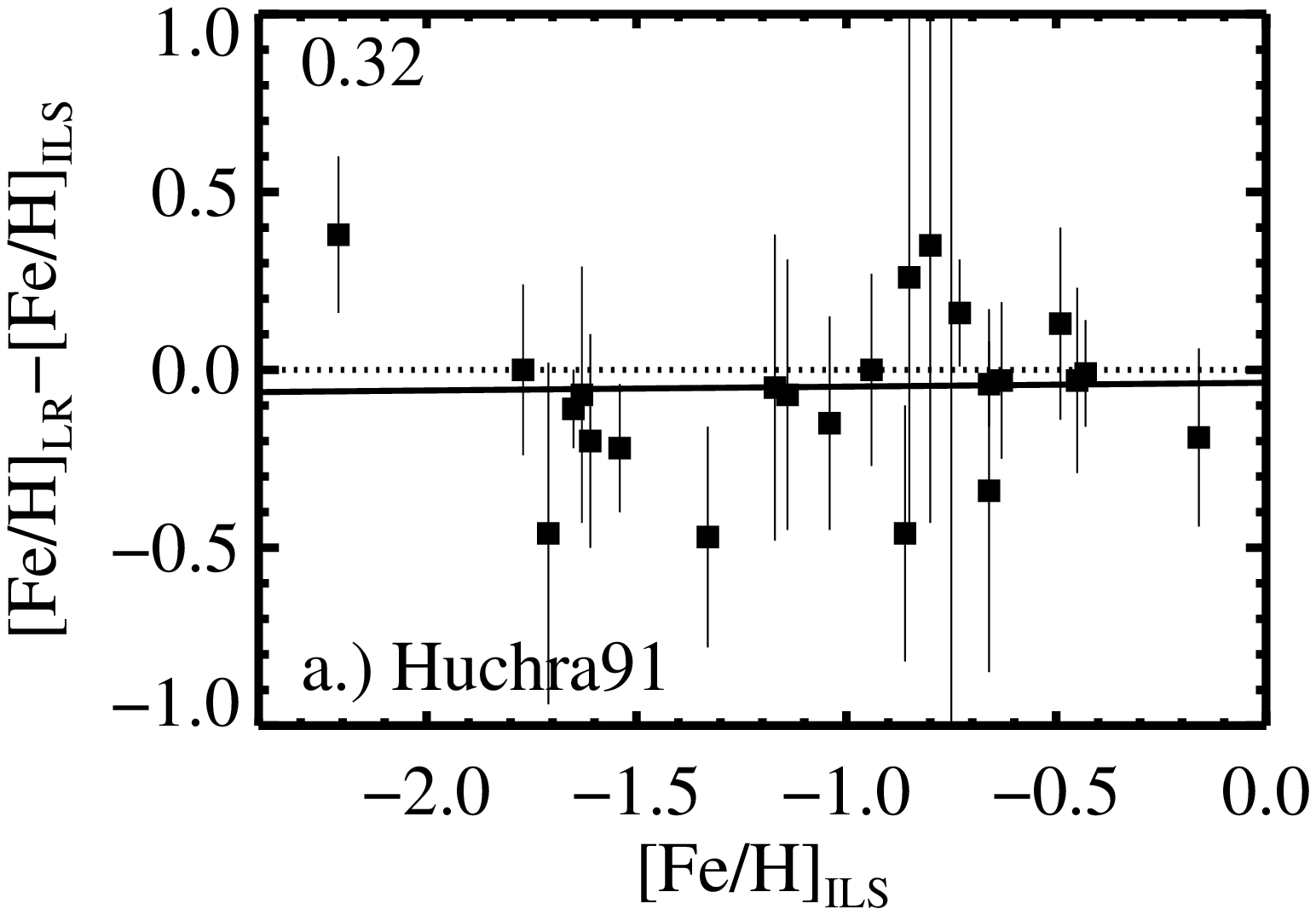}
\includegraphics[scale=0.450]{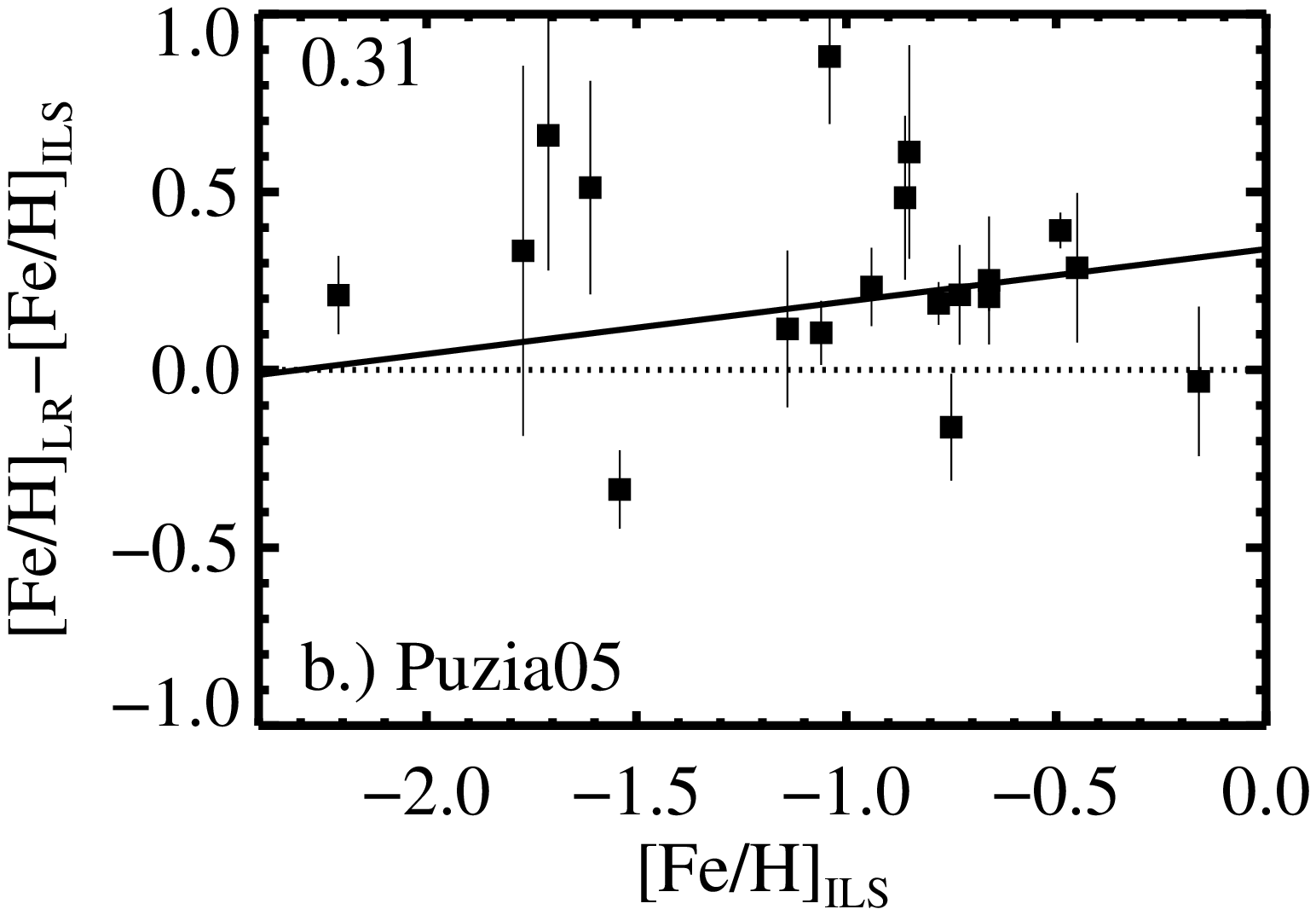}
\includegraphics[scale=0.450]{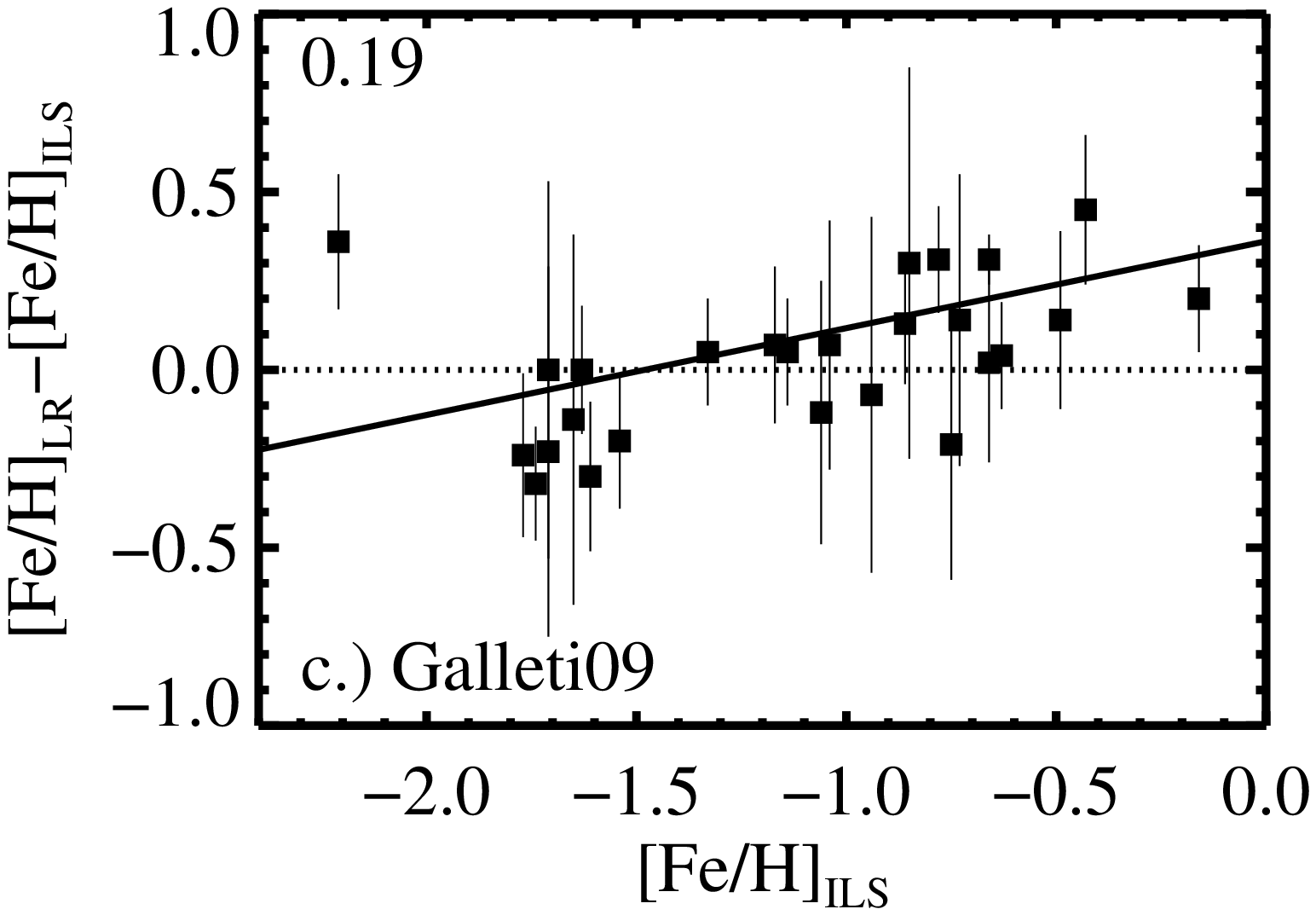}
\includegraphics[scale=0.450]{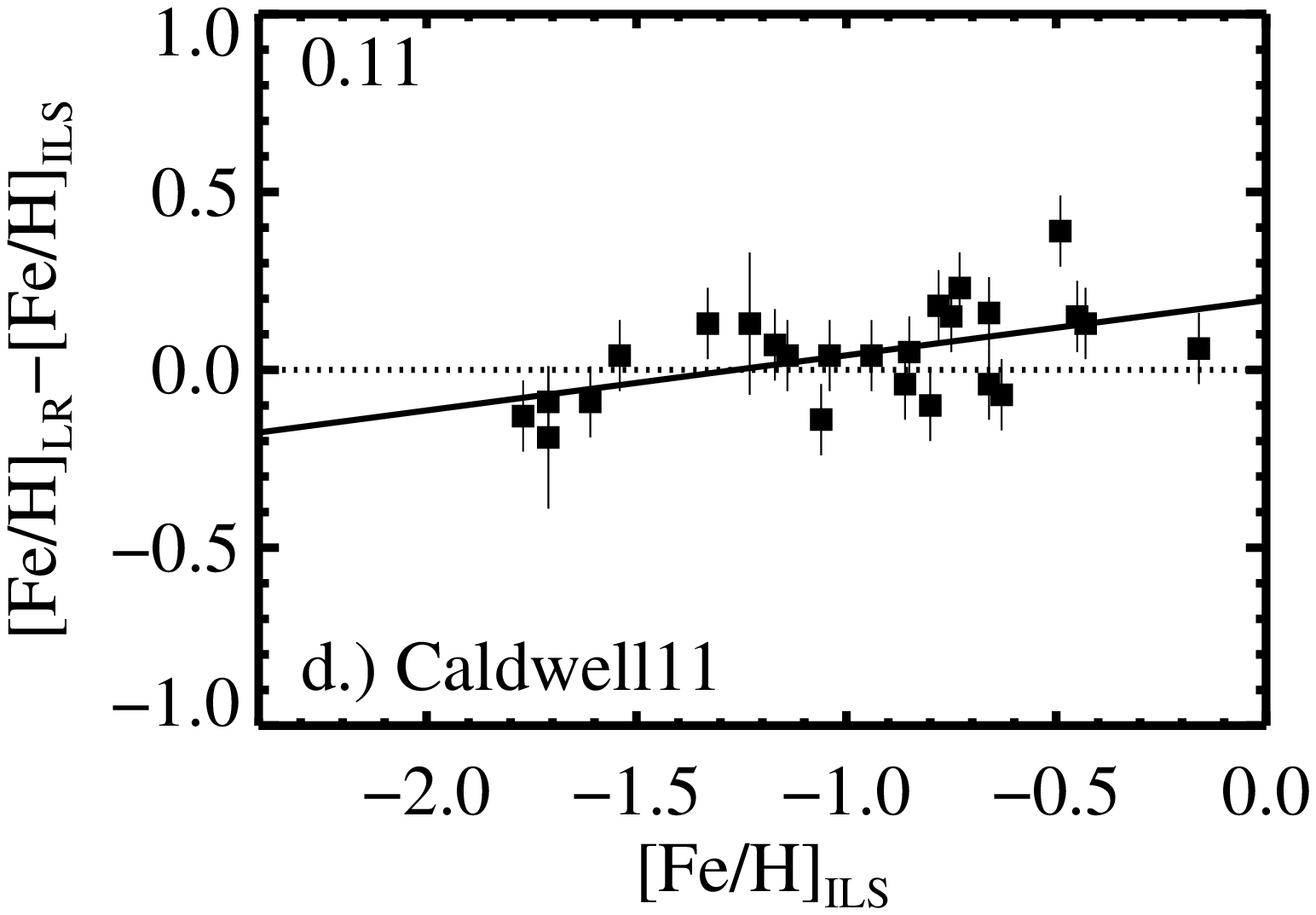}

\caption{ Comparison of [Fe/H] from low resolution studies to the high resolution [Fe/H] in this work.  a.)  Comparison for  \cite{huchra91} .  b.)  Comparison to  \cite{puzia05}, where we have accordingly  the used the relationship [Fe/H]=[Z/H]$-0.94$[$\alpha$/Fe] from \cite{2003MNRAS.339..897T}.  c.)  Comparison to \cite{galleti09}, and d.) Comparison to \cite{caldwell11}.  In each panel a solid  line shows a linear least squares fit, and the standard deviation of the residuals around the fit is noted in the upper left corner.}
\label{fig:lick-fe} 
\end{figure*}

There is  independent evidence  that  G002 is associated with a relic system.  \cite{mackey10} found that both G002 and G001 lie in an over density of GCs in the halo of M31, and classified the over density as   a substructure named ``Association 2.''  Follow up velocity work on the M31 outer halo GC system was performed by \cite{2014arXiv1406.0186V}, who found that the GCs in Association 2, including G002,  fall into two distinct kinematic subsets.  Their analysis showed a low probability that the GCs in this region inhabit an over density and have distinct  velocities by chance, and conclude that Association 2 may be a projection of two relic systems, possibly associated with the expected base of the prominent M31 substructure, the North-West Stream. According to \cite{2014arXiv1406.0186V}  there are 10 GCs associated with Association 2, including G001, G002, H2, H7, H8, PAndAS-18, 19, 21, 22 and 23.   Of these,  G001, H2, H7, PAndAS-21, and PAndAS-22 are luminous enough (M$_{V}<-6$) to be good candidates for high resolution ILS abundance analysis, and we therefore plan to target for follow-up. In the MW, metal-poor halo stars with low [Ca/Fe] have been tagged as an ``outlier" group \citep{cohenhalo}, so  it would be extremely interesting if some or all of the GCs in Association 2 shared the abundance pattern of G002, and could be chemically tagged as a distinct accretion event in M31.

\begin{figure}
\centering
\includegraphics[scale=0.50]{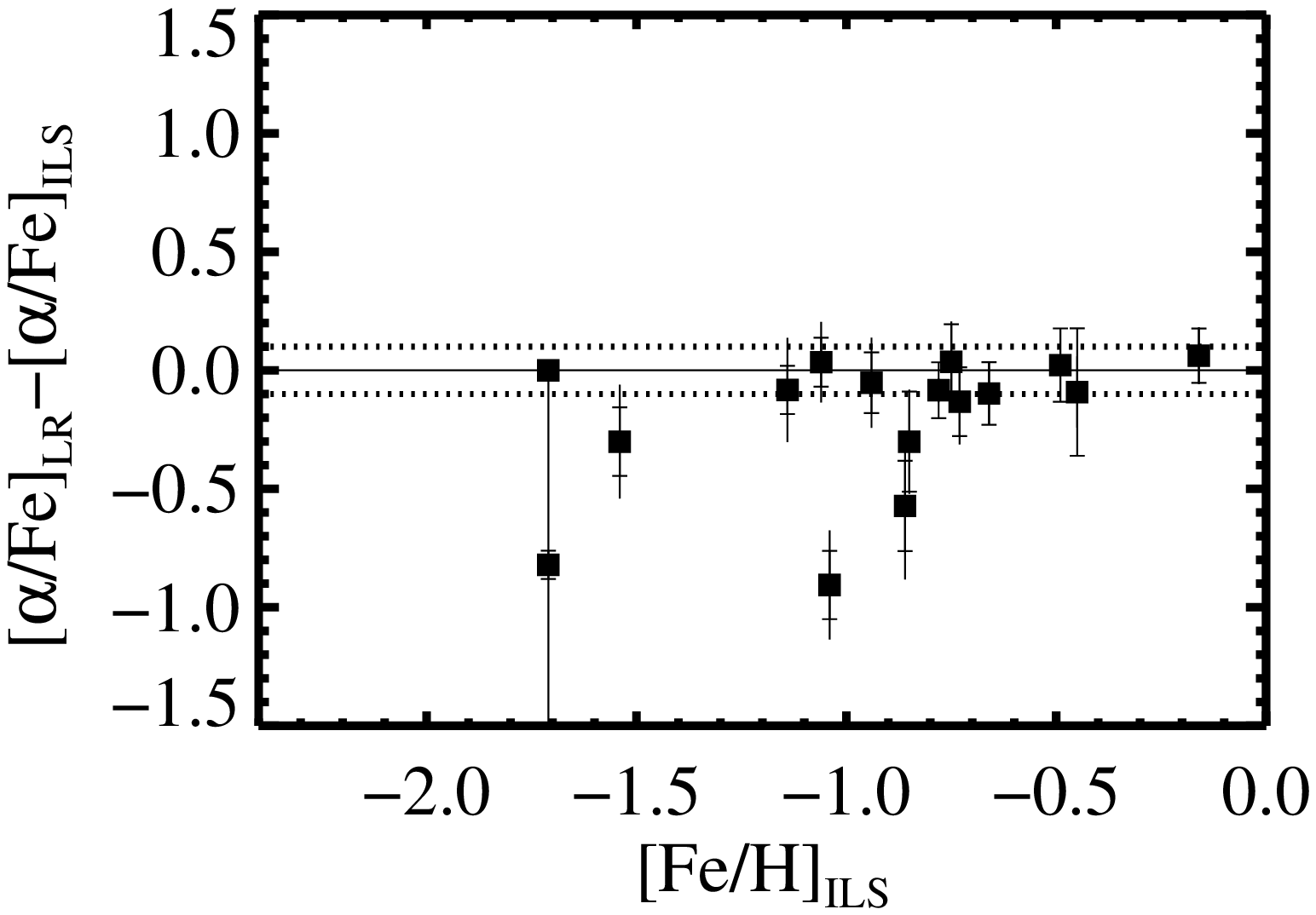}
\includegraphics[scale=0.50]{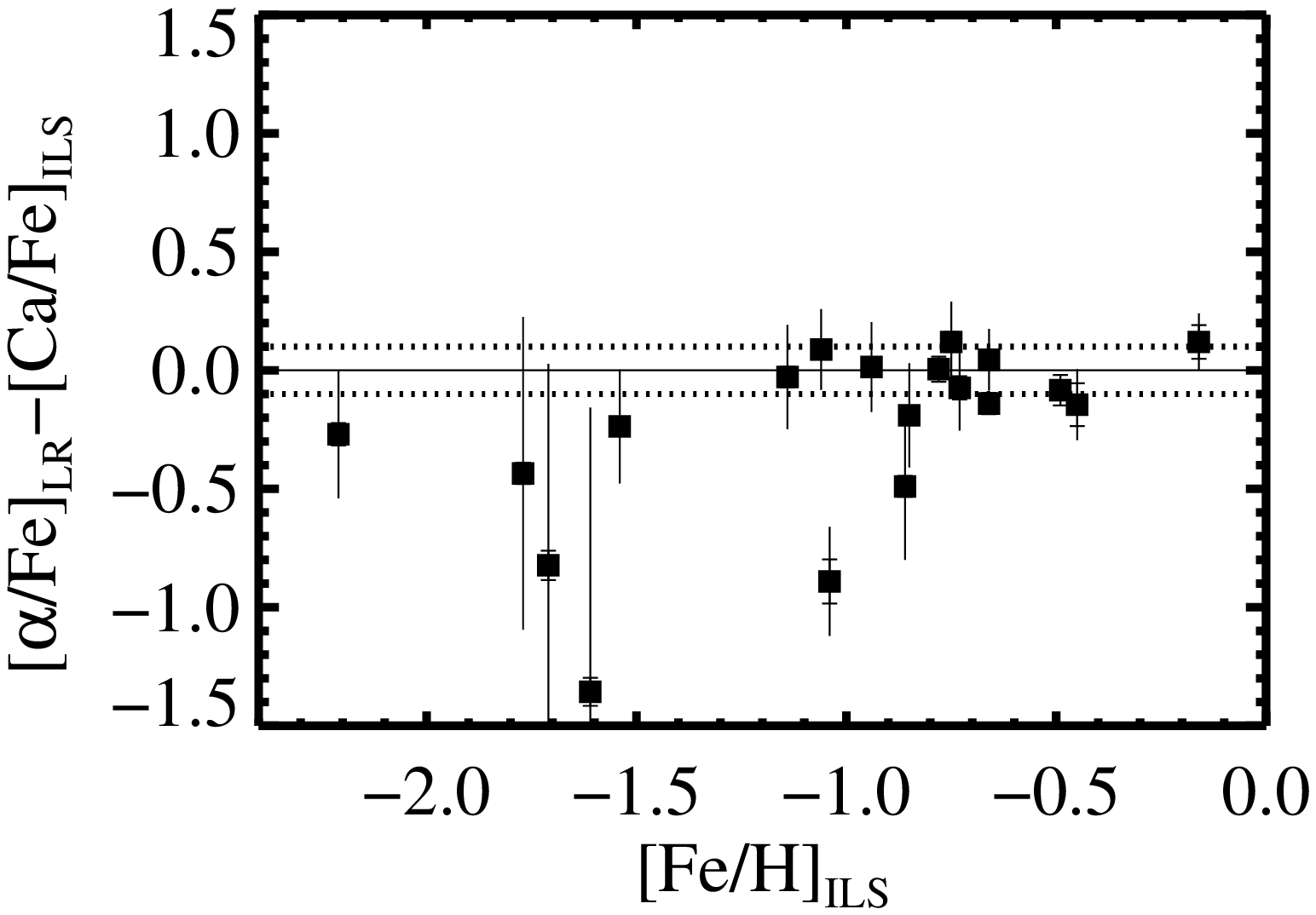}
\includegraphics[scale=0.50]{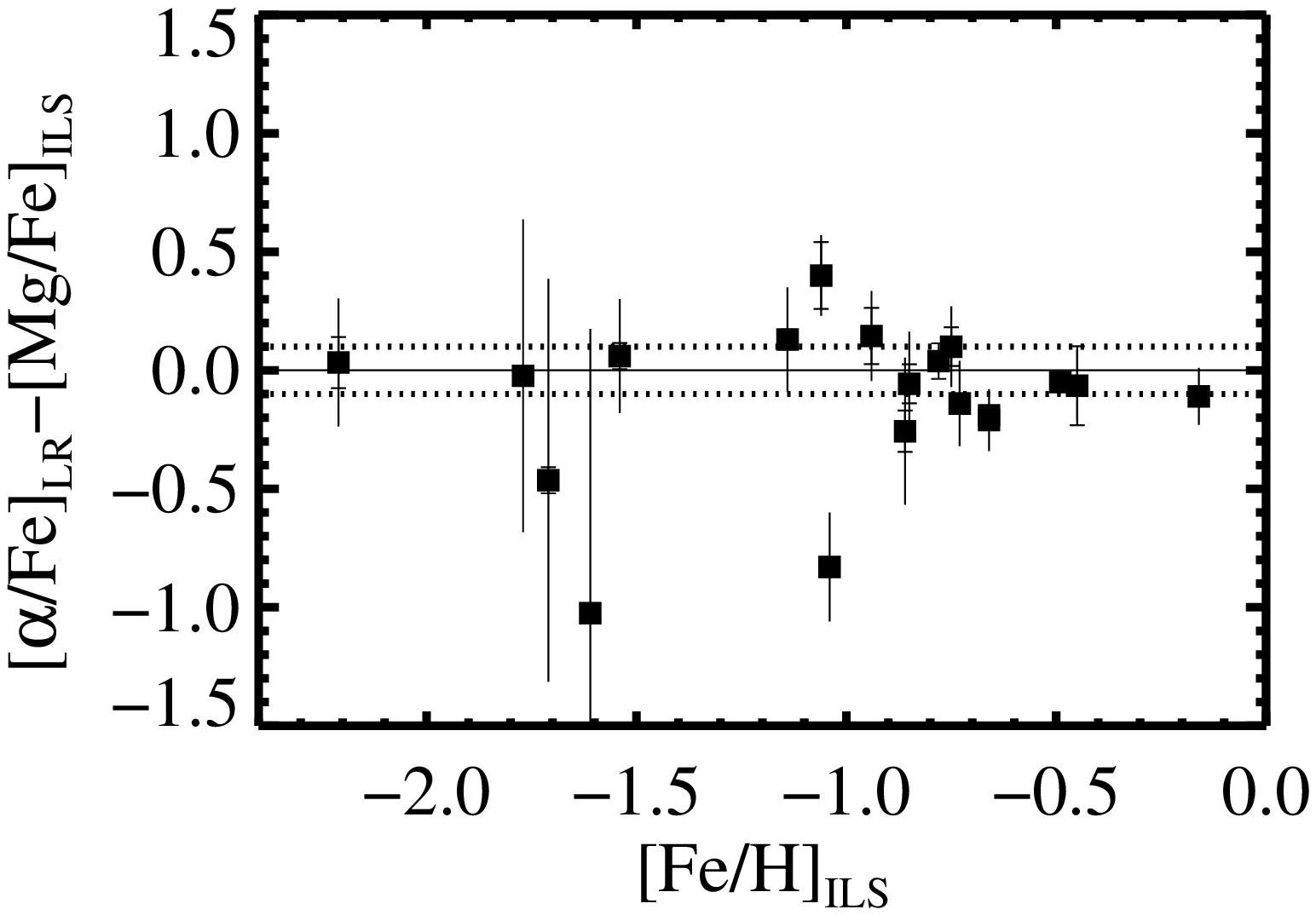}

\caption{ Comparison  to the [$\alpha$/Fe] estimates of \cite{puzia05}.  The top panel compares the average of our Ca, Si and Ti measurements,  and the middle panel compares only to Ca, which we have measured for all GCs in the sample.  The bottom panel compares to  [Mg/Fe], which we note does not track the other $\alpha$ elements in the GCs.  Dotted lines are shown at $\pm$0.10 dex to guide the eye, which is approximately the limit for meaningful distinctions between ``solar-scaled" and ``alpha-enhanced" abundances.  }
\label{fig:lick-alpha} 
\end{figure}

\subsection{Comparison with Previous Estimates: Fe and Alpha}
\label{sec:lick}

In this section we compare our precise high resolution abundance and age measurements with the wealth of previous data on M31 GCs that has previously been  obtained with lower resolution spectra and line index techniques.

First,  we compare the high resolution [Fe/H] to low resolution [Fe/H] in Figure \ref{fig:lick-fe} using measurements from \cite{huchra91}, \cite{puzia05}, \cite{galleti09}, and \cite{caldwell11}.  It's important to note that ``metallicity" measured from line index techniques isn't strictly an [Fe/H] in the same sense as the high resolution [Fe/H], which is measured directly from Fe lines. Instead, low resolution estimates  use a combination of metallicity sensitive indexes, which are largely sensitive to Fe, but also to Mg, for example \citep[e.g.][]{1984ApJ...287..586B,1985AJ.....90.1927R,1990ApJ...362..503B,1994ApJS...94..687W,1998ApJS..116....1T,2002MNRAS.333..383B,2003MNRAS.339..897T,2008ApJS..177..446G,2010MNRAS.404.1639V,2011MNRAS.412.2199T}. Our goal in this work is not to identify the reasons why metallicity
estimates vary between high resolution and low resolution
(e.g. different stellar and atmospheric models, assumed stellar
populations, or calibration samples), or vary between different low
resolution techniques themselves, since that level of detail is beyond
the scope of the current work.  Rather, we hope that our metallicity
scale can be used for future calibration of low resolution techniques
in abundance ranges that the MWÕs resolved GC system does not
include.

Following the fit to the differences in Figure \ref{fig:lick-fe}, it is clear that the low resolution solutions show a bias with metallicity that crosses over from under- to over-estimating the true metallicity at about [Fe/H]=$-1.5$ to $-1$.  Figure \ref{fig:lick-fe} also shows that after accounting for the bias,  the  overall dispersion in the residuals is smallest for the \cite{caldwell11} sample, although it has more GCs that deviate by more than 1 $\sigma$ because of the much smaller formal errors   than in the other samples.  
 While the claim of super-solar GCS in M31 has been around for some time \citep{huchra91,perrett02,puzia05,fan08,galleti09,caldwell11}, our IL spectra
and analysis do not suport that finding.  Moreover, we note that the
behavior seen in  Figure \ref{fig:lick-fe} casts doubt on the likelihood that there
are truly super-solar GCs in M31, as it is clear that abundances are
overestimated in metal-rich clusters.  A targeted follow up study of
the presumed solar and super-solar GCs in M31 is necessary to
determine if such clusters do exist; there are $\sim$10 GCs in the
\cite{caldwell11} sample that have [Fe/H$]>0$ and are bright enough for IL
abundance analysis (V$<18.5$), however the super-solar GCs tend to be
projected onto the bulge and inner disk of M31, so the analysis of
these GCs may be more difficult and special care would have to be
taken for background subtraction.

While low resolution bulk metallicity measurements are well established (and getting increasingly better according to Figure \ref{fig:lick-fe}), [$\alpha$/Fe] estimates from low resolution have only started to become available recently and are prone to greater systematics and inconsistencies \citep[see a longer discussion in ][]{brodierev06}.  With that in mind, we compare to the    [$\alpha$/Fe] estimates from \cite{puzia05} in Figure \ref{fig:lick-alpha}, which are currently the only [$\alpha$/Fe] estimates that are explicitly tabulated for individual GCs (however we anticipate a  much larger low resolution alpha element sample will soon be available for comparison in R. Schiavon et al. 2014).  Therefore,  while the [Fe/H] comparison discussed above spanned several different line index models and techniques, we are only able to compare to one set of [$\alpha$/Fe] measurements in this case, which were made from the specific models of \cite{2003MNRAS.339..897T,2004MNRAS.351L..19T}.

We compare both to the average of our Ca, Si, and Ti ratios, as well as the individual Ca ratios, and the individual Mg ratios.   A range of $\pm$0.10 dex is  marked in Figure \ref{fig:lick-alpha}, which is effectively the minimum scatter for meaningfully distinguishing between solar-scaled ratios of [$\alpha$/Fe]$=0.0$ and alpha-enhanced ratios of [$\alpha$/Fe]$=+0.3$.   
  Figure \ref{fig:lick-alpha} suggests that high resolution is necessary for useful IL [$\alpha$/Fe] measurements (e.g. precisions $<$0.1 dex), especially for metal poor GCs with  [Fe/H]$<-1$.   We find that low-resolution spectra can
produce more accurate measurements of [$\alpha$/Fe] 
at higher metallicities, but that the accuracy quickly
declines at metallicities below [Fe/H]$\sim -1.1$, at which [$\alpha$/Fe] is always
underestimated.This  can partly be explained by  the degeneracy of the SSP models at lower metallicities and the unavoidable disappearance of the spectral lines, as can be seen in \cite{puzia05}.  However, we find that the low resolution measurements appear to track the individual [Mg/Fe] measurements better than the other alpha elements.  This is not surprising, since strong Mg features probably dominate the low resolution alpha element indicators of the \cite{2003MNRAS.339..897T,2004MNRAS.351L..19T} models. However, it is a mixed blessing, since Mg doesn't  necessarily track the true alpha element abundances in the GCs because of the effect of  star-to-star abundance variations on the IL abundances.  Ideally, a better low resolution alpha element indicator would not be dominated by Mg features.

\section{Summary}
\label{sec:sum}

We have presented analysis of high resolution, high SNR, IL spectra of 31 GCs in M31. We report precise radial velocities and velocity dispersions using our high quality data; velocity dispersions of 
9 of the GCs are measured here for the first time. We have presented refinements of our original technique for detailed abundance analysis of IL spectra of GCs, which we have used  to obtain ages for the GCs  and abundances of Fe from Fe I and  Fe II lines,  Ca I, Si I, Ti from Ti I and  Ti II lines, Mg I, Na I, and Al I.   Below we summarize the key results from this work. 

1. For accurate age and abundance measurements, the spectra of clusters with large velocity dispersions (v$_{\sigma} \geq$ 15 \kms) and/or high metallicity ([Fe/H]$\geq -0.3$) must be analyzed using line synthesis, in order to perform accurate continuum placement and to account for line blending.

2. Of the 31 GCs analyzed in this work,  all of the GCs but B029 have ages consistent with being $\geq$ 10 Gyr.   For B029 we obtain an age of $\sim$2 Gyr, which is the first evidence for an intermediate GC in M31 that does not rely on integrated colors or Balmer line strengths.

3. The  mean, low metallicity  [$\alpha$/Fe] plateau values of the M31 GCs in our sample are similar to the MW GC stellar abundances of 
 \cite{pritzl05}. Subtle differences in the overall patterns of individual alpha elements are seen, just as in the MW.

4. The abundances of the light elements Mg I, Na I, and Al I show indirect evidence for star-to-star abundance variations within the GCs, which was first hinted at in \citetalias{m31paper}.

5.  We find  correlations of [Ca/Fe], [Na/Fe], and perhaps [Al/Fe] with proxies for cluster mass (M$_{v}$ and v$_{\sigma}$) when considering the sample as a whole, as well as when dividing the sample  into  metal-poor  and metal-rich subpopulations. This is the first evidence of mass-metallicity relationships in elements other than Fe.  We also  find a strong correlation of  [Mg/Fe] with proxies for GC mass in the metal-rich subpopulation.  This may indicate that the mechanism responsible for star-to-star abundance variations in GCs in dependent on cluster mass.

5. We find at least one GC, G002, that has a significantly different [$\alpha$/Fe] abundance pattern than other GCs at similar [Fe/H].  This abundance pattern may indicate that this GC, which is also associated with the Association 2 over-density, was accreted at late times into the M31 GC system.

6. We find a fairly constant, low  metallicity ([Fe/H]$=-1.6$) for the GCs in our sample that have projected galactocentric radii $>$20 kpc from M31. This resembles the ``old" halo population in the MW.

Detailed abundances of additional Fe-peak and heavy elements will be presented in the next papers in this series.

\acknowledgements
The authors thank the anonymous referee for a thoughtful report, which improved the clarity of the paper.  J.E.C. is supported by an NSF Astronomy and Astrophysics Postdoctoral Fellowship under award AST-1302710.  J.G.C. thanks NSF grant AST-0908139  for partial
support. J.E.C. and R.A.B. thank NSF grant AST-0507350 for partial support. Digitized Sky Surveys were produced at the Space Telescope Science Institute under U.S. Government grant NAG W-2166. The Second Palomar Observatory Sky Survey (POSS-II) was made by the California Institute of Technology with funds from the National Science Foundation, the National Geographic Society, the Sloan Foundation, the Samuel Oschin Foundation, and the Eastman Kodak Corporation.  The authors wish to recognize and acknowledge the very significant cultural role and reverence that the summit of Mauna Kea has always had within the indigenous Hawaiian community.  We are most fortunate to have the opportunity to conduct observations from this mountain.

\bibliographystyle{apj}

\end{document}